\documentclass[prd,preprintnumbers,nofootinbib,tightenlines,superscriptaddress]{revtex4}
\pdfoutput=1 

\usepackage{graphicx,epstopdf}
\usepackage{amssymb}
\usepackage{amsmath,mathrsfs,verbatim}
\usepackage[usenames, dvipsnames]{color}

\usepackage{times}
\usepackage{latexsym}

\DeclareGraphicsExtensions{.pdf,.png,.jpg}
\graphicspath{{RotationCurvePlots/}}

\usepackage{floatrow}

\usepackage{longtable}

\def\beq{\begin{equation}}
\def\eeq{\end{equation}}
\def\bey{\begin{eqnarray}}
\def\eey{\end{eqnarray}}

\def\lsim{\mathrel{\raise.3ex\hbox{$<$\kern-.75em\lower1ex\hbox{$\sim$}}}}
\def\gsim{\mathrel{\raise.3ex\hbox{$>$\kern-.75em\lower1ex\hbox{$\sim$}}}}

\newcommand{\vmax}{V_{\rm max}}

\newcommand{\rmax}{r_{\rm max}}

\newcommand{\MLdisk}{\Upsilon_{\star,{\rm disk}}}
\newcommand{\MLbulge}{\Upsilon_{\star,{\rm bulge}}}
\newcommand{\MLstar}{\Upsilon_{\star}}
\newcommand{\MLunits}{M_{\odot}/L_{\odot}}

\newcommand{\sigm}{\sigma/{\rm m}}
\newcommand{\micron}{\mu{\rm m}}

\begin{document}

\title{Reconciling the Diversity and Uniformity of Galactic Rotation Curves with Self-Interacting Dark Matter}

\author{Tao Ren}
\affiliation{Department of Physics and Astronomy, University of California, Riverside, California 92521, USA}
\author{Anna Kwa}
\affiliation{Department of Physics and Astronomy, University of California, Irvine, California 92697, USA}
\author{Manoj Kaplinghat}
\affiliation{Department of Physics and Astronomy, University of California, Irvine, California 92697, USA}
\author{Hai-Bo Yu}
\affiliation{Department of Physics and Astronomy, University of California, Riverside, California 92521, USA}
\date{\today}

\begin{abstract}
\vspace*{.0in}

Galactic rotation curves exhibit diverse behavior in the inner regions, while obeying an organizing principle, i.e., they can be approximately described by a radial acceleration relation or the Modified Newtonian Dynamics phenomenology. We analyze  the rotation curve data from the SPARC sample, and explicitly demonstrate that both the diversity and uniformity are naturally reproduced in a hierarchical structure formation model with the addition of dark matter self-interactions. The required concentrations of the dark matter halos are fully consistent with the concentration-mass relation predicted by the Planck cosmological model. The inferred stellar mass-to-light ($3.6~\micron$) ratios scatter around $0.5M_\odot/L_\odot$, as expected from population synthesis models, leading to a tight radial acceleration relation and baryonic Tully-Fisher relation. The inferred stellar-halo mass relation is consistent with the expectations from abundance matching. These results indicate that the inner dark matter halos of galaxies are thermalized due to the self-interactions of dark matter particles.

\end{abstract}

\maketitle

\section{Introduction}

Galactic rotation curves of spiral galaxies show a variety of behavior in the inner parts even across systems with similar halo and stellar masses, which lacks a self-consistent explanation in the standard cold dark matter (CDM) model~\cite{Flores:1994gz,Moore:1994yx,Burkert:1995yz,Persic:1995ru,deBlok:2001fe,Gentile:2004tb,KuziodeNaray:2007qi,deBlok:2008wp,Oh:2010ea,Oh:2015xoa,deNaray:2009xj,Oman:2015xda,Bullock:2017xww}. Along with this diversity, a long-standing observation is that many rotation curves can be understood in terms of Modified Newtonian Dynamics (MOND) phenomenology~\cite{Milgrom:1983ca,Milgrom:1983pn} (see~\cite{Famaey:2011kh} for a review), i.e., there exists a characteristic gravitational acceleration scale, $g_\dagger\approx 10^{-10}~{\rm m/s^2}\sim cH_0/7$ with $H_0$ being the present Hubble expansion rate, below which the observed acceleration can be approximated as $\sqrt{g_\dagger g_{\rm bar}}$ with ${g_{\rm bar}}$ being the baryonic acceleration (a.k.a. Milgrom's law). More recently, McGaugh et al.~\cite{McGaugh:2016leg} analyzed the Spitzer Photometry and Accurate Rotation Curves (SPARC) dataset~\cite{Lelli:2016zqa} and showed there is a tight relation between the total gravitational acceleration at any radius and the acceleration contributed by the baryons, assuming a constant stellar mass-to-light ratio $\MLdisk=0.5M_\odot/L_\odot$ and $\MLbulge=0.7M_\odot/L_\odot$ in the $3.6~\micron$ band. The scatter in this radial acceleration relation (RAR) is around $0.1$ dex, and the tightness of this relation has been interpreted as a signature of MOND~\cite{Li:2018tdo}. 

It has long been argued that the acceleration scale (including the $cH_0$ dependence) can emerge from hierarchical structure formation predicted in CDM~\cite{vandenBosch:1999dz,Kaplinghat:2001me}. Recent hydrodynamical simulations of galaxy formation with CDM have clearly shown that a RAR emerges~\cite{Keller:2016gmw,Ludlow:2016qzh,Navarro:2016bfs}. However, these simulated galaxies do not represent the full range of the diversity in the SPARC dataset and they cannot yet explain the rotation curves of low and high surface brightness galaxies simultaneously.

In this paper, we show that self-interacting dark matter (SIDM) provides a unified way to understand the diverse rotation curves of spiral galaxies, while reproducing the RAR with a small scatter. We analyze the SPARC dataset based on the SIDM halo model proposed in~\cite{Kaplinghat:2013xca,Kaplinghat:2015aga} and demonstrate three key observations leading to this result.

\begin{itemize}
\item For cross section per unit mass $\sigm \gtrsim 1~{\rm cm^2/g}$, dark matter self-interactions only thermalize the inner regions at distances less than about $10\%$ of the virial radius of galactic halos, while the outer regions remains unchanged. Thus, SIDM inherits essential features of the $\Lambda$CDM hierarchical structure formation model such as the halo concentration-mass relation, which sets the characteristic acceleration scale of halos.

\item In the inner halo, thermalization ties dark matter and baryon distributions together~\cite{Kaplinghat:2013xca,Vogelsberger:2014pda,Elbert:2016dbb}, and the SIDM halo can naturally accommodate the diverse range of `cored' and `cusped' central density profiles, depending on how the baryons are distributed. Combined with the scatter in the concentration-mass relation, this provides the diversity required to explain the rotation curves~\cite{Kaplinghat:2015aga,Kamada:2016euw,Creasey:2016jaq}. We will demonstrate the SIDM fits are systematically superior to the MOND ones.

\item For the same $\sigm$ that addresses the diversity problem, the baryon content of the galaxies and the mass model of their host halos also lead to the RAR with a scatter as small as the one in~\cite{McGaugh:2016leg}. In our SIDM fits, the inferred stellar $\MLdisk$ values for individual galaxies have a distribution peaked toward $0.5M_\odot/L_\odot$, as expected from stellar population synthesis models~\cite{Schombert:2013hga}.

\end{itemize}

The rest of the paper is organized as follows. In Sec.~\ref{sec:diversity}, we present the SIDM fits to 135 galaxies from the SPARC sample, which exemplify the full range of the diversity. In Sec.~\ref{sec:accel}, we show the radial acceleration relation and the distribution of the stellar mass-to-light ratios from our SIDM fits, compared to the MOND fits. In Sec.~\ref{sec:cosmo}, we discuss the host halo properties and the origin of the acceleration scale. In Sec.~\ref{sec:baryons}, we show the predicted stellar -- halo mass relation and the baryonic Tully-Fisher relation (BTFR). We comment on future directions and conclude in Sec.~\ref{sec:con}. In the appendix, {\bf Methods}, we provide detailed information about the model and the fitting procedure. In {\bf Supplementary Materials}, we present SIDM and MOND fits to $135$ individual galaxies from the SPARC sample and additional results that support the main text, including model fits to simulated halos.

\section{The diversity of galactic rotation curves}
\label{sec:diversity}

\begin{figure}[!t]
\centering
\includegraphics[scale=0.37]{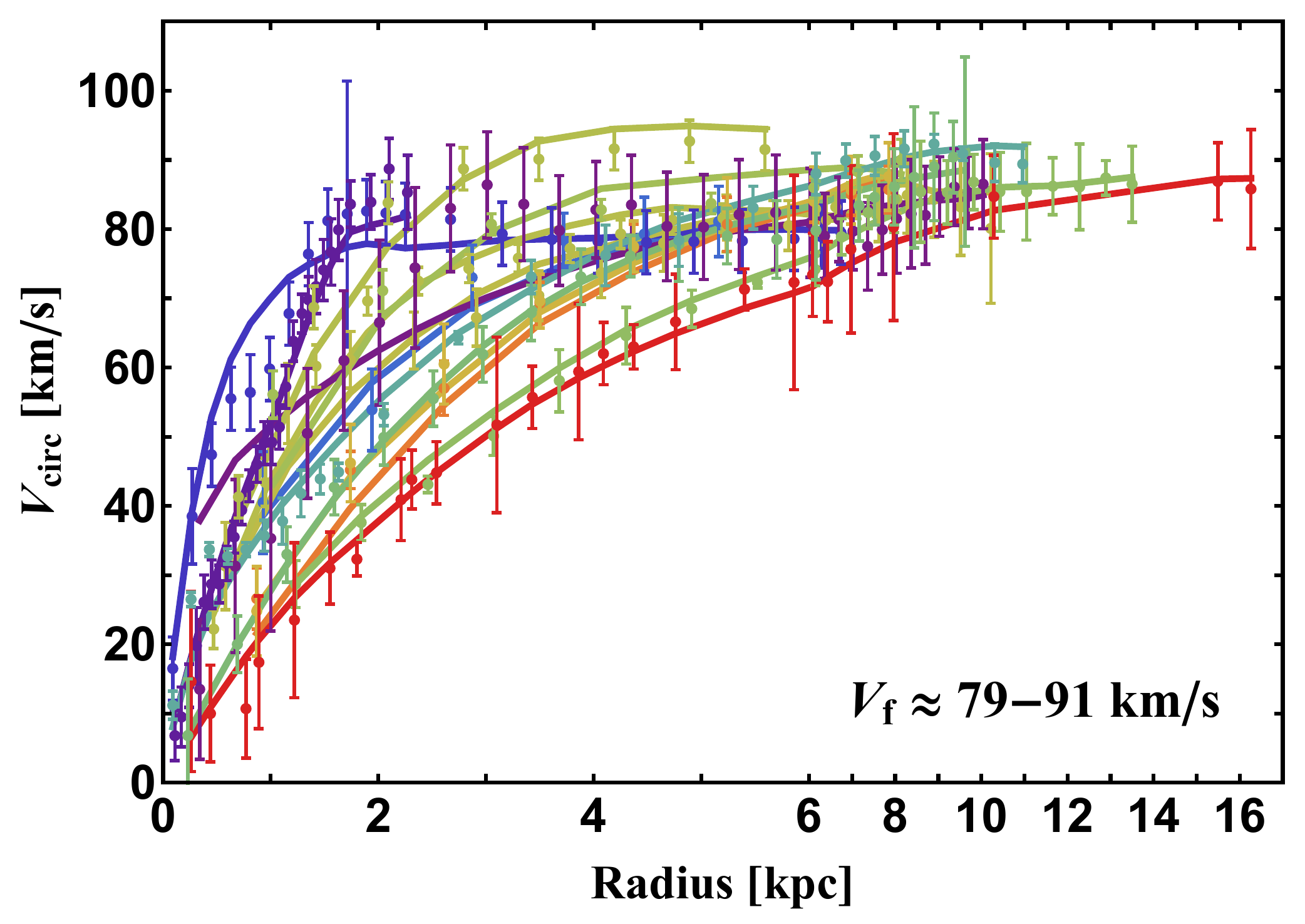} 
\includegraphics[scale=0.37]{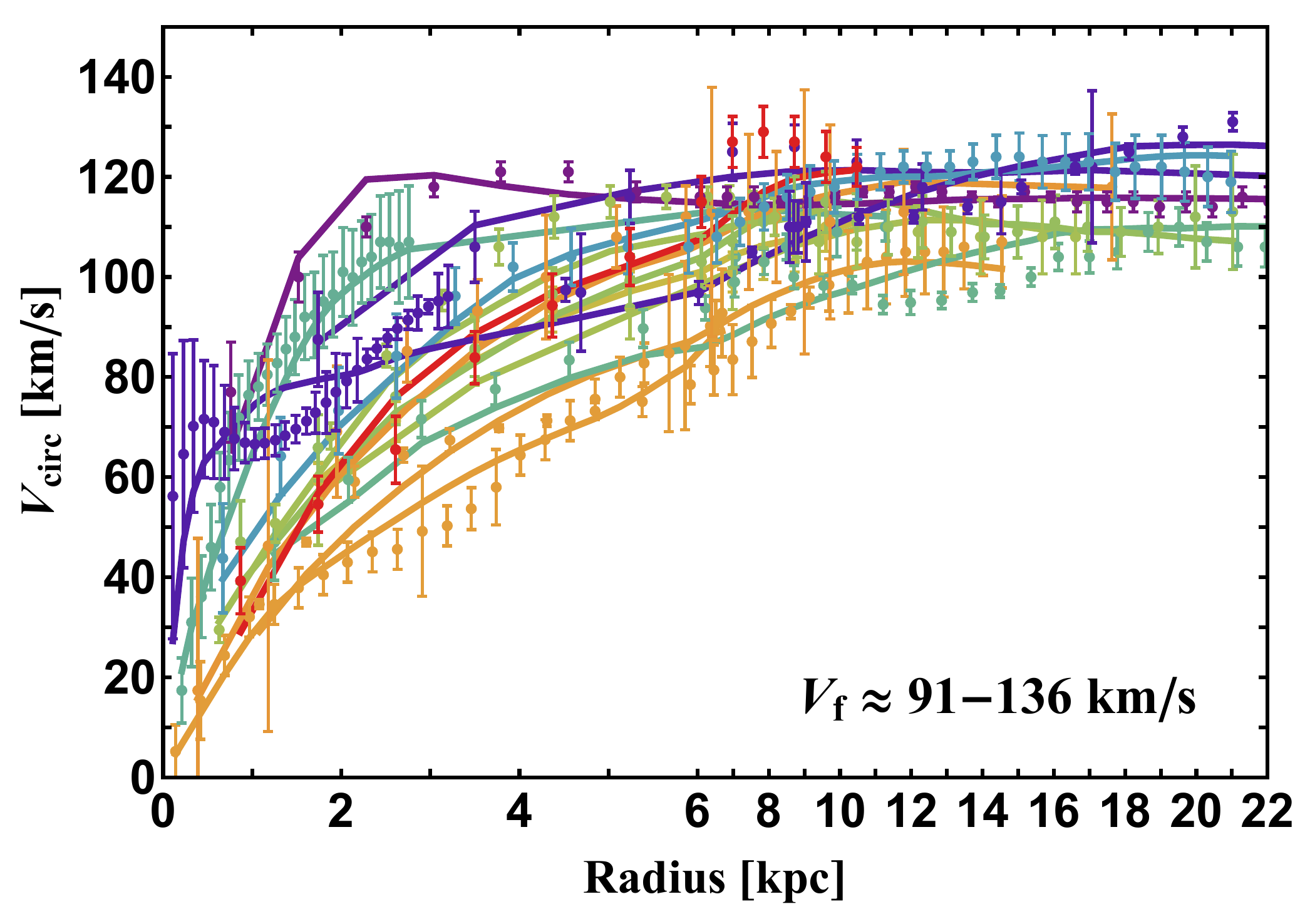} \\
\includegraphics[scale=0.37]{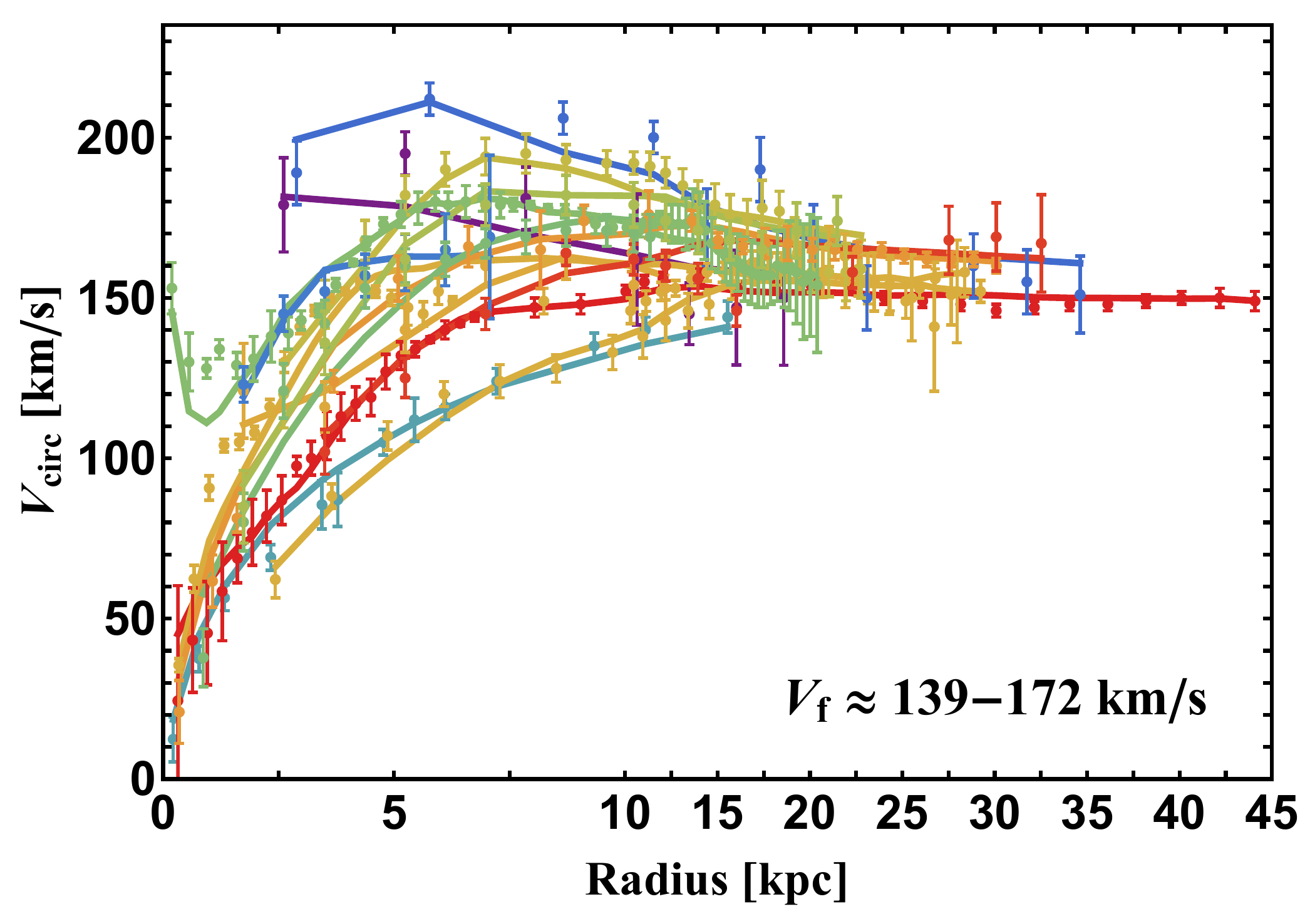}
\includegraphics[scale=0.37]{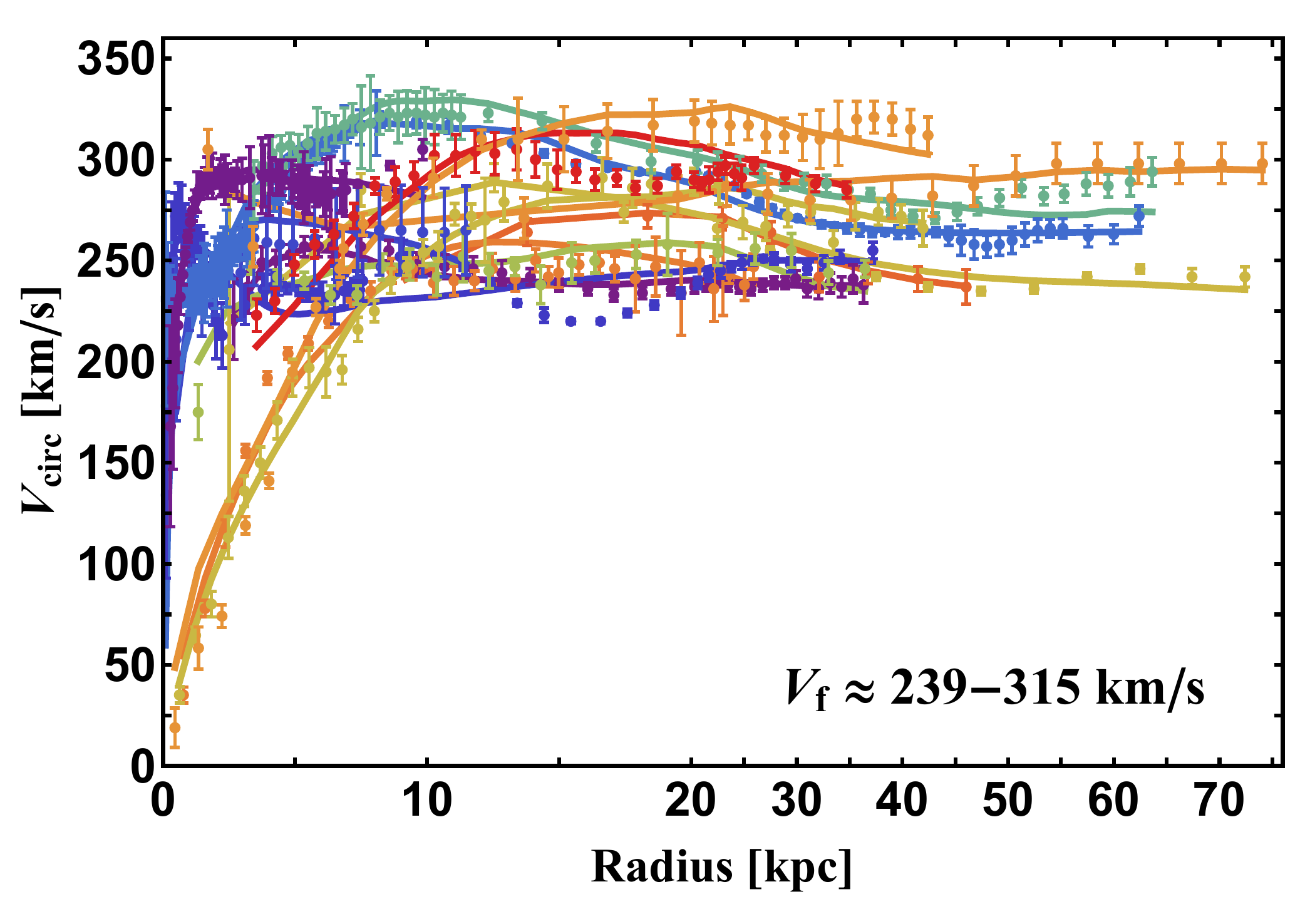}
\caption{SIDM fits (solid) to the diverse rotation curves across a range of spiral galaxy masses, where we take $\sigm=3~{\rm cm^2/g}$. The data points with error bars are from the SPARC dataset~\cite{Lelli:2016zqa}. Each panel contains $14$ galactic rotation curves that are selected to have similar flat rotation velocities at their furthest radial data points, and the corresponding $V_{\rm f}$ bins are $79\textup{--}91$, $91\textup{--}126$, $139\textup{--}172$ and $239\textup{--}315~{\rm km/s}$, spanning the mass range of the galaxies considered in this work. The galaxies are colored according to their relative surface brightness in each panel from low (red) to high (violet).}
\label{fig:Diverse}
\end{figure}

We select $135$ out of $175$ galaxies in the full SPARC sample based on the criteria that they must have a recorded value for the flat part of the rotation curve, $V_{\rm f}$. In our sample, $87$, $42$ and $6$ galaxies have quality flags $1$, $2$ and $3$, respectively. It spans a wide range of galaxy masses and inner shapes of rotation curves with $V_{\rm f}$ ranging from $20~{\rm km/s}$ to $300~{\rm km/s}$. In fitting to the data, we utilize the analytical SIDM halo model~\cite{Kaplinghat:2015aga,Kamada:2016euw}, where we assume the dark matter distribution in the inner halo follows the isothermal density profile,
\begin{equation}
\label{eq:rho_iso_axi}
\rho_{\text{iso}}(R,z)=\rho_{0} \exp \left(\left[\Phi_{\text{tot}}(0,0)-\Phi_{\text{tot}}(R,z)\right]/\sigma_{\text{v0}}^{2}\right),
\end{equation}
where $\rho_0$ is the central dark matter density, $\sigma_{\text{v0}}$ is the one-dimensional dark matter velocity dispersion, $\Phi_{\text{tot}}(R,z)$ is the total gravitational potential and $R,z$ are cylindrical coordinates aligned with the stellar disk. We match this isothermal profile to a Navarro-Frenk-White (NFW) form~\cite{Navarro:1995iw,Navarro:1996gj} at $r_1$, where a dark matter particle has scattered ${\cal O}(1)$ times over the age of the galaxy, assuming continuity in both the density and the enclosed mass at $r_1$. In this way, the isothermal parameters ($\rho_0$, $\sigma_{\rm v0}$) directly map on to the NFW parameters ($r_s$, $\rho_s$) or ($r_{\rm max}$, $V_{\rm max}$). This model provides an approximate way to calculate the SIDM distribution in a halo if its CDM counterpart is known, and vice versa. It correctly predicts the halo central density and its scalings with the outer halo properties, stellar profiles and cross section, as confirmed in both isolated and cosmological N-body simulations with and without baryons, see, e.g.,~\cite{Kaplinghat:2015aga,Elbert:2016dbb,Creasey:2016jaq,Robertson:2017mgj,Sameie:2018chj}. See {\bf Methods} and {\bf Supplementary Materials} for a detailed description of the model and additional comparisons between model predictions and cosmological simulations. 

We adopt two independent but complementary approaches to perform the analysis. In the controlled sampling (CS) approach, we demand that the host halos follow the concentration-mass relation within a $2\sigma$ range predicted in cosmological simulations~\cite{Dutton:2014xda}. We model the stellar distribution as an axisymmetric thin disk as in~\cite{Kamada:2016euw}, which directly enters into the calculation of the density profile of SIDM through the gravitational potential $\Phi(R,z)$. In the CS fits, we start with the outer NFW halo and find the SIDM density profile that matches its mass and density at $r_1$. In the second approach, we use the Markov Chain Monte Carlo (MCMC) sampling (MS) to explore the full likelihood. To save computational time, we assume spherical symmetry by spreading the mass within the disk at radius $R$ into a sphere of the same radius~\cite{Kaplinghat:2013xca, Elbert:2016dbb}. The rotation curves generated from two approaches agree well and the differences in the fits are small (see {\bf Supplementary Materials}). For our main results, we show inferences from both of the approaches. 

In Fig.~\ref{fig:Diverse}, we show the SIDM fits to the diverse rotation curves from the controlled sampling with $\sigm=3~{\rm cm^2/g}$. In each panel, galaxies are selected to have similar flat rotation velocities at their outermost data points. The rise up to $V_{\rm f}$ within their central regions displays a wide variety of slopes and the SIDM halo model provides equally good fits to the shallow and steeply rising rotation curves. The fits for the other galaxies in the sample are as good as those in Fig.~\ref{fig:Diverse} (see {\bf Supplementary Materials}).

The success of the SIDM halo model stems from a combination of the following effects. First, SIDM thermalization ties the baryon and dark matter distributions together. For low surface brightness galaxies, thermalization leads to a shallow density core and a circular velocity profile that rises mildly with radius~\cite{Spergel:1999mh,Dave:2000ar,Rocha:2012jg,Peter:2012jh,Vogelsberger:2012ku,Zavala:2012us,Vogelsberger:2015gpr}. While, for high surface brightness ones, the core shrinks in response to the deeper baryonic potential and the central SIDM density increases accordingly~\cite{Kaplinghat:2013xca,Elbert:2016dbb,Creasey:2016jaq,Sameie:2018chj}. The galaxies in our sample have a variety of central surface brightnesses, resulting in diverse central dark matter densities. Second, scatter in the cosmological halo concentration-mass relation leads to scatter in the characteristic SIDM core density and radius, which is reflected in the rotation curves~\cite{Kaplinghat:2015aga}. Ref.~\cite{Kamada:2016euw} fitted $30$ galaxies and illustrated the importance of these effects in explaining the diverse rotation curves. In this work, we fit a larger sample of galaxies and demonstrate that the observed galaxies are fully consistent with the SIDM predictions. 

We have assumed a constant cross section to fit the SPARC sample because it is hard to pin down the cross section for individual galaxies. For low surface brightness galaxies with a large core, a large cross section, such as $\sigm=3~{\rm cm^2/g}$ is preferred~\cite{Kamada:2016euw}. However, since the central SIDM density varies mildly with the cross section in range of $1\textup{--}10~{\rm cm^2/g}$~\cite{Elbert:2014bma,Sokolenko:2018noz}, a feature that is well-captured in our analytical model~\cite{Kaplinghat:2015aga}, an even larger cross section may work as well. For high surface brightness galaxies, to which most of galaxies with high $V_{\rm f}$ belong, the fits are insensitive to the cross section because of the degeneracy between $\sigm$ and $\MLstar$~\cite{Kamada:2016euw}. The effect in the SIDM fits induced by varying $\sigm$ can be compensated by a minor change in the stellar mass-to-light ratio, and many of these systems are actually compatible with an NFW profile. The cross section may have a mild velocity dependence over the sample, as implied by the constraint from galaxy clusters~\cite{Kaplinghat:2015aga,Firmani:2000ce,Yoshida:2000uw}, but it is impossible to extract it from the SPARC dataset given the reasons discussed above. In this work, we present the results for fixed $\sigm=3~{\rm cm^2/g}$ and they remain the same qualitatively for other values larger than $\sim1~{\rm cm^2/g}$ on galaxy scales.

An important consequence of the large cross section is that the SIDM profile is driven quickly to be isothermal in the inner regions. This implies that the resultant SIDM fits will not depend sensitively on the formation history of individual galaxies~\cite{Kamada:2016euw}, but the {\em final} stellar and gas distributions~\cite{Kaplinghat:2013xca}. This has been explicitly confirmed in recent hydrodynamical SIDM simulations~\cite{Robertson:2017mgj} and those with idealized disk growth~\cite{Elbert:2016dbb}. Furthermore, in our fits $r_1$ is close to $r_s$, which is well outside the stellar disk or budge in the galaxies. It is unlikely that a viable baryonic feedback process could change the halo mass profile significantly at that far distance. Thus, our analytical model takes into account the realistic baryon distribution for individual galaxies and encodes this effect on the SIDM halo profile through the matching procedure.

\section{The Radial Acceleration Relation in SIDM}
\label{sec:accel}

\begin{figure}[t!]
\centering
\begin{tabular}{@{}ccc@{}}
\includegraphics[scale=0.37]{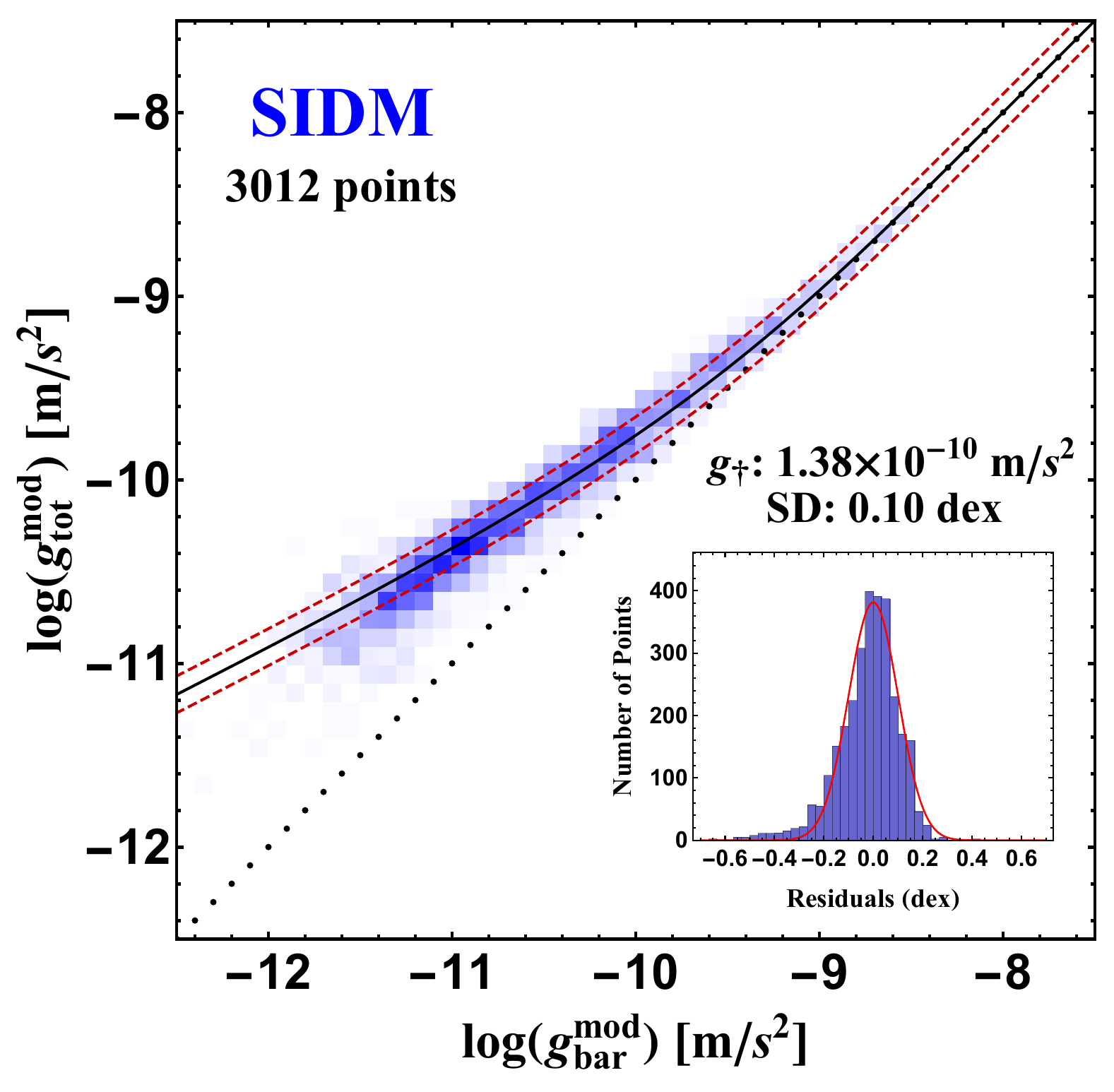} &
\includegraphics[scale=0.27]{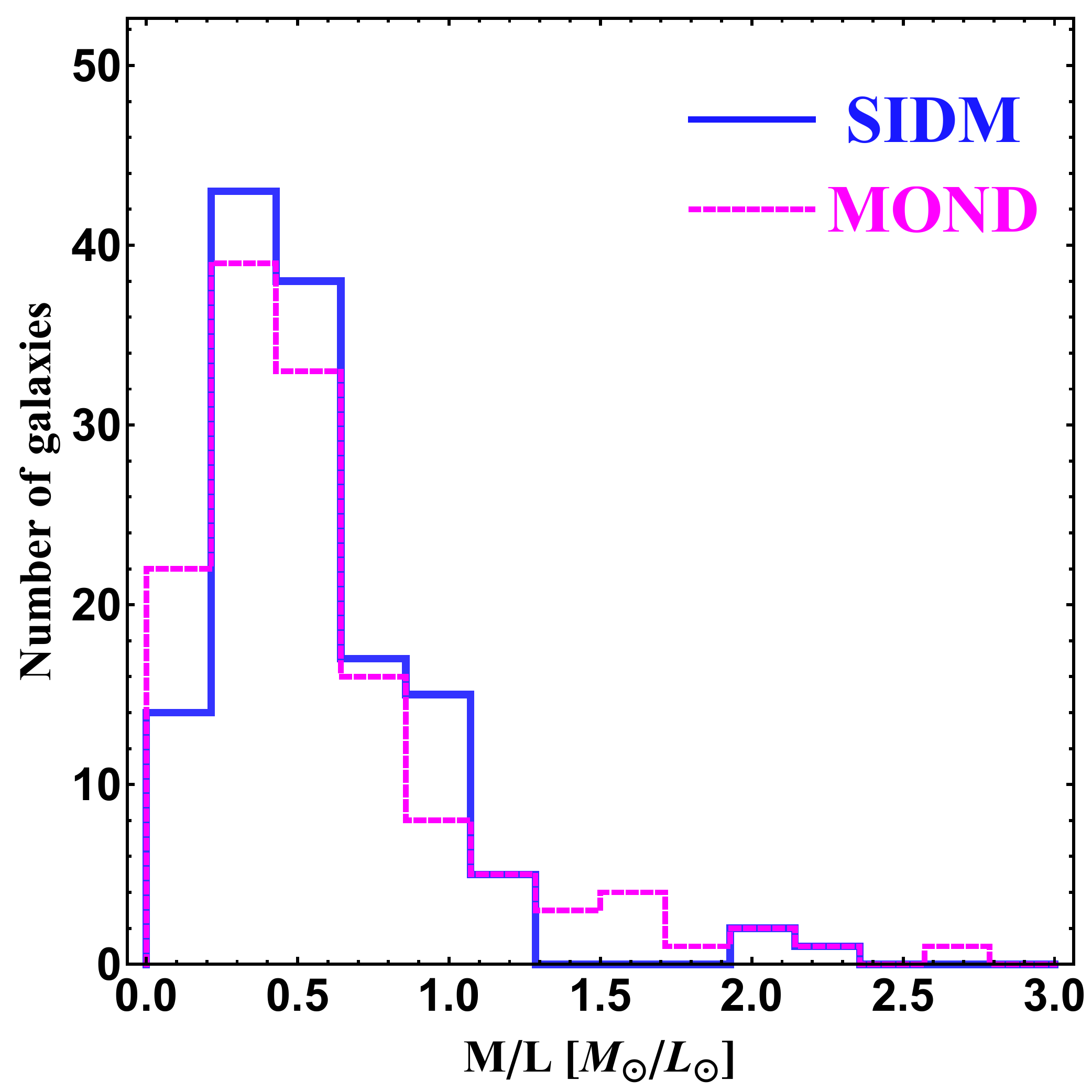}  &
\includegraphics[scale=0.28]{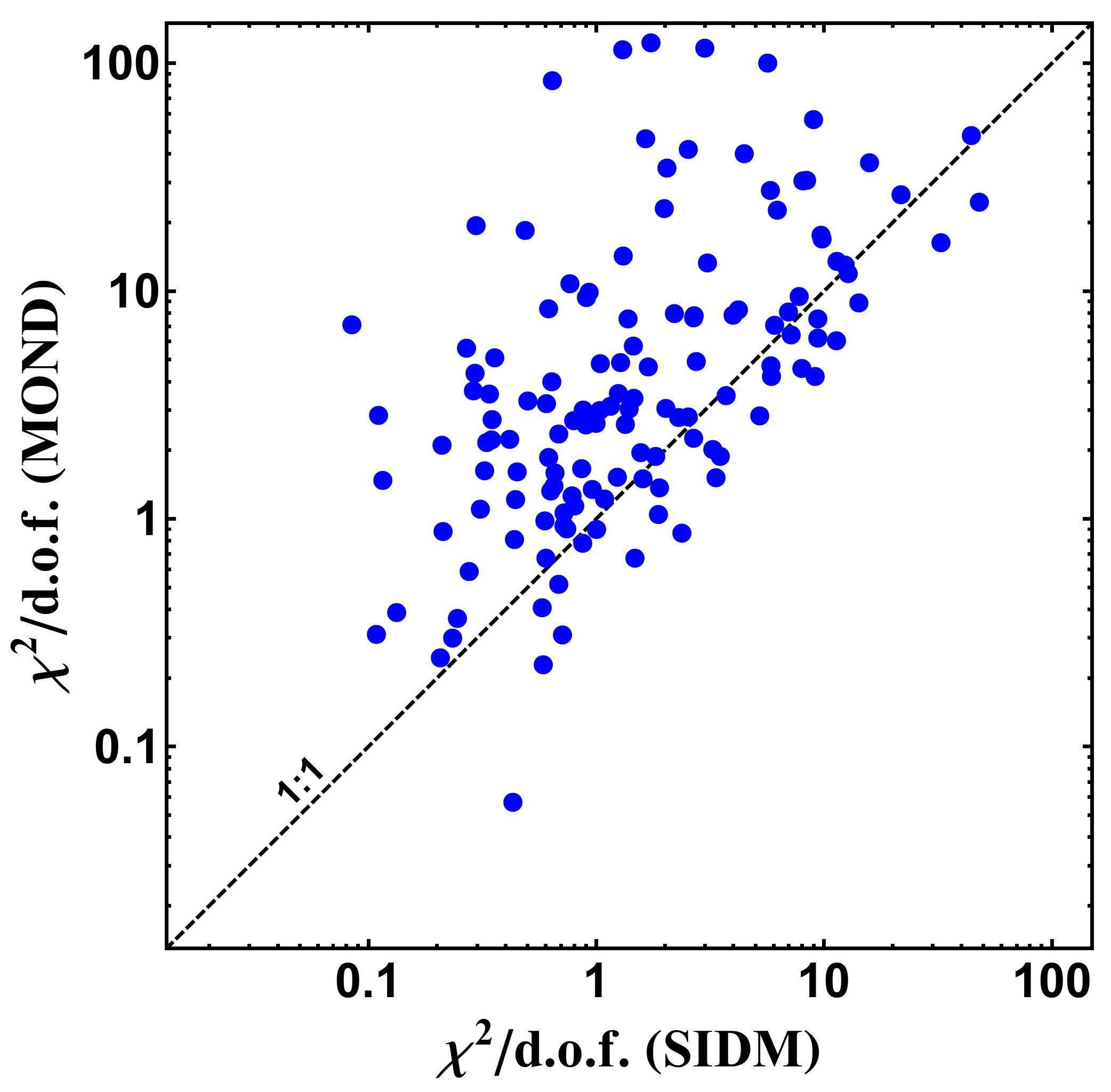}  
\end{tabular}
\caption{{\it Left}: The radial acceleration relation from the SIDM fits, where $g^{\rm mod}_{\rm tot}$ and $g^{\rm mod}_{\rm bar}$ are inferred from the $\sigm=3~{\rm cm^2/g}$ fits. The black solid line is the best fit to Eq.~\ref{eqn:radialAccel}; the two red dashed curves correspond to the $1\sigma$ deviation from this fit. The black dotted line is the one-to-one reference line. {\bf Insets}: Corresponding histograms of residuals after subtracting the fit function with the best-fitting scale parameter $g_\dagger=1.38\times10^{-10}~{\rm m/s^2}$, together with the Gaussian fits to the residuals, which have $1\sigma$ widths of $0.10$ dex. {\it Middle}: Inferred $\MLdisk$ distributions for the SIDM and MOND fits. {\it Right}: distribution of $\chi^2/{\rm d.o.f.}$ values for individual galaxies from the SIDM and MOND fits.}
\label{fig:Acceleration}
\end{figure}

In the RAR described in Ref.~\cite{McGaugh:2016leg}, the gravitational acceleration $g_{\rm tot}$ at radius $r$ is found to be related to the acceleration $g_{\rm bar}$ at the same radius. This relation can be fit to a functional form with a single parameter $g_\dagger$:
\begin{equation}
g_{\rm tot}(r) = g_{\rm bar}(r)\left(1-e^{-\sqrt{ g_{\rm bar}(r)/g_\dagger}} \right)^{-1}.
 \label{eqn:radialAccel}
\end{equation}
Their best-fit value of $g_\dagger=1.2\times 10^{-10}~{\rm m/s^2}$ is the oft-quoted MOND acceleration scale.

In the left panel of Fig.~\ref{fig:Acceleration}, we show the inferred total and baryonic acceleration values from the controlled sampling, where $g^{\rm mod}_{\rm{tot}}$ and $g^{\rm mod}_{\rm{bar}}$ are {\it calculated} from the SIDM fits, using the halo parameters and the best-fit $\MLstar$ values for each galaxy. The intensity of color in Fig.~\ref{fig:Acceleration} (left) reflects the density of points. After fitting the data with the empirical relation given in Eq.~\ref{eqn:radialAccel}, we find the best-fit value of $g_\dagger$ is $1.38\times10^{-10}~{\rm cm^2/g}$ and the resulting dispersion in the residuals is $0.10$ dex. Fig.~\ref{fig:Acceleration} (middle) shows $\MLdisk$ distribution from the SIDM fits (solid). It is peaked toward $\MLdisk=0.5 M_{\odot}/L_{\odot}$, in good agreement with predictions from stellar population synthesis models~\cite{Schombert:2013hga}. This is remarkable because {\it no} priors based on the stellar population synthesis models were used. We have also reproduced the analysis in Ref.~\cite{McGaugh:2016leg} with $\MLdisk$ and $\MLbulge$ were fixed to $0.5 M_{\odot}/L_{\odot}$ and $0.7 M_{\odot}/L_{\odot}$, respectively. For this fixed $\MLstar$ case, we obtained $g_\dagger=1.19\times10^{-10}~{\rm m/s^2}$ and dispersion $0.12~\text{dex}$, both in agreement with previous work~\cite{McGaugh:2016leg}. 

For a more detailed comparison, we also fit the sample of $135$ SPARC galaxies using the MOND relation in Eq.~\ref{eqn:radialAccel}, where we fixed $g_\dagger=1.2\times10^{-10}~{\rm m/s^2}$, but varied $\MLdisk$ and $\MLbulge$ using MCMC sampling (see also~\cite{Frandsen:2018ftj}). The results look similar if we set $g_\dagger$ to $1.0\times10^{-10}$ or $1.4\times10^{-10}~\rm{m/s^2}$. The middle panel of Fig.~\ref{fig:Acceleration} shows the $\MLdisk$ distribution from the MOND fits (dotted), which closely tracks the one from the SIDM fits. The right panel shows the distribution of minimum $\chi^2/{\rm d.o.f.}$ values for individual galaxies from the SIDM and MOND fits. The SIDM model provides a better fit than MOND for most of the galaxies ($\sim77\%$), while maintaining a tight RAR. In fact, $72\%~(45\%)$ of them have $\chi^2/{\rm d.o.f.}\leq3~(1)$ in the controlled SIDM fits and those with a large $\chi^2/{\rm d.o.f.}$ value have either tiny errors or wiggles in the observed rotation curves that cannot be reproduced by MOND either. 

\begin{figure}[!t]
\centering
\includegraphics[scale=0.27]{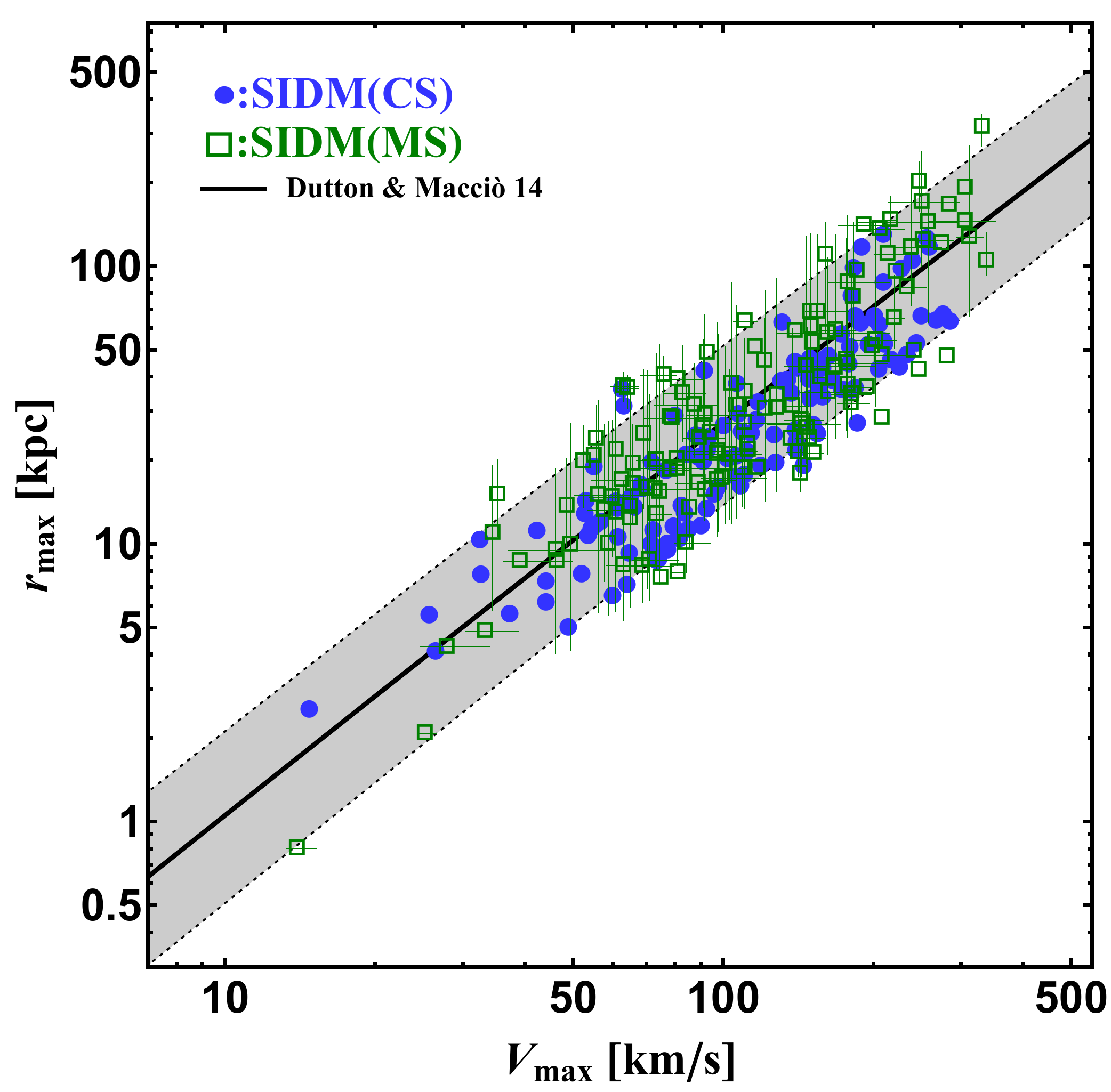}  
\includegraphics[scale=0.27]{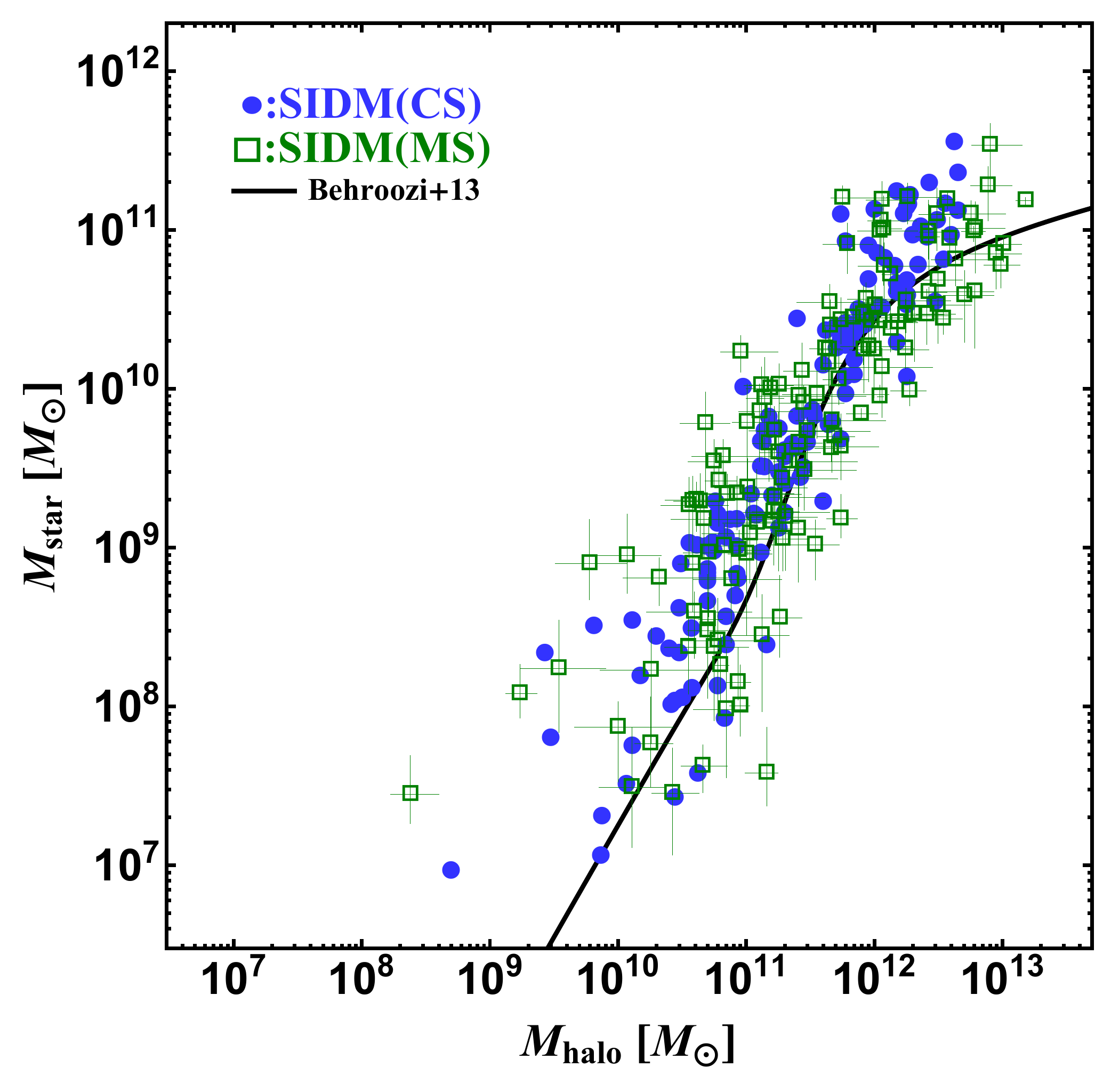}   
\includegraphics[scale=0.27]{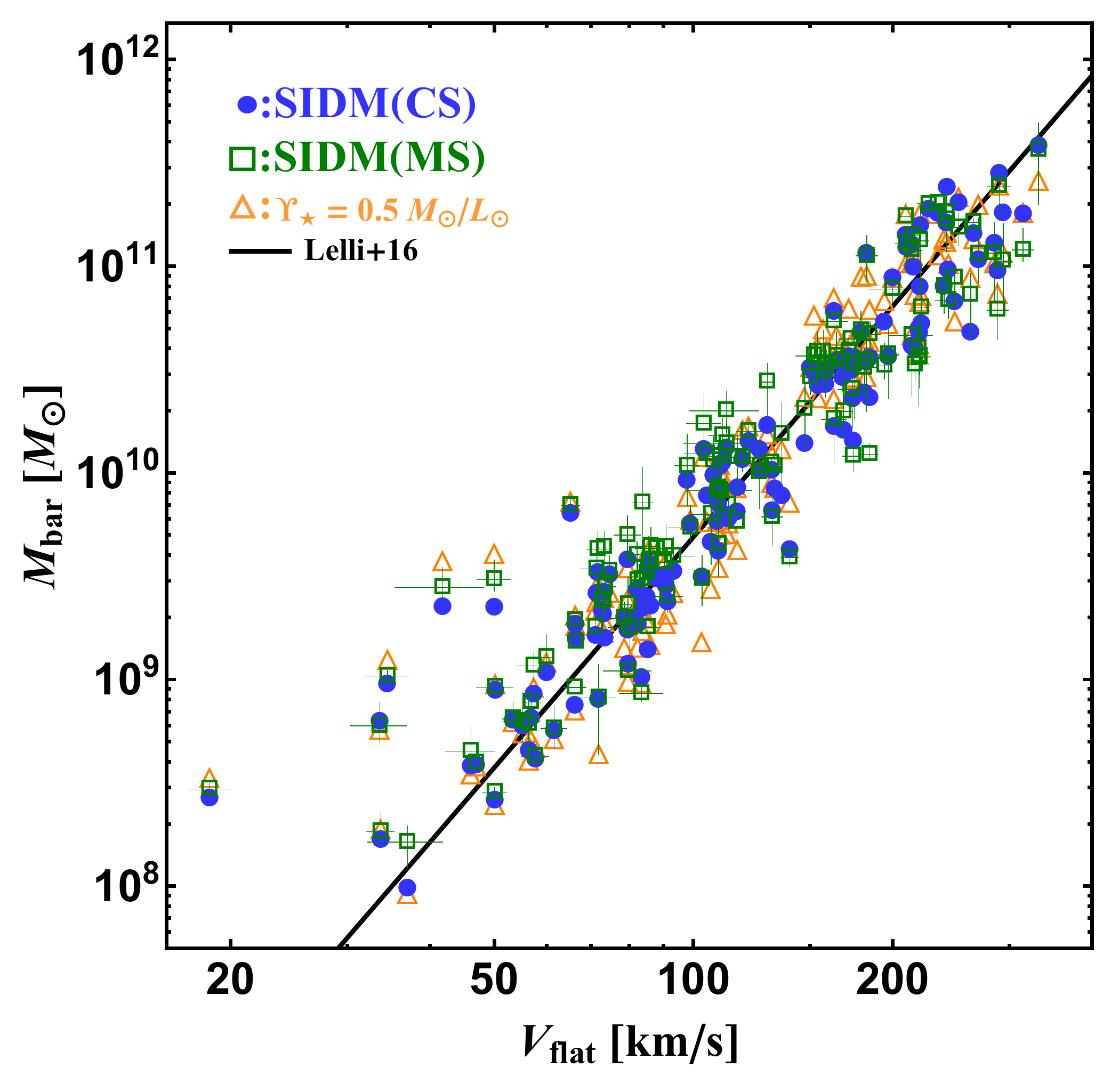} 
\caption{{\it Left}: $r_{\rm max}\textup{--}V_{\rm max}$ distributions of the host halos in the SIDM fits with controlled (circles) and MCMC (squares) samplings. We also show the mean relation (black solid) and $2\sigma$ scatter (gray shaded) predicted in cosmological CDM simulations~\cite{Dutton:2014xda}. {\it Middle}: Halo virial mass vs galaxy stellar mass from the SIDM fits. The black solid line corresponds to the abundance matching inference from~\cite{Behroozi:2012iw}. {\it Right}: total baryonic mass vs flat circular velocity for the $135$ galaxies, where $M_{\rm bar}$ is inferred from our SIDM fits (circles and squares). For comparison, we also show the case (triangles) when $\MLdisk$ and $\MLbulge$ are fixed to $0.5{M_\odot/L_\odot}$~\cite{Lelli:2015wst}. The black solid line is the mean baryonic Tully-Fisher relation from~\cite{Lelli:2015wst}, derived from $118$ SPARC galaxies with $\MLdisk=\MLbulge=0.5{M_\odot/L_\odot}$, at which the scatter is minimized.}
\label{fig:relations}
\end{figure}

We emphasize that the diversity in the inner rotation curves is also reflected in the $g_{\rm tot}\textup{--}g_{\rm bar}$ plane, as explicitly demonstrated in {\bf Supplementary Materials}, where we show the $g_{\rm tot}$ vs $g_{\rm bar}$ plot, but now split the sample into two sets: radii outside and inside $2R_{\rm d}$ with $R_{\rm d}$ being the scale radius of the stellar disk. The scatter is relatively large for radii $<2R_{\rm d}$, and this is due to the different shapes in the inner rotation curves and not just the result of random errors (see also~\cite{Desmond:2016azy}). On the other hand, there is a clear ordered behavior of $g_{\rm tot}$ vs $g_{\rm bar}$ curves for radii $> 2R_{\rm disk}$, which is a reflection of the BTFR: the tight correlation between the flat circular velocity, $V_{\rm f}$, and the total baryonic mass, $M_{\rm bar}$ for spiral galaxies~\cite{McGaugh:2000sr}. In this regime, $g_{\rm tot}\approx\sqrt{g_\dagger g_{\rm bar}}$, where $g_{\rm tot}\approx V^2_{\rm f}/r$ and $g_{\rm bar}\approx GM_{\rm bar}/r^2$, hence we have $V^4_{\rm f}/(GM_{\rm bar})\approx g_\dagger$. This is the success of MOND, i.e., if one assumes $M_{\rm bar} \propto V_{\rm f}^4$, then the normalization of the BTFR also predicts the rotation curve, which in many cases is a good fit to the observed one. Many studies do find $M_{\rm bar} \propto V_{\rm f}^s$ with $3<s<4$~\cite{Lelli:2015wst,2016ApJ...832...11B,2018MNRAS.474.4366P}, as we will also show in Sec.~\ref{sec:baryons}; $s=4$ is not forced upon us by the data, but it is not ruled out either. However, the MOND relation (Eq.~\ref{eqn:radialAccel})  cannot explain the full range of the diversity in the inner rotation curves, while the success of SIDM is deeply rooted to hierarchical structure formation, as we discuss in the next section. 

\section{The concentration-mass relation and origin of the characteristic acceleration scale}
\label{sec:cosmo}

We have demonstrated that SIDM explains {\em both} the diversity and the tight RAR exhibited in the rotation curves, as dark matter self-interactions thermalize the inner halo in the presence of the baryonic potential. Here, we show the host halos in the SIDM fits are consistent with predictions in the hierarchical structure formation model, see, e.g.,~\cite{Prada:2011jf,2014MNRAS.441..378L,Dutton:2014xda}. Since the outer halo ($r\gtrsim r_1$) remains unchanged for $\sigm=3~{\rm cm^2/g}$, we parameterize an SIDM halo using the concentration and mass or, equivalently, the maximal circular velocity ($\vmax$) and the associated radius ($\rmax$) of its CDM counterpart. Ideally, one would measure these halo parameters directly from the kinematics data and compare them with simulations. Unfortunately, most rotation curves do not have the radial extent needed to sufficiently constrain them. In this work, we impose the cosmological concentration-mass relation~\cite{Dutton:2014xda} as a prior similar to Ref.~\cite{Katz:2016hyb} and examine the consistency between its consequences and observations. 

In Fig.~\ref{fig:relations} (left), we show the $\rmax\textup{--}\vmax$ distributions from our controlled (circles) and MCMC (squares) samplings. For the former, we intend to seek the best SIDM fits to the rotation curves following the mean relation (solid) from simulations. For the sample we consider, $97\%$ galaxies can be fitted within the $2\sigma$ band (gray shaded), calculated from the relation $\log_{10}c_{200}=0.905-0.101\log_{10}(M_{200}/10^{12}h^{-1}M_\odot)$ with an intrinsic scatter of $0.11$ ($1\sigma$)~\cite{Dutton:2014xda}. For the latter, we impose the $c_{200}\textup{--}M_{200}$ relation as a top-hat prior within the $3\sigma$ range in our MCMC sampling, together with an additional constrain on $\vmax$, $1/\sqrt{2}<\vmax/V_{\rm f}<\sqrt{2}$. The resulting inferences (median and $1\sigma$ error) are shown in the figure. The two results agree well with each other. It is remarkable that even with the stringent constraints on $\vmax$ and $\rmax$ (through the $c_{200}\textup{--}M_{200}$ relation), the SIDM halo model is able to fit the diverse rotation curves, as illustrated in Figs.~\ref{fig:Diverse} and \ref{fig:Acceleration} (left). Indeed, with the concentration-mass relation, we find the $\MLdisk$ distribution is peaked toward $0.5 \MLunits$ in the fits, shown in Fig.~\ref{fig:Acceleration} (middle).

To see the MOND acceleration scale emerging from the hierarchical structure formation model, we parametrize a CDM halo with its  gravitational acceleration at $r=0$ as $g_{\rm NFW}(0)= GM/r^2|_{r\rightarrow0}\approx2\pi G\rho_s r_s\approx2\pi\vmax^2/(1.26\rmax)$. Taking the mean cosmological $\vmax\textup{--}\rmax$ relation, $\rmax=27~{\rm kpc}(\vmax/{100~{\rm km/s}})^{1.4}$, we have $g_{\rm NFW}(0)\approx1.0\times10^{-10}~{\rm m/s^2}\left({\vmax}/{240~{\rm km/s}}\right)^{0.6}$, which is close to the MOND acceleration parameter $g_\dagger$. This is the underlying reason why the empirical MOND relation captures the overall stellar kinematics of spiral galaxies well. In the presence of dark matter self-interactions and baryons, the actual central acceleration deviates from $g_{\rm NFW}({0})$, but the general argument still holds. For example, we can characterize a halo with the acceleration at the scale radius $r_s$, where the impact of dark matter self-interactions and influence of baryons tend to be small, $g_{\rm NFW}(r_s)\approx0.39g_{\rm NFW}(0)$, slightly smaller than $g_{\rm NFW}$ at the center. The characteristic halo acceleration has a mild dependence on $\vmax$, ranging from $20$ to $300~{\rm km/s}$ in the sample, and it also varies with the scatter in the cosmological relation. This variation is important, as shown in Fig.~\ref{fig:relations} (left). Since MOND does not have such flexibility, its overall fits are worse than the SIDM ones, as illustrated in Fig.~\ref{fig:Acceleration} (right). We emphasize that $g_\dagger=1.38\times10^{-10}~{\rm m/s^2}$ inferred from our SIDM fits in Sec.~\ref{sec:accel} is an average quantity over the sample after fitting to Eq.~\ref{eqn:radialAccel}, not a universal value for all the galaxies as in MOND. 

The calculation of the acceleration due to dark matter toward the center is more subtle. Inside a constant density core $g_{\rm SIDM}(r) \propto r$, and we need to specify the radius where the acceleration is being computed. On average, the stellar half-light radius is empirically observed to track the virial radius as $r_{1/2}\approx 0.015 r_{\rm vir}$~\cite{Kravtsov:2012jn}, and we have $r_{1/2}\approx1.7 R_{\rm d}$ for an exponential disk model. Without a significant contribution from baryons to the gravitational potential, SIDM predicts that $g_{\rm SIDM}(r_{1/2})=10^{-11}~{\rm m/s^2} (V_{\rm max}/100~{\rm km}/{\rm s})^{0.2}$ for the median halo concentration, and its dependence on the halo mass is extremely mild. When baryons contribute, $g_{\rm tot}$ does not increase linearly with $g_{\rm bar}$ since both the central SIDM density and the core radius depend on the gravitational potential contributed by the baryons. The net result is a strong correlation between $g_{\rm tot}$ and $g_{\rm bar}$, which is clearly evident in Fig.~\ref{fig:Acceleration}. The model predictions have a definite width in the $g_{\rm tot}$ vs $g_{\rm bar}$ plane and we have shown clearly that this scatter is required to fully explain the diversity in the rotation curve data.

\section{The correlations between the total luminous and dark matter masses}
\label{sec:baryons}

We have seen the SIDM fits to the rotation curves require values for the halo concentration parameter that are completely in line with the expectations from the Planck experiment~\cite{Ade:2013zuv,Ade:2015xua}. In addition, the stellar mass-to-light ratios are consistent with the results from stellar population models~\cite{Schombert:2013hga}.

This leads to a natural question: what is the predicted halo mass for a given stellar mass in the SIDM model? Since we assume the primordial matter power spectrum is unchanged from the CDM one for the scales we are interested in, there should be a relation consistent with the abundance matching results in the literature. In the middle panel of Fig.~\ref{fig:relations}, we show the stellar mass vs halo mass relation derived using the mass-to-light ratios from controlled (circles) and MCMC (squares) samplings. The error bars on the MCMC points denote the $1\sigma$ widths from the posteriors ($16^{\rm th}$ and $84^{\rm th}$ percentiles). Our results are consistent with the overall trend in the relation from abundance matching (solid)~\cite{Behroozi:2012iw} (see~\cite{Desmond:2016azy} for the CDM case).

We have already alluded to the importance of the BTFR in our discussion of the RAR. Lelli et al.~\cite{Lelli:2015wst} selected $118$ SPARC galaxies and found that their $V_{\rm f}\textup{--}M_{\rm bar}$ inferences can be fitted with a simple relation: $\log(M_{\rm bar})=s\log(V_{\rm f})+\log(A)$, where $s=3.71\pm0.08$ and $\log(A)=2.27\pm0.18$ for $\MLdisk=\MLbulge=0.5M_\odot/L_\odot$. The right panel of Fig.~\ref{fig:relations} shows the $V_{\rm f}\textup{--}M_{\rm bar}$ inferences with the $\MLdisk$ and $\MLbulge$ values from the controlled (circles) and MCMC (squares) fits. The error bars in $M_{\rm bar}$ on the MCMC points denote the $1\sigma$ widths in the stellar mass-to-light ratios from the posteriors, and the errors in $V_{\rm f}$ are taken directly from the SPARC dataset~\cite{Lelli:2016zqa}. We also show the fit from~\cite{Lelli:2015wst} as the solid line of Fig.~\ref{fig:relations} (right). Note that this fit used $118$ galaxies and a few outliers at the low $V_{\rm f}$ end were not included. For comparison, we plot the $135$ galaxies in our sample as triangles by fixing $\MLdisk=\MLbulge=0.5M_\odot/L_\odot$. We see that their distribution in the $V_{\rm f}\textup{--}M_{\rm bar}$ plane is almost identical to the one from our SIDM fits. This is not surprising, as the $\MLdisk$ values inferred from the SIDM fits are peaked toward $0.5~M_\odot/L_\odot$ as shown in Fig.~\ref{fig:Acceleration} (middle). Thus, we conclude that the SIDM fits also lead to a tight BTFR relation. For our fits, we find $s\approx3.46$ (CS), $3.27$ (MS) and $3.58$ ($0.5 M_{\odot}/L_{\odot}$), excluding six obvious outliers on the left side of the black line. 

We note that there is no evidence in the data for $s=4$, i.e., $M_{\rm bar}\propto V_{\rm f}^4$, the motivation for MOND, either in the constant $\MLstar$ fits or in the SIDM fits. Many of the recent CDM simulations with efficient baryonic feedback seem to get something akin to the BTFR with $s \approx 3.6\textup{--}3.8$~\cite{Dutton:2012jh,Chan:2015tna,Sales:2016dmm,Dutton:2016qfg}, but it is fair to say that this is still not well understood theoretically, in particular, the smallness of the scatter in the BTFR, equivalent to the one seen in the RAR~\cite{2018arXiv180301849W}. We expect that there will be interplay between dark matter self-interactions and baryonic feedback in changing the halo potential, and understanding how the BTFR emerges in SIDM is fertile territory for research in galaxy formation.


\section{Discussion and Conclusions}
\label{sec:con}

In this work, we have investigated SIDM as a solution to two puzzles that are present in galactic rotation curves: (1) the diversity of inner rotation curves in galaxies that have similar baryon content and similar flat circular velocities, and (2) the small scatter in the radial acceleration relation between the total gravitational acceleration and the one inferred from the baryonic mass content, i.e., uniformity. 

We have fitted our SIDM halo model to the rotation curves of $135$ SPARC galaxies, and found that it reproduces the observed diversity in the inner regions. The distribution of resulting $3.6~\micron$ stellar disk mass-to-light ratios for the sample peaks at $\MLdisk\approx0.5~\MLunits$, in good agreement with the stellar population models. Our fits lead to a radial acceleration relation described by the characteristic acceleration scale $\sim 10^{-10} {\rm m/s}$, with tight scatter of $0.10$ dex. The host halos are fully consistent with the Planck cosmology. The inferred stellar mass-halo mass relation agrees with the result from the abundance matching method, and the fits also predict a tight BTFR. These results provide compelling arguments in favor of the idea that the inner halos of galaxies are kinematically thermalized due to dark matter self-interactions.

The SIDM model automatically inherits all of the successes of the CDM model on large scales, as the predictions are indistinguishable at distances larger than about $10\%$ of the virial radius of galactic halos. The required cross section is similar to the proton-neutron elastic scattering cross section and this may be a strong hint that the dark matter sector replicates some elements of the standard model. The large cross section keeps the inner halo isothermal and this makes the predictions for the central halo profile at later times insensitive to the star formation history, as confirmed in recent hydrodynamical N-body simulations~\cite{Robertson:2017mgj,2017MNRAS.472.2945R}. This implies that a large variety of feedback models, e.g,~\cite{Governato:2012fa,Hopkins:2013vha,Onorbe:2015ija,Wang:2015jpa,Read:2017lvq,2018MNRAS.473.4392S}, can be compatible with the SIDM model we have discussed here. The predictions are quantitatively the same for $\sigm\gtrsim 1~{\rm cm^2/g}$. This makes our results robust, but it makes hard to precisely determine the cross section from kinematic datasets on galaxy scales~\cite{Kamada:2016euw}.

There are a number of promising directions that can further test SIDM and explore galaxy formation and evolution in this framework. Here, we highlight a few of them. SIDM simulations predict a correlation between the half-light radius of the stars and the dark matter core size in dwarf and low surface brightness galaxies~\cite{Vogelsberger:2014pda}, which should be further explored and may provide an observational test of SIDM. Similarly, the ultra-diffuse galaxies in the clusters could be a test laboratory~\cite{2018arXiv180506896C}. A related issue is the origin of the large spread in the surface brightness of galaxies, which remains poorly understood. Interestingly, hydrodynamical simulations of galaxy clusters show that the stellar density profiles in SIDM are more diverse than in their CDM counterparts~\cite{Robertson:2017mgj}. Is this a more general feature in SIDM due to the dynamical interplay between core formation and feedback? How does this interplay impact the emergence of the BTFR? Finally, at the lowest mass end, the dwarf spheroidal galaxies, including the so-called ultra-faint dwarfs, in the Local Group could provide a key test of SIDM (see~\cite{Valli:2017ktb,Read:2018pft}). Dedicated SIDM simulations with the baryons will be required to explore these exciting topics.

The predictive power of the SIDM model, the clear connection to cosmology, and its rich implications for other astrophysical observations and particle physics phenomenology~\cite{Feng:2009mn,Feng:2009hw,Buckley:2009in,Loeb:2010gj,Frandsen:2010yj,Frandsen:2011kt,Aarssen:2012fx,Cline:2013zca,Kahlhoefer:2013dca,Tulin:2013teo,Schutz:2014nka,Boddy:2014yra,Hochberg:2014dra,Hochberg:2014kqa,Cyr-Racine:2015ihg,Blennow:2016gde,Boddy:2016bbu,Tulin:2017ara,Kamada:2018hte,Braaten:2018xuw}, all taken together make a clear case that it should be treated on the same footing as the CDM model. The economical explanation, with the addition of just one parameter, for the diverse rotation curves across the entire range of observed galaxies argues in favor of the idea that the dark matter particles have a large affinity for the self-interactions.

\acknowledgments
We thank the authors of~\cite{Rocha:2012jg,Elbert:2014bma,Creasey:2016jaq} for the simulation data, which are used for comparison with the analytical model, shown in {\bf Supplementary Materials}. This work was supported by the National Science Foundation Grant No.~PHY-1620638 (MK), the U. S. Department of Energy under Grant No.~DE-SC0008541 (HBY), the Hellman Fellows Fund (HBY), UCR Academic Senate Regents Faculty Development Award (HBY), and the National Science Foundation under Grant No.  NSF PHY-1748958 as part of the KITP  ``The Small-Scale Structure of Cold (?) Dark Matter" and ``High Energy Physics at the Sensitivity Frontier" workshops (MK and HBY). HBY also acknowledges Tsung-Dao Lee Institute, Shanghai, for its hospitality during the completion of this work.

\section*{Methods}
\label{sec:Methods}

We provide a detailed description of the analytical model developed previously~\cite{Kaplinghat:2015aga,Kamada:2016euw} and the fitting procedure in this section. We divide the halo into an inner and an outer region~\cite{Rocha:2012jg} with the aim that the outer halo is not significantly changed by the self-scattering process. In the inner region, dark matter self-interactions thermalize the halo in the presence of the baryonic potential, and we model the dark matter distribution using the  isothermal density profile, $\rho_{\rm iso}\propto\exp(-\Phi_{\rm tot}(R,z)/\sigma^2_{\rm v0})$. Poisson's equation relates $\Phi_{\text{tot}}$ to the dark matter and baryon profiles as 
\begin{equation}
\label{eq:poisson}
\nabla^{2}\Phi_{\text{tot}}(R,z)=4\pi G[\rho_{\text{iso}}(R,z)+\rho_{\text{b}}(R,z)].
\end{equation}
For the outer halo, where the self-scattering effect becomes negligible, we model the dark matter distribution with an NFW profile $\rho_{\rm NFW}(r)=\rho_s r^3_s/r(r+r_s)^2$. To construct the full SIDM halo profile, we define a radius $r_1$, where dark matter particles had one interaction on average over the age of the galaxy. We join the spherically-averaged isothermal ($\rho_{\rm iso}$) and spherical NFW ($\rho_{\rm NFW}$) profiles at $r=r_1$ such that the mass and density are continuous at $r_1$. Thus, the isothermal parameters ($\rho_0$, $\sigma_{\rm v0}$) directly map on to the NFW parameters ($r_s$, $\rho_s$) or ($r_{\rm max}$, $V_{\rm max}$).

The value of $r_1$ is determined by the following condition, 
\begin{equation}
\left<\sigma v_{\rm rel}\right>\rho_{\rm NFW}(r_1)t_{\rm age}/m=N_{\rm sc}, 
\label{eq:r1}
\end{equation}
where $\sigma$ is the self-scattering cross section, $m$ is the dark matter particle mass, $v_{\rm rel}$ is the dark matter relative velocity in the halo, $\left<...\right>$ denotes averaging over the Maxwellian velocity distribution, $t_{\rm age}$ is the age of the galaxy, and $N_{\rm sc}$ is a factor of order unity, to be determined by calibrating to simulations. In this work, we have set $t_{\rm age}=10~{\rm Gyr}$ and $N_{\rm sc}=1$, which reproduce simulation results well; see {\bf Supplementary Materials}. In principle, we should use different ages for each galaxy, say between $10~{\rm Gyr}$ and $13~{\rm Gyr}$. However, our model can only constrain the combination of the cross section and the age. More importantly, we have set $\sigm$ to a large enough value that the SIDM density profile is insensitive to small changes in the cross section. We assume that this cross section is a constant over the SPARC sample, so $\left<\sigma v_{\rm rel}\right>=\sigma(4/\sqrt{\pi})\sigma_{\rm v0}$. 

We take two independent but complementary approaches. In the first one, we assume a thin-disk profile for the stellar disk in solving Eq.~(\ref{eq:poisson}), $\rho_{\rm b}(R,z)=\Sigma_0\exp(-R/R_{\rm d})\delta(z)$, where $\Sigma_{0}$ is the central surface density and $R_{\rm d}$ is the scale radius. For each galaxy, we reconstruct the $\Sigma_{0}$ and $R_{\rm d}$ values by fitting the profile to the disk contribution of the rotation curve as in~\cite{Kamada:2016euw}. We neglect the baryonic influence on the SIDM halo from the gas and bulge potentials, but include all the mass components in modeling the total circular velocity. This is a reasonable approximation for the following reasons: (1) the gas is less centrally concentrated and so its impact on the SIDM density profile is smaller, (2) the bulge (when present) mainly affects the innermost region, while the disk contributes in this region as well as at farther radii. Ref.~\cite{Kamada:2016euw} solved Eq.~(\ref{eq:poisson}) with the thin-disk approximation and created numerical templates for the isothermal density profile on the grid of $a \equiv 8\pi G \rho_{0}R_{\text{d}}^{2}/(2\sigma_{\text{v0}}^{2})$ and $b \equiv 8\pi G \Sigma_{0}R_{\text{d}}/(2\sigma_{\text{v0}}^{2})$. When the stellar profile is known, the parameters $a$ and $b$ give the central density and dispersion of the isothermal dark matter halo, which completely specify the inner density profile. We interpolate the templates to generate rotation curves for any set of $(\rho_0, \sigma_{\text{v0}}, \Sigma_{0}, R_{\text{d}})$. The fixed value of the cross section allows us to match this density profile to the outer NFW density profile.

In fitting to the SPARC sample with the templates, we take a controlled sampling approach. For a given galaxy, we start with the mean $\rmax\textup{--}\vmax$ relation from cosmological $\Lambda$CDM simulations~\cite{Dutton:2014xda} and an NFW profile that matches the flat part of the rotation curve. Then, we choose an appropriate $\MLdisk$ ($\MLbulge$) value to reproduce the inner rotation curve. We calculate a $\chi^2/{\rm d.o.f.}$ value for each fit and iterate this process manually by adjusting the parameters until a good fit is achieved. For most galaxies, the very first step provides decent fits, showcasing the simplicity of the model and its ability to fit the observed data simultaneously. For each galaxy, we demand the ($\rmax$, $\vmax$) values to be within the $\sim2\sigma$ band. In this way, we have good control over the halo parameters in the fits. The goal is to see to what degree are the galaxy halos of the SPARC sample consistent with predictions of the hierarchical structure formation scenario, and the extracted $\MLstar$ values consistent with stellar population synthesis model results~\cite{Schombert:2013hga}.       

In our second approach, we perform a MCMC sampling of the SIDM model parameter space. Since it is computationally expensive to use the templates, we use a spherical approximation to model the baryon distribution~\cite{Kaplinghat:2013xca,Elbert:2016dbb}. We create a spherical baryonic mass profile from the stellar and gas masses, such that the baryonic mass within a sphere of radius $r$ is $M_{\text{b}}(r)=(V_{\rm disk}^2+V_{\rm bulge}^2+V_{\rm gas}^2)r/G$, where $V_{\rm disk}$ is the contribution to the rotation curve from the disk at radius $r$ and similarly for the bulge and gas. We solve Eq.~(\ref{eq:poisson}) in the spherical limit by taking $r=\sqrt{R^2+z^2}$. We compute $\rho_{\text{iso}}(r)$ starting at a small radius assuming a core and integrate the equation to larger radii. The behavior of the baryonic density profile as $r \rightarrow 0$ is chosen so that a core at small radii is physical~\cite{Kaplinghat:2013xca}. We compared the isothermal halos from this spherical approximation to those from the axisymmetric case (templates) and found agreement within $10\textup{--}20\%$. Thus, while we expect some variance in the inferred parameters between the two methods, the overall features should be very similar. This expectation is borne out by our final fits. 

We match the isothermal density profile $\rho_{\text{iso}}$, parameterized by $(\rho_{0},\sigma_{\text{v0}})$, to the NFW density profile at $r_1$, and this determines $(\vmax,\rmax)$. Thus, the spherical model has four parameters, two for the entire halo and two for the mass-to-light ratios: $(\rho_{0}, \sigma_{\text{v0}}, \MLdisk, \MLbulge)$. We use the \texttt{emcee} implementation of the Affine invariant MCMC ensemble sampler \cite{ForemanMackey:2012ig} to infer the posteriors of these four model parameters. To streamline the calculation of $r_1$ at each point in parameter space for matching onto the outer NFW radius, we use the rate of scatterings, $\Gamma_0=\rho_0(\sigm)(4/\sqrt{\pi})\sigma_{\rm v0}$, within the isothermal core as the MCMC parameter in lieu of the core density $\rho_0$.

The prior distributions used for the halo parameters and the mass-to-light ratios in the MCMC scan are as follows:
\begin{itemize}
\item $\Gamma_0$: Uniform prior in the range of $\log_{10} 2<\log_{10}(\Gamma_0\times 10~{\rm Gyr})<5$. 
  
\item $\sigma_0$: Uniform prior in range of $\log_{10} 2<\log_{10}\sigma_0 /({\rm km/s}) <\log_{10} 500$. 
 
\item $\MLstar$: Uniform prior in $0.1 < \MLdisk, \MLbulge < 10~\MLunits$. The parameter $\MLbulge$ is only included for galaxies whose surface brightness profiles have a stellar bulge decomposition provided in the SPARC dataset. All galaxies have $\MLdisk$ as a parameter describing their stellar disk.
 
\item $c_{200}\textup{--}M_{200}$ relation: We adopt the mean concentration-mass relation $\log_{10}c_{200}=0.905-0.101\log_{10}(M_{200}/10^{12}h^{-1}M_\odot)$~\cite{Dutton:2014xda}, then use a top-hat prior with $0.33$ dex spread in $\log_{10}c_{200}$, corresponding to a $\pm 3\sigma$ scatter.

\end{itemize}

Additionally, we also impose two regularization priors.
\begin{itemize}
\item We add $5\%$ of $V_{\rm f}$ in quadrature for calculating the likelihood function. This allows the code to disregard the points deep within the central regions and those with tiny errors. We have checked that it doesn't change the inference of cores/cusps. We {\em do not} include this regularization error when quoting $\chi^2$ values. 
 \item We impose a uniform regulation prior on $\vmax$: $1/\sqrt{2}<\vmax/V_{\rm f}<\sqrt{2}$. For most of the galaxies ($\sim80\%$), our MCMC program can find physical fits without this prior. However, in some cases, the MCMC sampler tends to pick up fits not consistent with hierarchical structure formation predictions --- either the baryonic contribution dominates the total rotation velocity at all radii and the halo concentration is unreasonable low (for some high surface brightness galaxies), or the opposite (for some low surface brightness galaxies). This is typically due to the lack of an extended rotation curve to fully constrain the halo parameters. The additional regularization prior fixes this issue. We have also checked that the results are similar if we consider a more generous range $1/2<\vmax/V_{\rm f}<2$ (see {\bf Supplementary Materials}).
 
 \end{itemize}

\bibliography{sidm}

\newpage

\floatsetup[table]{capposition=top}
\floatsetup[figure]{capposition=top}
\renewcommand{\thetable}{S\arabic{table}}
\renewcommand{\thefigure}{S\arabic{figure}}

{\Huge \bf Supplementary Materials}\\

We provide additional information and results to supplement the results in the main text.

\begin{itemize}
\item{In Table S1, we list the galaxies that are shown in Fig. 1 of the main text. }
\item{We show the total acceleration vs the baryonic acceleration for the inner and outer regions in Fig.~S1.}
\item{Fig.~S2 shows $r_{\rm max}\textup{--}V_{\rm max}$, $M_{\rm star}\textup{--}M_{\rm halo}$, and $M_{\rm bar}\textup{--}V_{\rm flat}$ relations, similar to Fig. 3 of the main text, but we impose the top-hat prior on the concentration-mass relation with a wider $V_{\rm max}$ regulation, $1/2<V_{\rm max}/V_{\rm f}<2$. In addition, we show the results with a Gaussian prior on the concentration-mass relation and $1/\sqrt{2}<V_{\rm max}/V_{\rm f}<\sqrt{2}$.}
\item{Fig. S3 shows the SIDM density profiles predicted in the analytical model, compared to cosmological N-body simulations from Elbert et al. MNRAS 453 (2015) no. 1, 29-37, and Creasey et al., MNRAS 468 (2017) no. 2, 2283-2295.}
\item{Fig. S4 shows the SIDM density profiles predicted in the analytical model, compared to cosmological N-body simulations from Rocha et al. MNRAS 430 (2013) no. 1, 81-104.}
\item{Fig. S5 shows detailed SIDM and MOND fits to individual $135$ SPARC galaxies.}
\item {Table S2 contains the model parameters and $\chi^2/{\rm d.o.f.}$ values for the SIDM fits shown in Fig. S5.}
\end{itemize}

\clearpage

\addtocounter{table}{-1}

\centering
\begin{table}[t]
\begin{longtable}{cc|cc|cc|cc}
\hline
Name & ${V}_{\rm{f}}$ [km/s] & Name & ${V}_{\rm{f}}$ [km/s] & Name & ${V}_{\rm{f}}$ [km/s] & Name & ${V}_{\rm{f}}$ [km/s]\\
\hline
UGC06923 & 79.6 & UGC04278 & 91.4 & F571-8 & 139.7 & NGC7331 & 239.0\\
UGC05721 & 79.7 & NGC0247 & 104.9 & NGC4138 & 147.3 & NGC3992 & 241.0\\
UGC06446 & 82.2 & NGC0024 & 106.3 & NGC3198 & 150.1 & NGC6674 & 241.3\\
UGC08286 & 82.4 & UGC06930 & 107.2 & UGC09037 & 152.3 & IC4202 & 242.6\\
NGC2915 & 83.5 & UGC06917 & 108.7 & NGC2683 & 154.0 & UGC06787 & 248.1\\
UGC06667 & 83.8 & NGC1003 & 109.8 & NGC6015 & 154.1 & NGC6195 & 251.7\\
UGC06399 & 85.0 & NGC4183 & 110.6 & NGC4051 & 157.0 & NGC5005 & 262.2\\
NGC2976 & 85.4 & F568-V1 & 112.3 & NGC4100 & 158.2 & UGC02953 & 264.9\\
NGC0055 & 85.6 & UGC05986 & 113.0 & NGC6946 & 158.9 & UGC11455 & 269.4\\
F583-1 & 85.8 & NGC6503 & 116.3 & NGC3949 & 163.0 & NGC2841 & 284.8\\
UGC02259 & 86.2 & NGC3769 & 118.6 & NGC1090 & 164.4 & UGC11914 & 288.1\\
NGC0100 & 88.1 & NGC4559 & 121.2 & NGC3726 & 168.0 & UGC02885 & 289.5\\
NGC5585 & 90.3 & NGC4010 & 125.8 & NGC3877 & 168.4 & NGC5985 & 293.6\\
UGC04325 & 90.9 & UGC03580 & 126.2 & NGC4088 & 171.7 & ESO563-G021 & 314.6\\
\hline
\end{longtable}
\caption{Galaxies shown in Fig. 1 of the main text.}
\end{table}

\clearpage

\addtocounter{figure}{-3}

\begin{figure}[H]
\centering
\begin{tabular}{@{}cc@{}}
\includegraphics[scale=0.4]{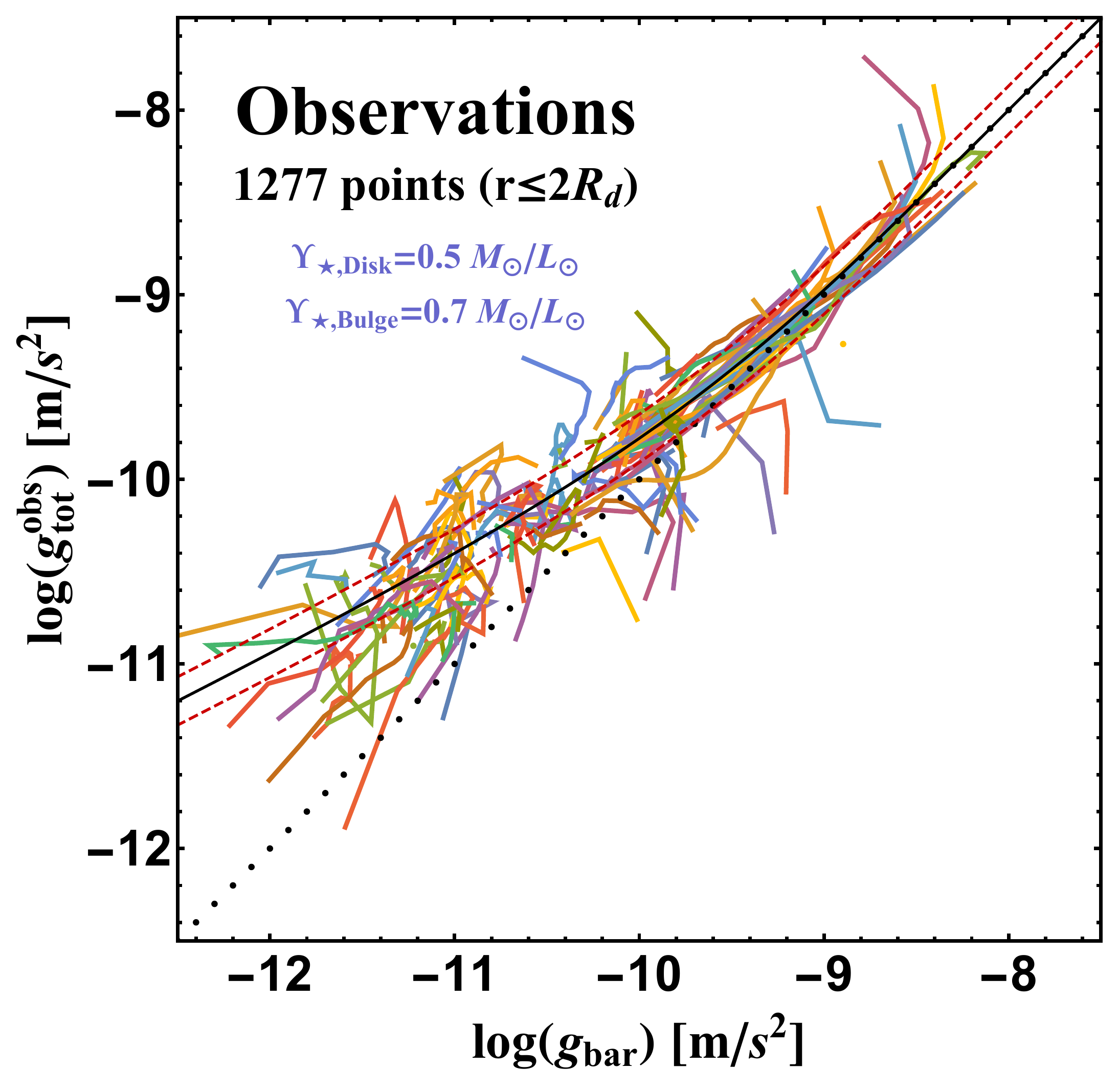} & \includegraphics[scale=0.4]{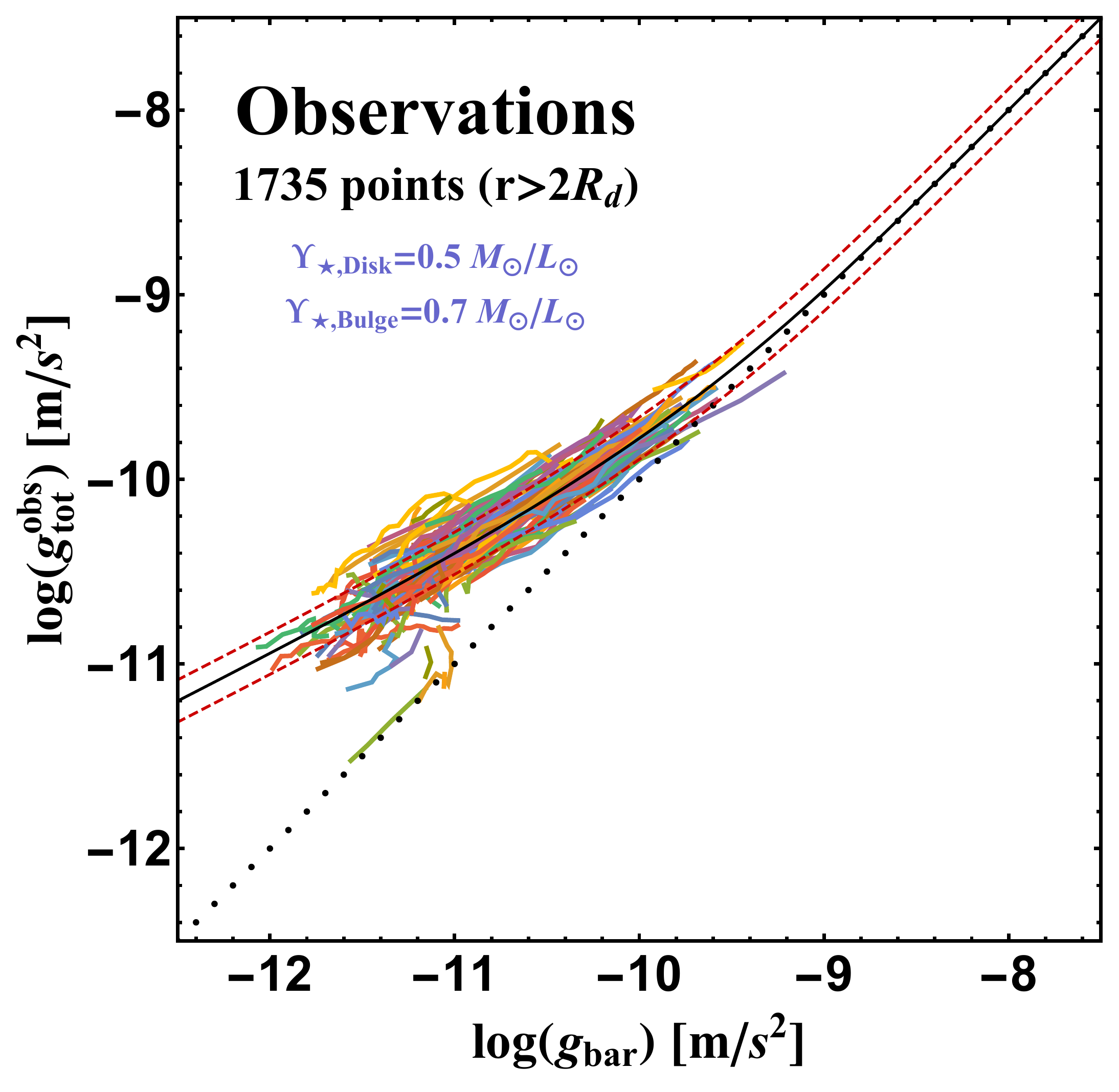} \\
\includegraphics[scale=0.56]{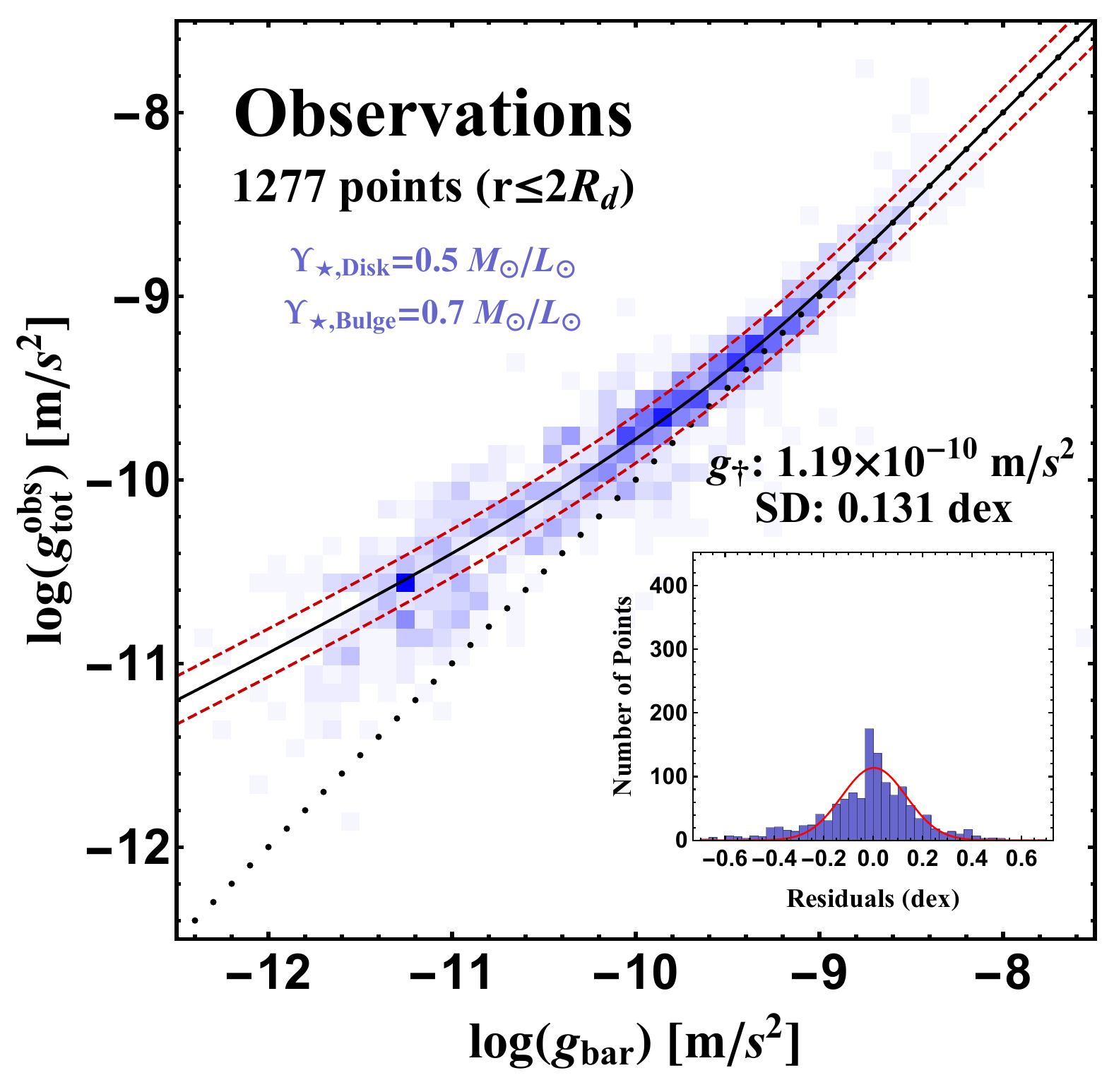} &  \includegraphics[scale=0.56]{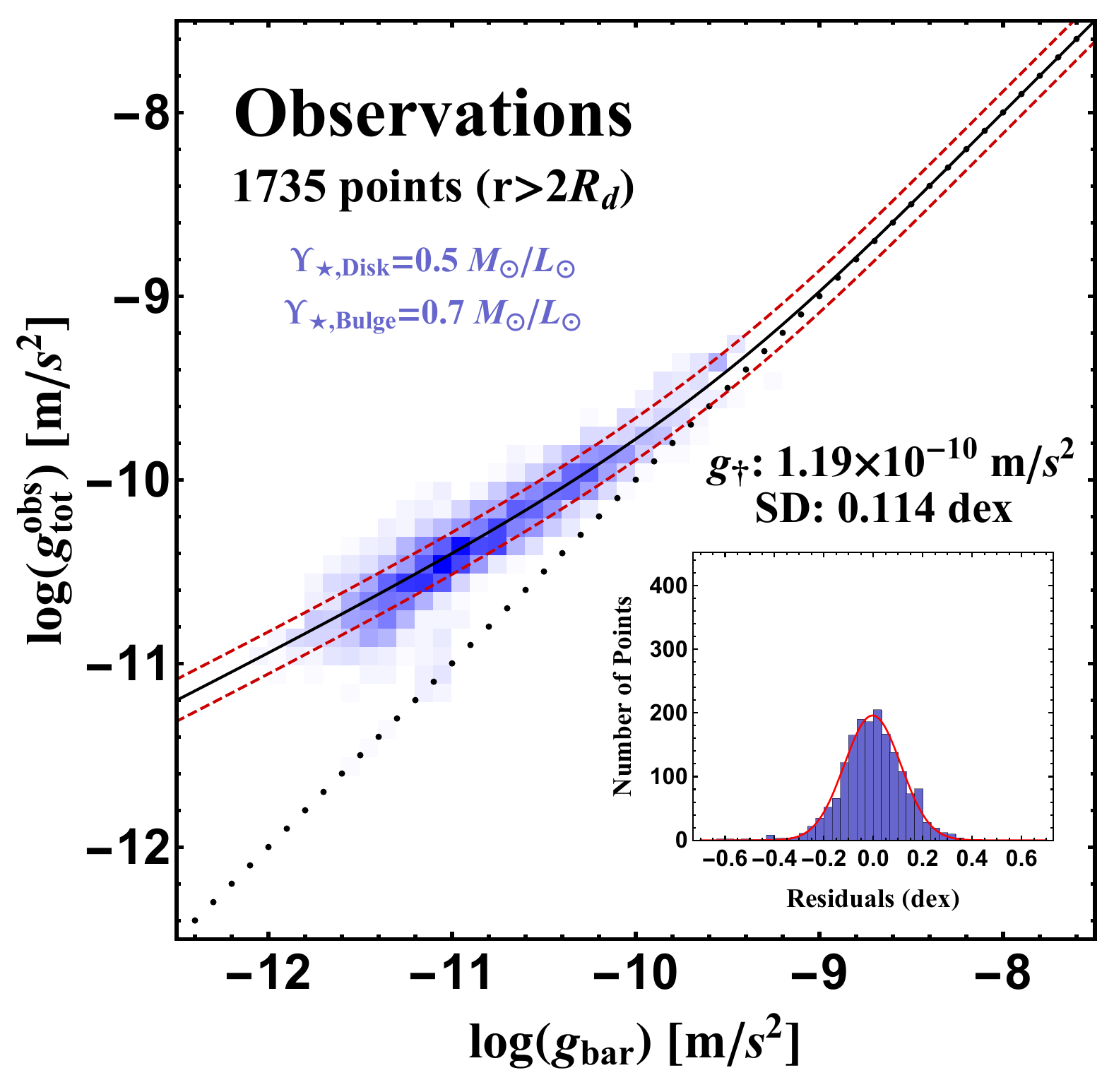}\\
\end{tabular}
\caption{{\it Upper}: the total acceleration vs the baryonic acceleration (colored) for the inner ($r\leq 2 R_d$, left) and outer ($r>2 R_d$, right) regions, where $R_d$ is the scale radius of the stellar disk. {\it Lower:} The $g_{\rm tot}\textup{--}g_{\rm bar}$ relation with a different color scheme, where the intensity is proportional to the density of points. The scatter in the $g_{\rm tot}\textup{--}g_{\rm bar}$ relation of the inner regions is visibly larger (black solid).}
\end{figure}

\begin{figure}[H]
\centering
\label{fig:c200gac}
\begin{tabular}{@{}ccc@{}}
\includegraphics[scale=0.28]{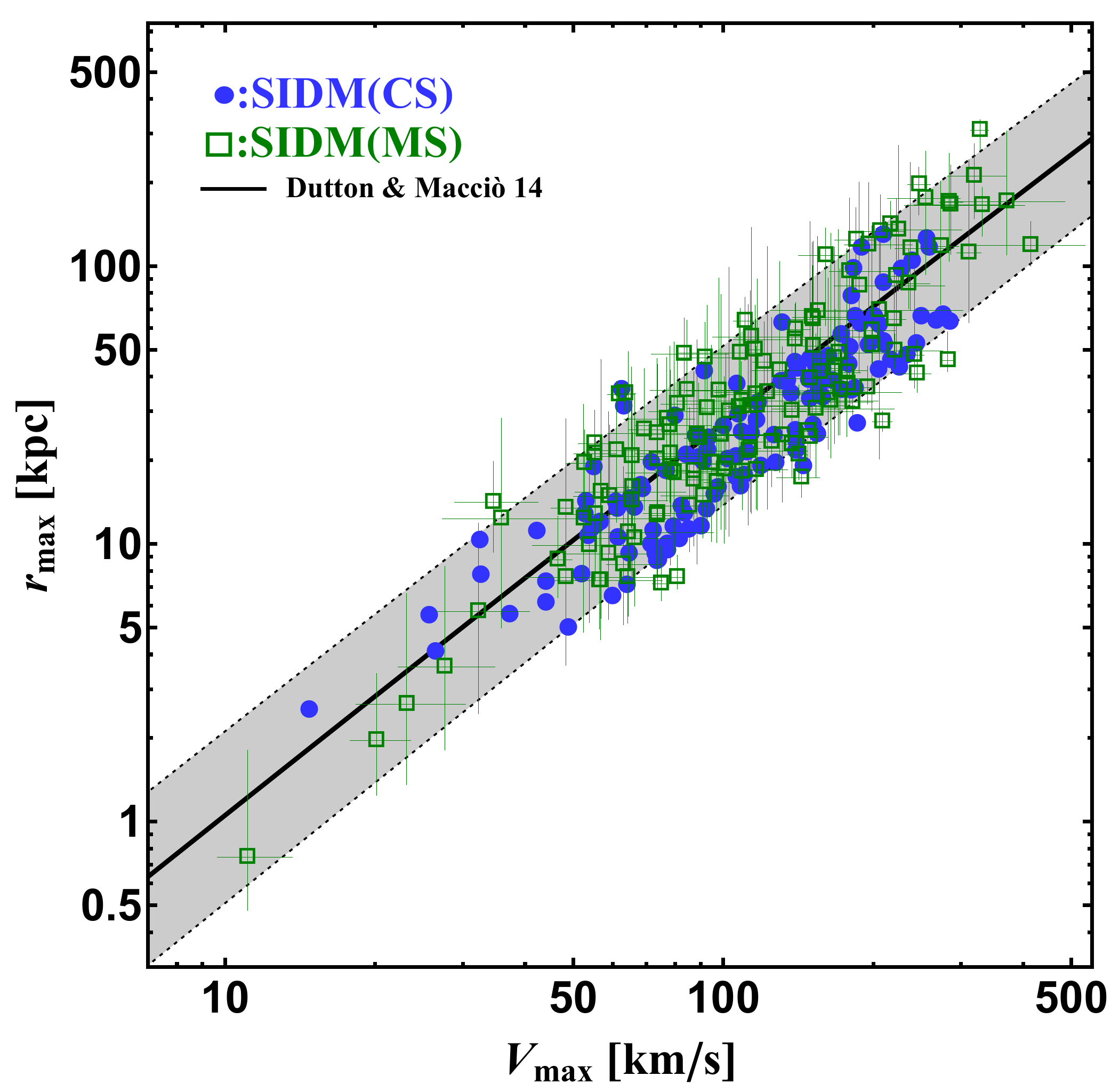} & \includegraphics[scale=0.28]{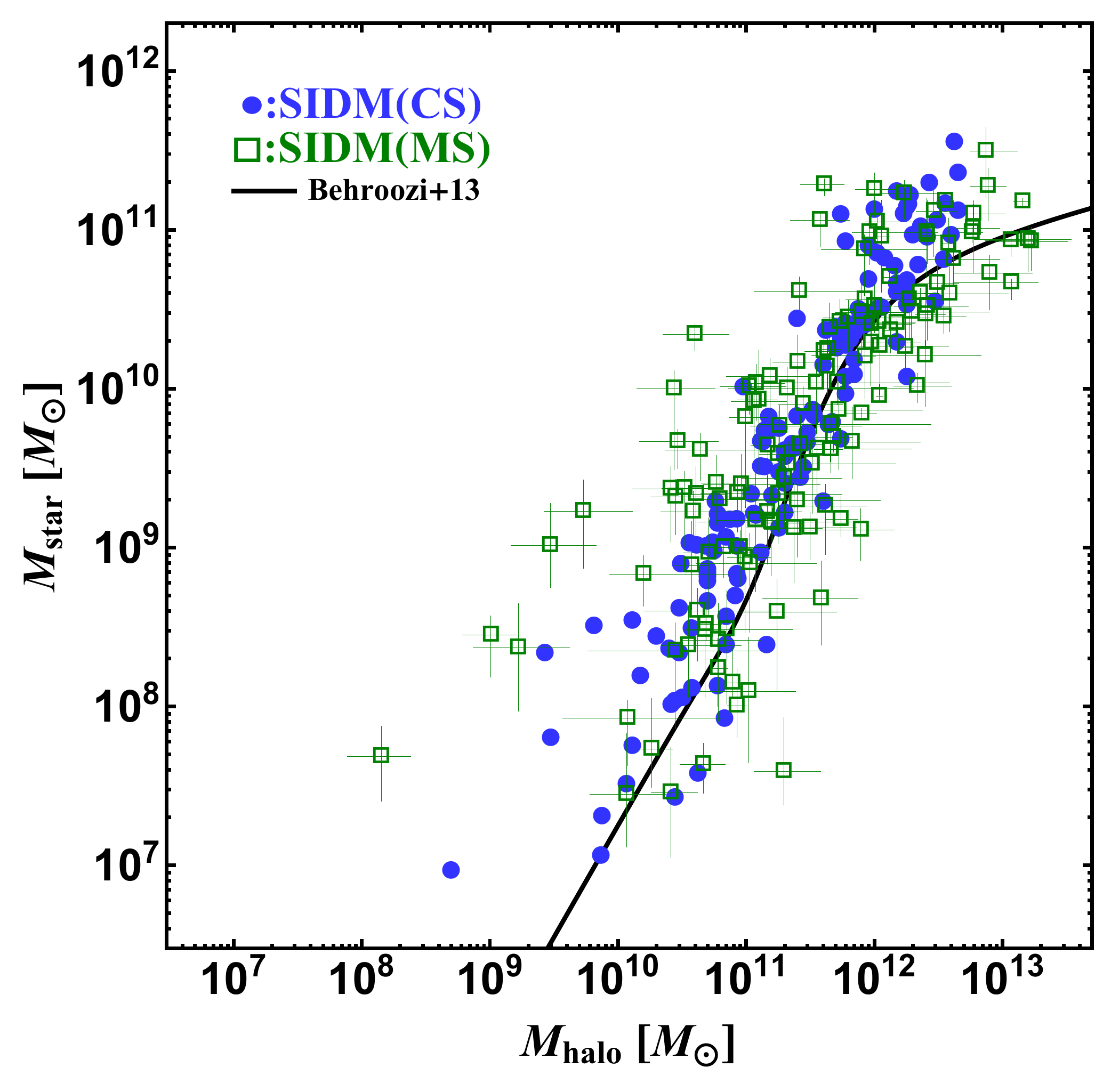}  & \includegraphics[scale=0.28]{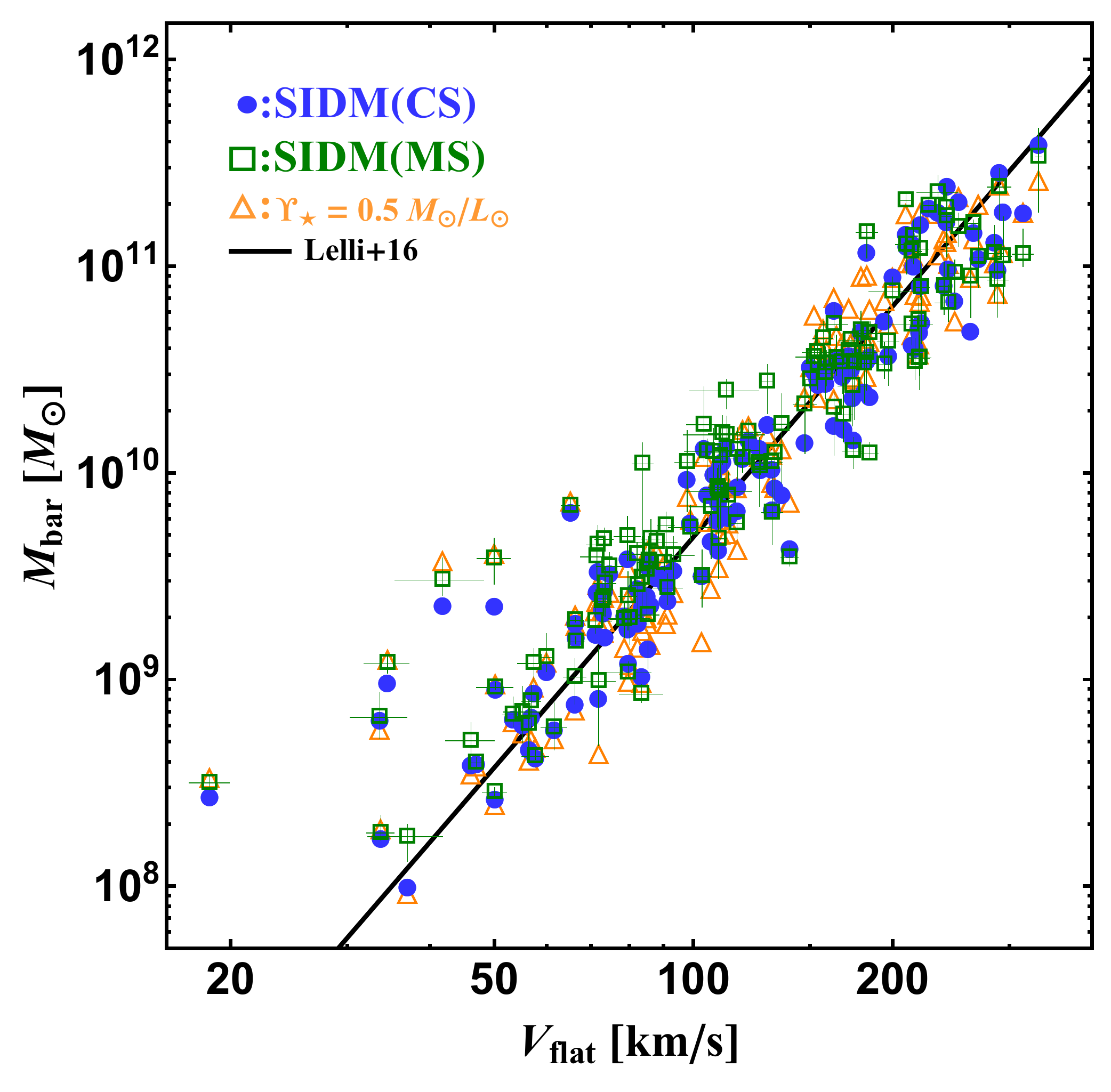} \\
\includegraphics[scale=0.28]{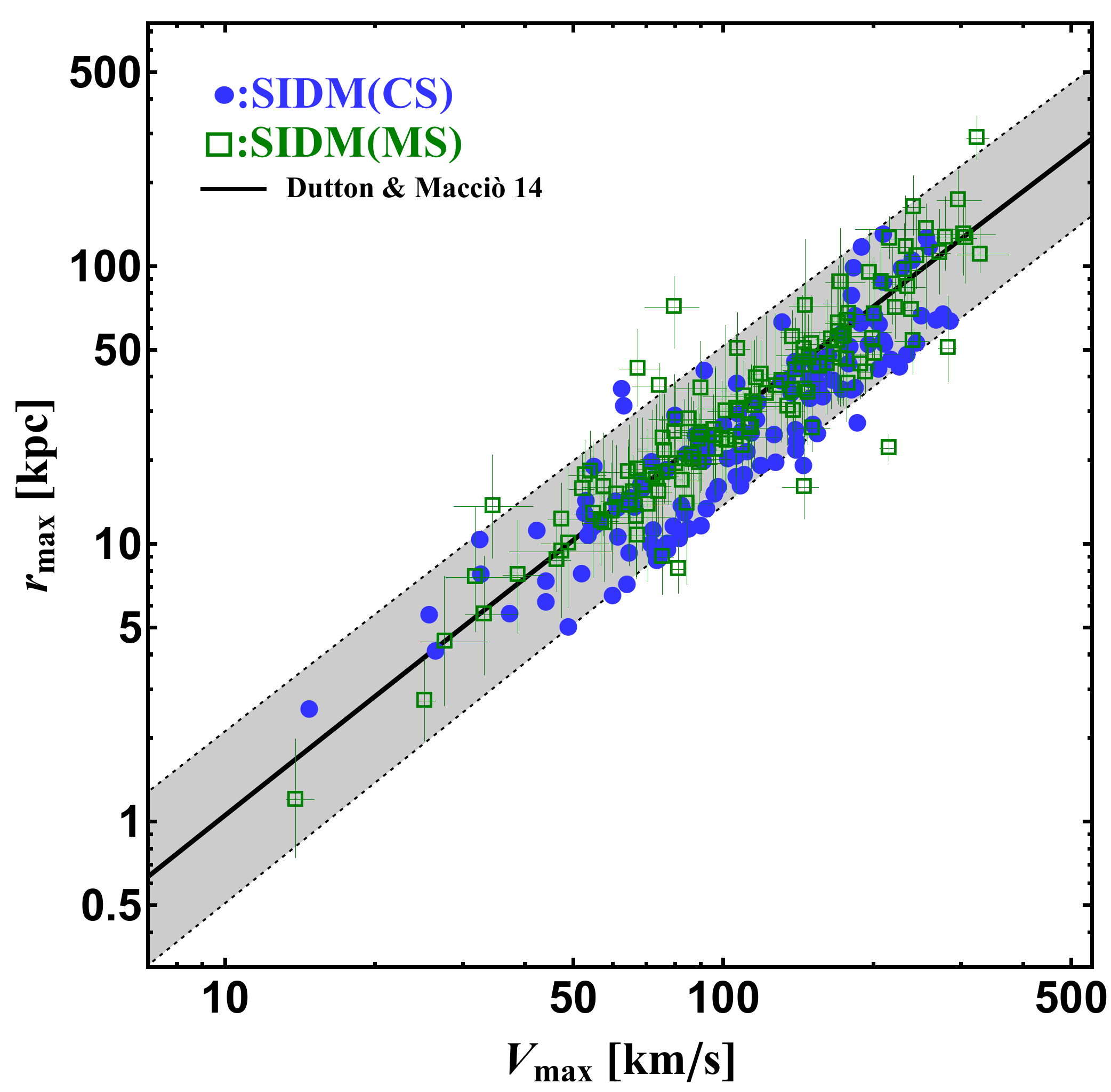} & \includegraphics[scale=0.28]{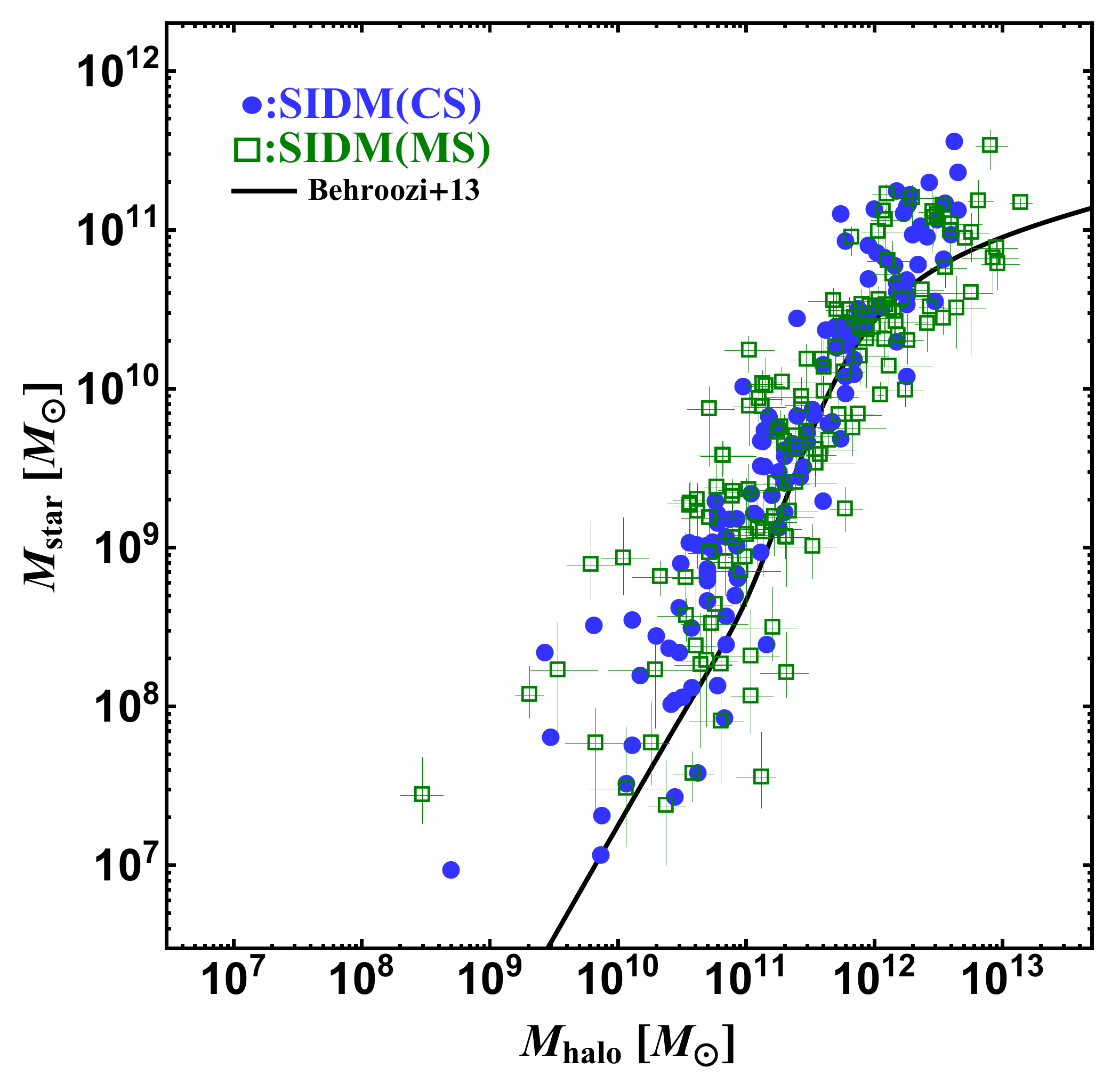}  & \includegraphics[scale=0.28]{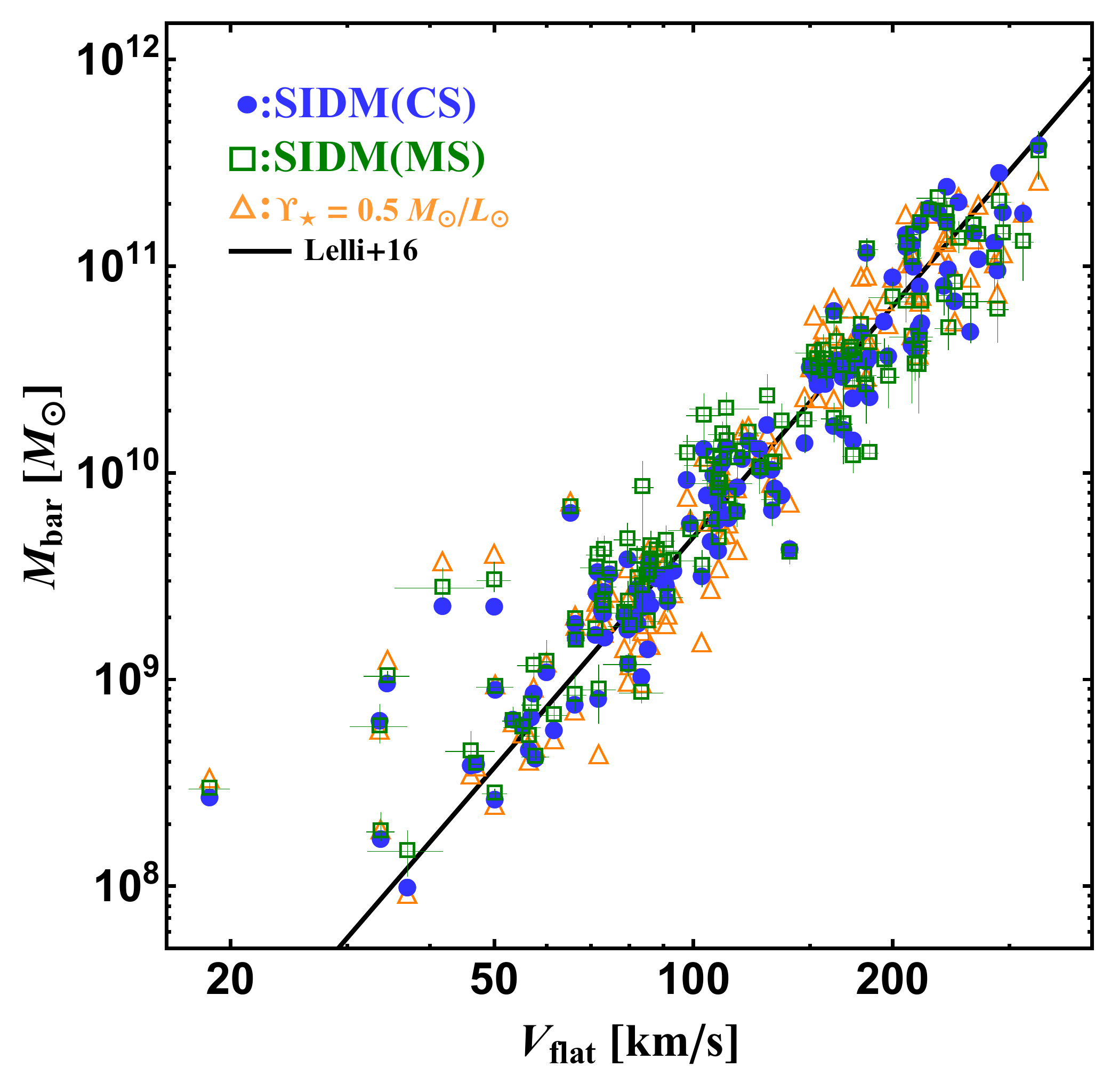} 
\end{tabular}
\caption{{\it Upper}: Similar to Fig. 3 of the main text, but we impose the top-hat prior on the concentration-mass relation with a wider $V_{\rm max}$ regulation, $1/2<V_{\rm max}/V_{\rm f}<2$. {\it Lower}: Similar to Fig. 3 of the main text, but with a Gaussian prior on the concentration-mass relation (with width 0.11 dex) and $1/\sqrt{2}<V_{\rm max}/V_{\rm f}<\sqrt{2}$.}
\end{figure}

\begin{figure}[H]
\centering
\label{fig:comparision}
\begin{tabular}{@{}ccc@{}}
\includegraphics[scale=0.7]{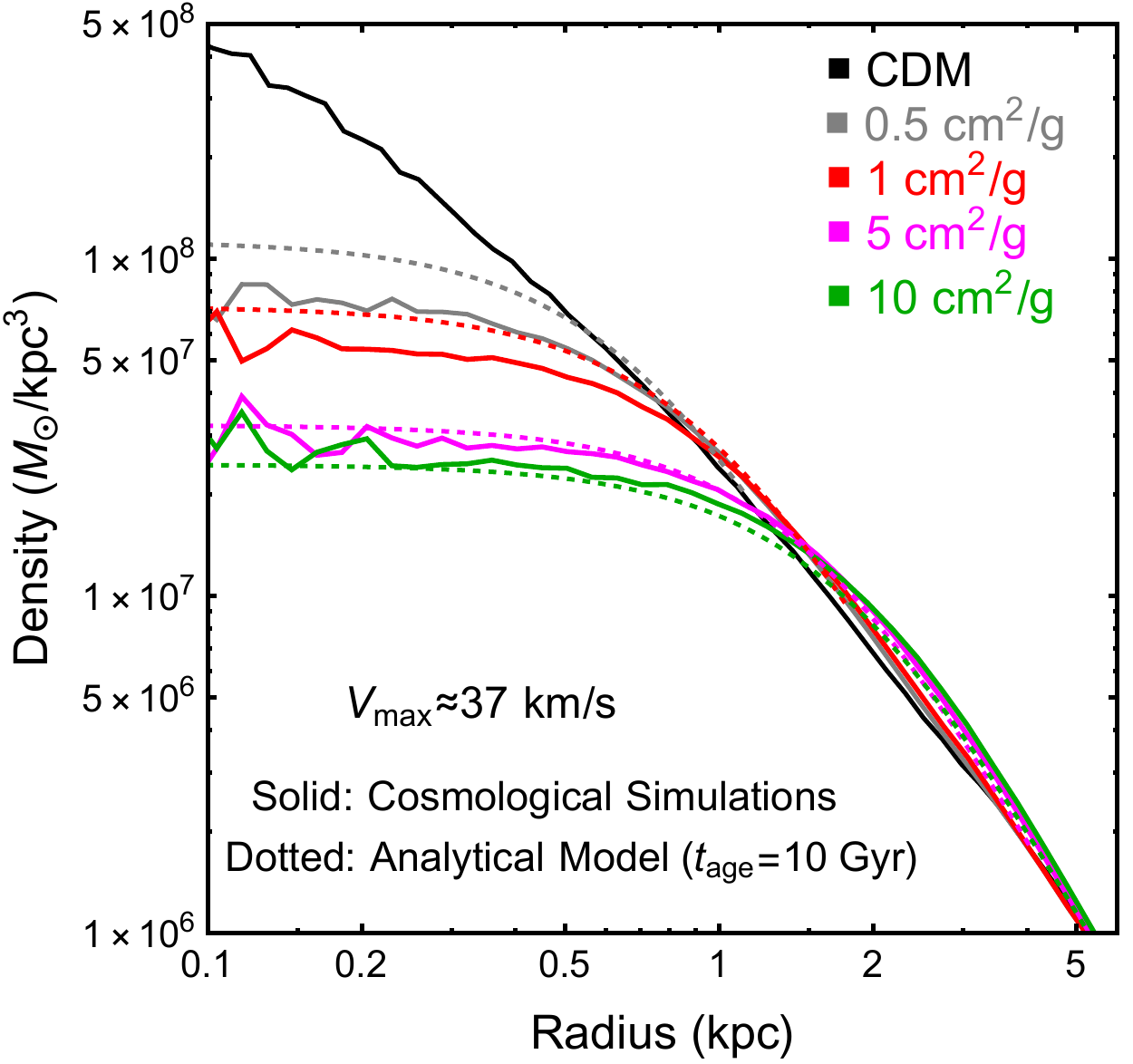} & \includegraphics[scale=0.7]{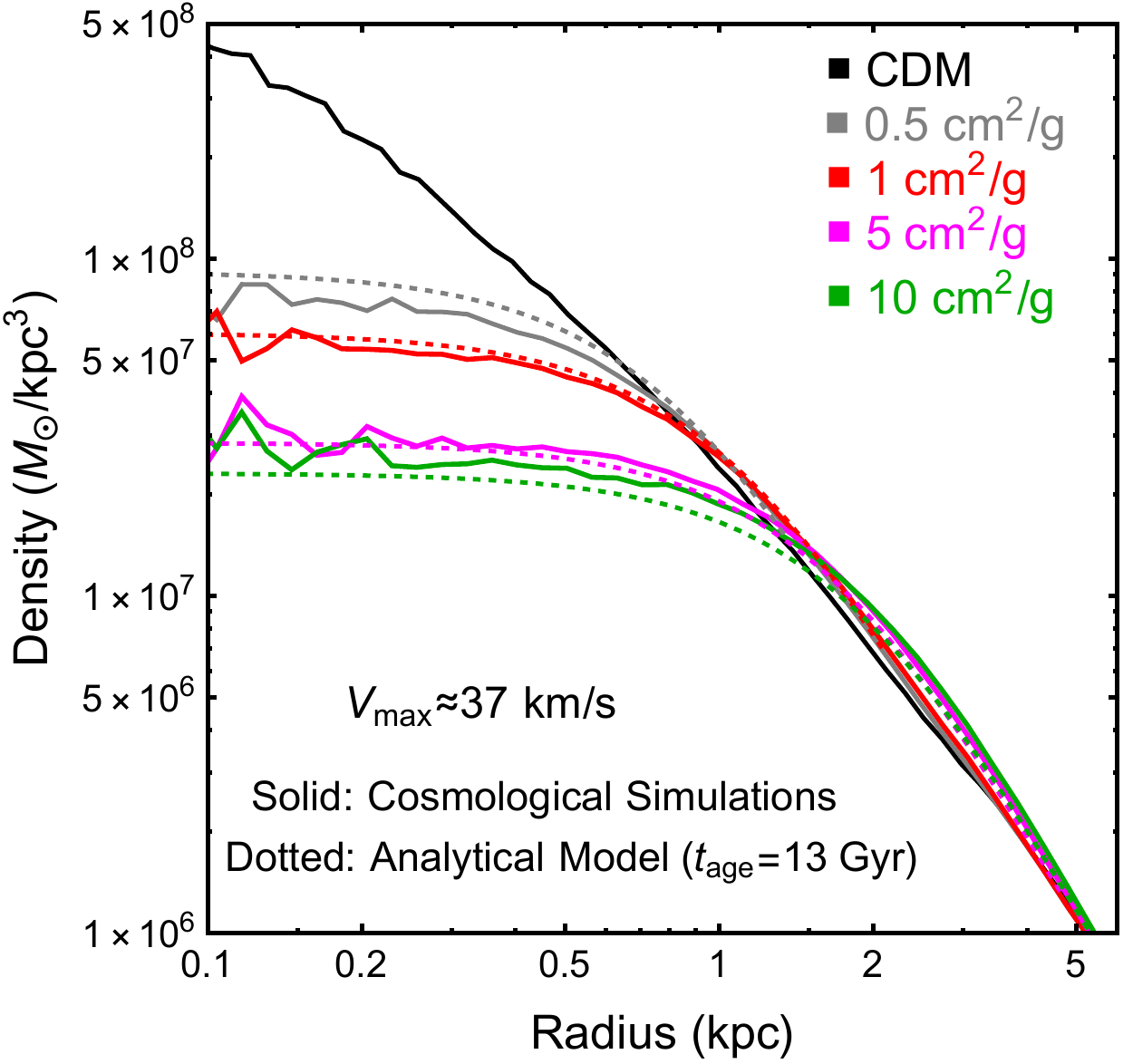}  \\
\includegraphics[scale=0.7]{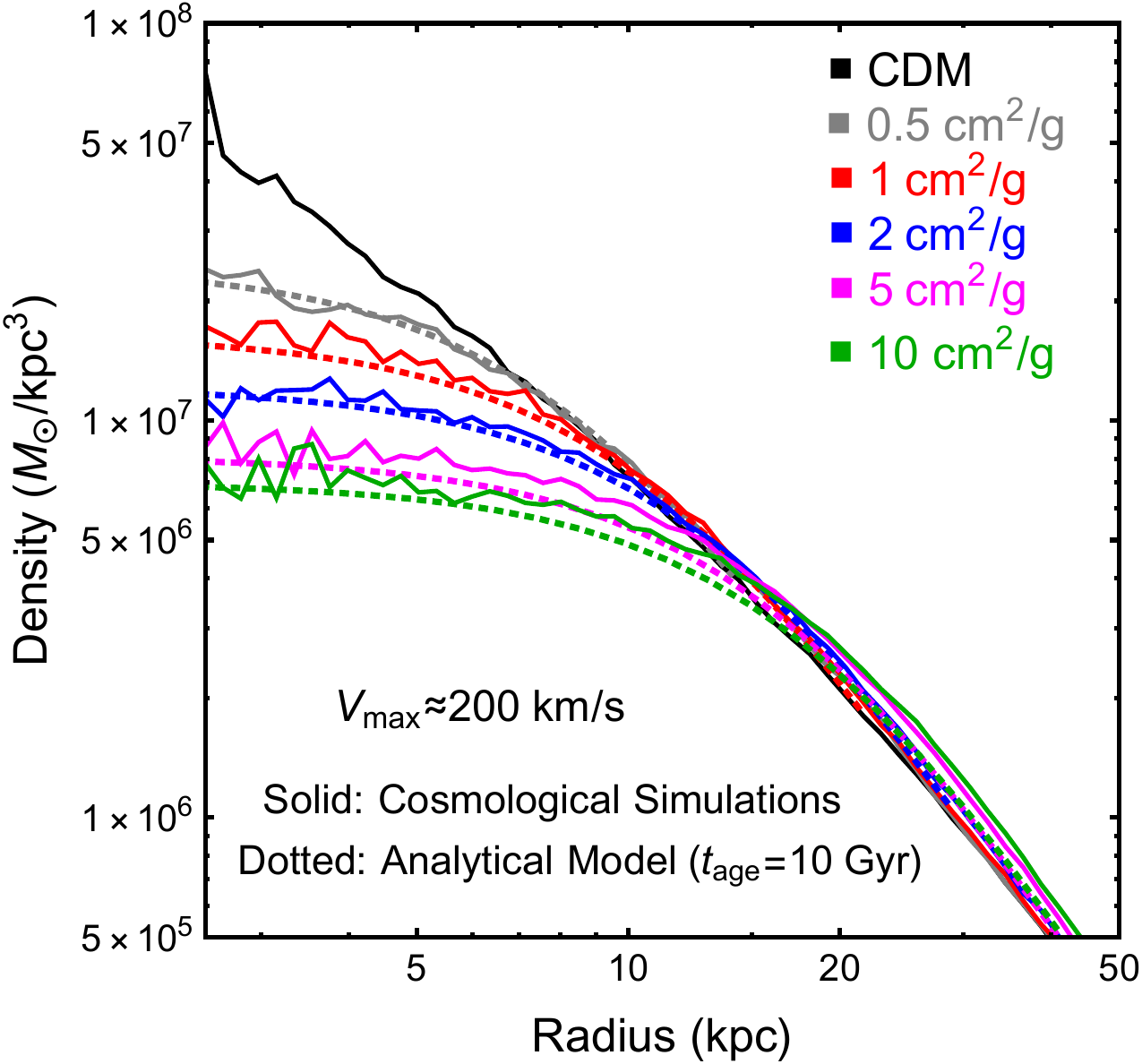}& \includegraphics[scale=0.7]{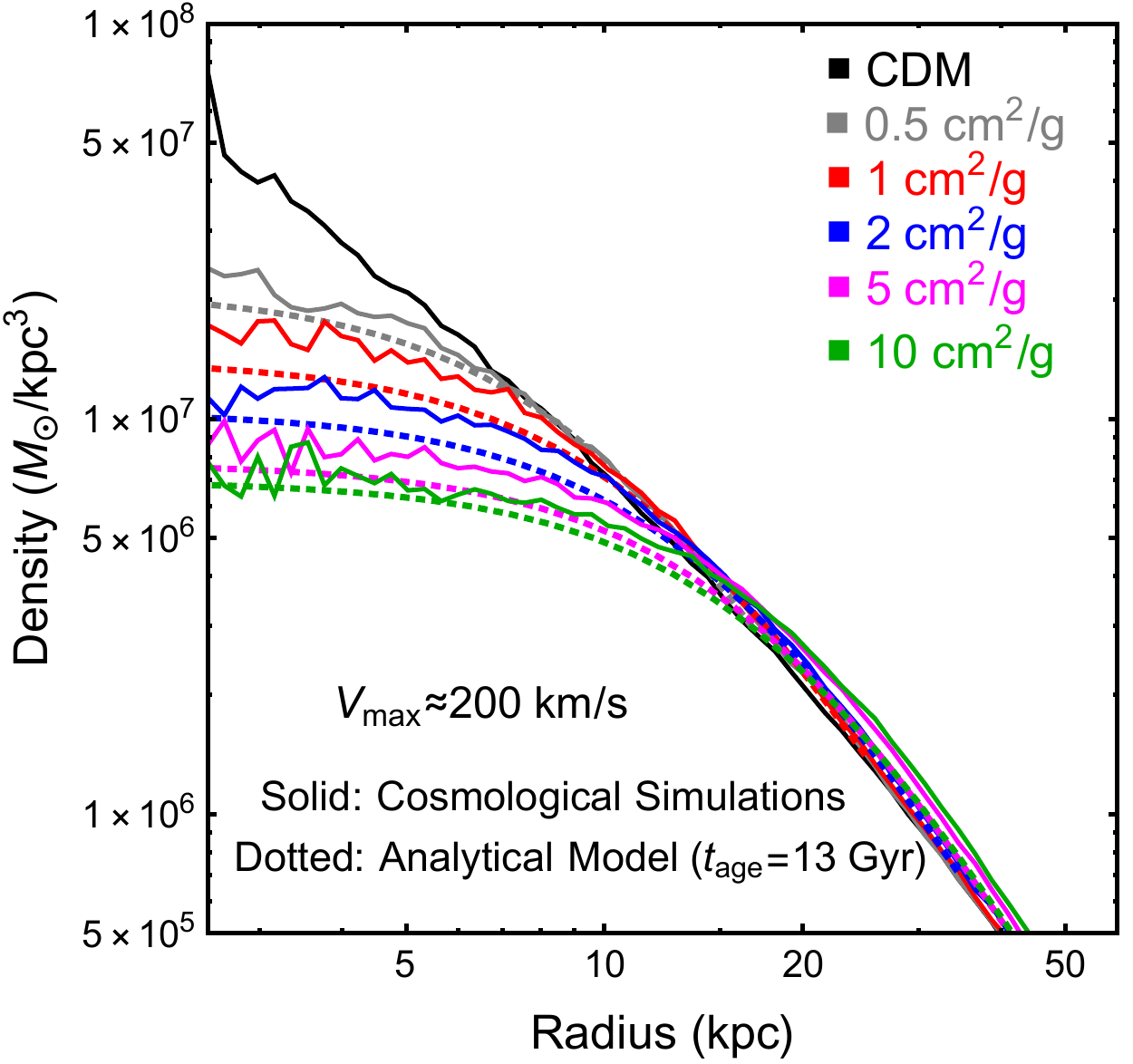}
\end{tabular}
\caption{{\it Upper:} Density profiles predicted in the analytical model (dotted), compared with simulations (solid) from Elbert et al., MNRAS 453 (2015) no. 1, 29-37, based on the SIDM code developed Rocha et al. MNRAS 430 (2013) no. 1, 81-104. {\it Lower:} A similar comparison with simulations from Creasey et al., MNRAS 468 (2017) no. 2, 2283-2295, which used the code developed in Vogelsberger et al., MNRAS 423 (2012) no. 4, 3740-3752. Despite the fact that we impose the exact matching condition at $r_1$, i.e., $\rho_{\rm iso}=\rho_{\rm NFW}$ and $M_{\rm iso}=M_{\rm NFW}$, and the agreement is better than $\sim5\textup{--}20\%$ for $\sigm\geq1~{\rm cm^2/g}$ and the results change very mildly from $t_{\rm age}=10~{\rm Gyr}$ to $13~{\rm Gyr}$. Sokolenko et al., 1806.11539, also showed the core sizes predicted in this analytical model are consistent with their simulations. The agreement can be further improved through tweaks to this model by including small halo mass or cross section dependence in the $r_1$ definition or allowing freedom in the matching at the level of $\sim 5\%$. In the paper, we take $\sigm=3~{\rm cm^2/g}$, $t_{\rm age}=10~{\rm Gyr}$ and the exact matching condition.}
\end{figure}

\begin{figure}[H]
\centering
\label{fig:comparision2}
\begin{tabular}{@{}ccc@{}}
\includegraphics[scale=0.47]{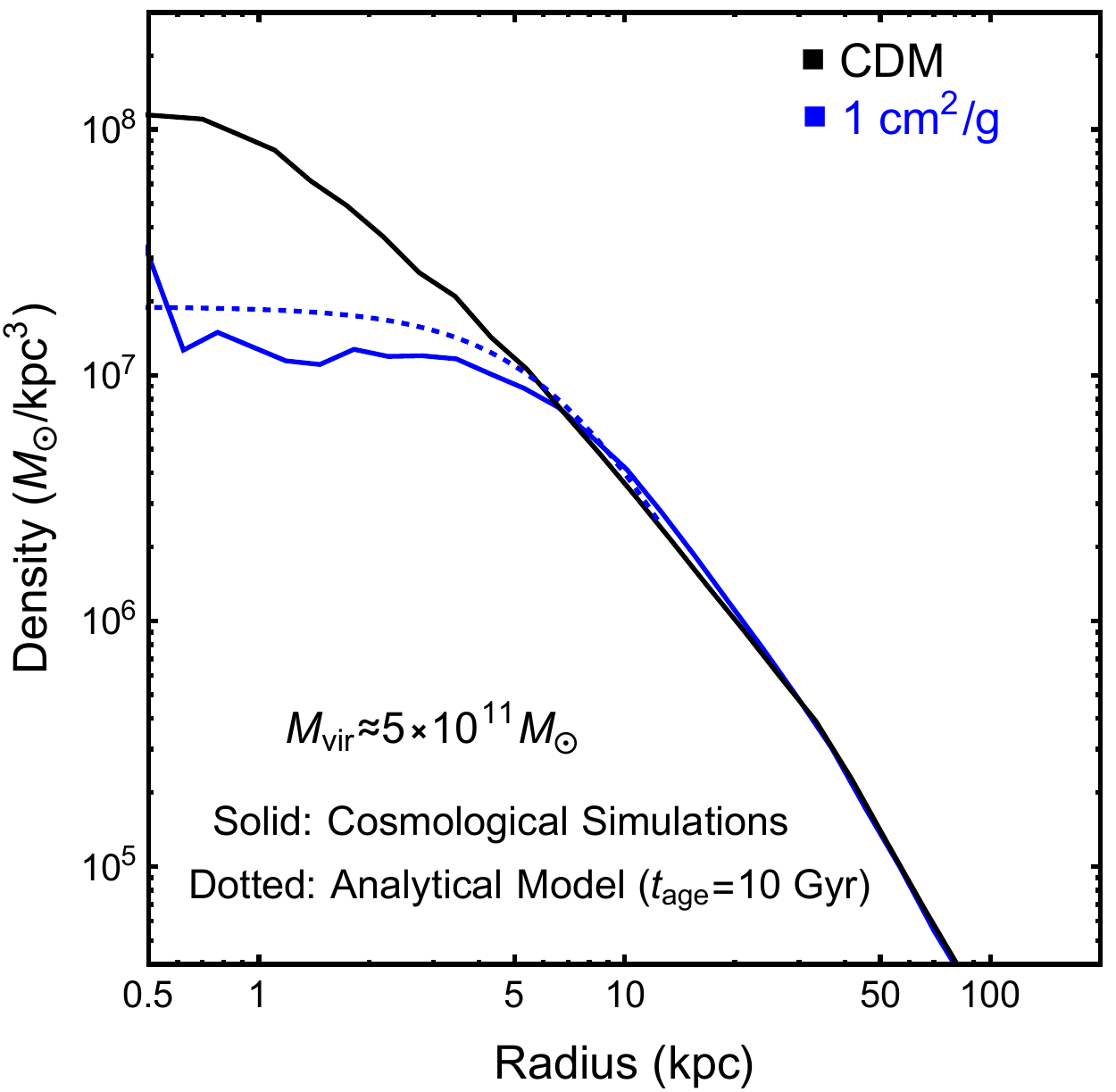} & \includegraphics[scale=0.47]{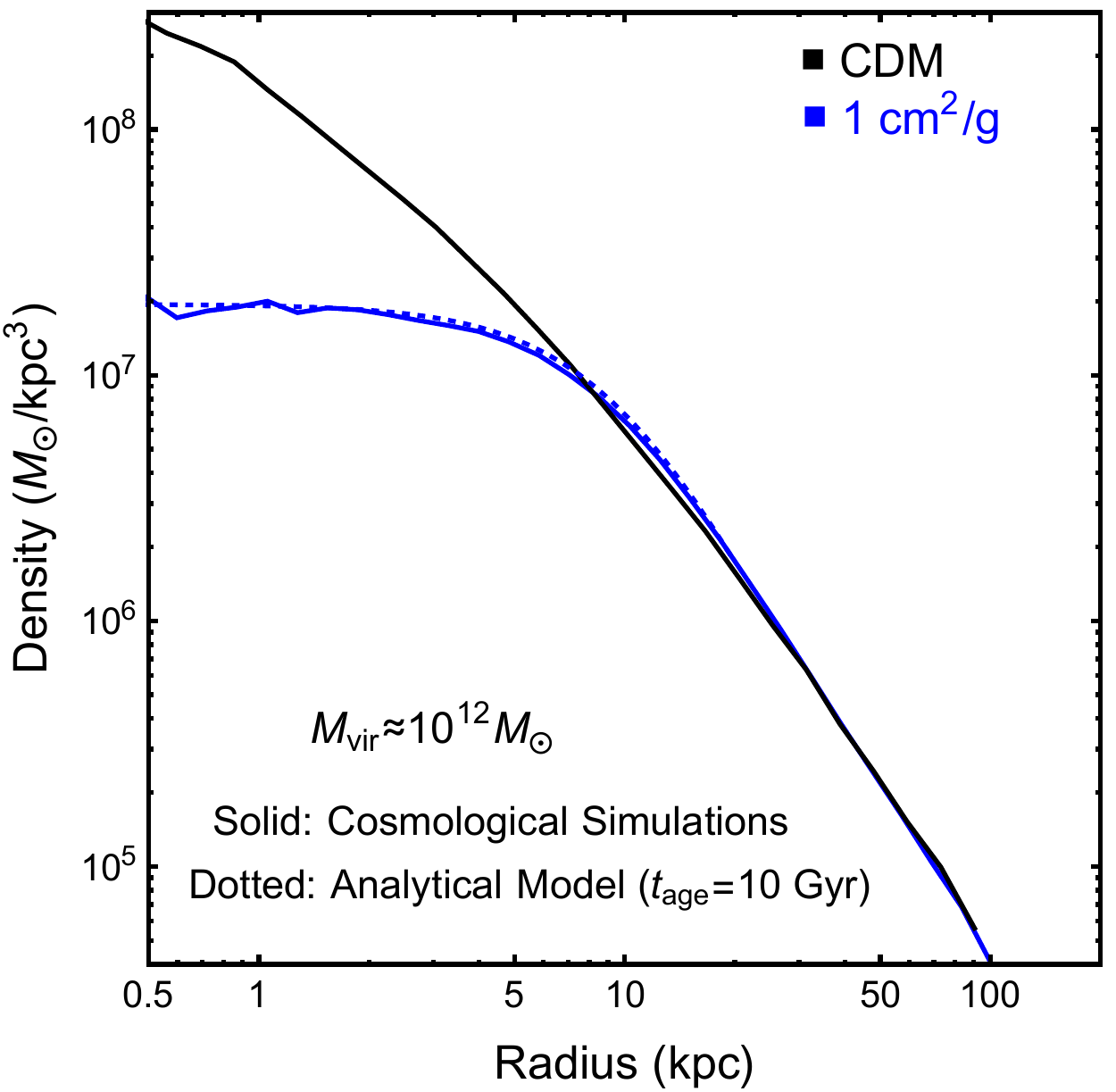}  & \includegraphics[scale=0.47]{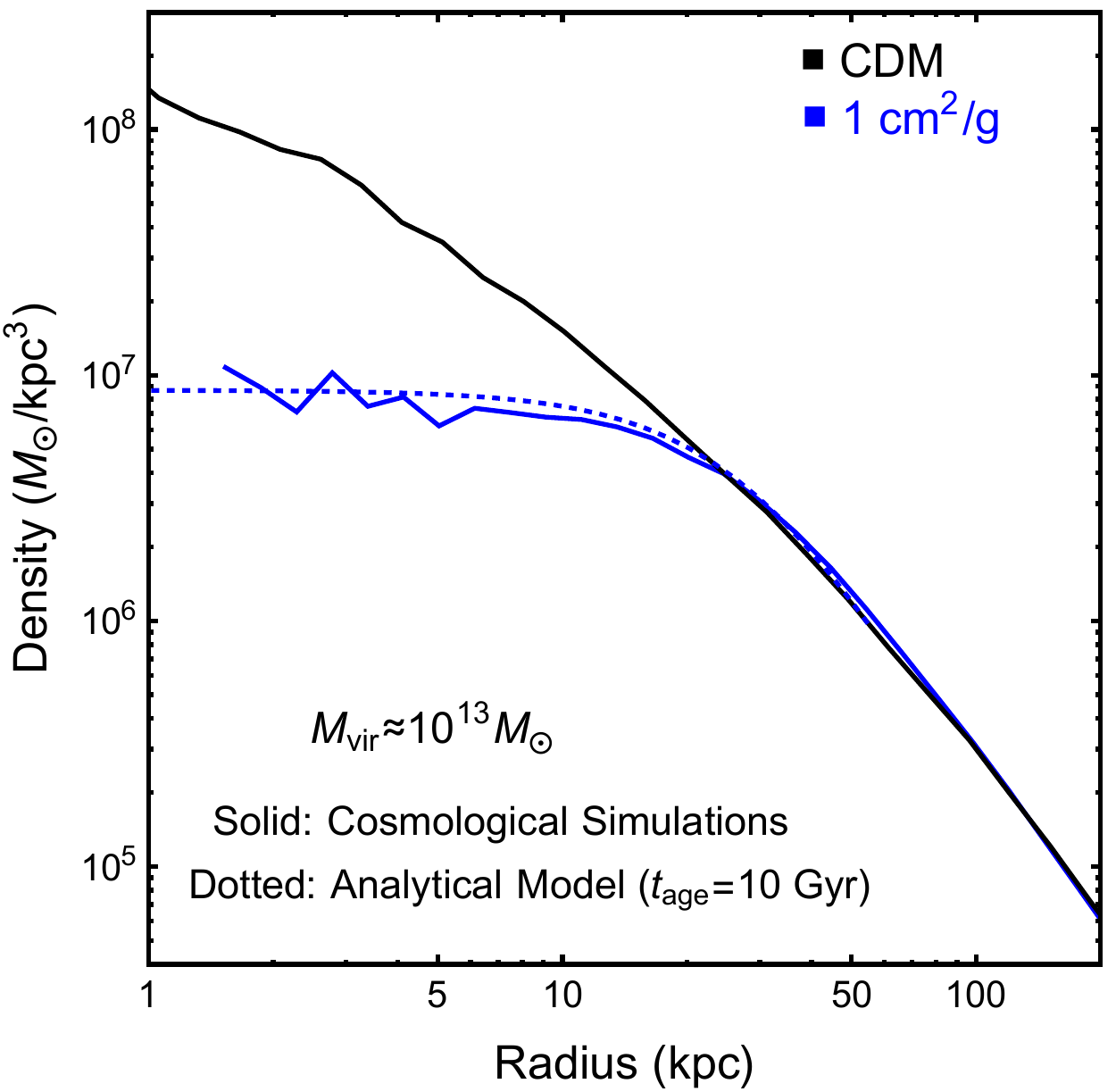} \\
\includegraphics[scale=0.47]{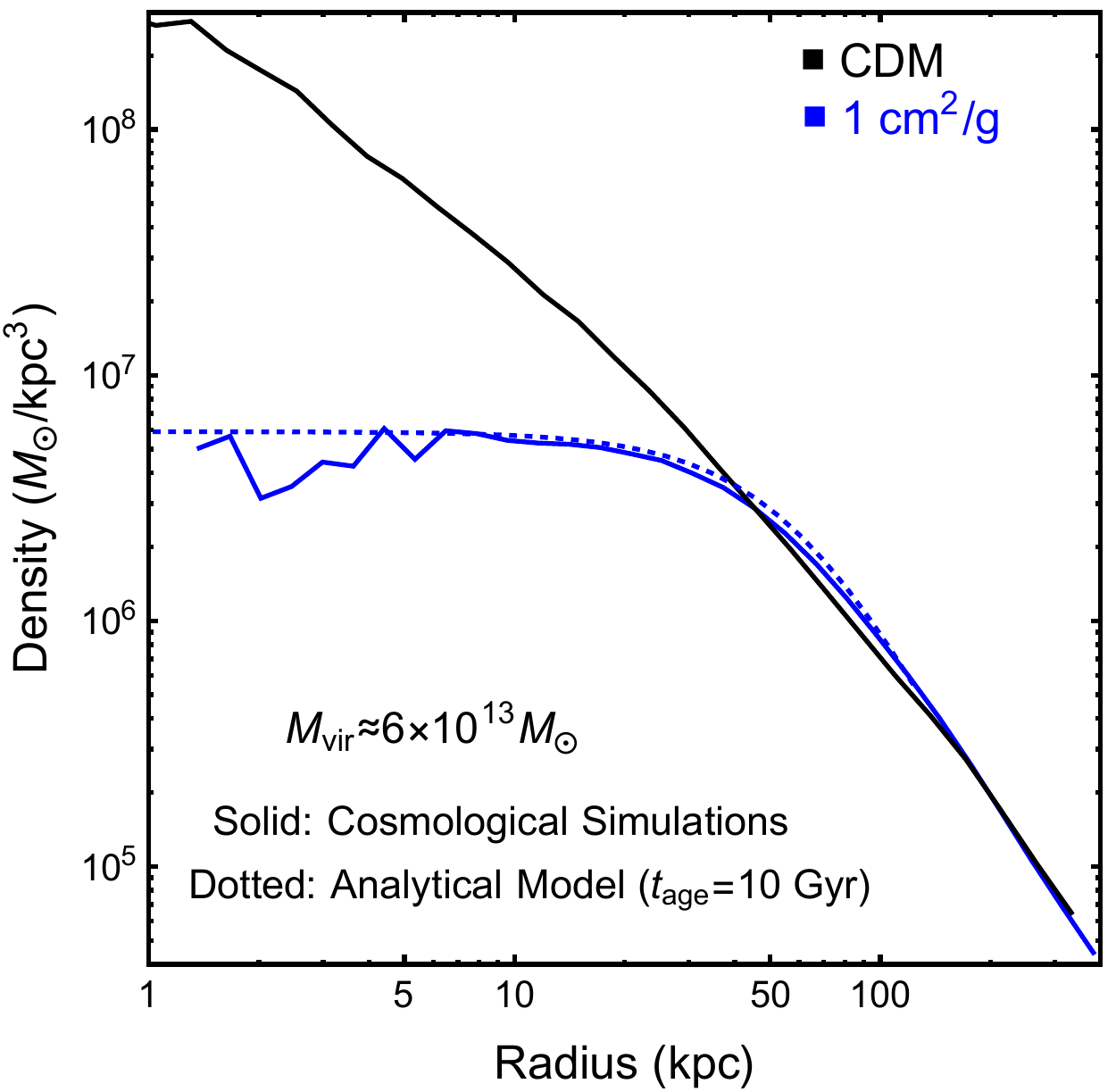} & \includegraphics[scale=0.47]{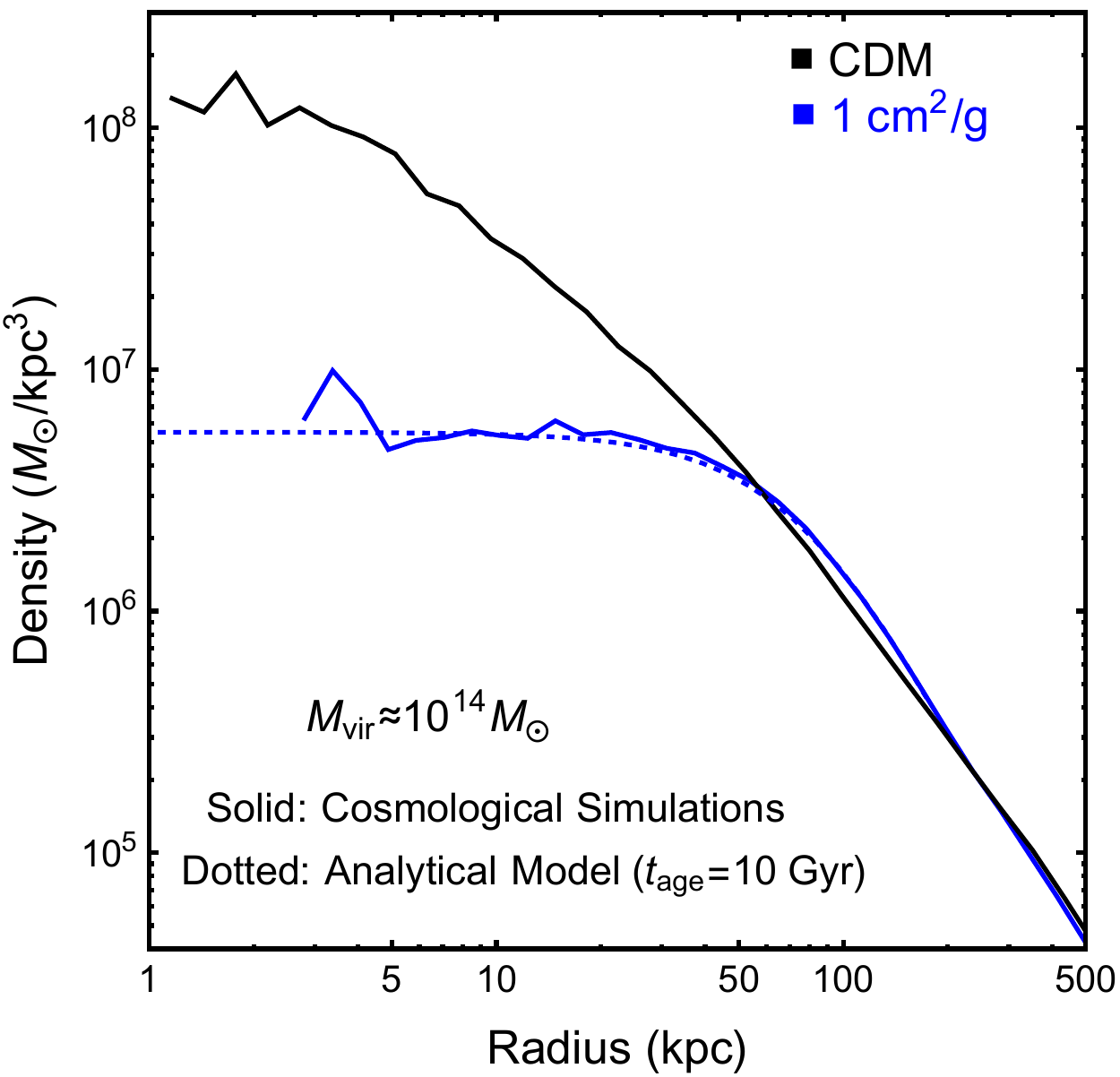}  & \includegraphics[scale=0.47]{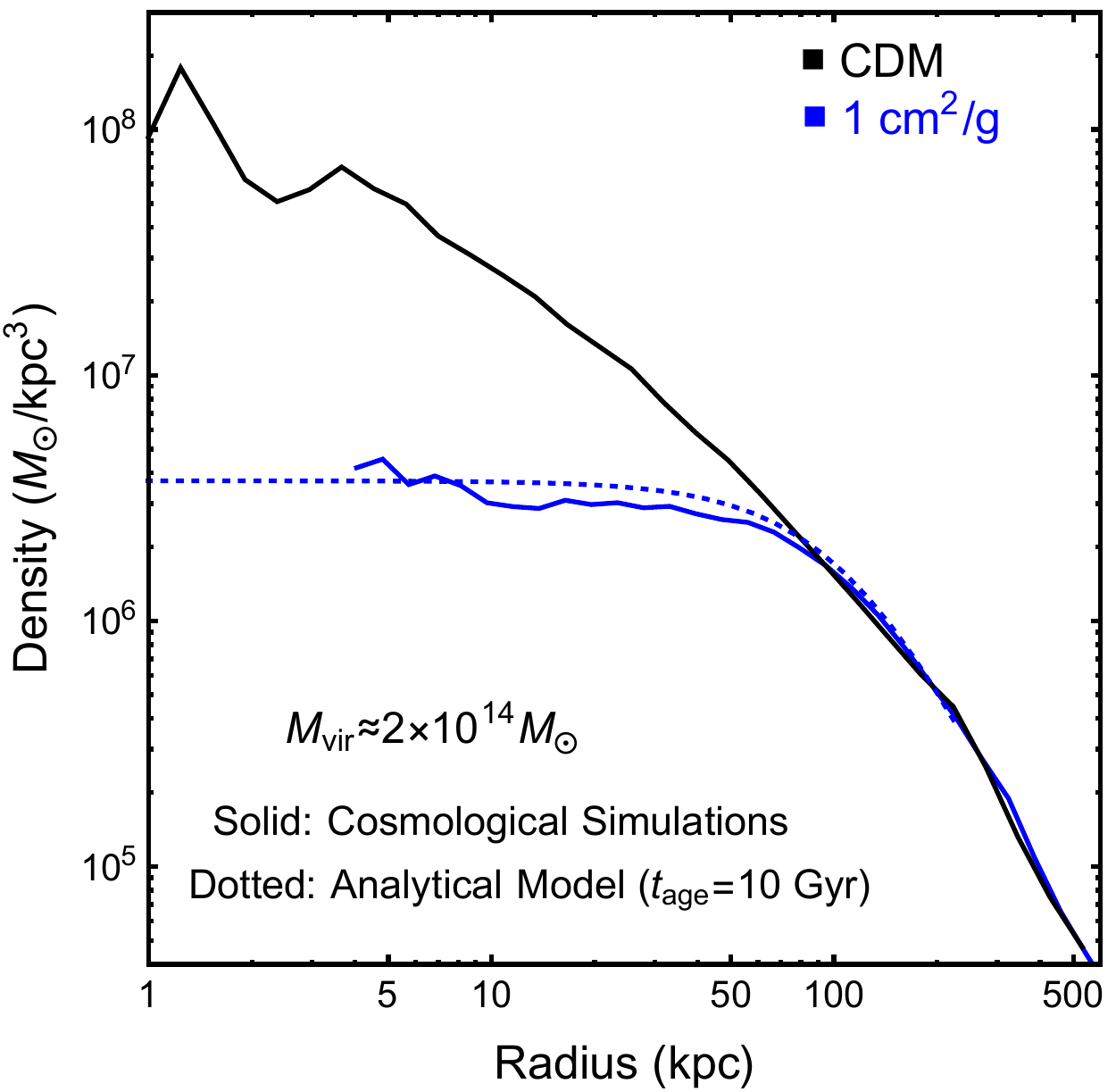} 
\end{tabular}
\caption{Density profiles predicted in the analytical model (dotted), compared with simulations (solid) from Rocha et al. MNRAS 430 (2013) no. 1, 81-104. Our model reproduces the simulation results over a wide range of halo masses.}
\end{figure}

\begin{figure}[H]
\centering
\begin{tabular}{@{}ccc@{}}
\includegraphics[scale=0.25]{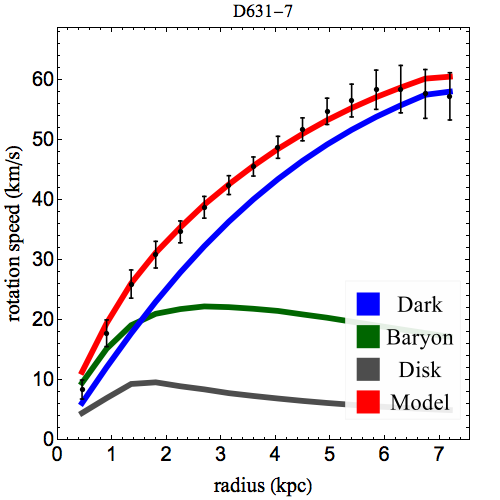} & \includegraphics[scale=0.192]{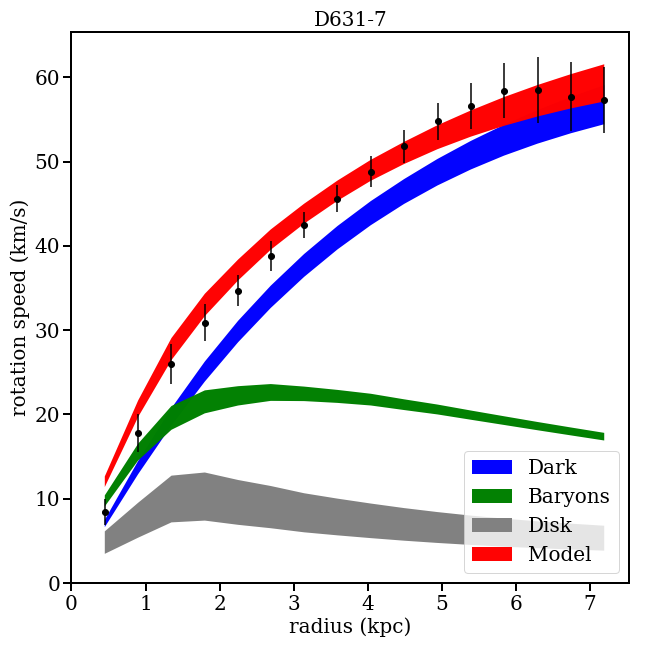} & \includegraphics[scale=0.215]{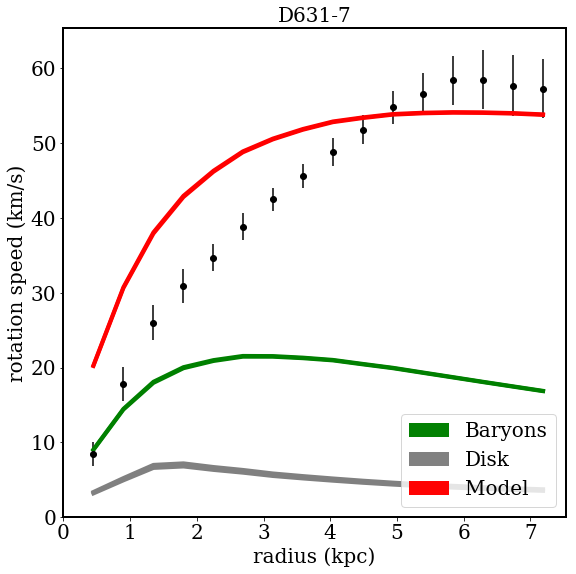} \\
\includegraphics[scale=0.25]{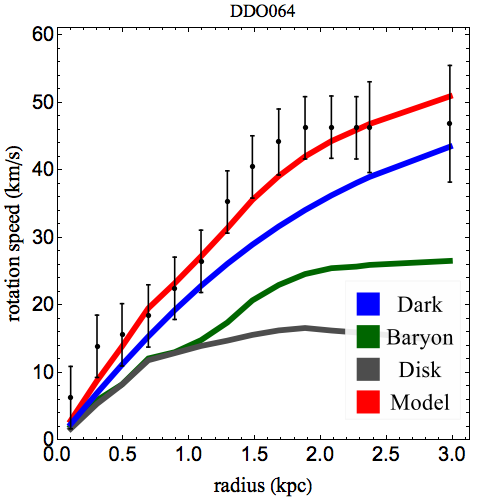} & \includegraphics[scale=0.192]{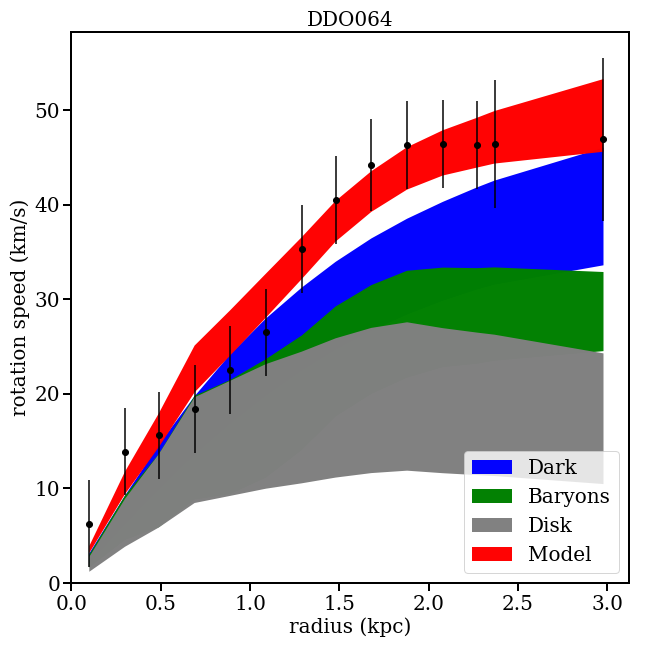} & \includegraphics[scale=0.215]{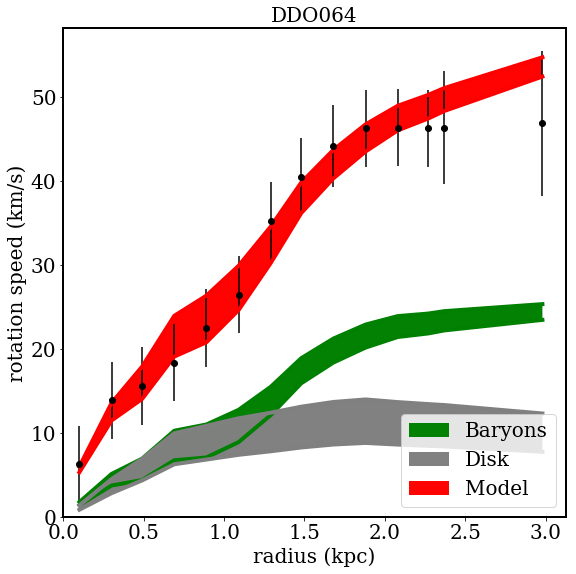} \\
\includegraphics[scale=0.25]{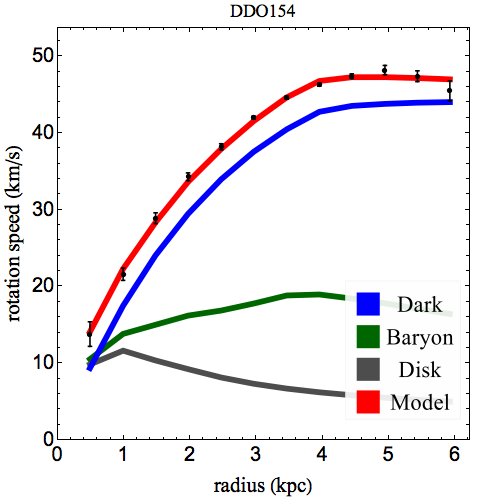} & \includegraphics[scale=0.192]{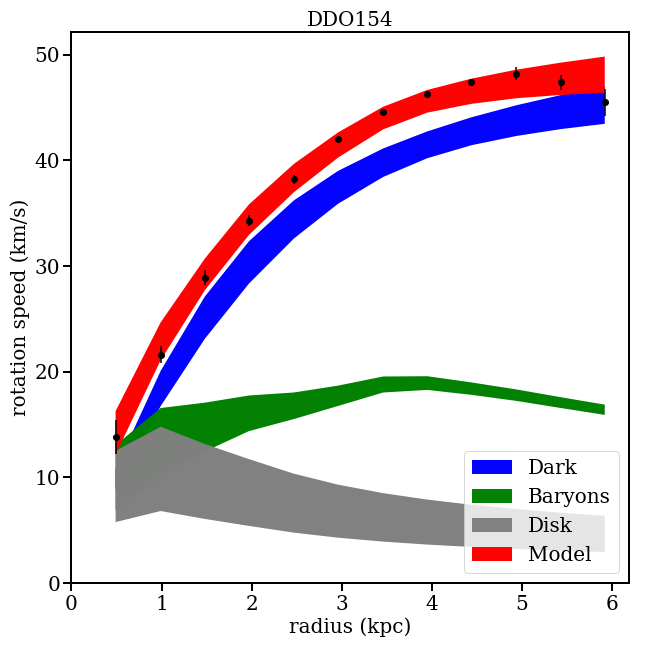} & \includegraphics[scale=0.215]{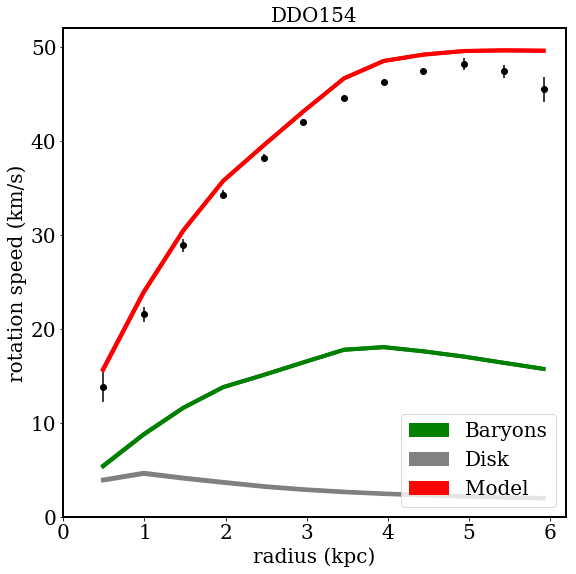} \\
\includegraphics[scale=0.25]{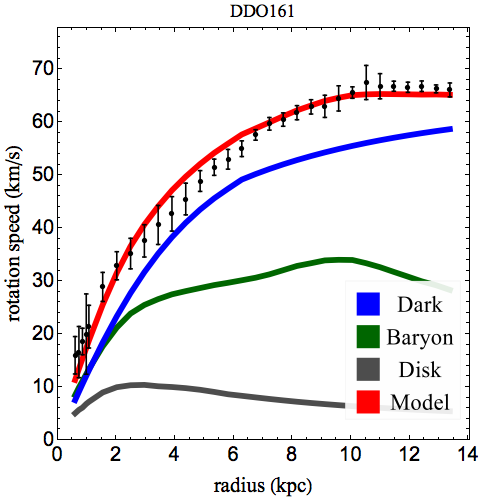} & \includegraphics[scale=0.192]{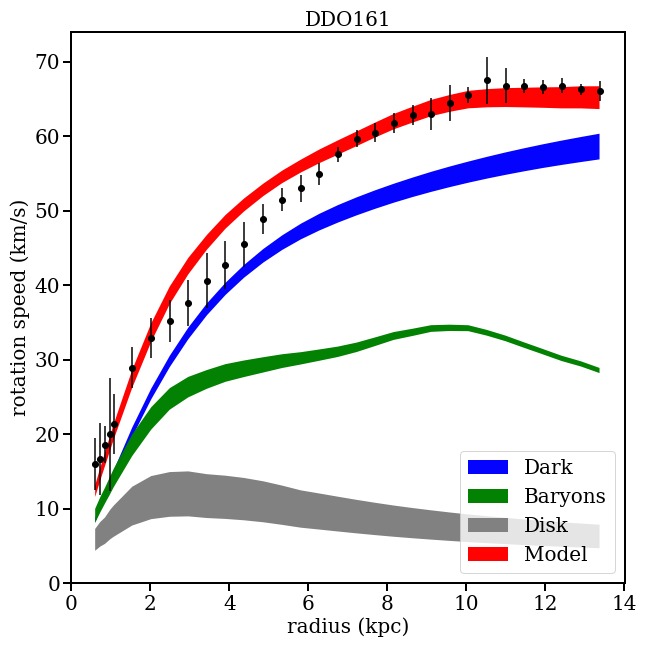} & \includegraphics[scale=0.215]{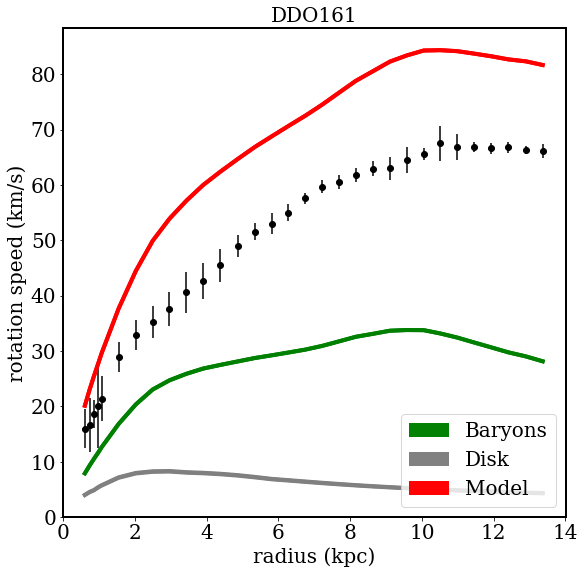} \\
\includegraphics[scale=0.25]{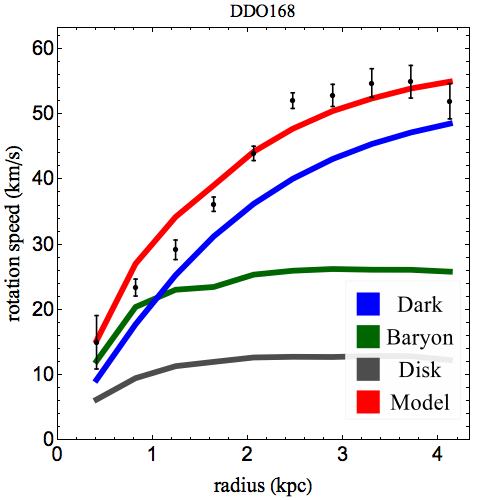} & \includegraphics[scale=0.192]{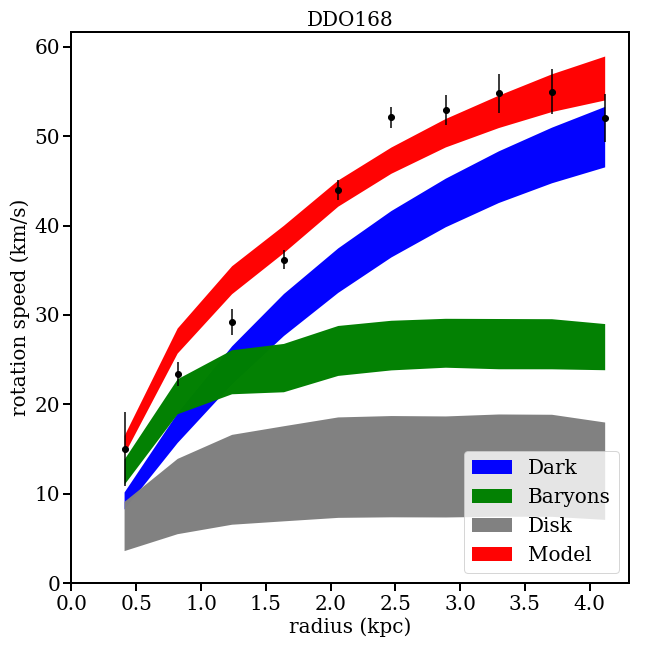} & \includegraphics[scale=0.215]{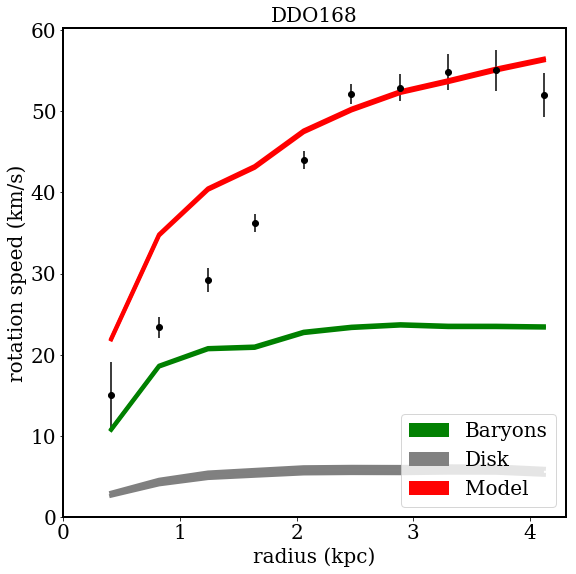} \\
\end{tabular}
\caption{Detailed SIDM fits to the $135$ SPARC galaxies with the controlled (left) and MCMC (tophat prior, middle) sampling methods, with ${\rm \sigm}=3~{\rm cm^2/g}$. The model parameters and $\chi^2/{\rm d.o.f.}$ values are collected in Table S2. The MOND fits (right) are also shown for comparison. The observational data are taken from Lelli et al. Astron. J. {\bf 152}, 157 (2016), 1606.09251.}
\end{figure}

\addtocounter{figure}{-1}

\begin{figure}[H]
\centering
\begin{tabular}{@{}ccc@{}}
\includegraphics[scale=0.25]{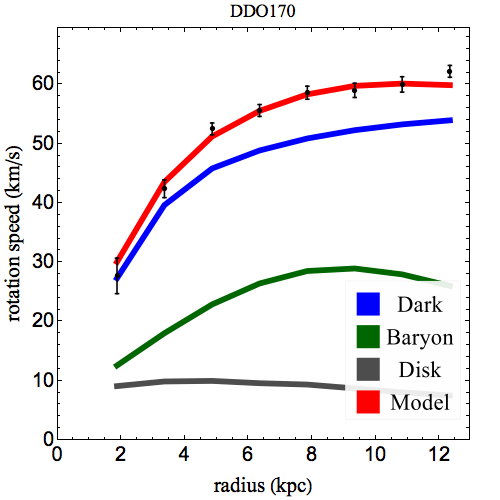} & \includegraphics[scale=0.192]{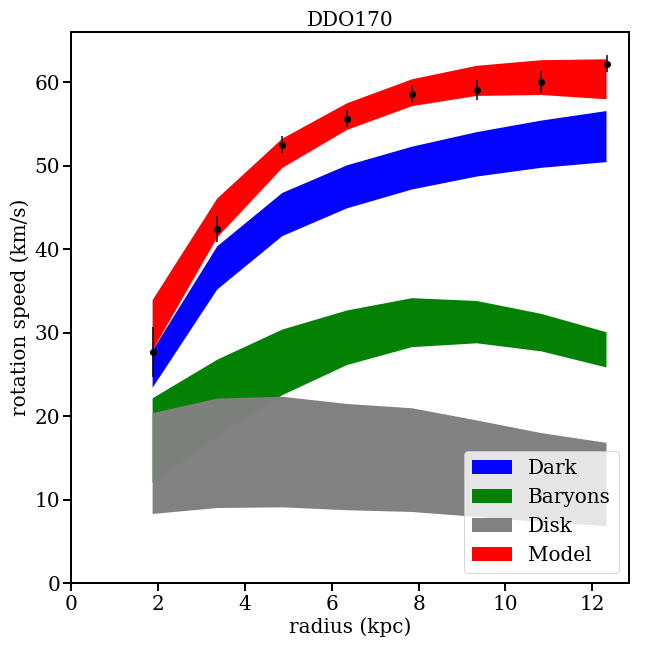} & \includegraphics[scale=0.215]{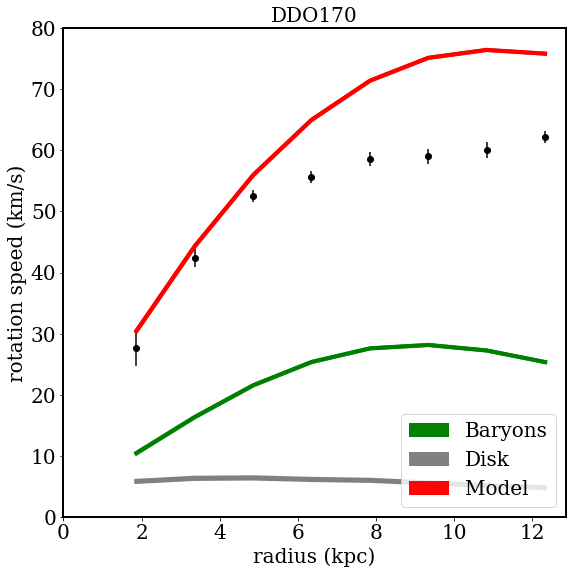} \\
\includegraphics[scale=0.25]{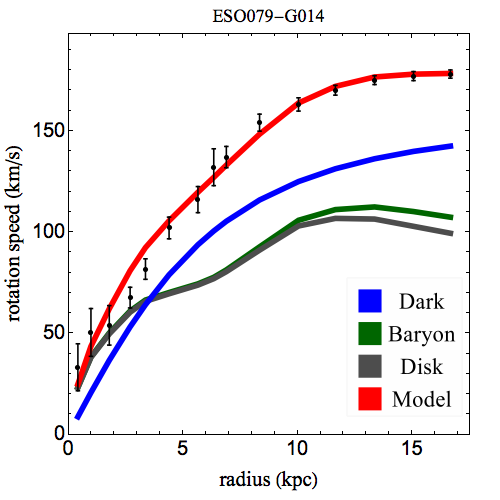} & \includegraphics[scale=0.192]{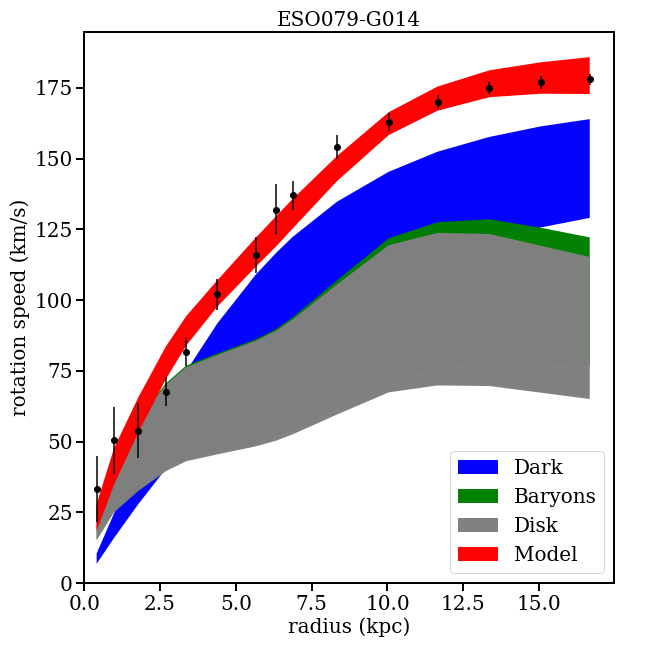} & \includegraphics[scale=0.215]{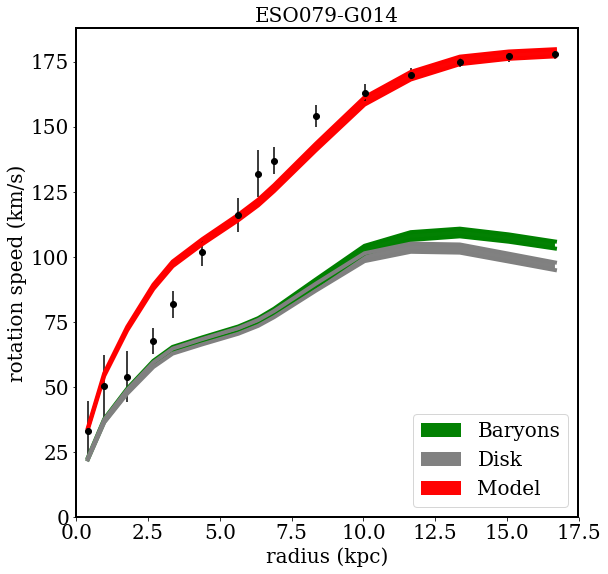} \\
\includegraphics[scale=0.25]{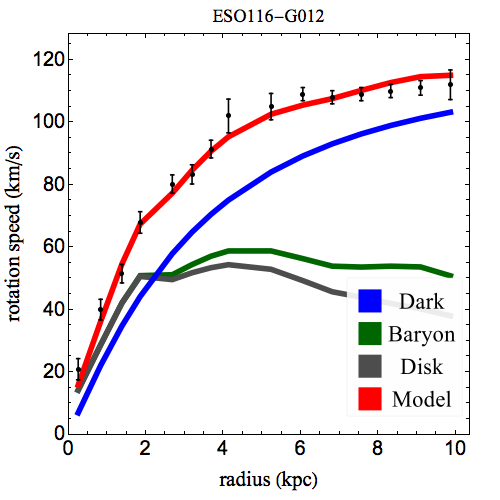} & \includegraphics[scale=0.192]{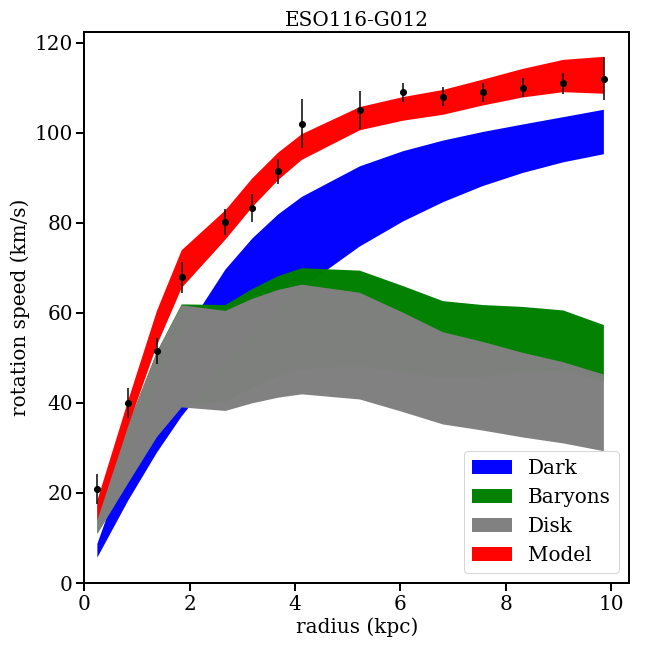} & \includegraphics[scale=0.215]{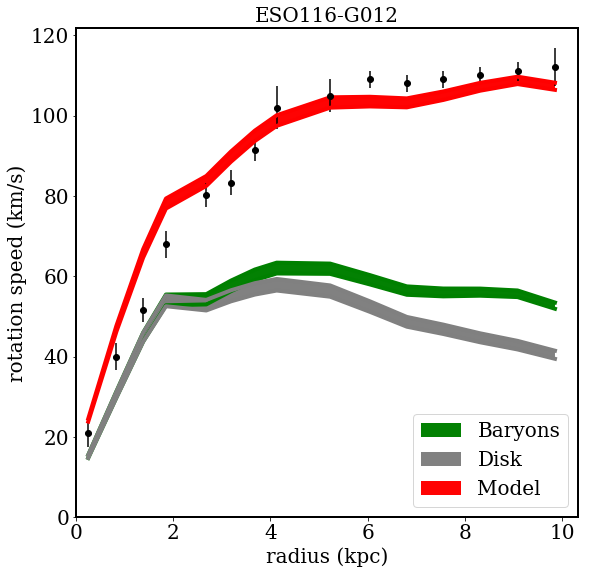} \\
\includegraphics[scale=0.25]{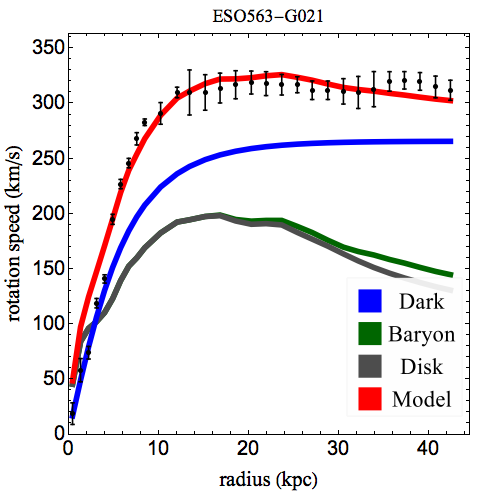} & \includegraphics[scale=0.192]{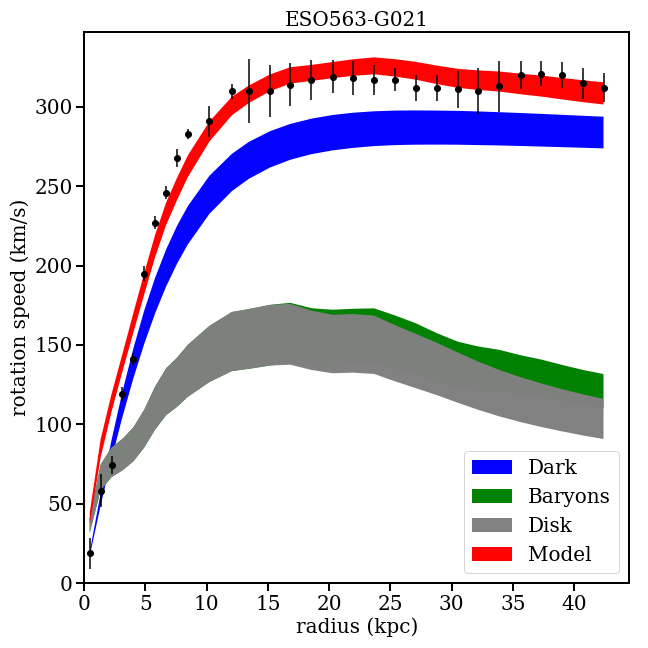} & \includegraphics[scale=0.215]{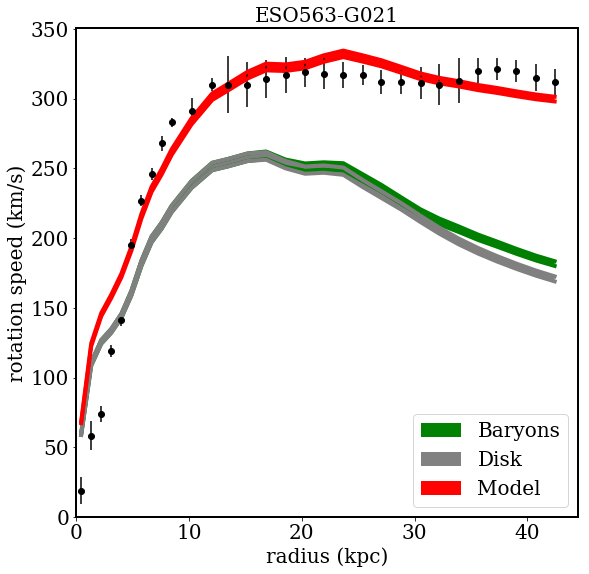} \\
\includegraphics[scale=0.25]{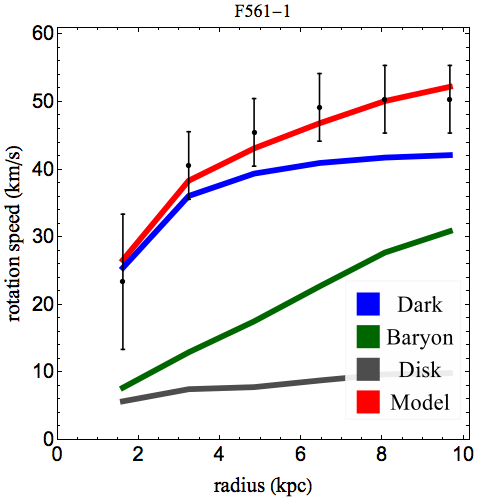} & \includegraphics[scale=0.192]{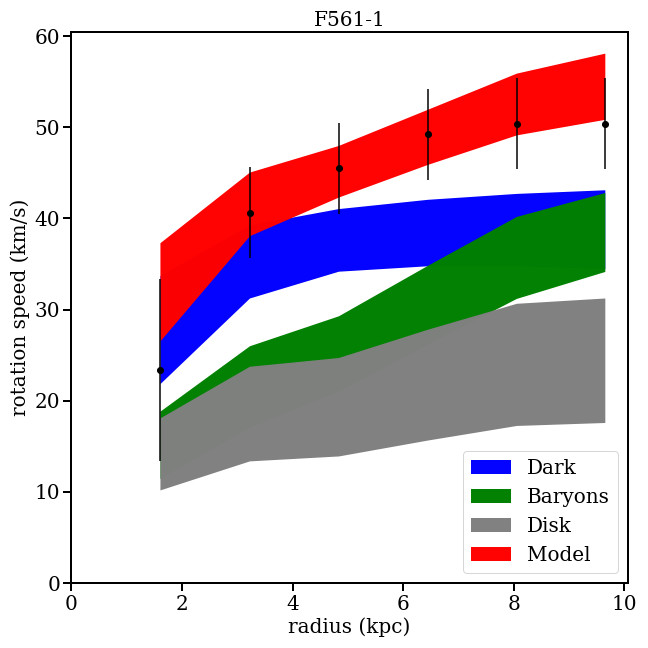} & \includegraphics[scale=0.215]{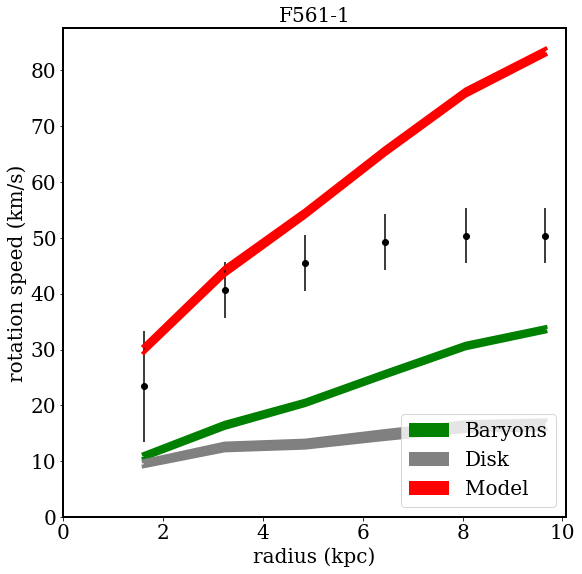} \\
\end{tabular}
\caption{Continued.}
\end{figure}

\addtocounter{figure}{-1}
\begin{figure}[H]
\centering
\begin{tabular}{@{}ccc@{}}
\includegraphics[scale=0.25]{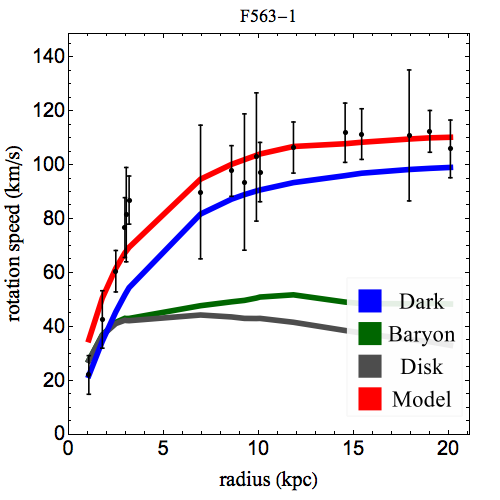} & \includegraphics[scale=0.192]{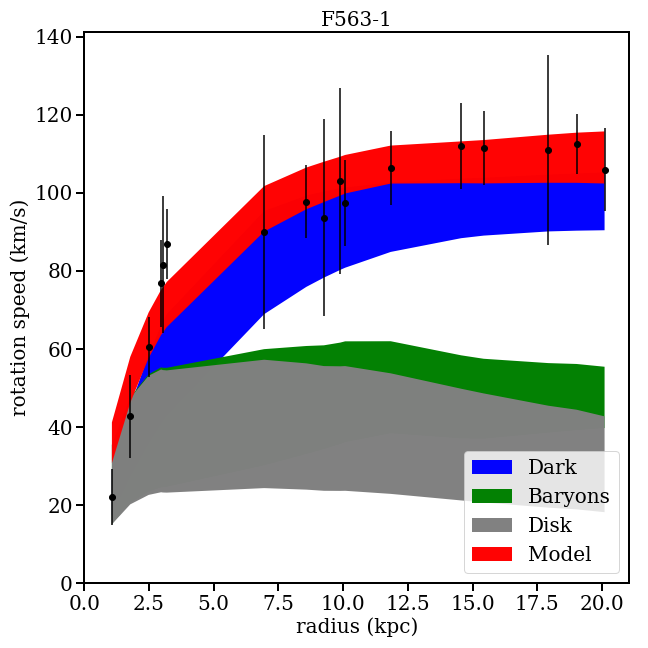} & \includegraphics[scale=0.215]{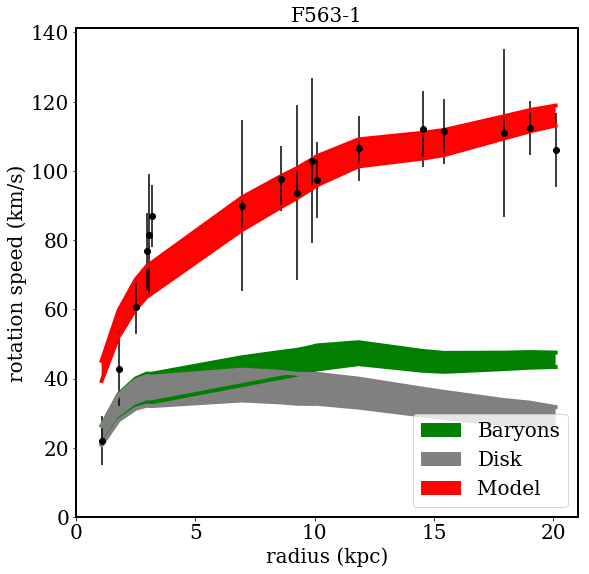} \\
\includegraphics[scale=0.25]{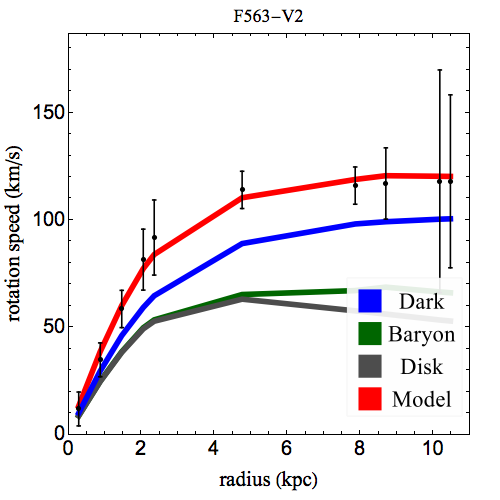} & \includegraphics[scale=0.192]{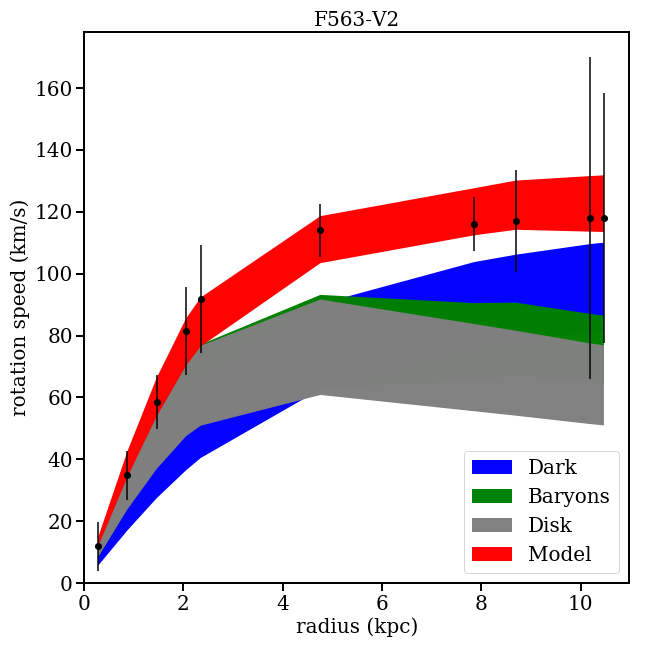} & \includegraphics[scale=0.215]{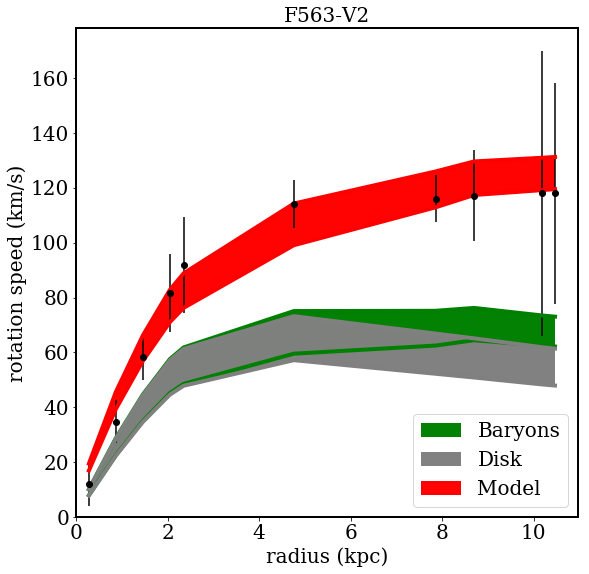} \\
\includegraphics[scale=0.25]{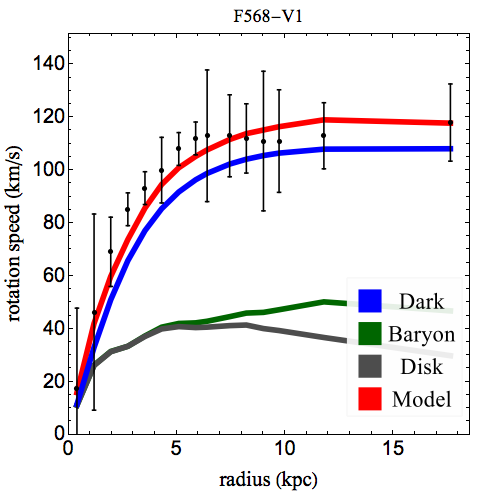} & \includegraphics[scale=0.192]{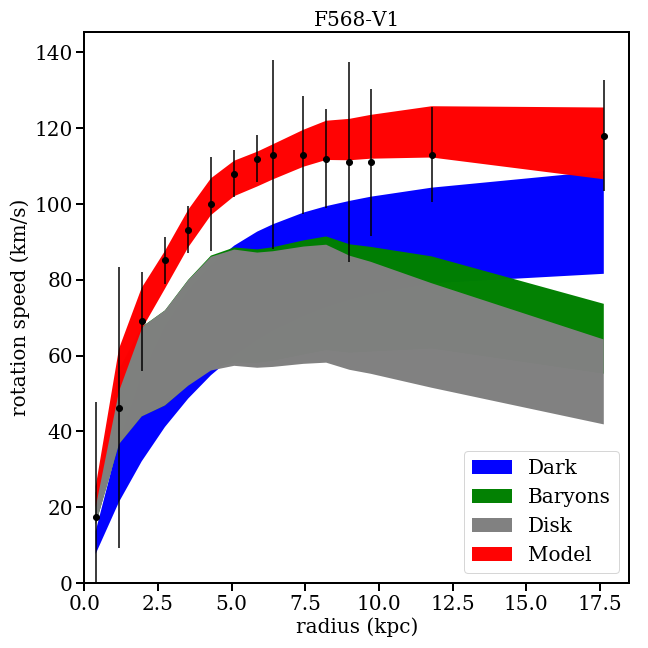} & \includegraphics[scale=0.215]{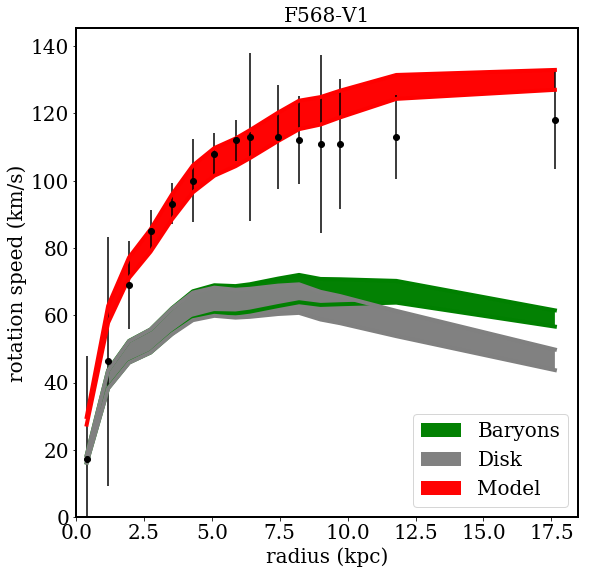} \\
\includegraphics[scale=0.25]{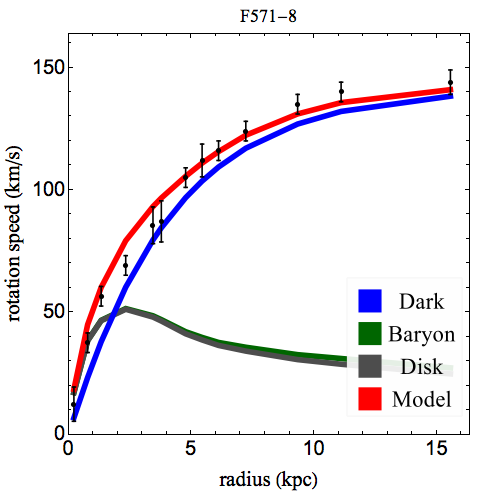} & \includegraphics[scale=0.192]{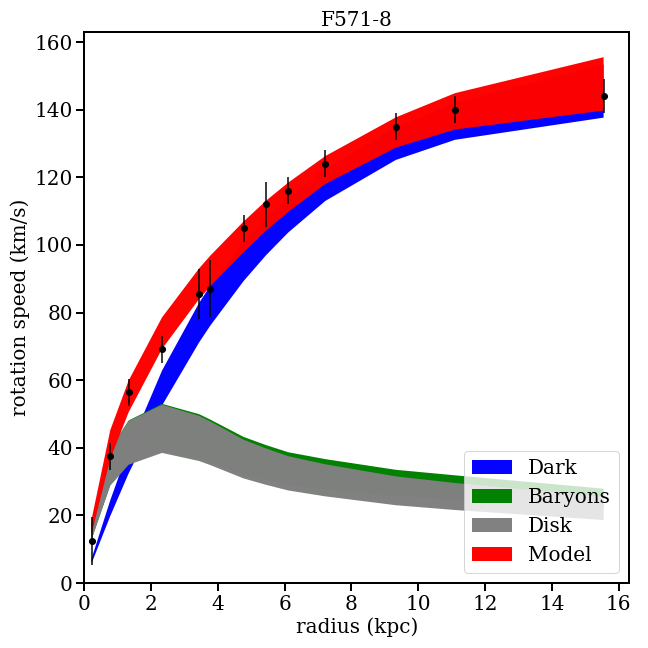} & \includegraphics[scale=0.215]{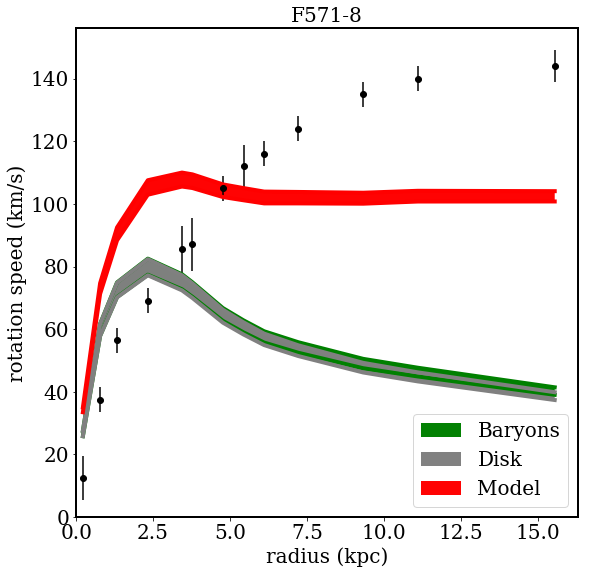} \\
\includegraphics[scale=0.25]{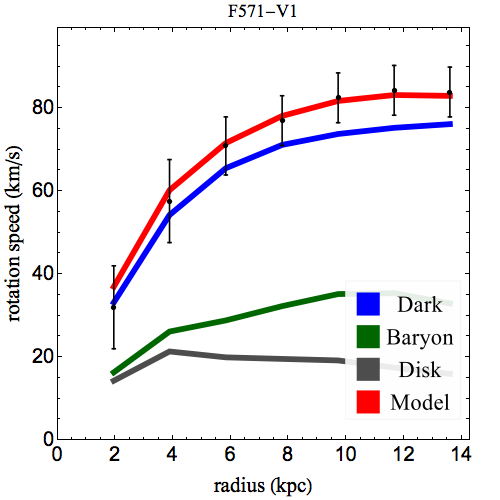} & \includegraphics[scale=0.192]{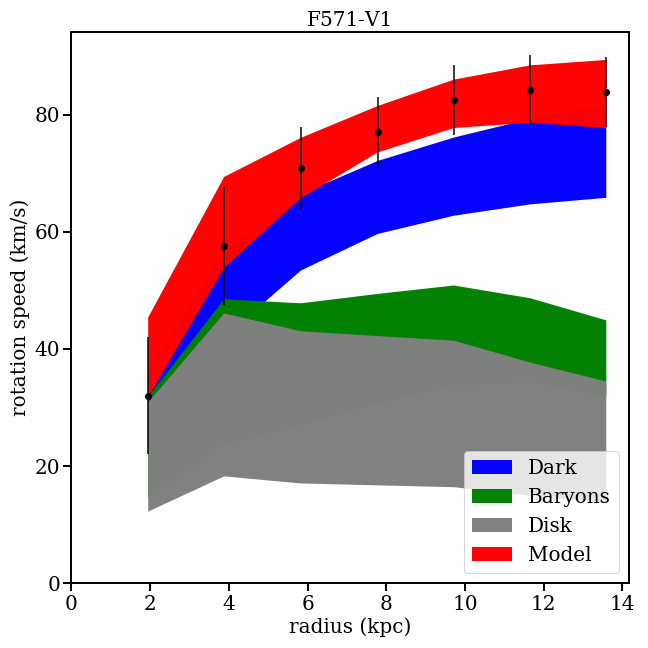} & \includegraphics[scale=0.215]{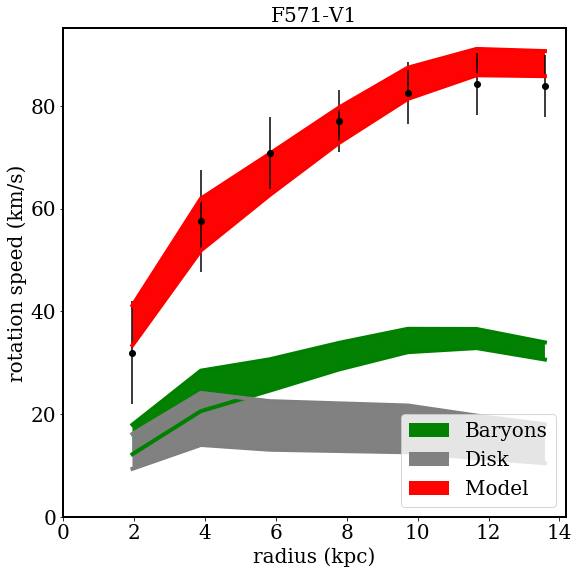} \\
\end{tabular}
\caption{Continued.}
\end{figure}

\addtocounter{figure}{-1}
\begin{figure}[H]
\centering
\begin{tabular}{@{}ccc@{}}
\includegraphics[scale=0.25]{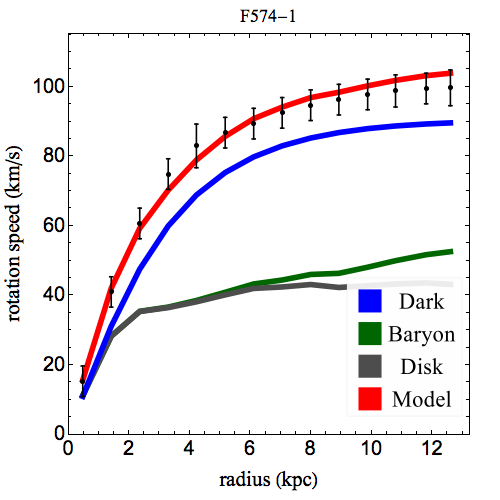} & \includegraphics[scale=0.192]{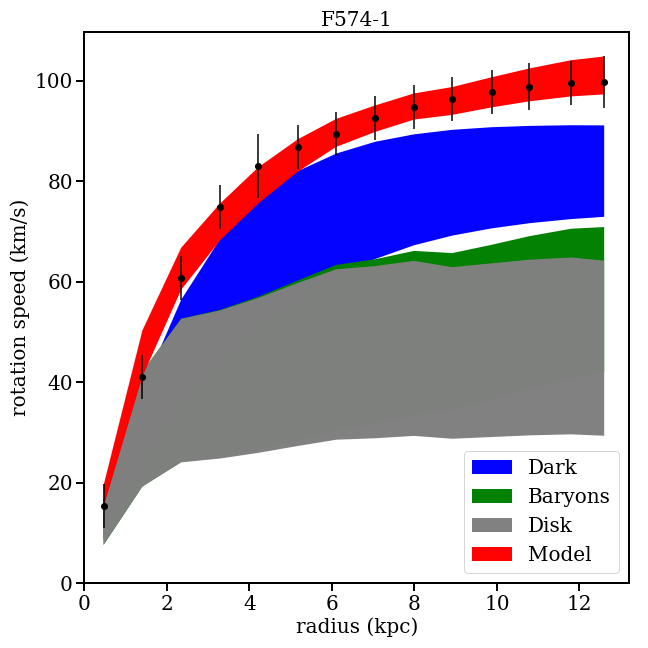} & \includegraphics[scale=0.215]{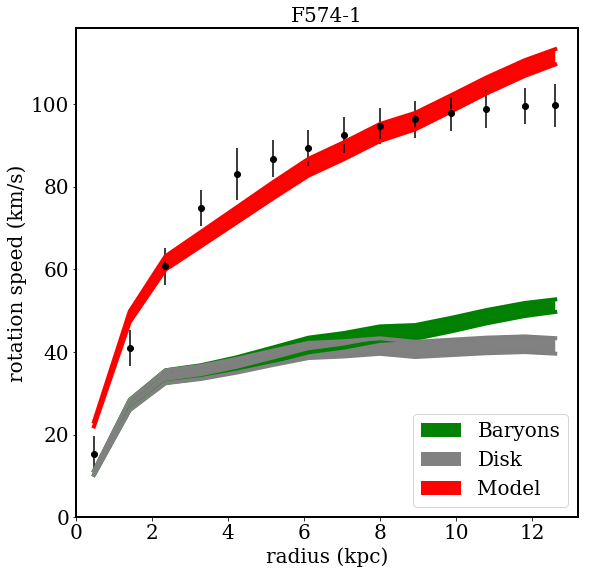} \\
\includegraphics[scale=0.25]{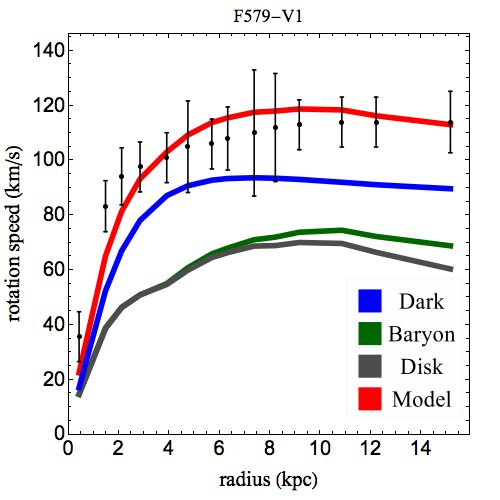} & \includegraphics[scale=0.192]{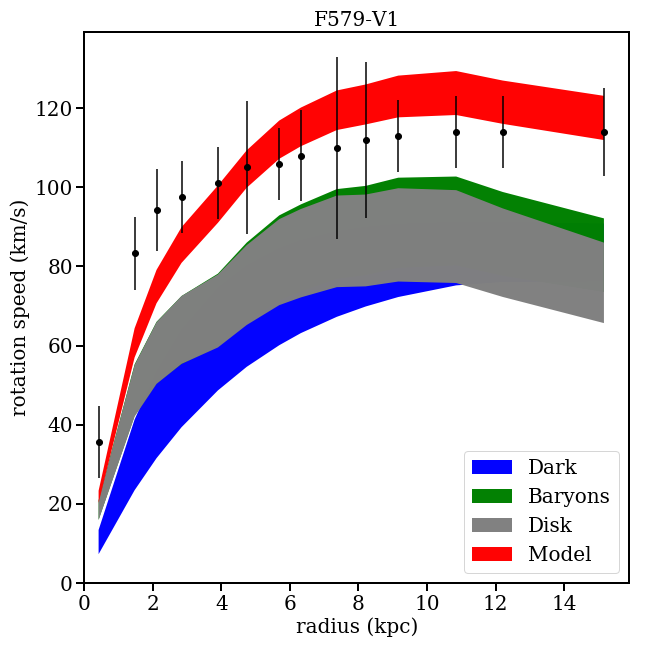} & \includegraphics[scale=0.215]{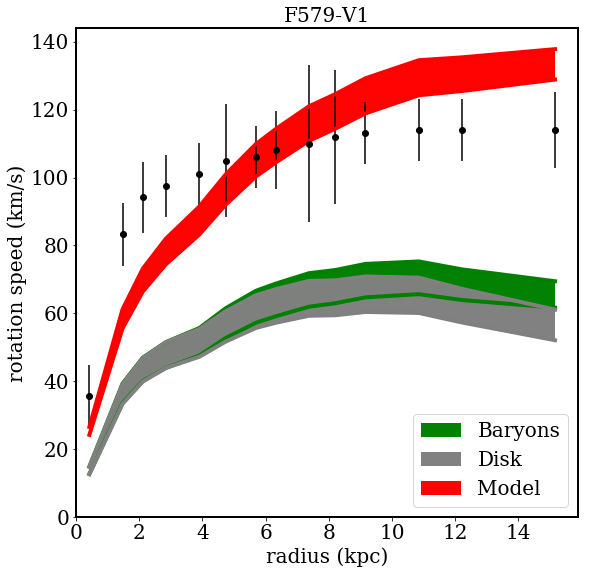} \\
\includegraphics[scale=0.25]{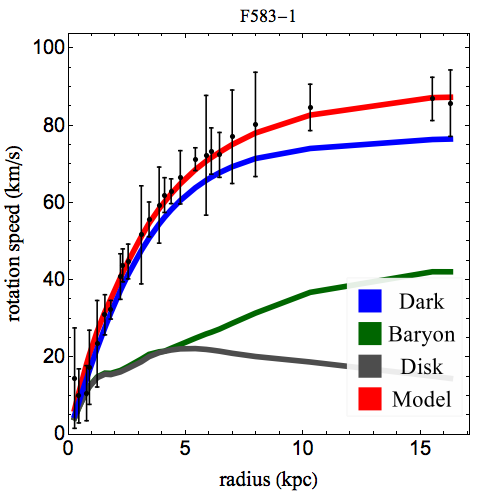} & \includegraphics[scale=0.192]{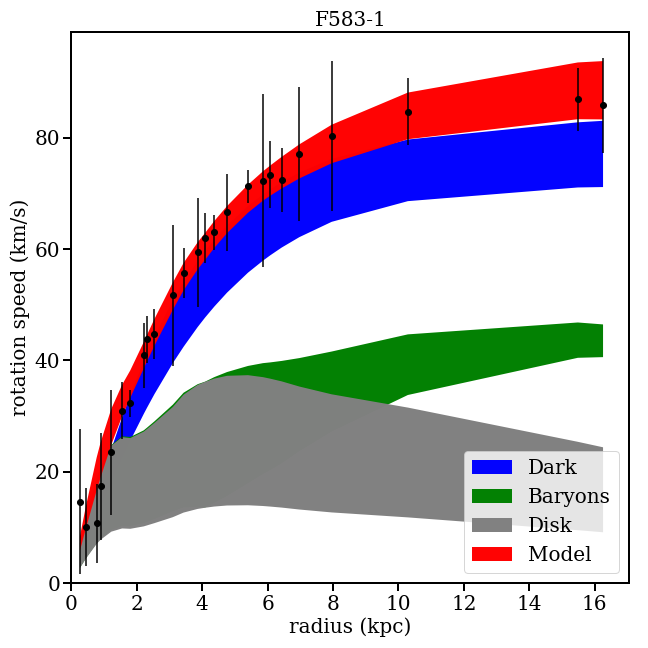} & \includegraphics[scale=0.215]{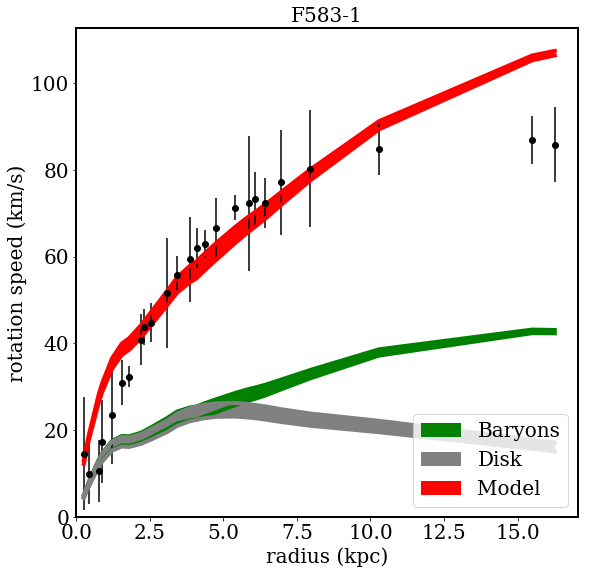} \\
\includegraphics[scale=0.25]{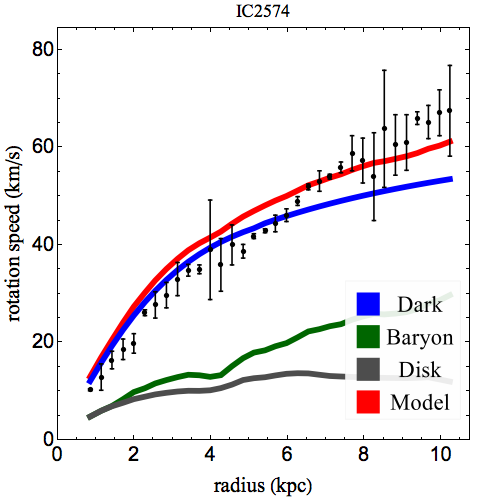} & \includegraphics[scale=0.192]{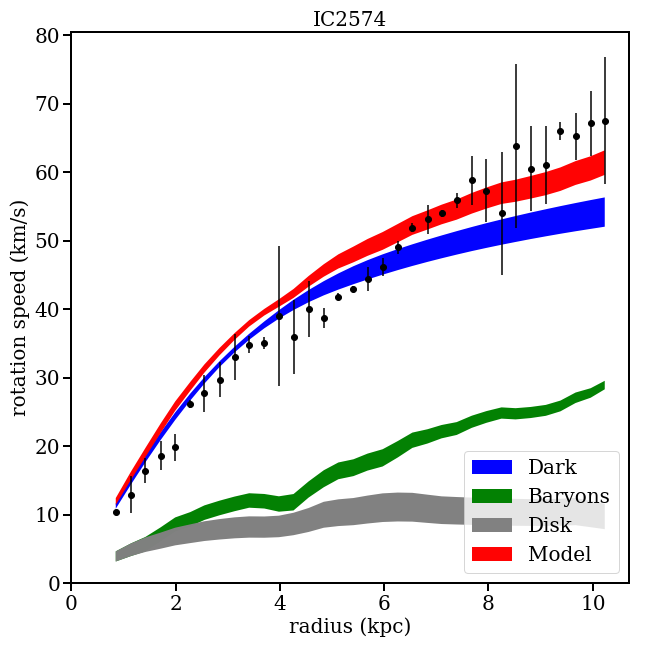} & \includegraphics[scale=0.215]{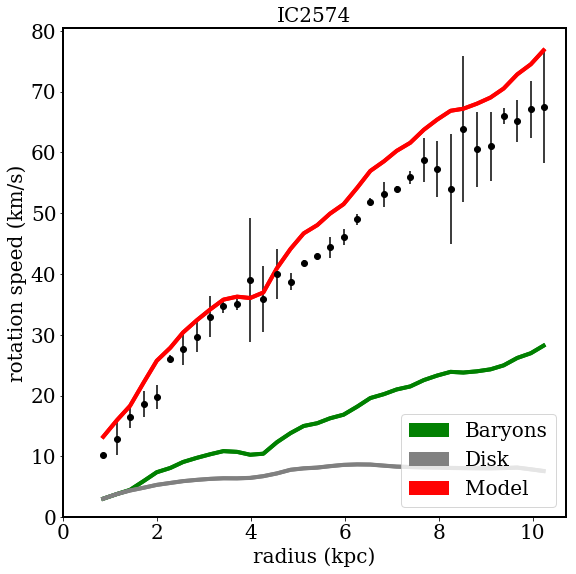} \\
\includegraphics[scale=0.25]{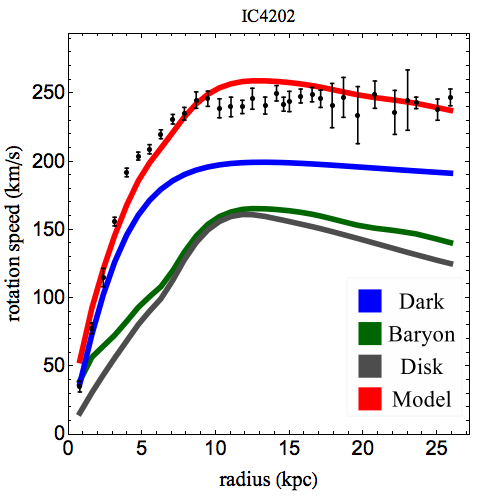} & \includegraphics[scale=0.192]{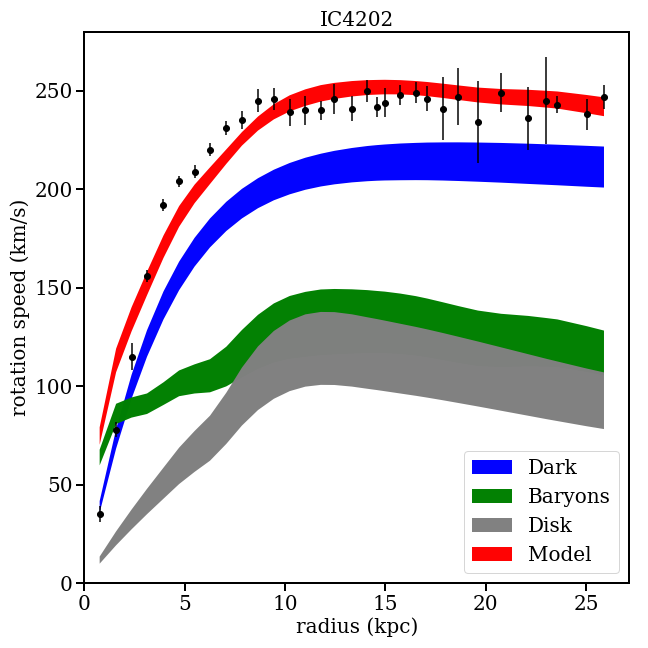} & \includegraphics[scale=0.215]{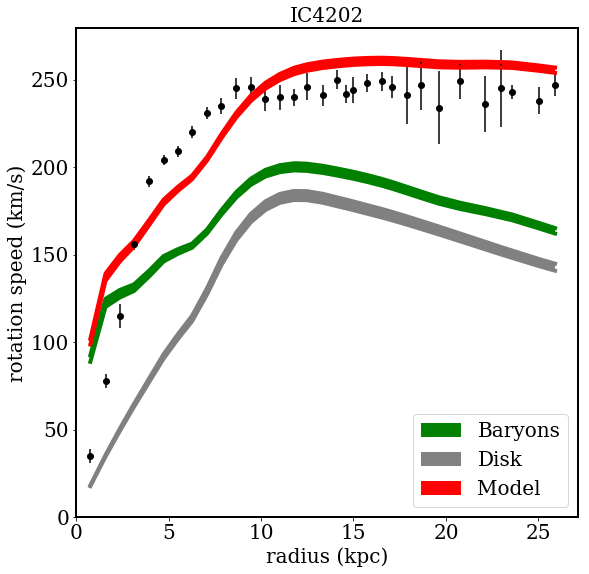} \\
\end{tabular}
\caption{Continued.}
\end{figure}

\addtocounter{figure}{-1}
\begin{figure}[H]
\centering
\begin{tabular}{@{}ccc@{}}
\includegraphics[scale=0.25]{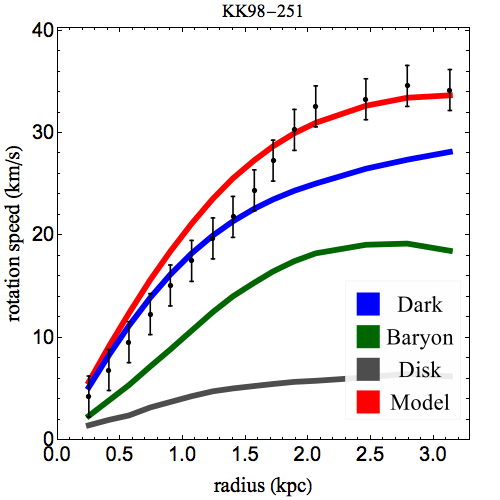} & \includegraphics[scale=0.192]{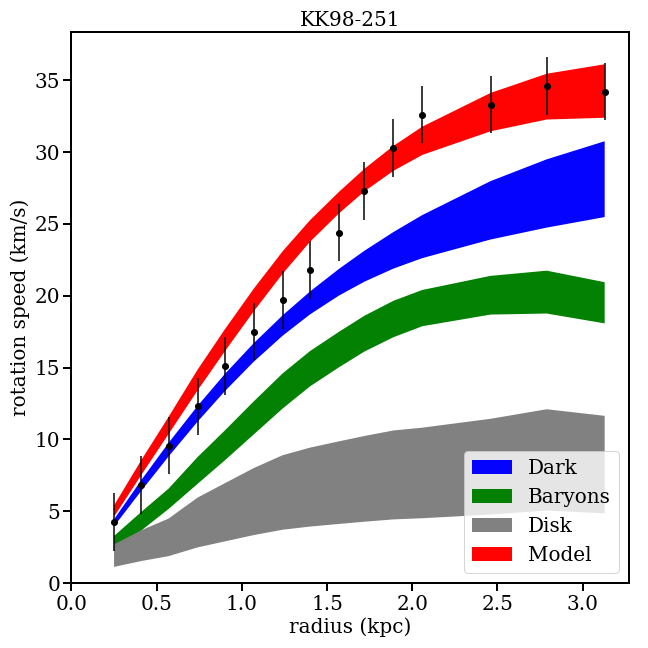} & \includegraphics[scale=0.215]{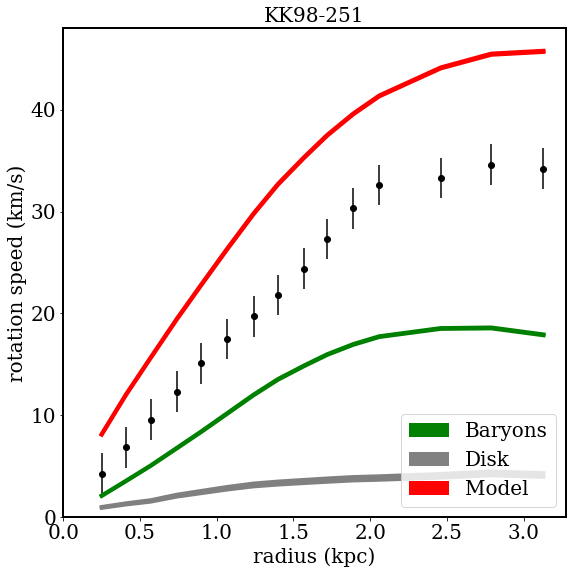} \\
\includegraphics[scale=0.25]{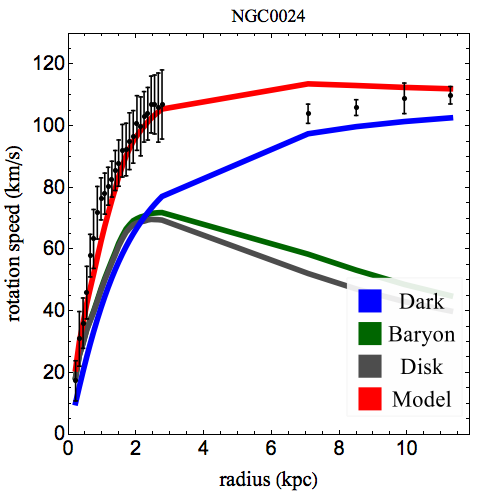} & \includegraphics[scale=0.192]{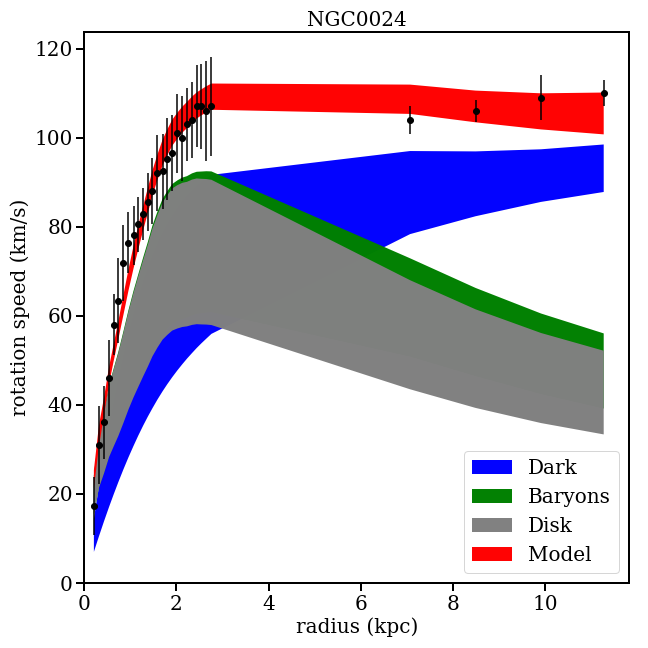} & \includegraphics[scale=0.215]{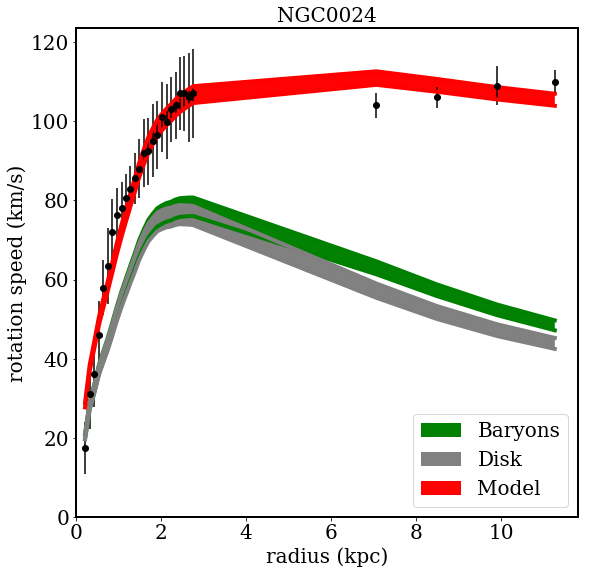} \\
\includegraphics[scale=0.25]{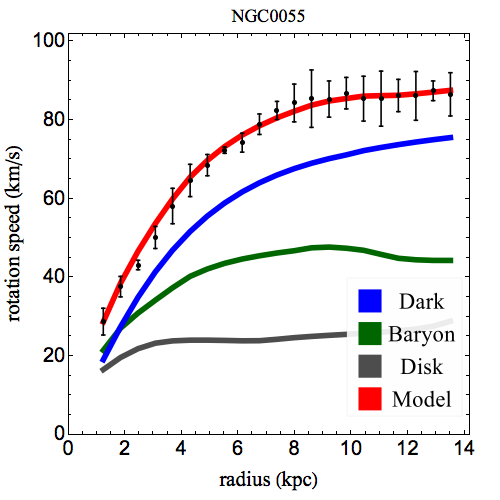} & \includegraphics[scale=0.192]{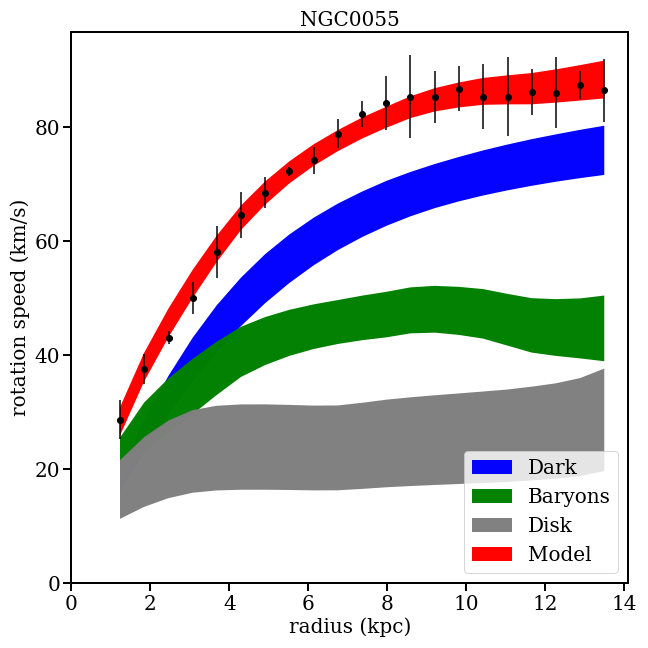} & \includegraphics[scale=0.215]{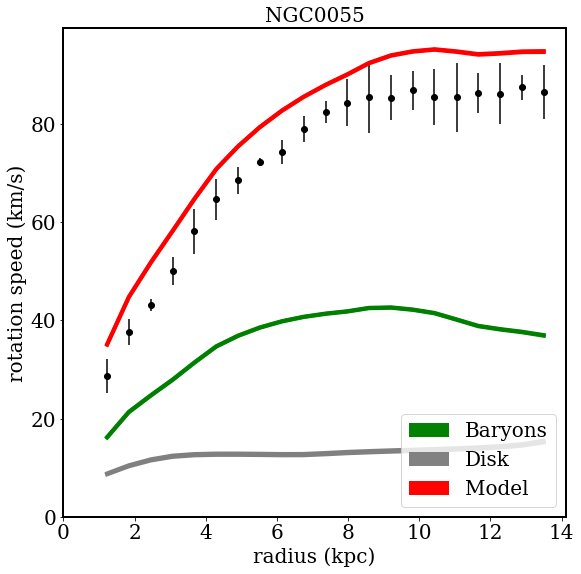} \\
\includegraphics[scale=0.25]{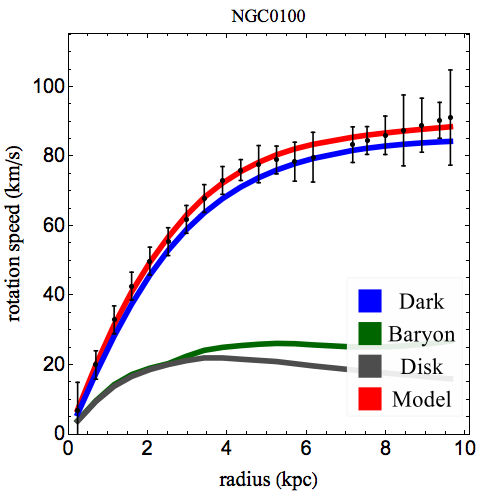} & \includegraphics[scale=0.192]{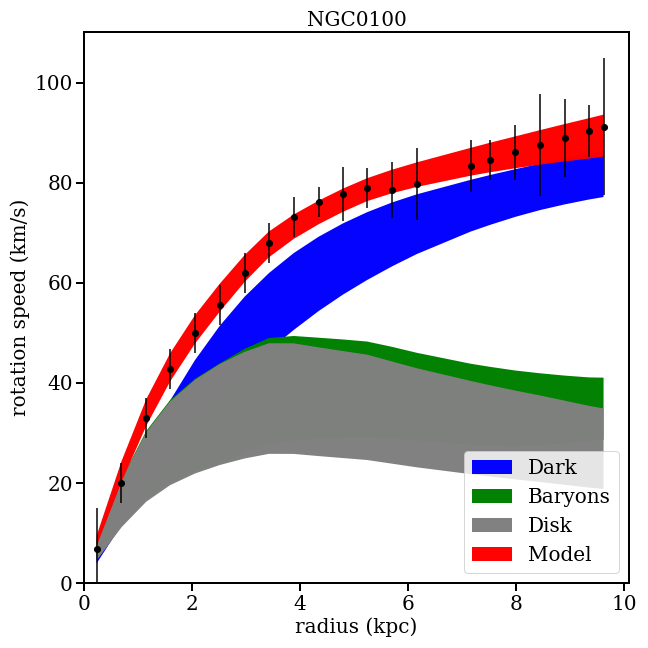} & \includegraphics[scale=0.215]{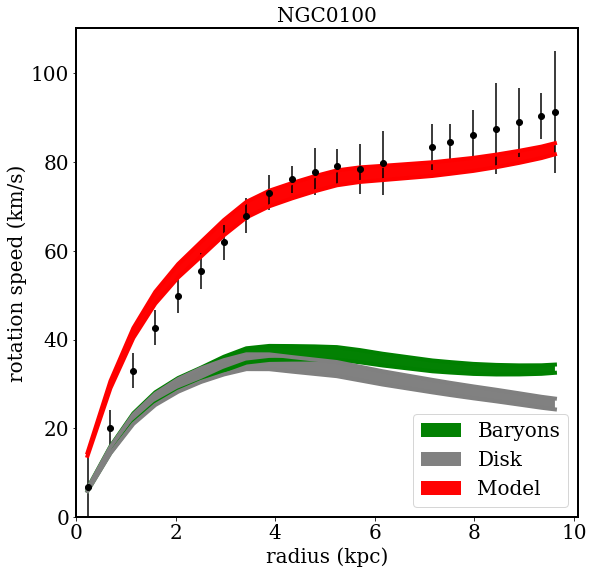} \\
\includegraphics[scale=0.25]{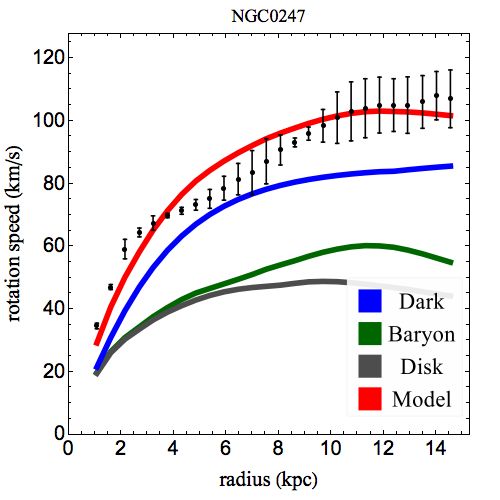} & \includegraphics[scale=0.192]{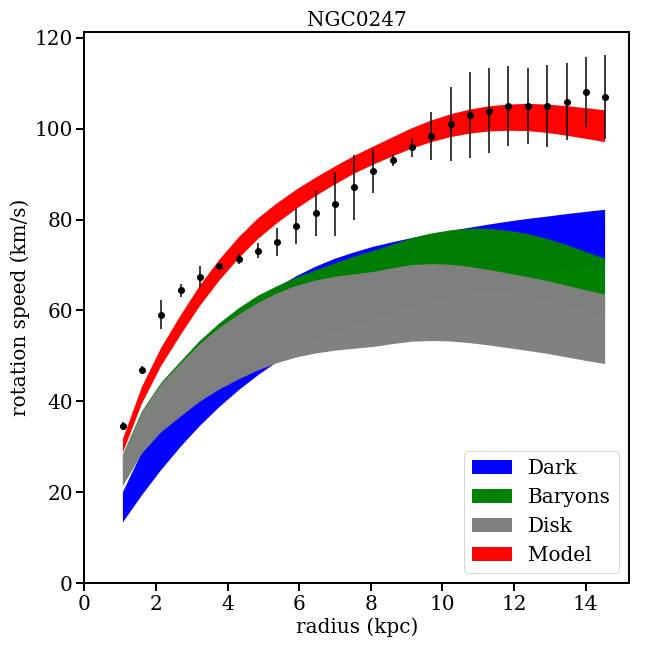} & \includegraphics[scale=0.215]{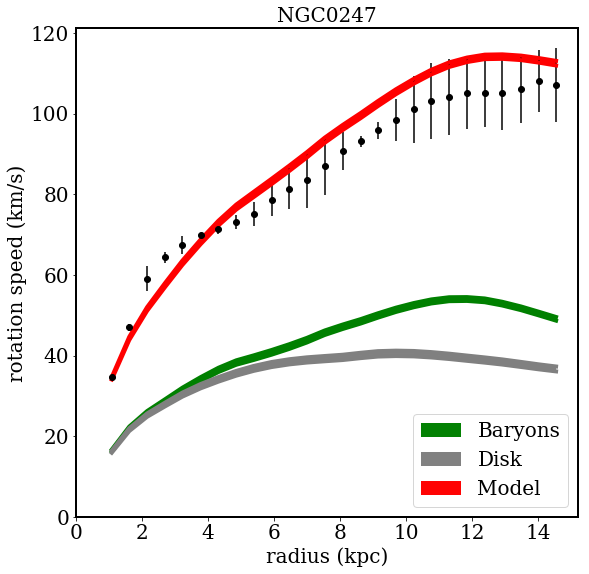} \\
\end{tabular}
\caption{Continued.}
\end{figure}

\addtocounter{figure}{-1}
\begin{figure}[H]
\centering
\begin{tabular}{@{}ccc@{}}
\includegraphics[scale=0.25]{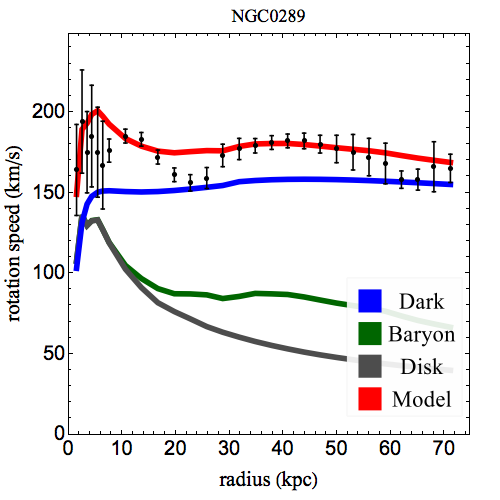} & \includegraphics[scale=0.192]{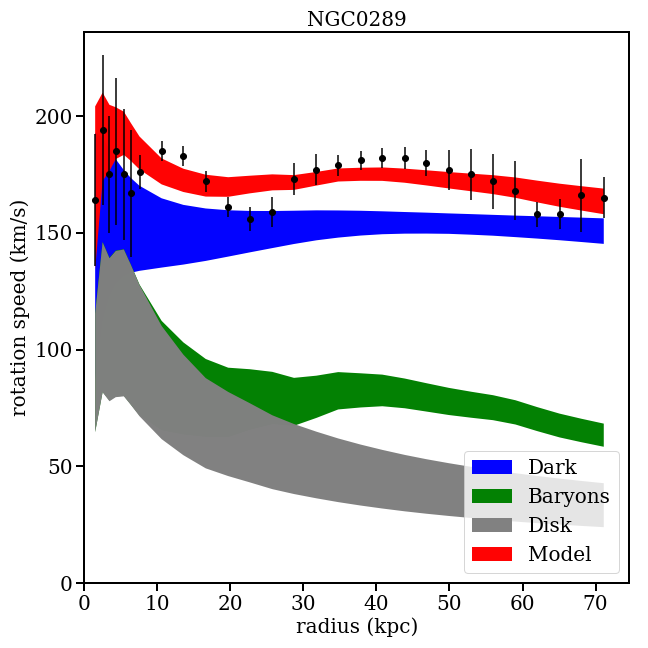} & \includegraphics[scale=0.215]{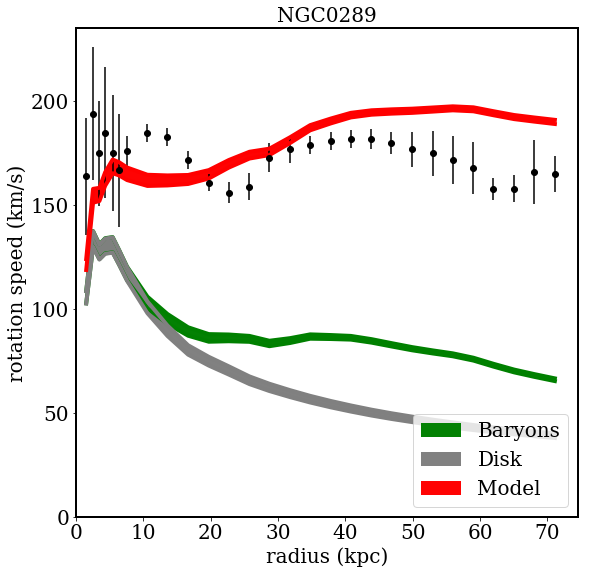} \\
\includegraphics[scale=0.25]{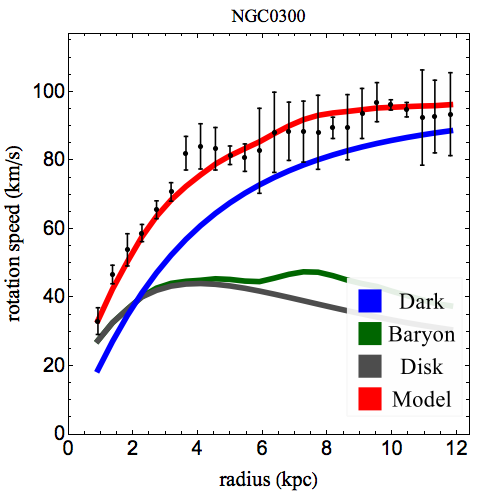} & \includegraphics[scale=0.192]{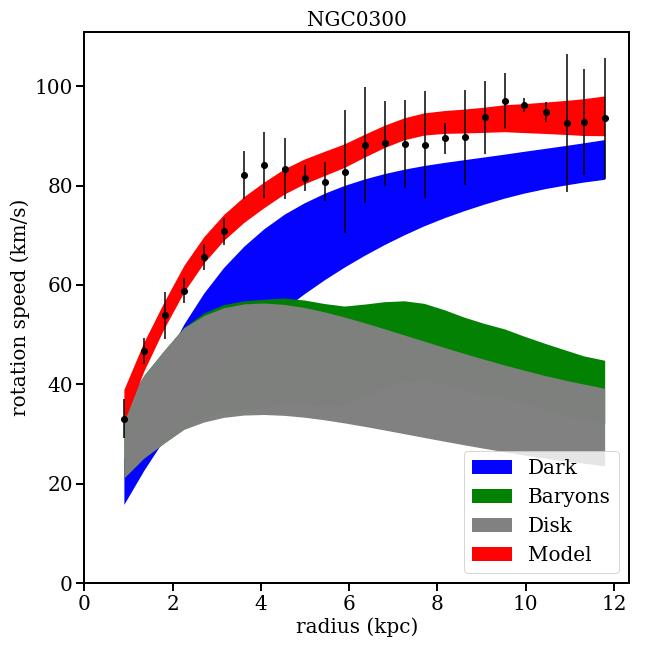} & \includegraphics[scale=0.215]{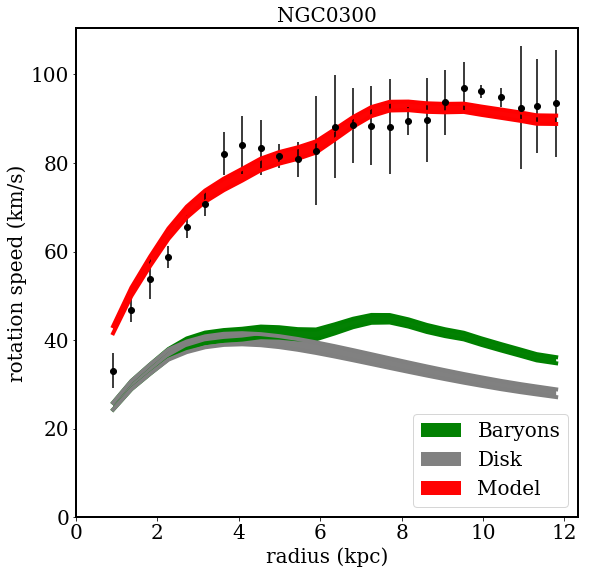} \\
\includegraphics[scale=0.25]{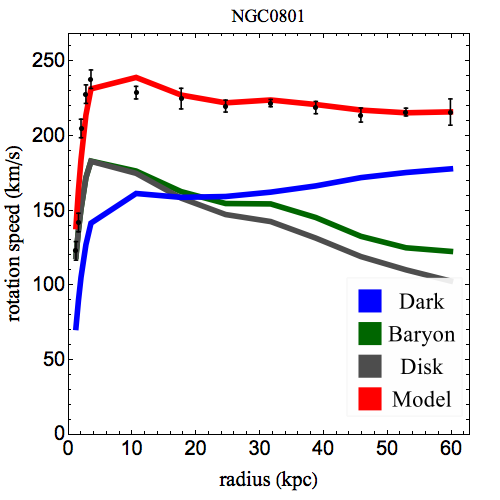} & \includegraphics[scale=0.192]{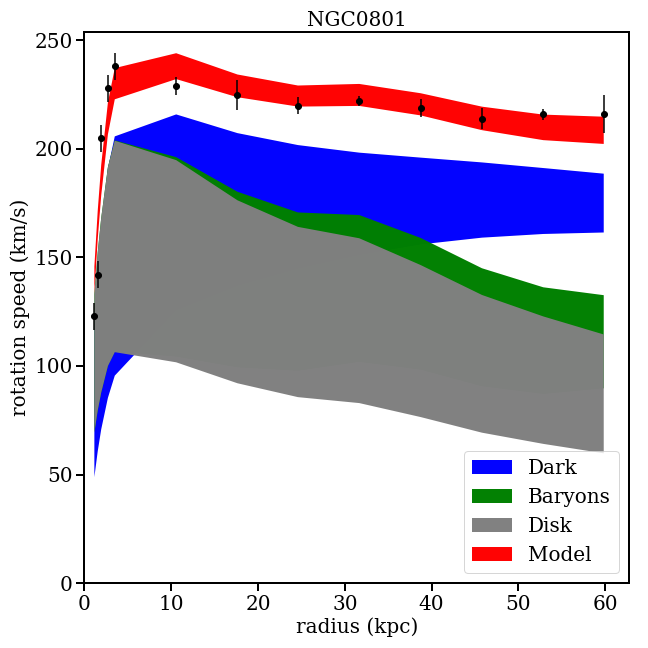} & \includegraphics[scale=0.215]{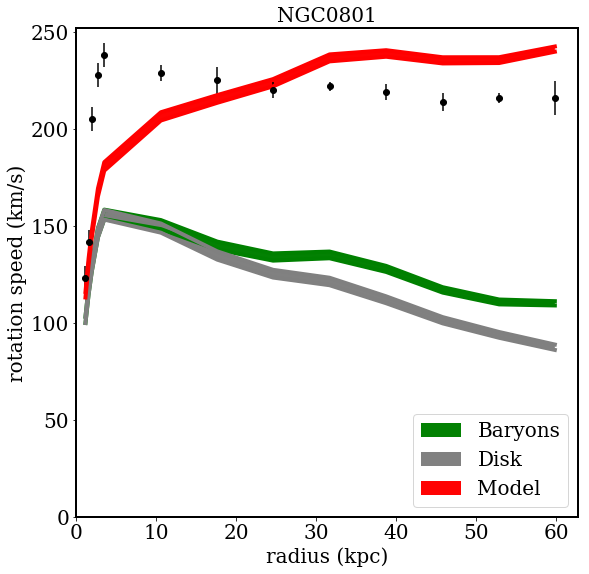} \\
\includegraphics[scale=0.25]{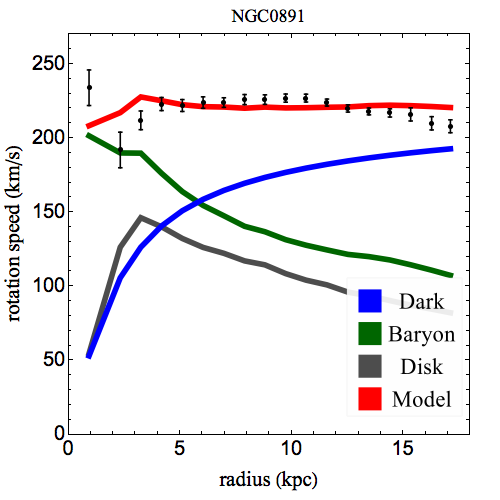} & \includegraphics[scale=0.192]{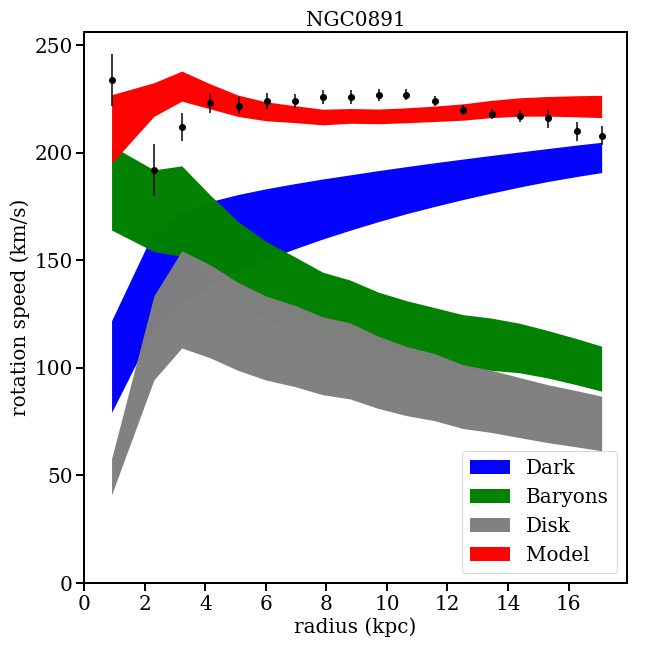} & \includegraphics[scale=0.215]{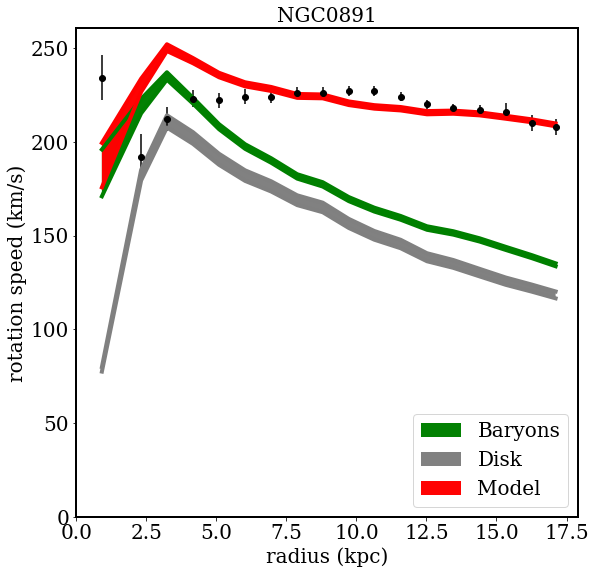} \\
\includegraphics[scale=0.25]{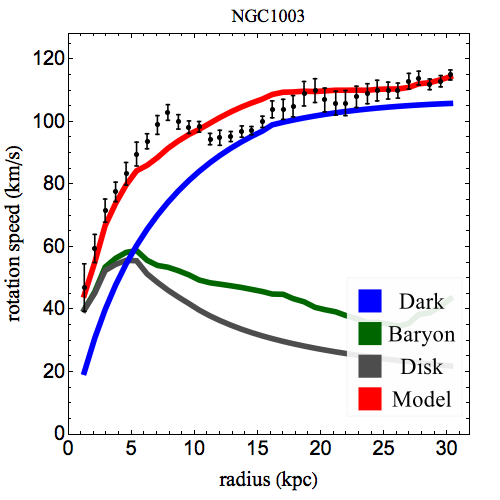} & \includegraphics[scale=0.192]{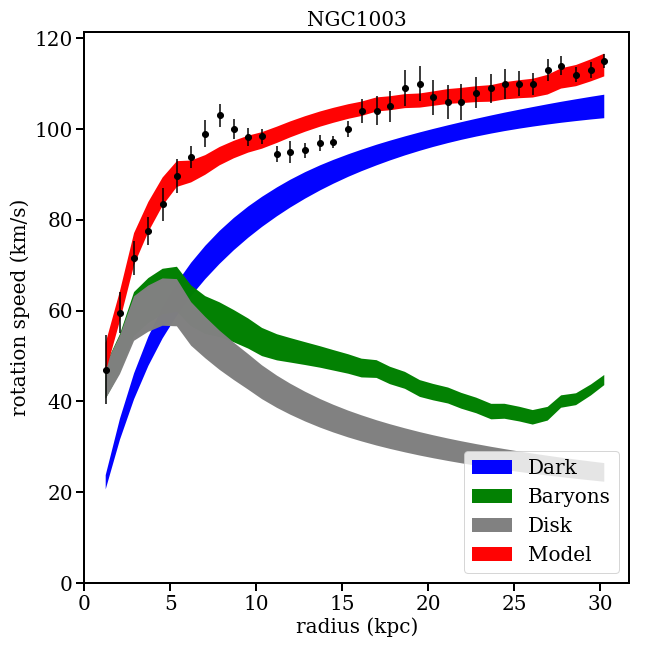} & \includegraphics[scale=0.215]{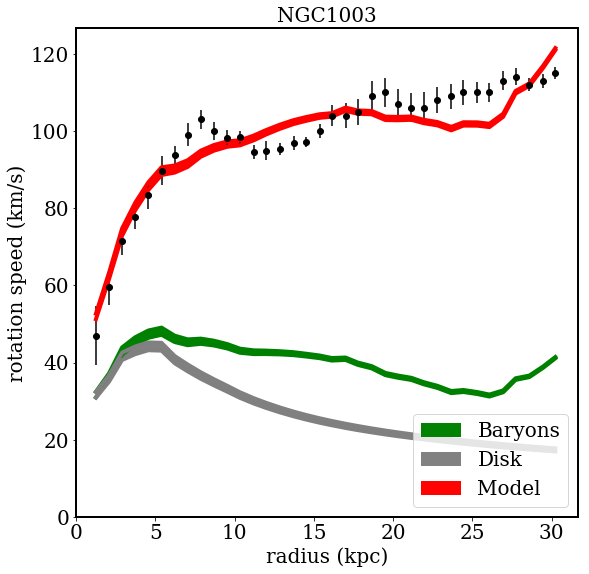} \\
\end{tabular}
\caption{Continued.}
\end{figure}

\addtocounter{figure}{-1}
\begin{figure}[H]
\centering
\begin{tabular}{@{}ccc@{}}
\includegraphics[scale=0.25]{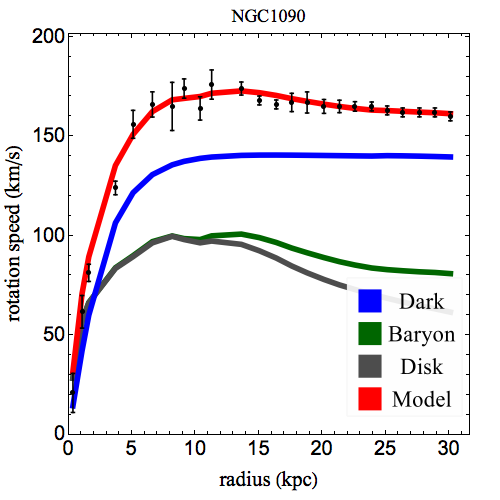} & \includegraphics[scale=0.192]{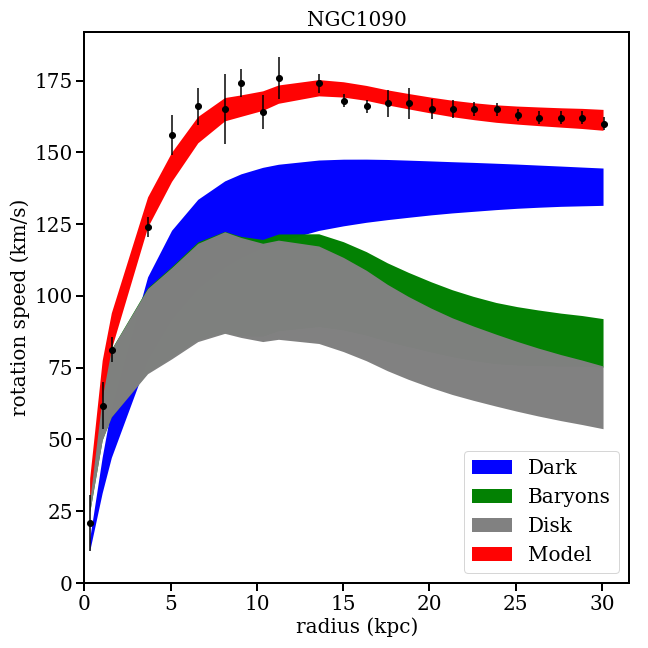} & \includegraphics[scale=0.215]{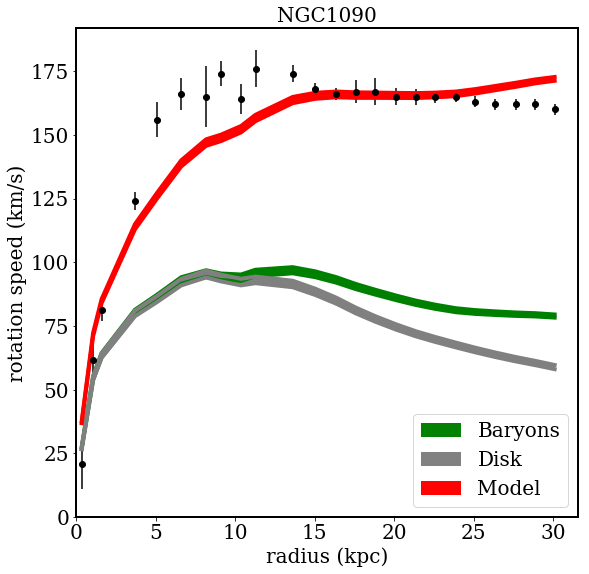} \\
\includegraphics[scale=0.25]{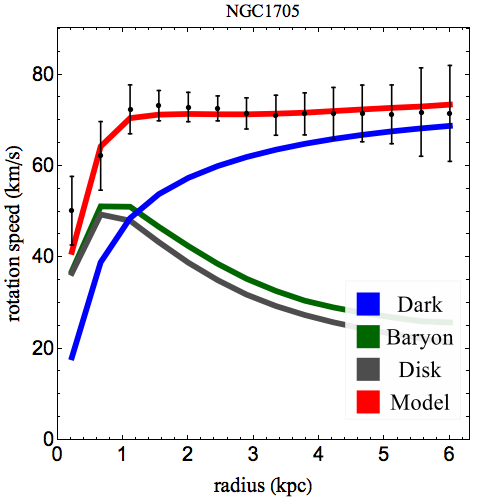} & \includegraphics[scale=0.192]{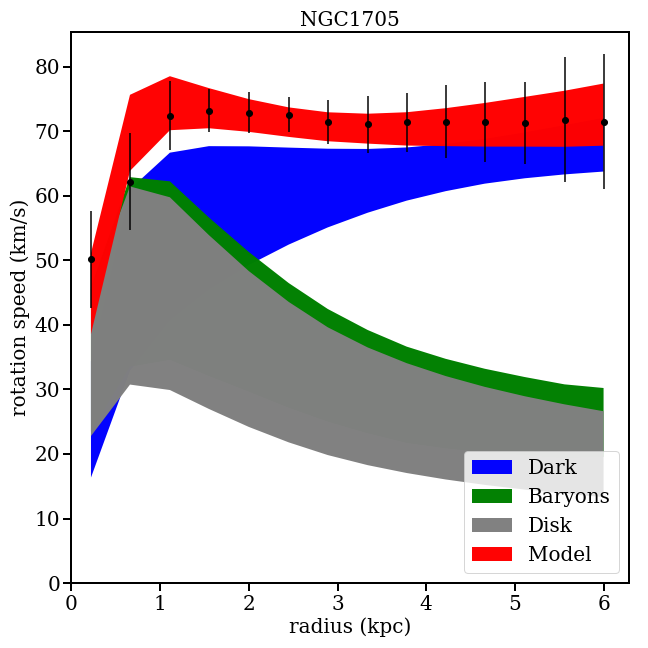} & \includegraphics[scale=0.215]{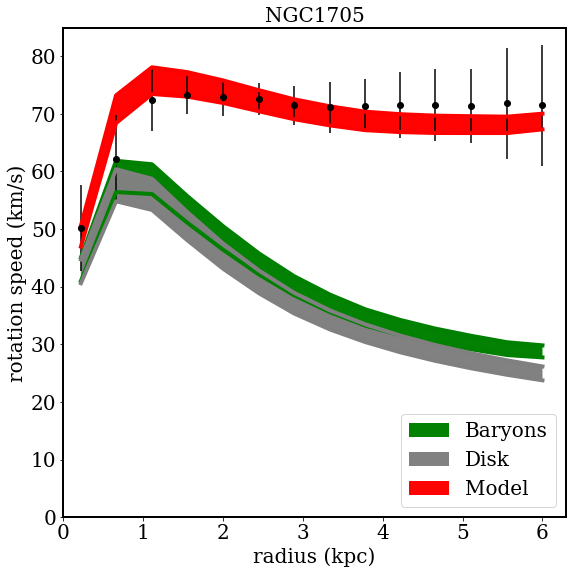} \\
\includegraphics[scale=0.25]{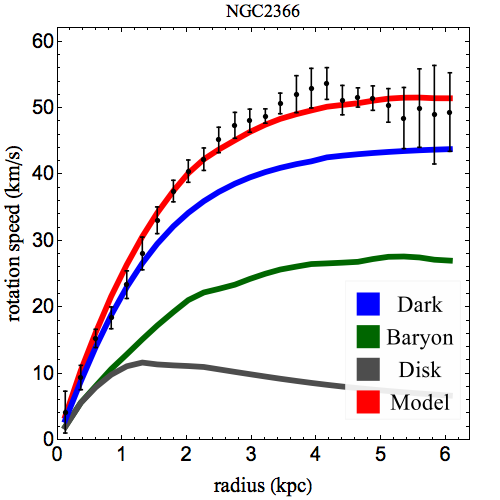} & \includegraphics[scale=0.192]{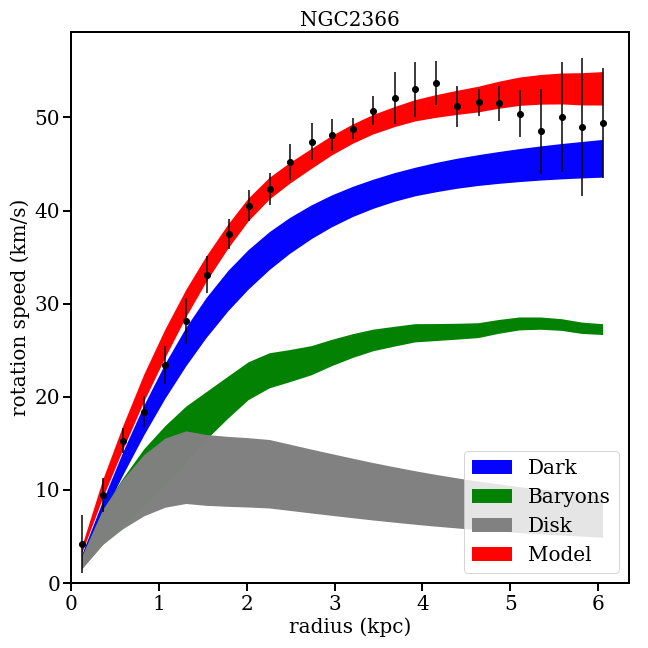} & \includegraphics[scale=0.215]{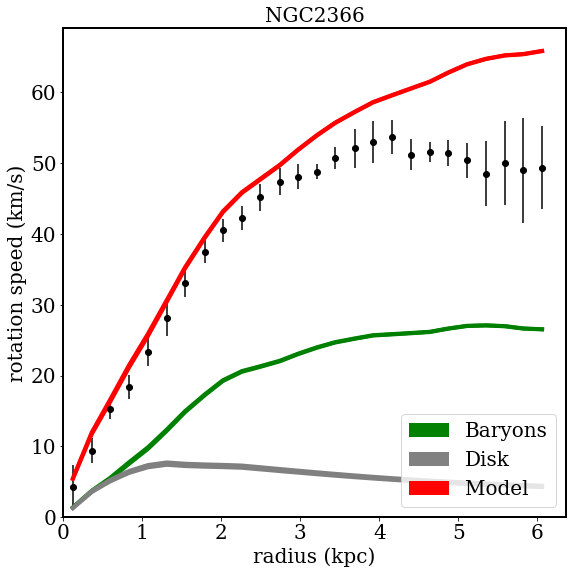} \\
\includegraphics[scale=0.25]{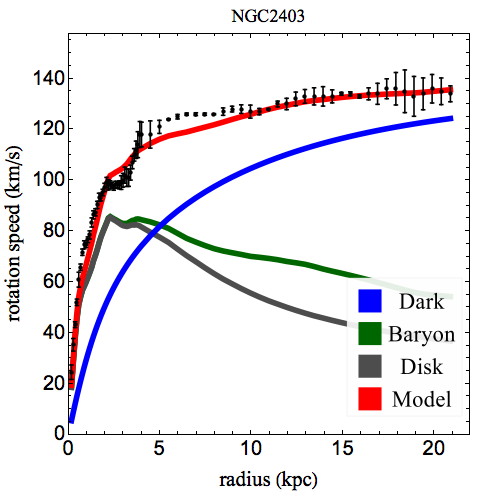} & \includegraphics[scale=0.192]{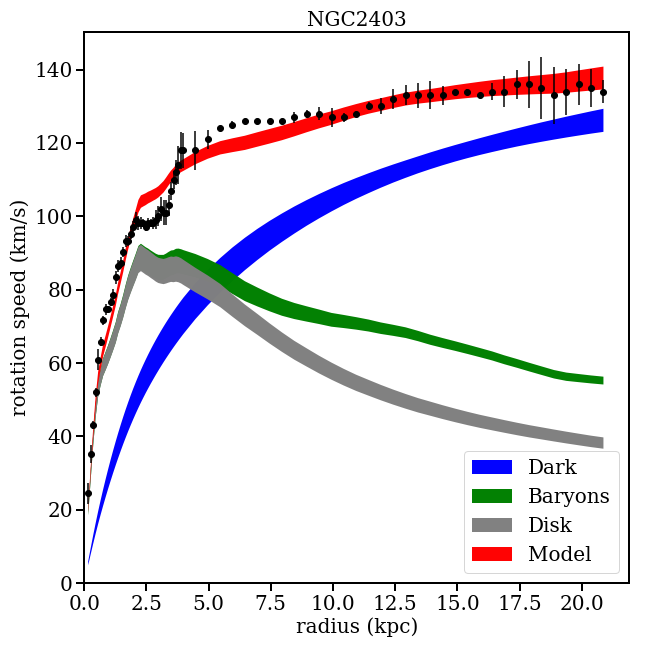} & \includegraphics[scale=0.215]{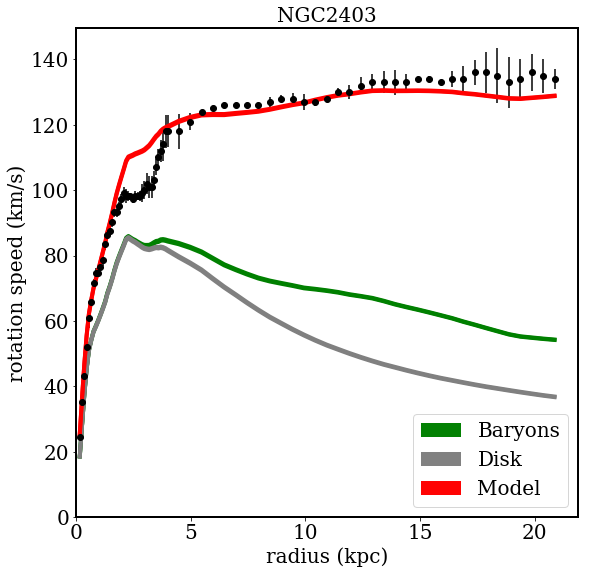} \\
\includegraphics[scale=0.25]{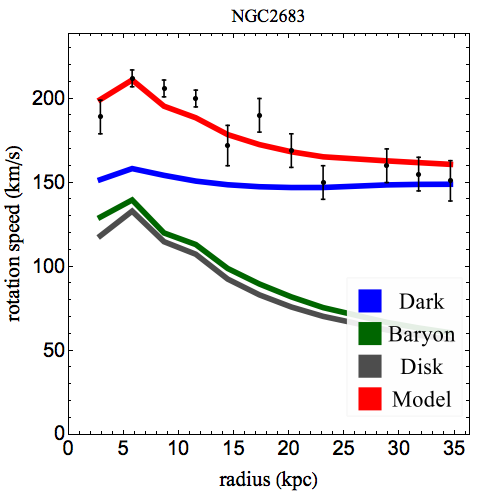} & \includegraphics[scale=0.192]{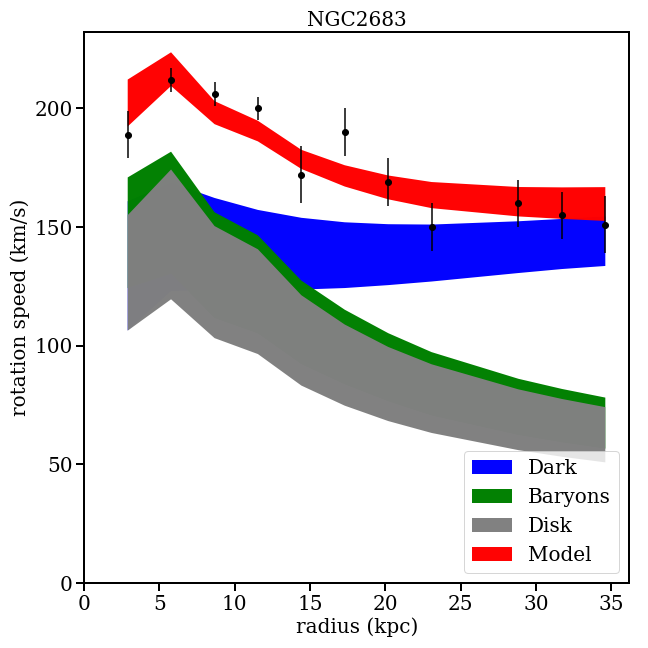} & \includegraphics[scale=0.215]{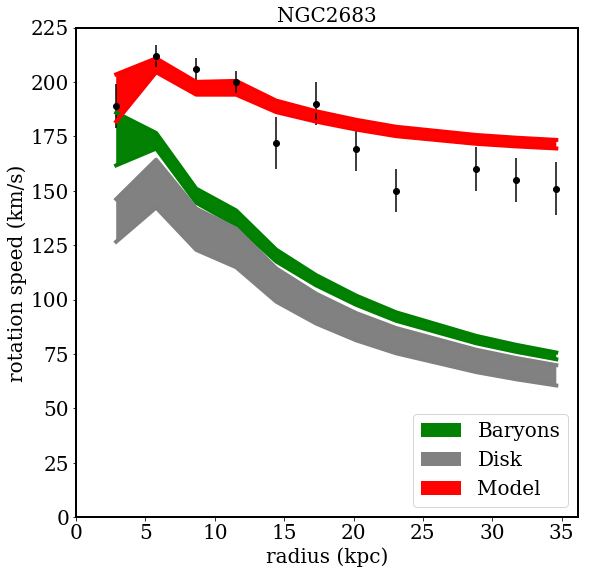} \\
\end{tabular}
\caption{Continued.}
\end{figure}

\addtocounter{figure}{-1}
\begin{figure}[H]
\centering
\begin{tabular}{@{}ccc@{}}
\includegraphics[scale=0.25]{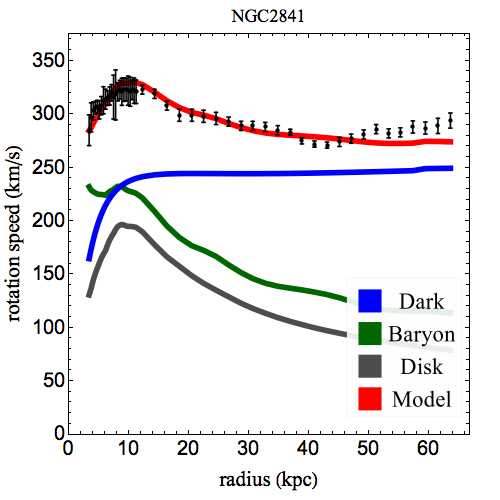} & \includegraphics[scale=0.192]{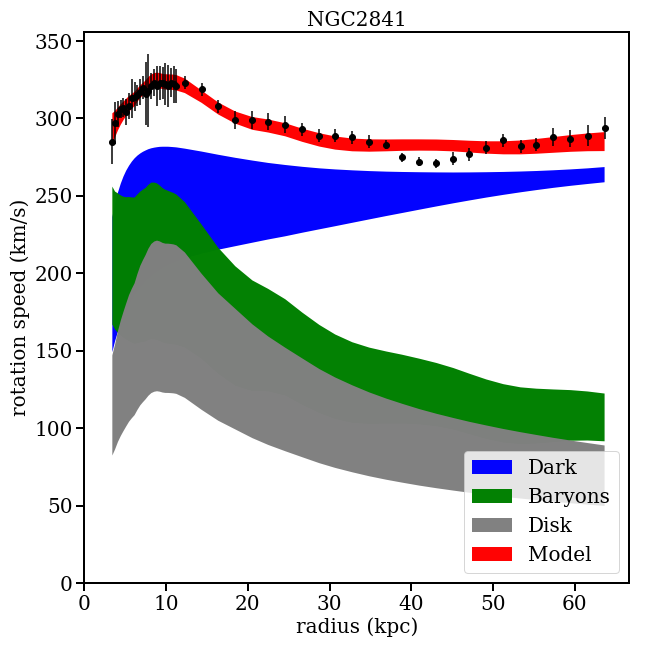} & \includegraphics[scale=0.215]{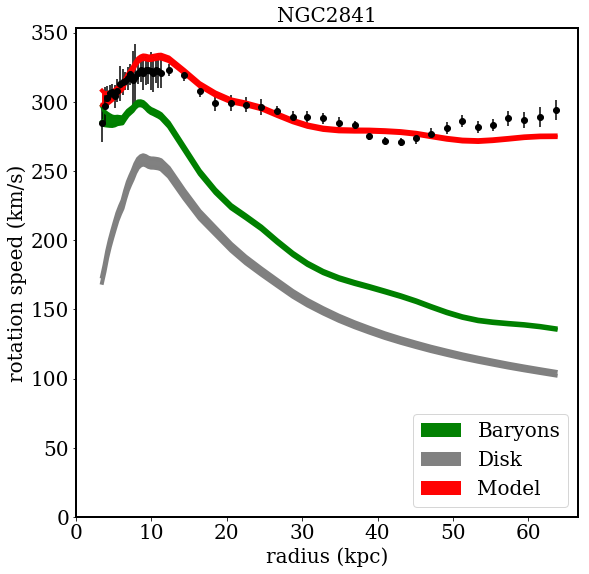} \\
\includegraphics[scale=0.25]{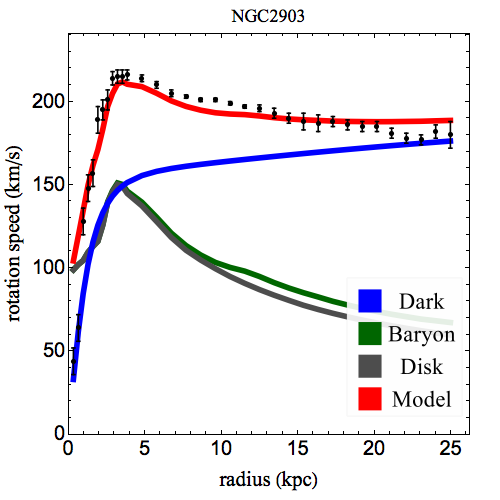} & \includegraphics[scale=0.192]{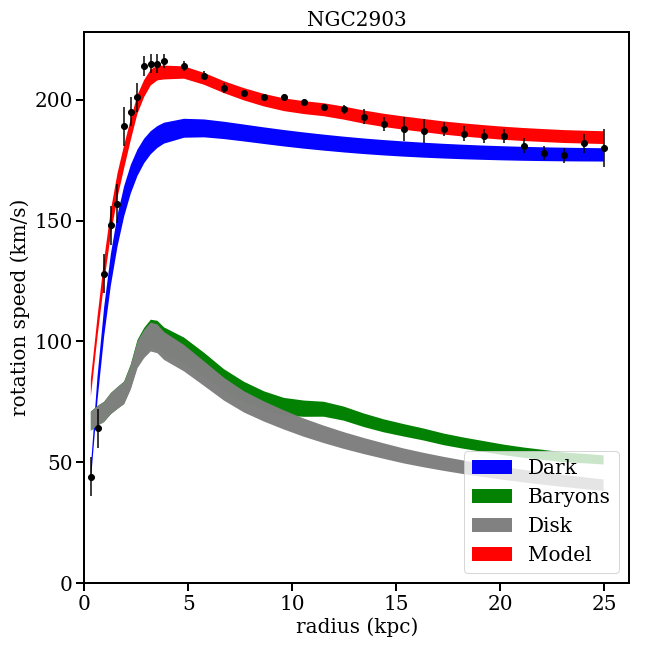} & \includegraphics[scale=0.215]{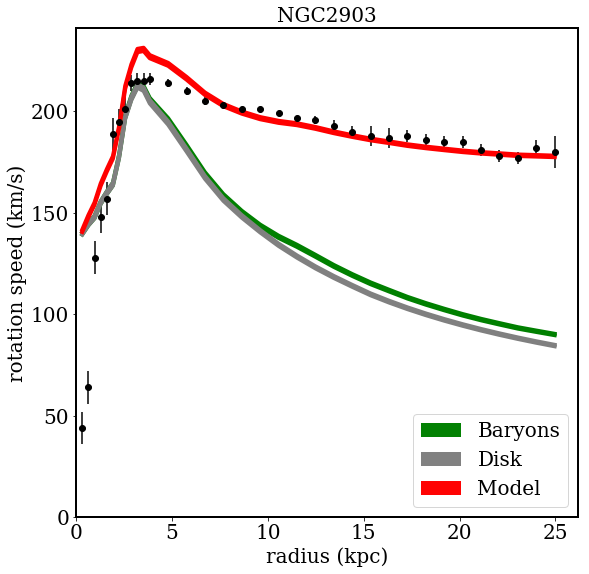} \\
\includegraphics[scale=0.25]{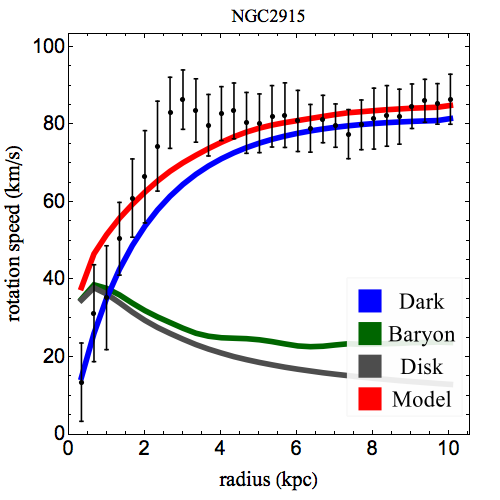} & \includegraphics[scale=0.192]{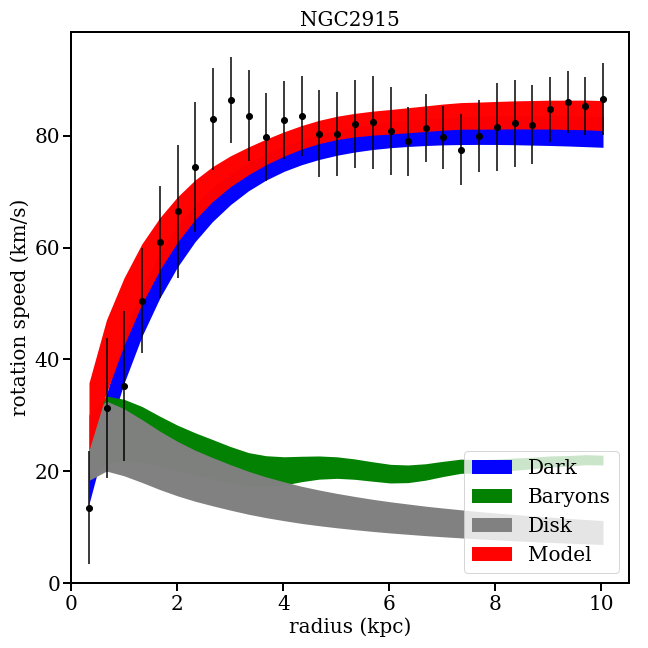} & \includegraphics[scale=0.215]{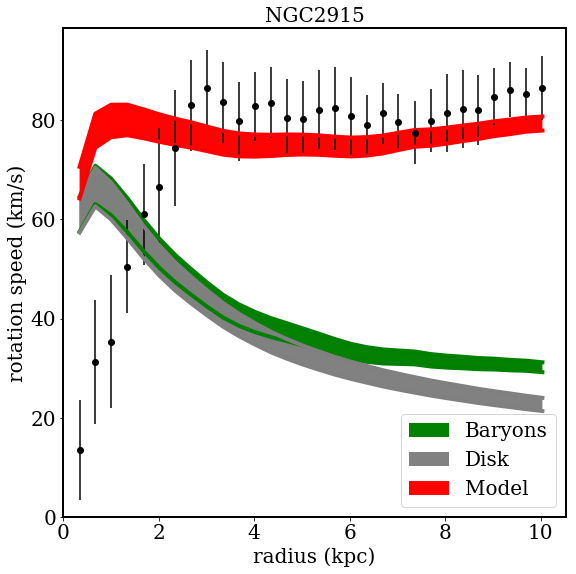} \\
\includegraphics[scale=0.25]{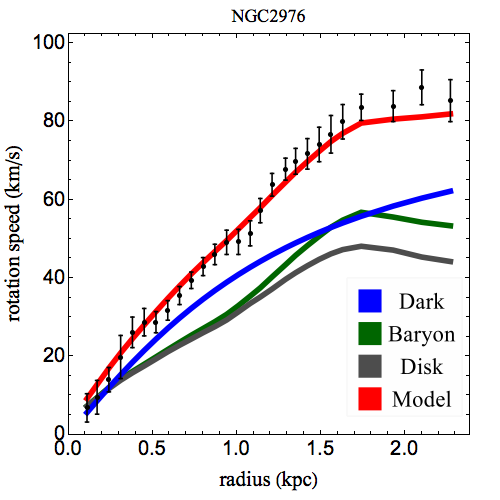} & \includegraphics[scale=0.192]{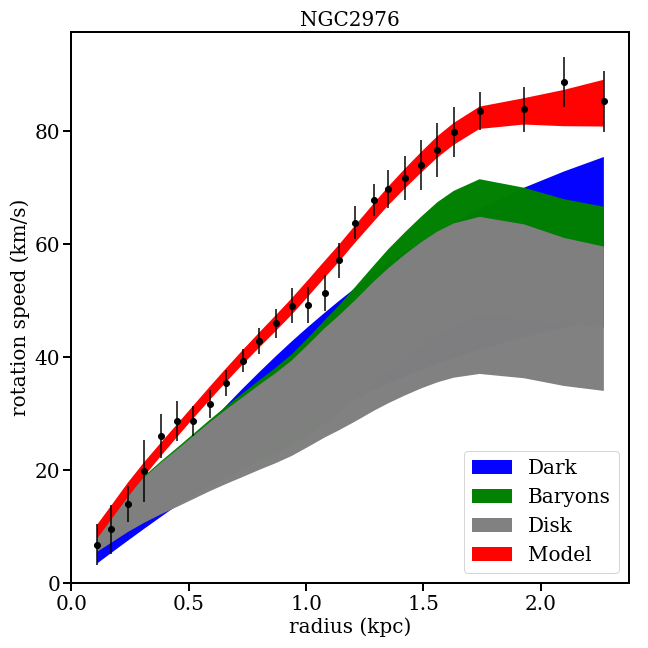} & \includegraphics[scale=0.215]{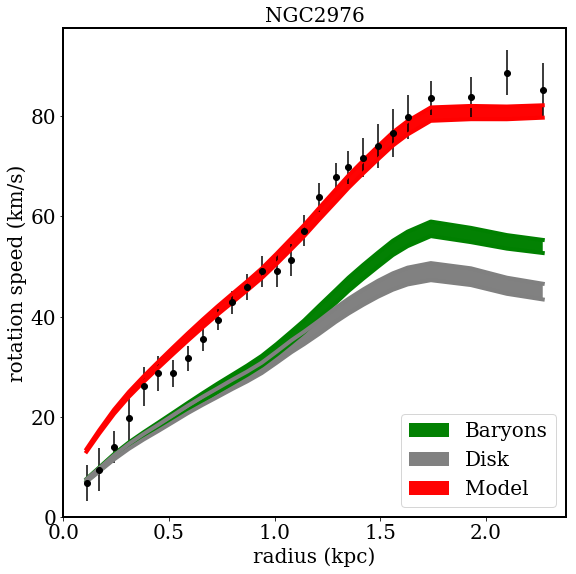} \\
\includegraphics[scale=0.25]{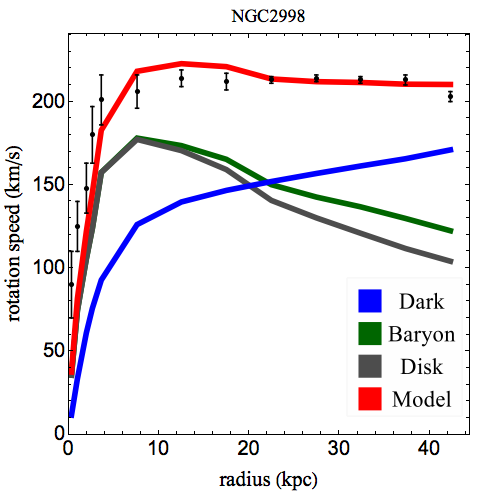} & \includegraphics[scale=0.192]{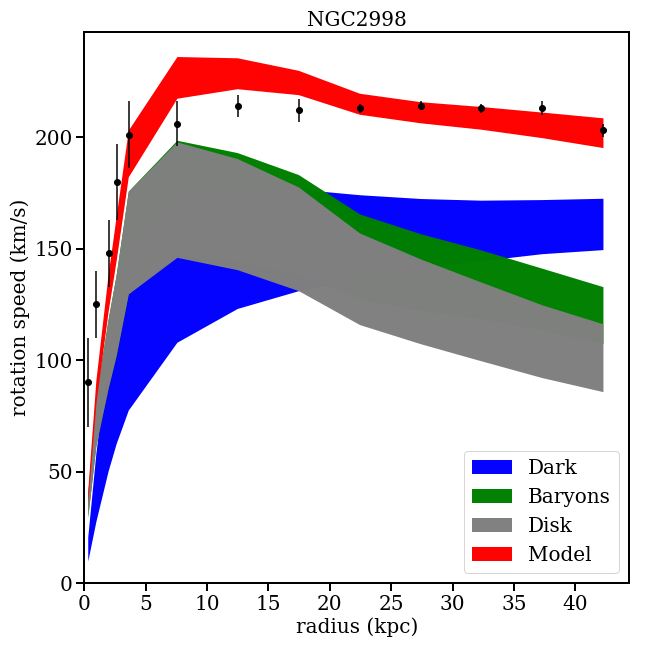} & \includegraphics[scale=0.215]{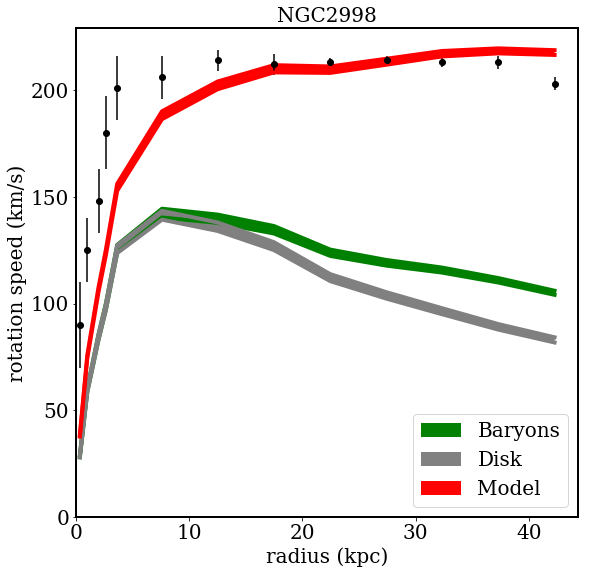} \\
\end{tabular}
\caption{Continued.}
\end{figure}

\addtocounter{figure}{-1}
\begin{figure}[H]
\centering
\begin{tabular}{@{}ccc@{}}
\includegraphics[scale=0.25]{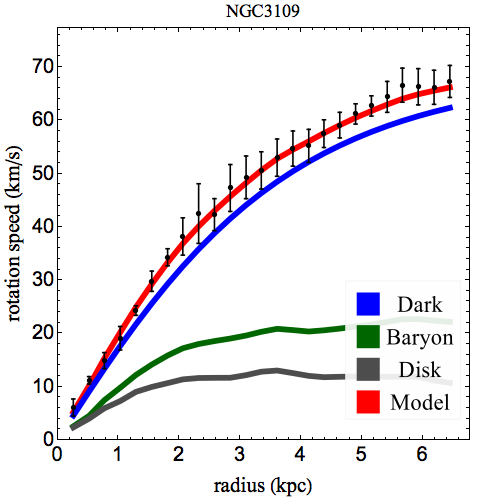} & \includegraphics[scale=0.192]{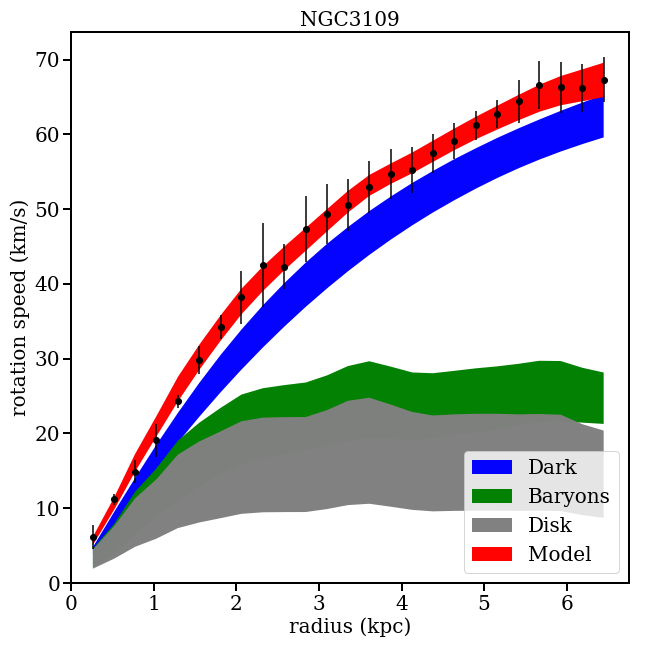} & \includegraphics[scale=0.215]{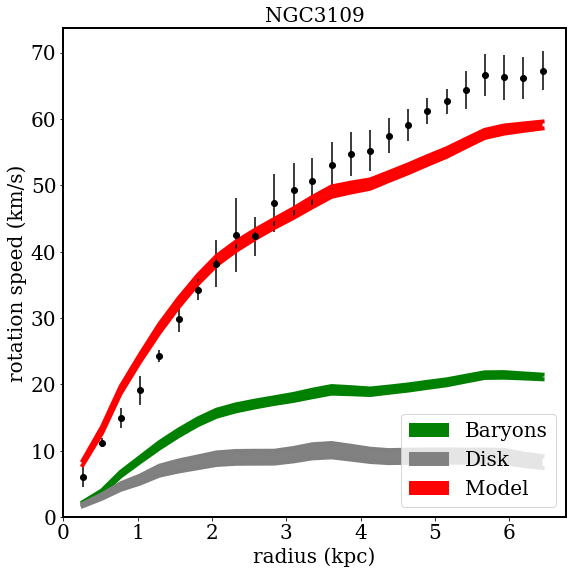} \\
\includegraphics[scale=0.25]{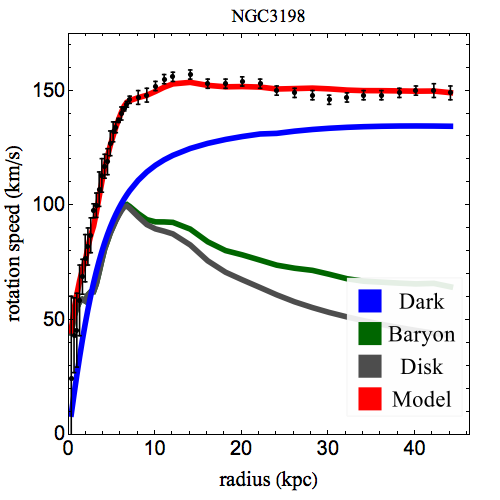} & \includegraphics[scale=0.192]{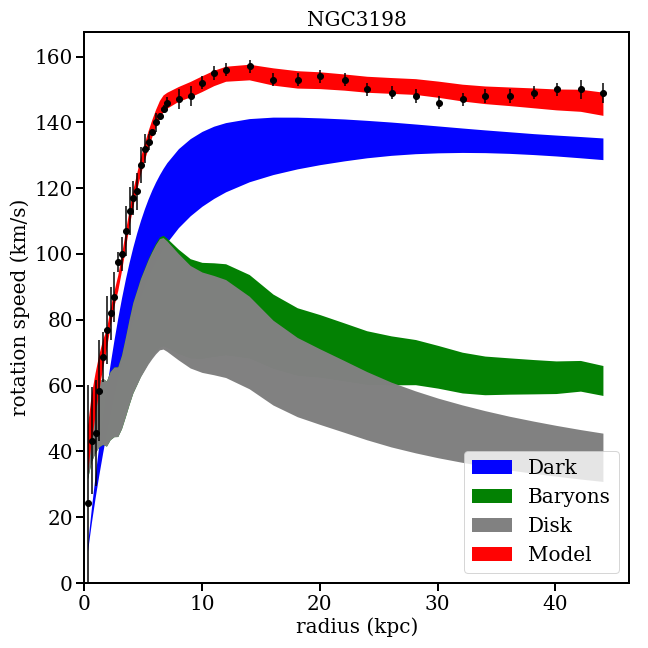} & \includegraphics[scale=0.215]{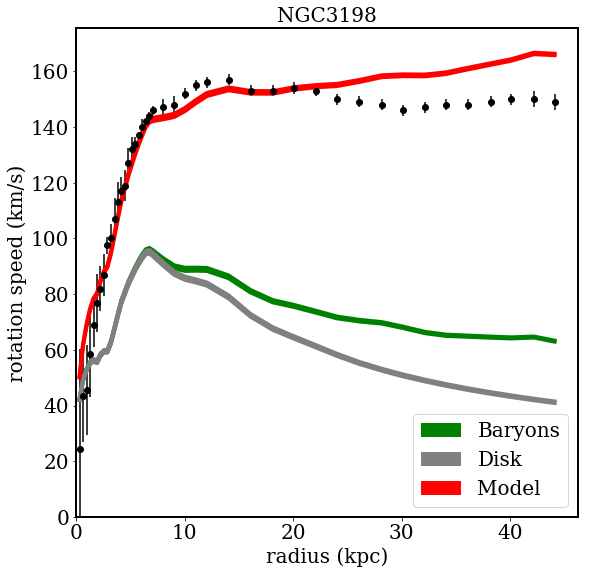} \\
\includegraphics[scale=0.25]{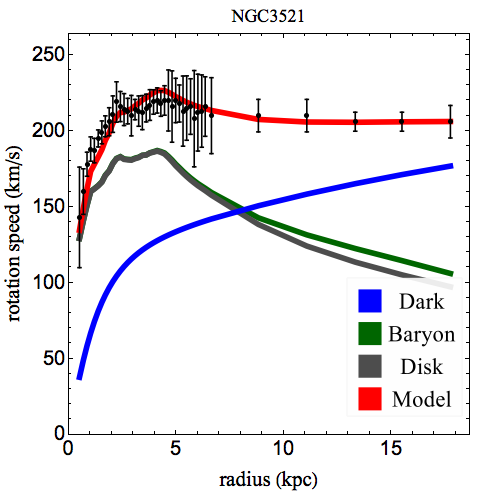} & \includegraphics[scale=0.192]{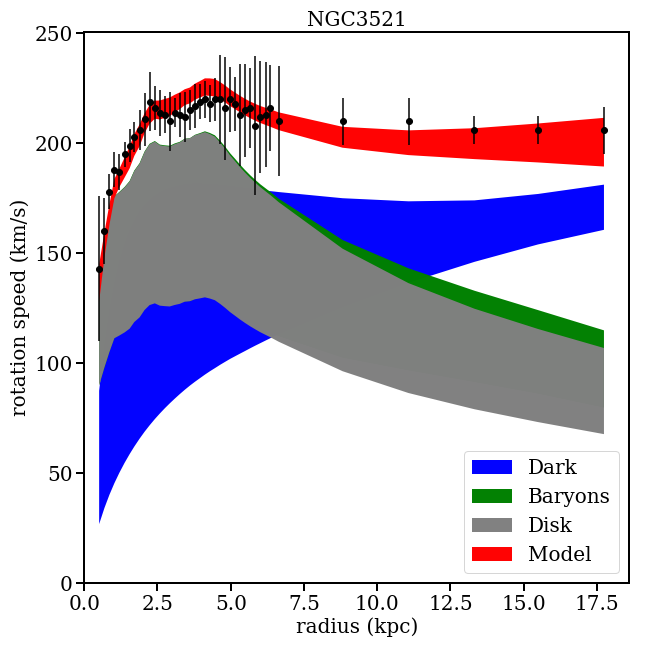} & \includegraphics[scale=0.215]{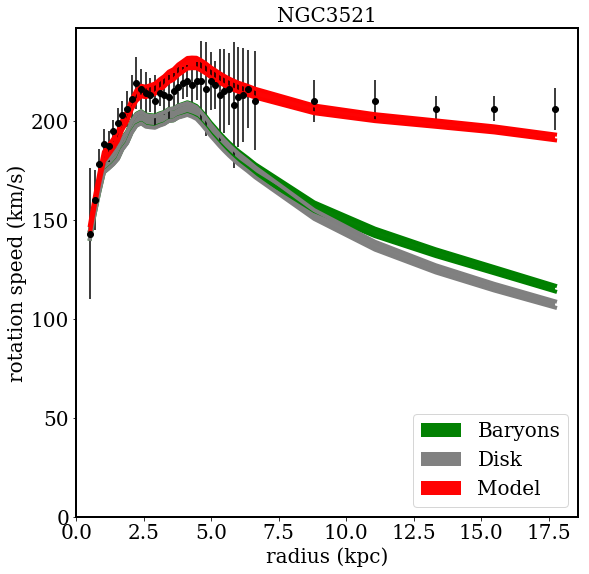} \\
\includegraphics[scale=0.25]{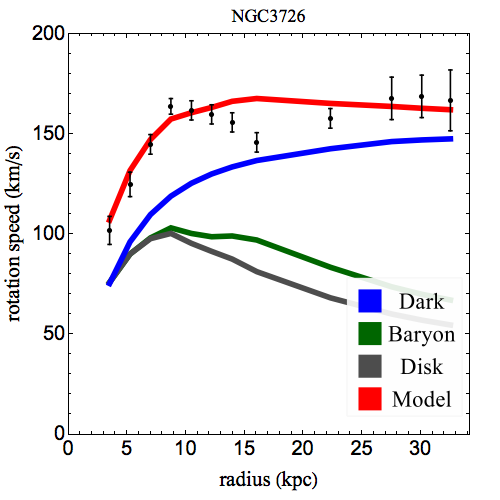} & \includegraphics[scale=0.192]{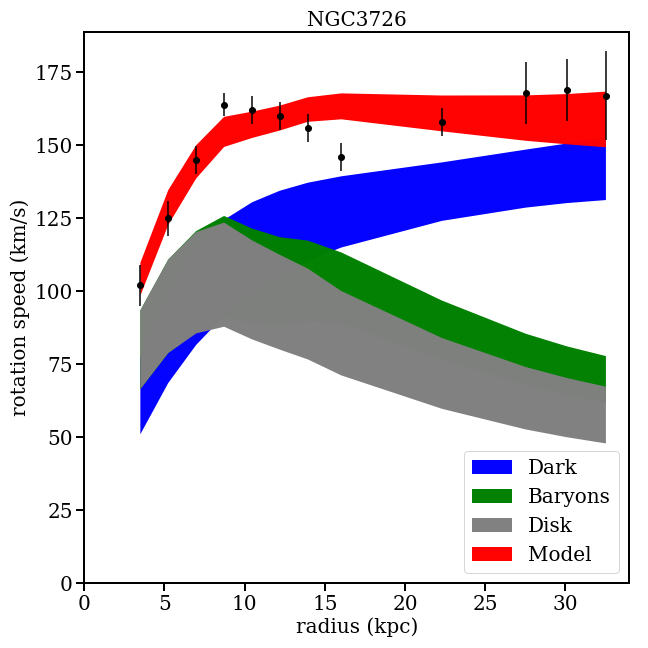} & \includegraphics[scale=0.215]{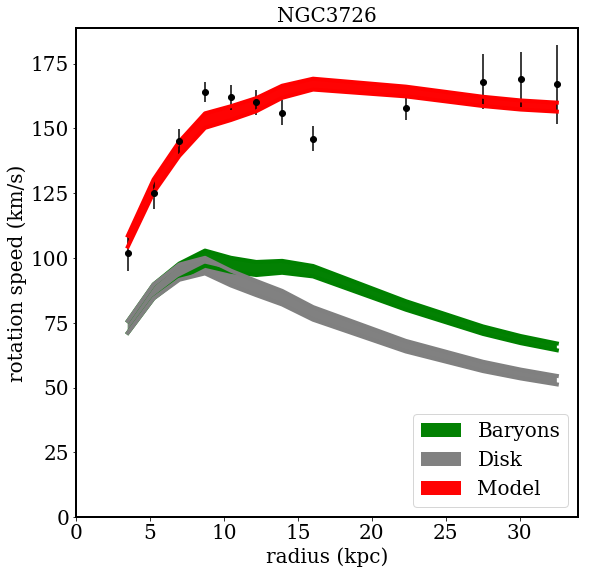} \\
\includegraphics[scale=0.25]{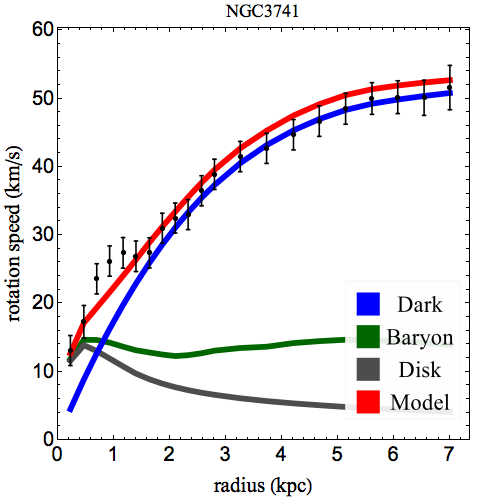} & \includegraphics[scale=0.192]{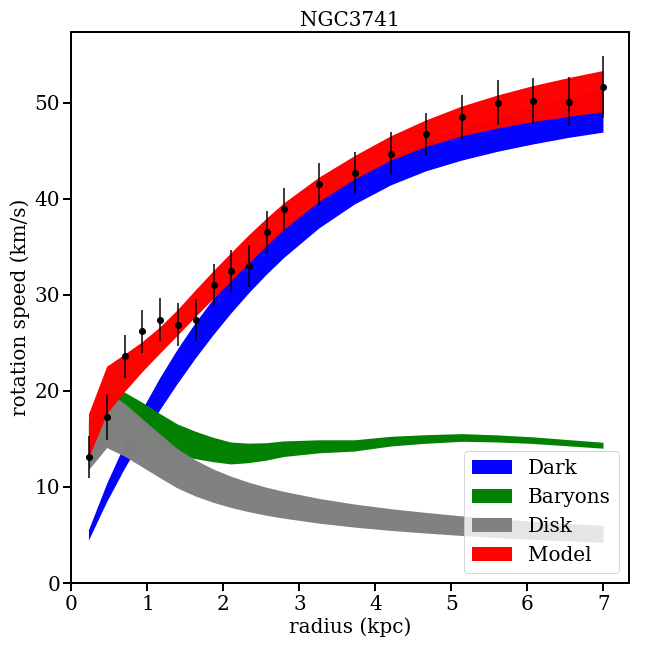} & \includegraphics[scale=0.215]{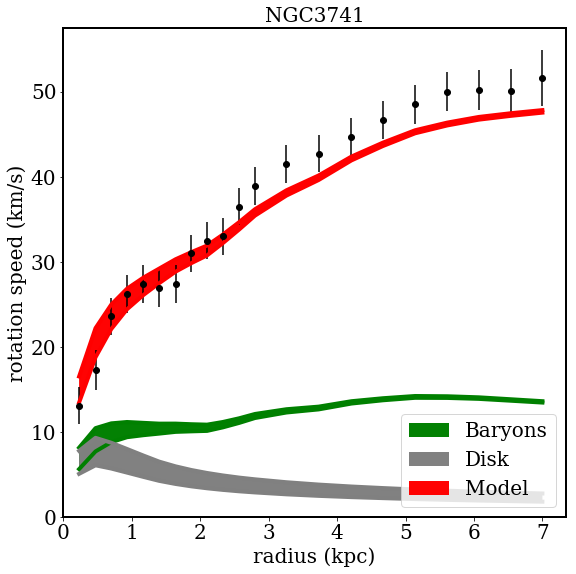} \\
\end{tabular}
\caption{Continued.}
\end{figure}

\addtocounter{figure}{-1}
\begin{figure}[H]
\centering
\begin{tabular}{@{}ccc@{}}
\includegraphics[scale=0.25]{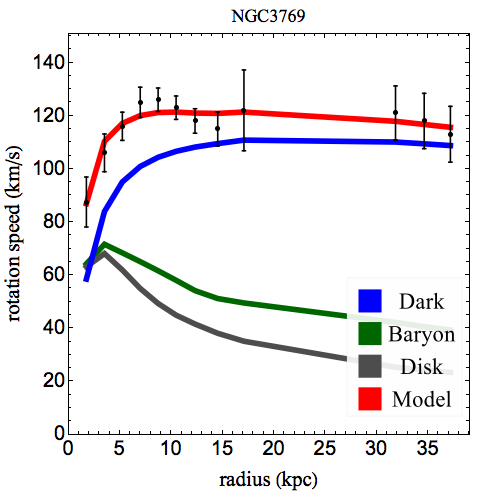} & \includegraphics[scale=0.192]{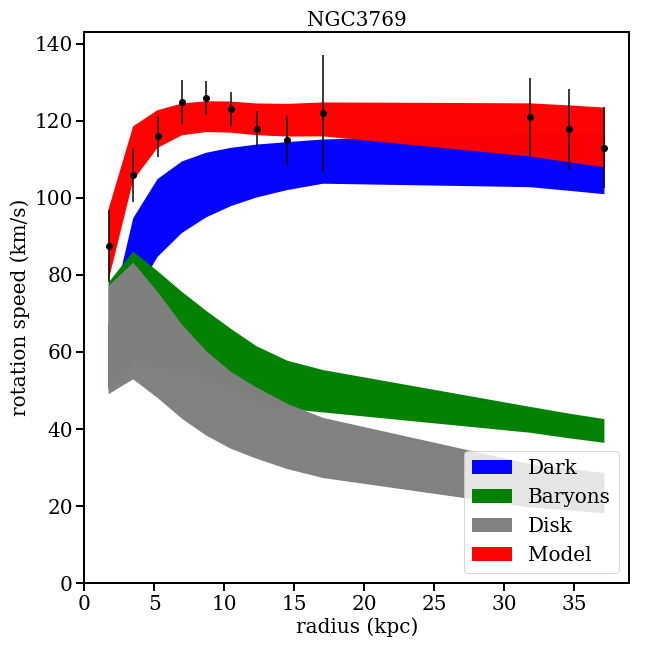} & \includegraphics[scale=0.215]{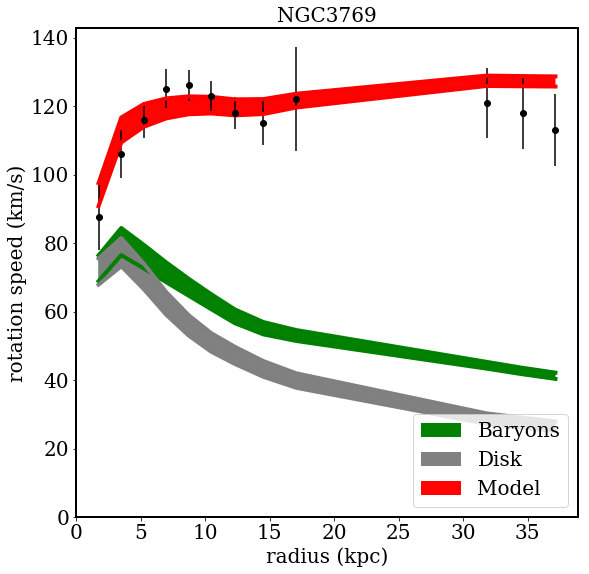} \\
\includegraphics[scale=0.25]{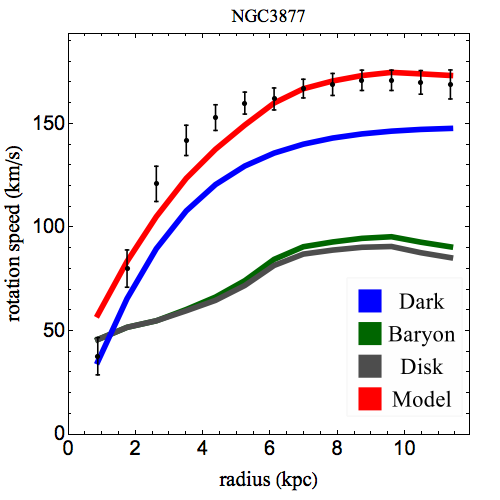} & \includegraphics[scale=0.192]{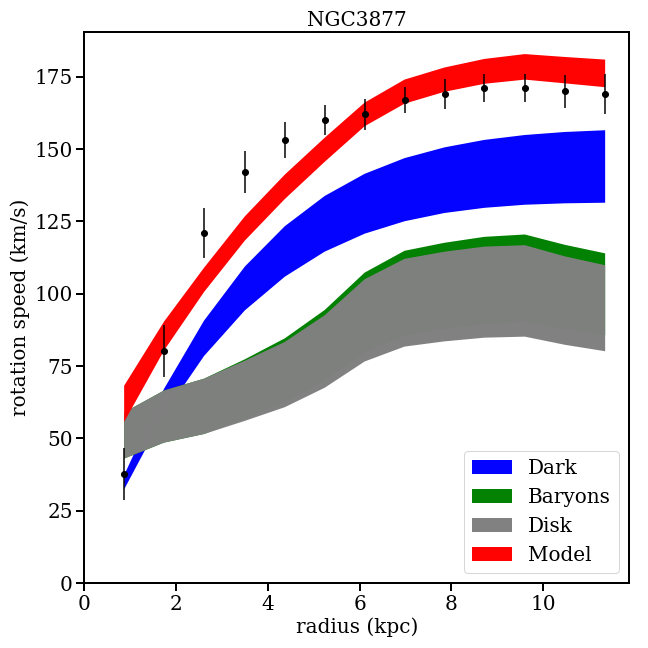} & \includegraphics[scale=0.215]{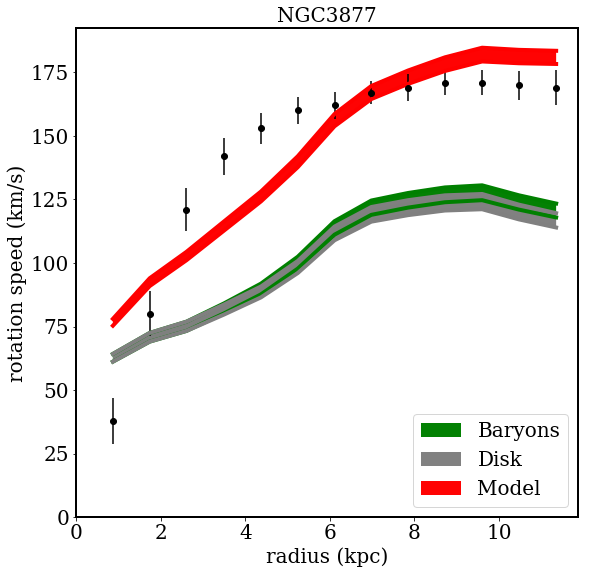} \\
\includegraphics[scale=0.25]{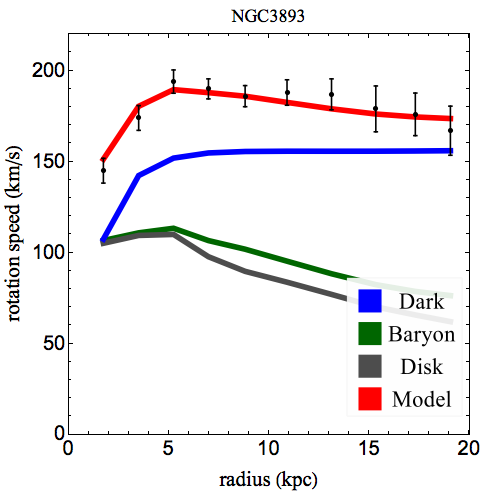} & \includegraphics[scale=0.192]{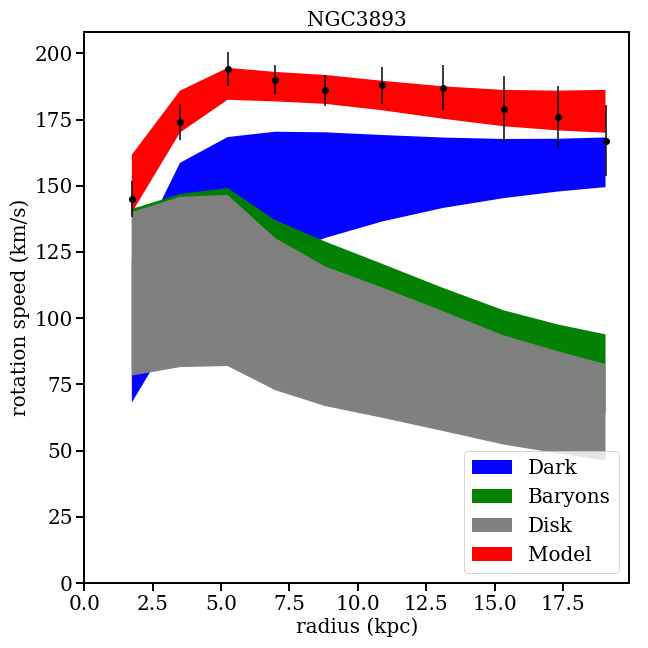} & \includegraphics[scale=0.215]{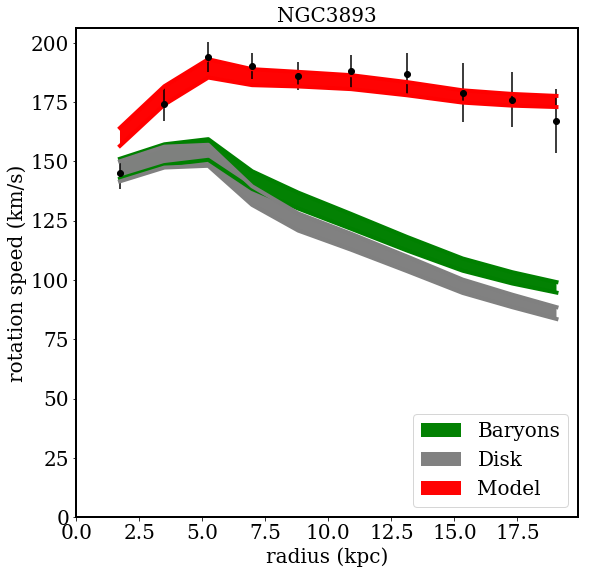} \\
\includegraphics[scale=0.25]{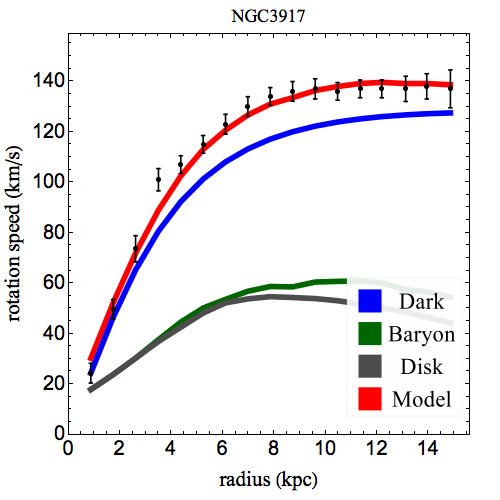} & \includegraphics[scale=0.192]{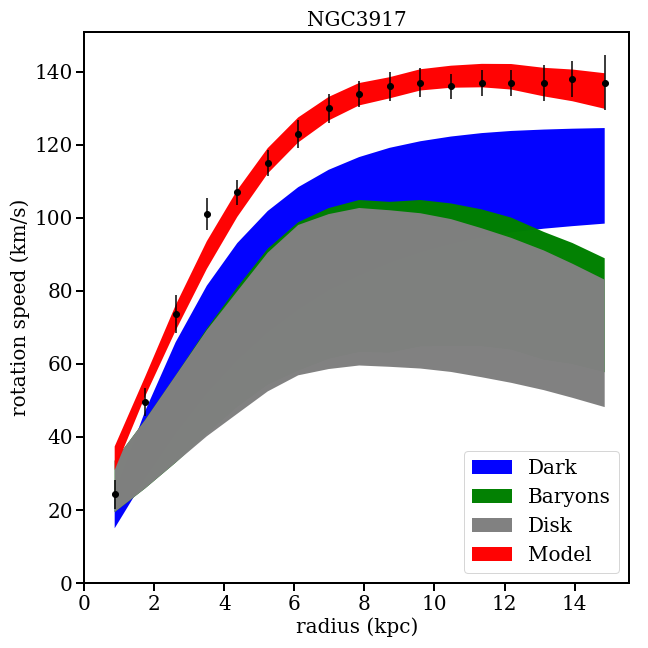} & \includegraphics[scale=0.215]{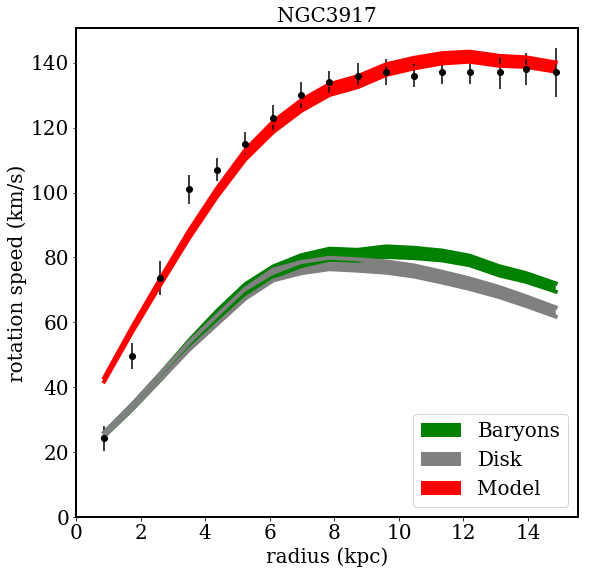} \\
\includegraphics[scale=0.25]{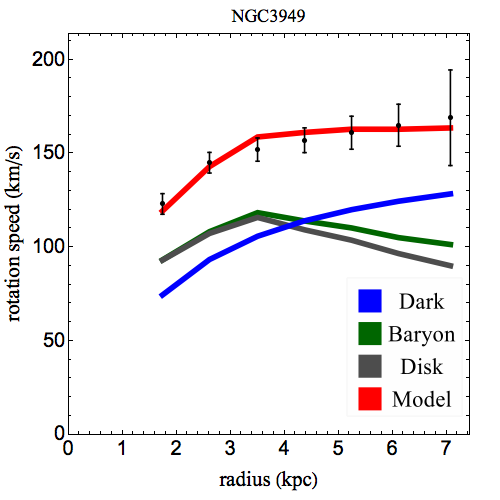} & \includegraphics[scale=0.192]{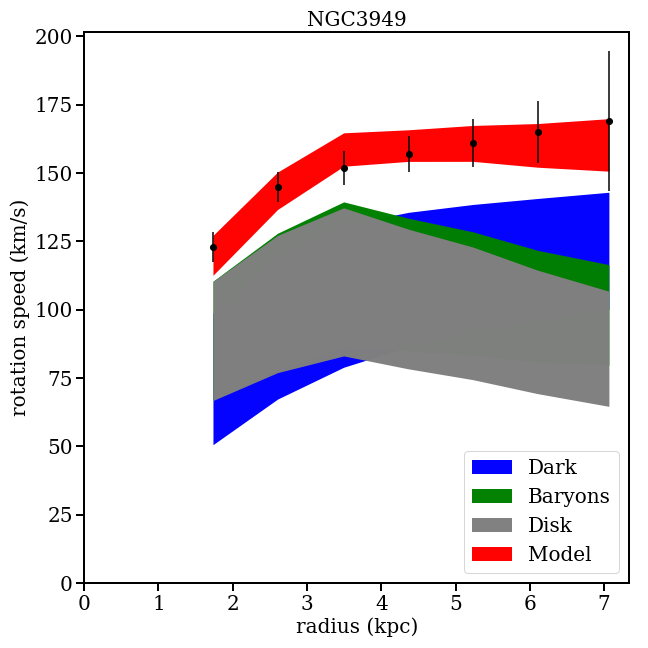} & \includegraphics[scale=0.215]{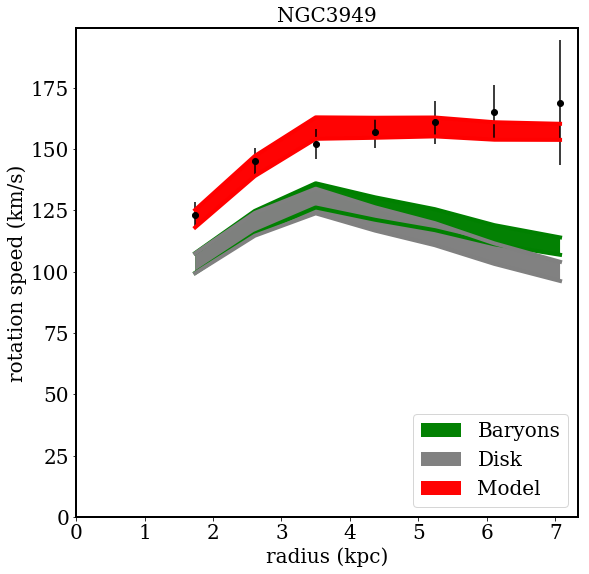} \\
\end{tabular}
\caption{Continued.}
\end{figure}

\addtocounter{figure}{-1}
\begin{figure}[H]
\centering
\begin{tabular}{@{}ccc@{}}
\includegraphics[scale=0.25]{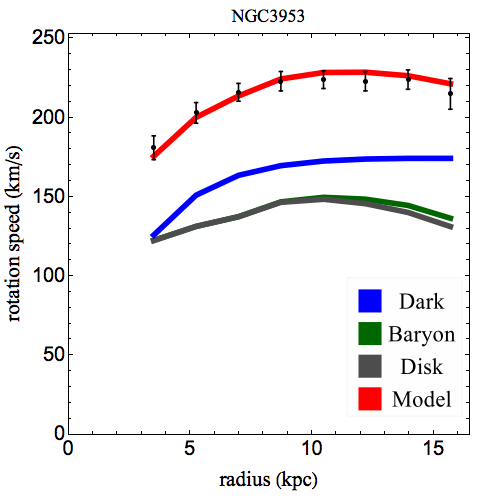} & \includegraphics[scale=0.192]{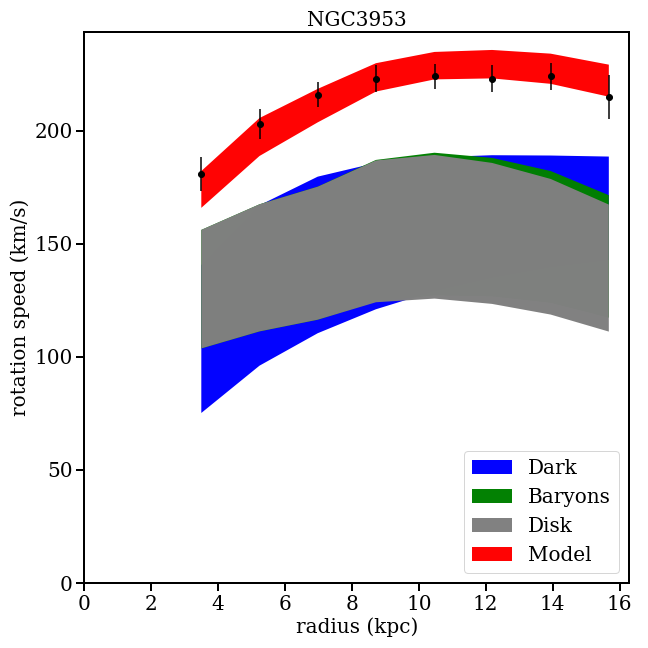} & \includegraphics[scale=0.215]{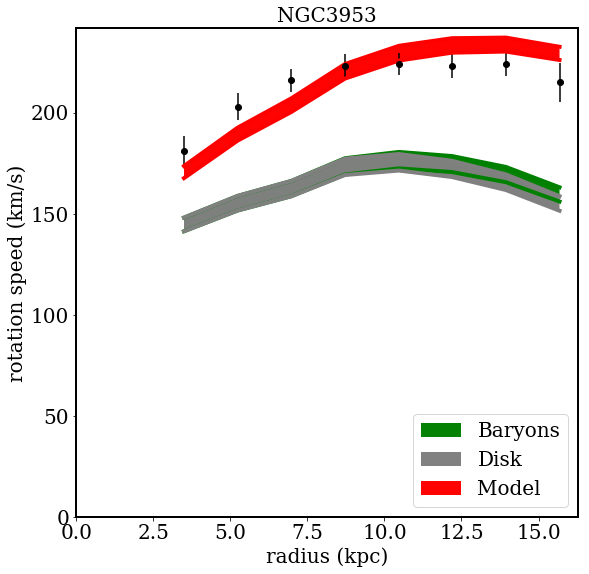} \\
\includegraphics[scale=0.25]{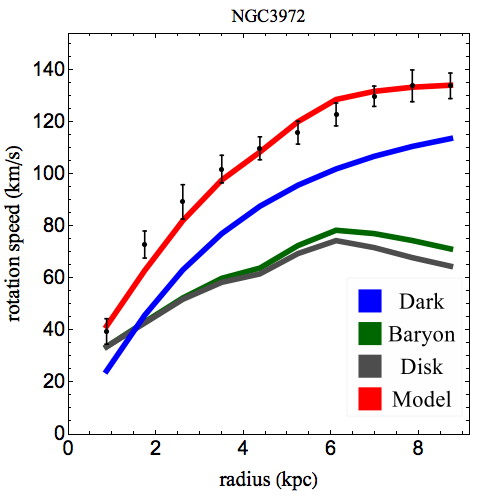} & \includegraphics[scale=0.192]{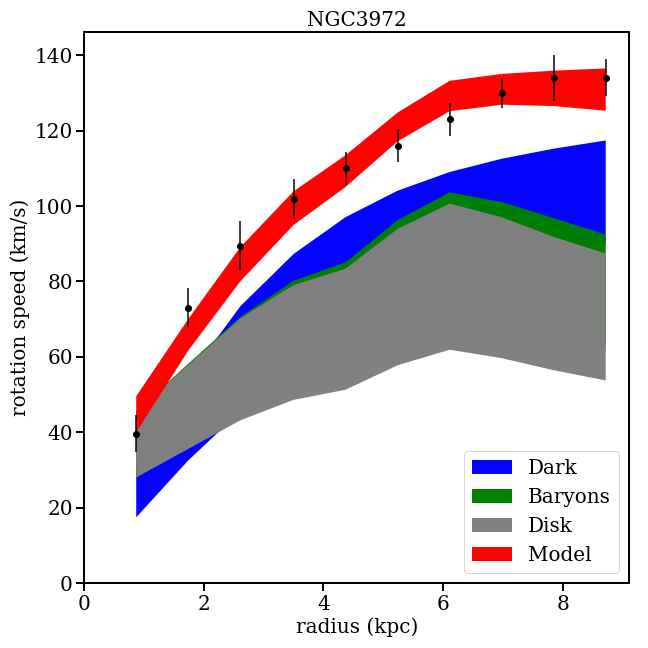} & \includegraphics[scale=0.215]{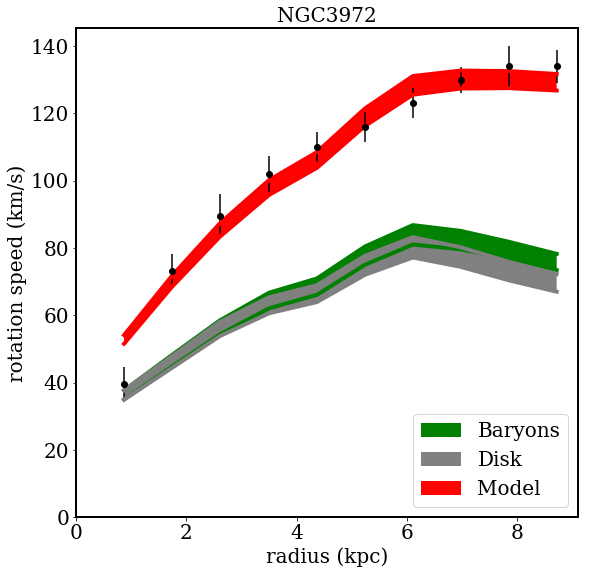} \\
\includegraphics[scale=0.25]{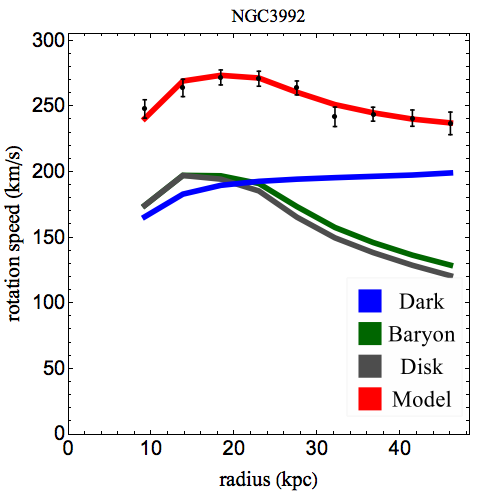} & \includegraphics[scale=0.192]{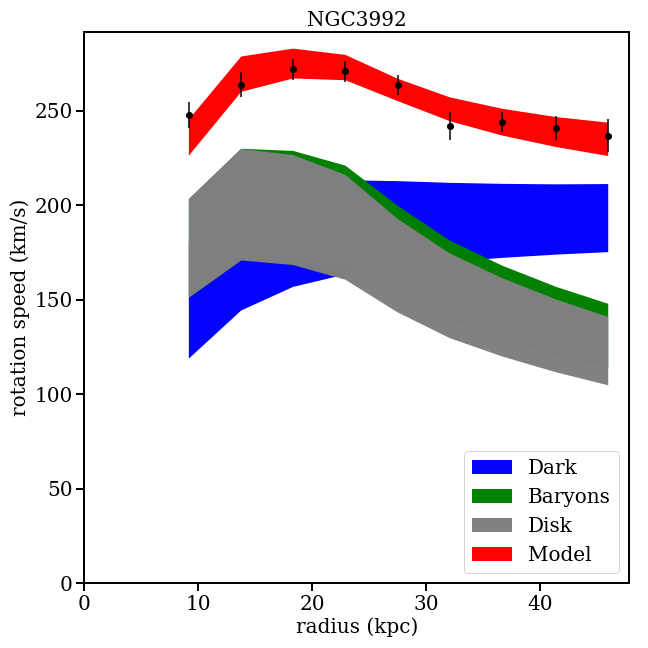} & \includegraphics[scale=0.215]{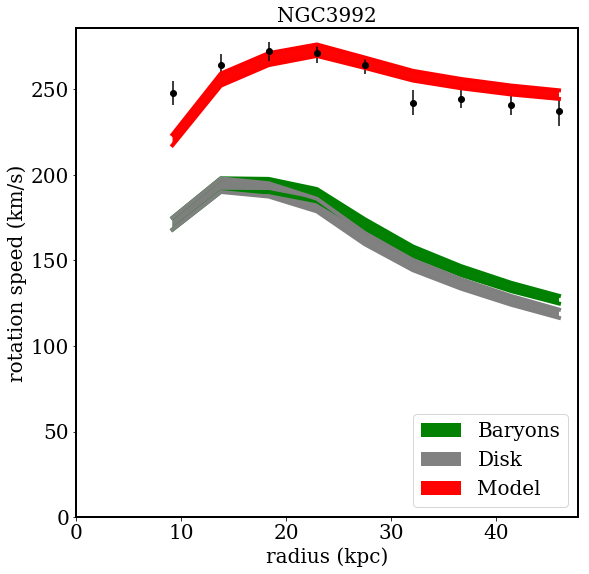} \\
\includegraphics[scale=0.25]{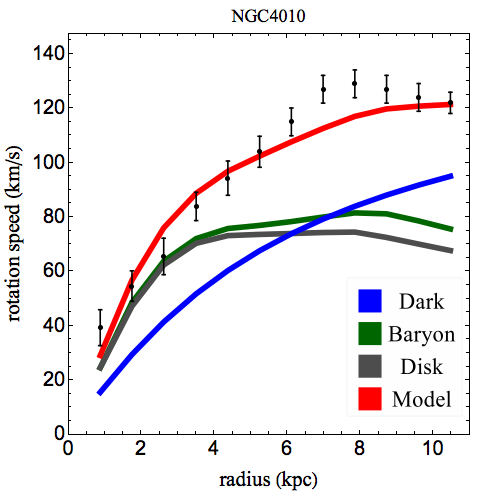} & \includegraphics[scale=0.192]{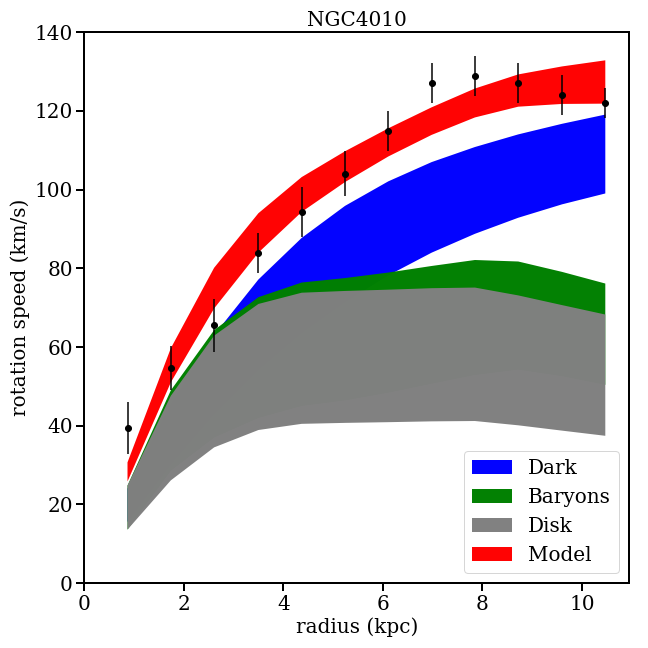} & \includegraphics[scale=0.215]{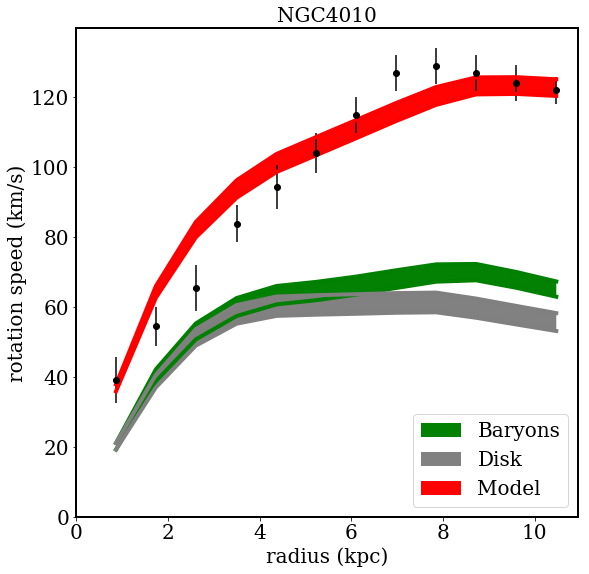} \\
\includegraphics[scale=0.25]{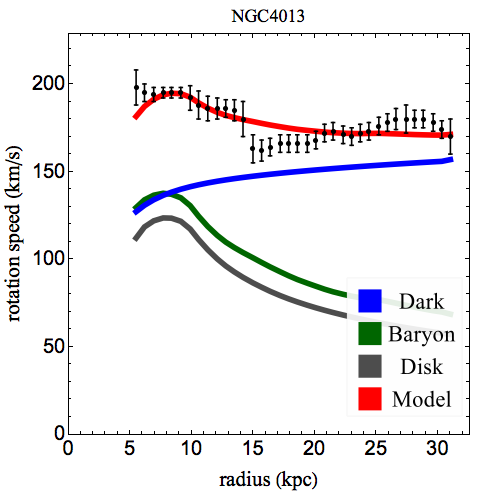} & \includegraphics[scale=0.192]{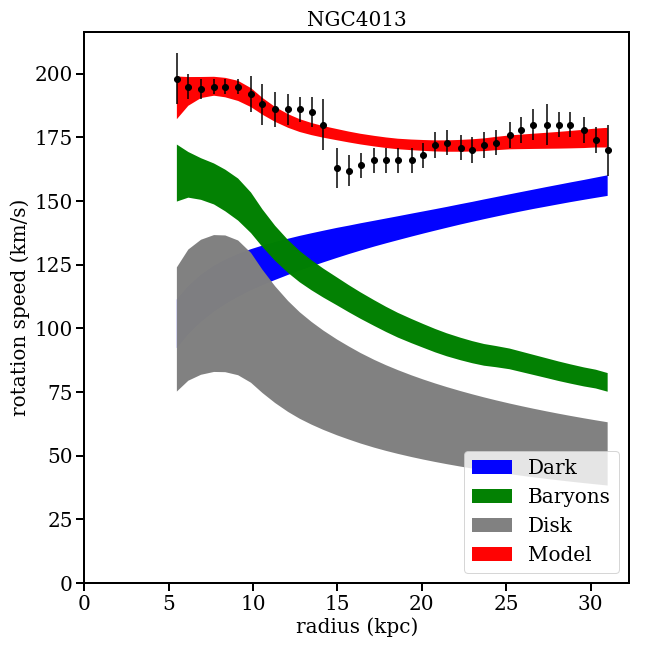} & \includegraphics[scale=0.215]{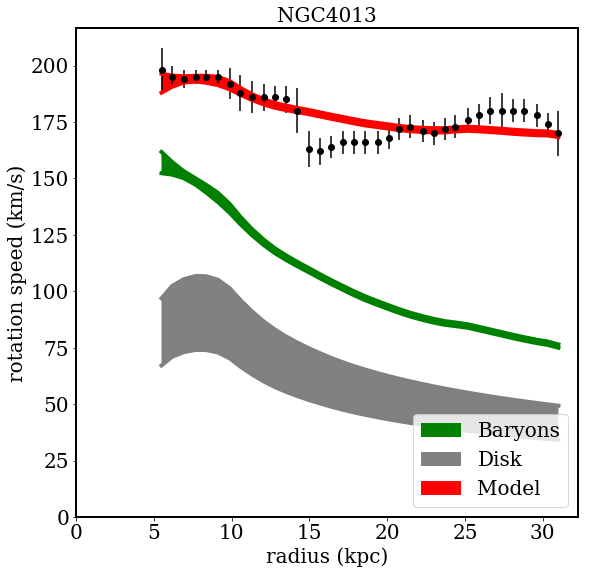} \\
\end{tabular}
\caption{Continued.}
\end{figure}

\addtocounter{figure}{-1}
\begin{figure}[H]
\centering
\begin{tabular}{@{}ccc@{}}
\includegraphics[scale=0.25]{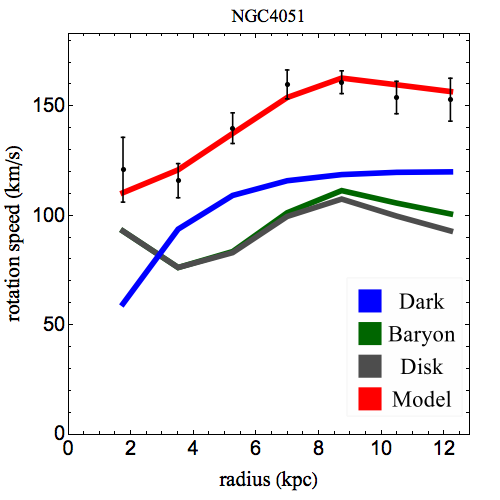} & \includegraphics[scale=0.192]{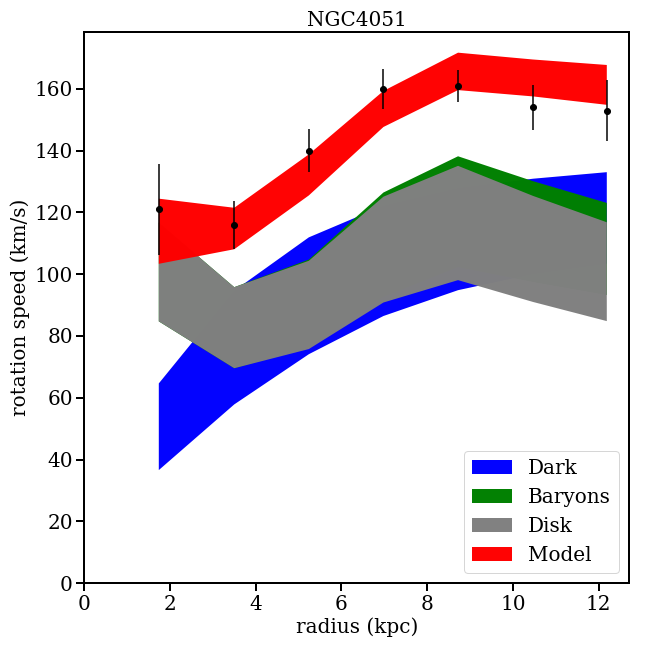} & \includegraphics[scale=0.215]{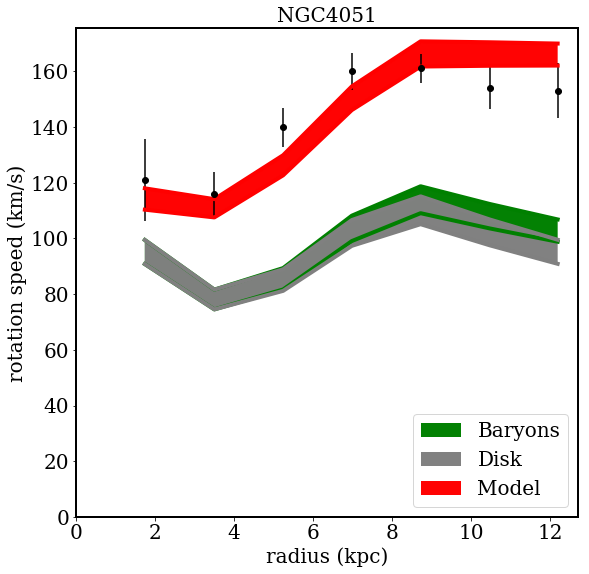} \\
\includegraphics[scale=0.25]{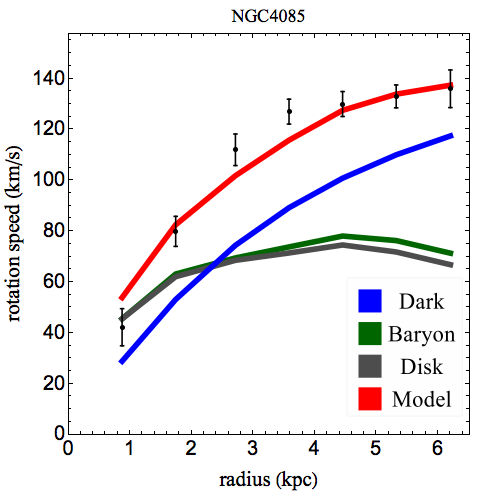} & \includegraphics[scale=0.192]{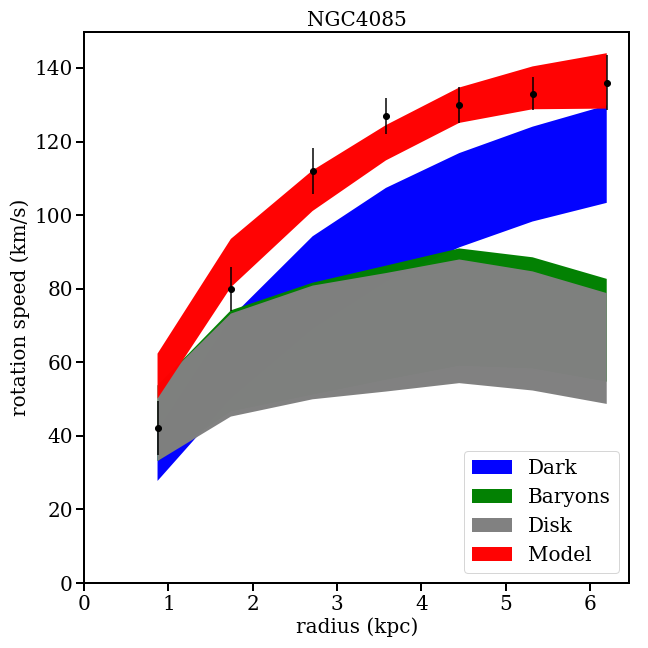} & \includegraphics[scale=0.215]{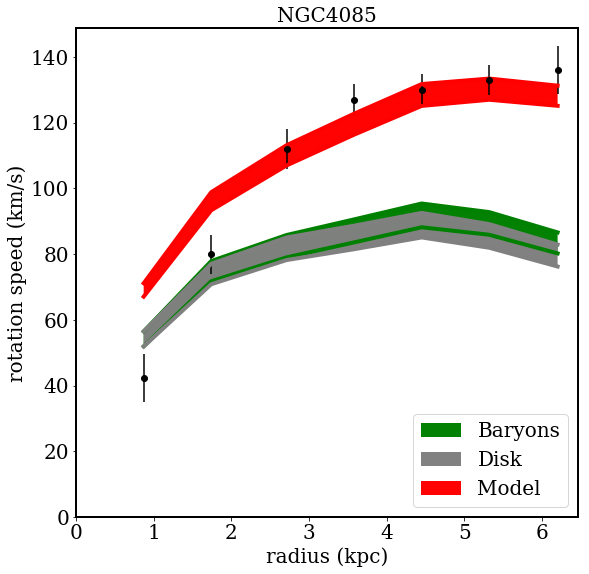} \\
\includegraphics[scale=0.25]{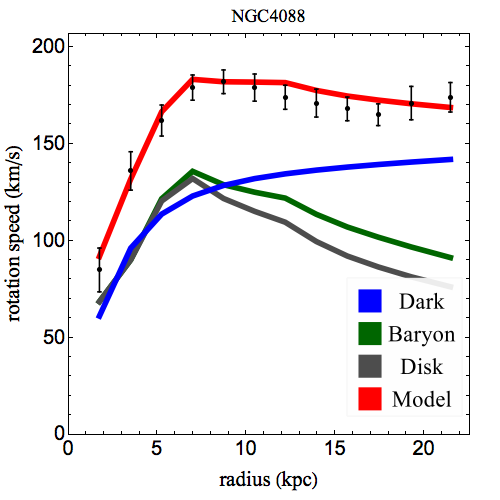} & \includegraphics[scale=0.192]{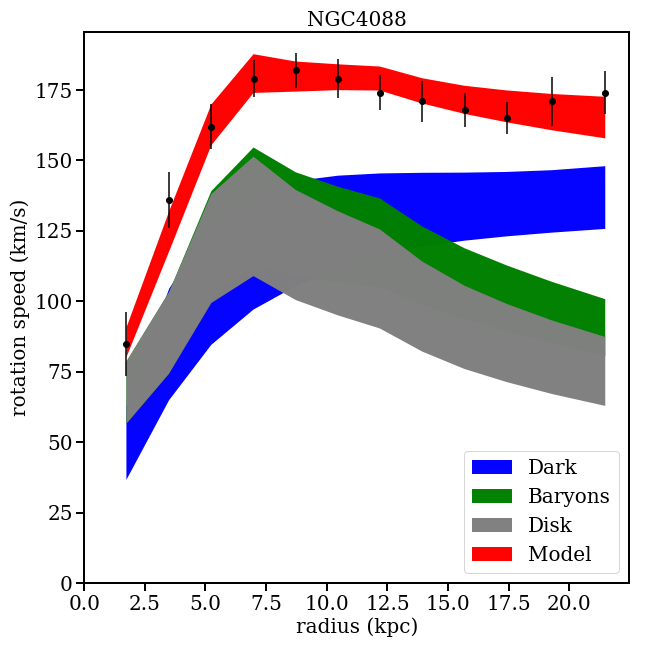} & \includegraphics[scale=0.215]{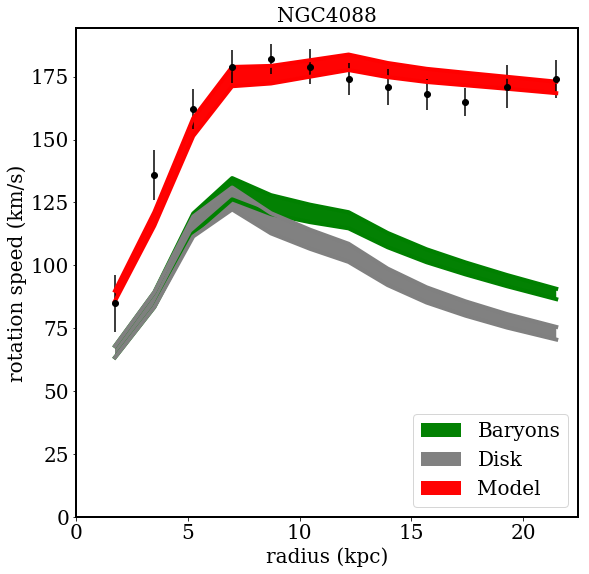} \\
\includegraphics[scale=0.25]{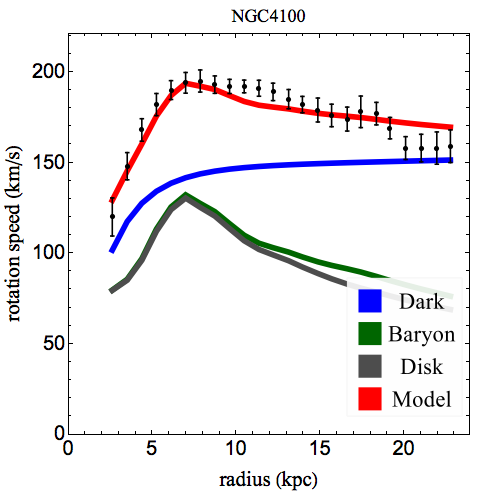} & \includegraphics[scale=0.192]{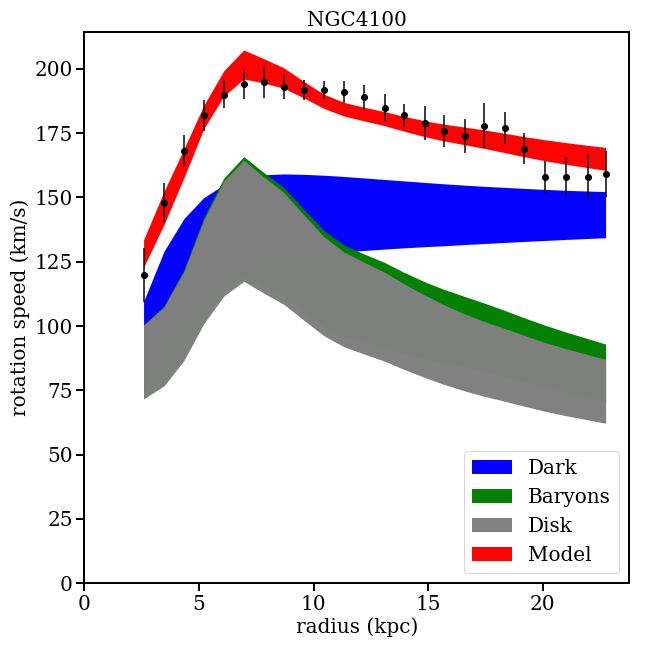} & \includegraphics[scale=0.215]{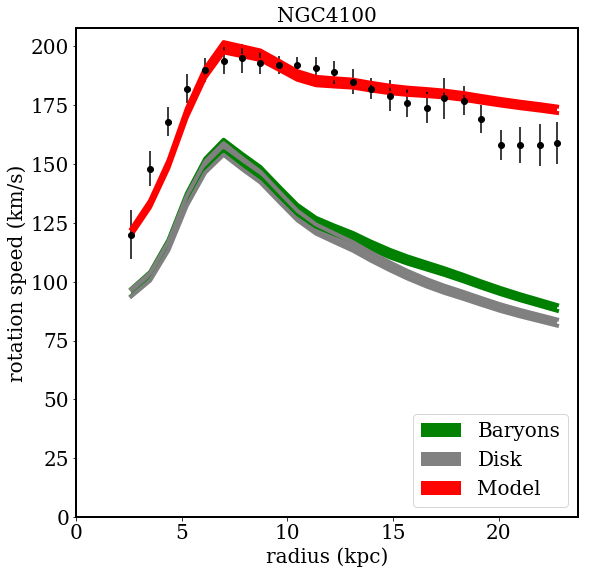} \\
\includegraphics[scale=0.25]{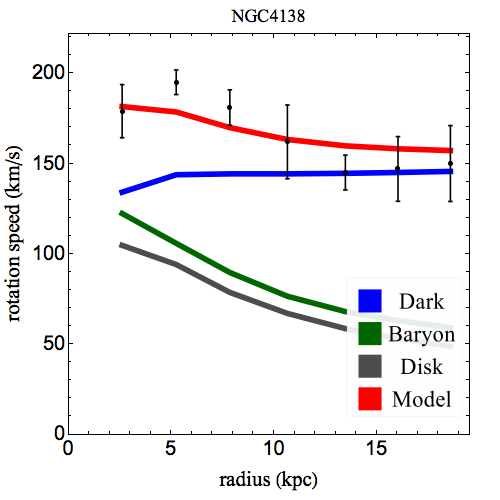} & \includegraphics[scale=0.192]{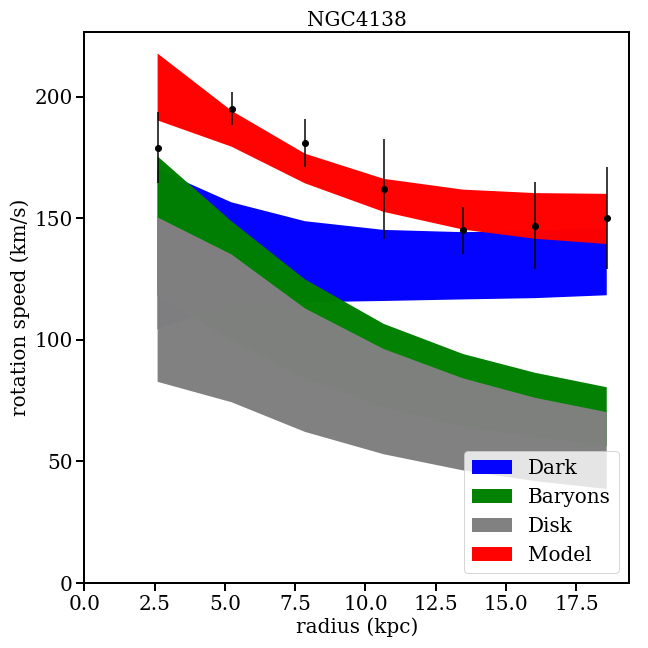} & \includegraphics[scale=0.215]{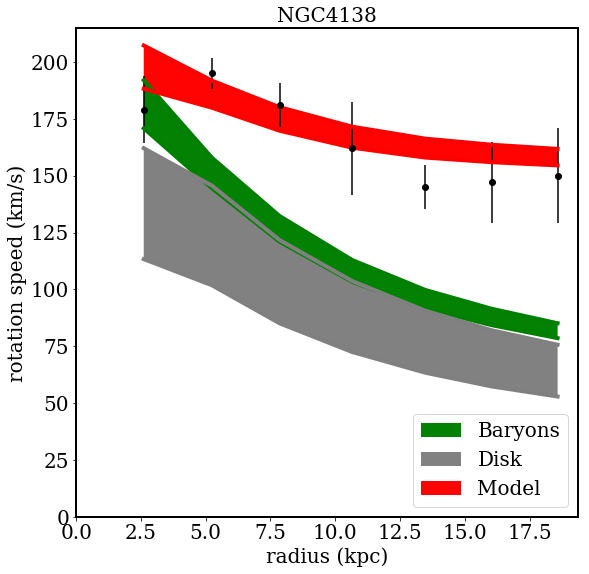} \\
\end{tabular}
\caption{Continued.}
\end{figure}

\addtocounter{figure}{-1}
\begin{figure}[H]
\centering
\begin{tabular}{@{}ccc@{}}
\includegraphics[scale=0.25]{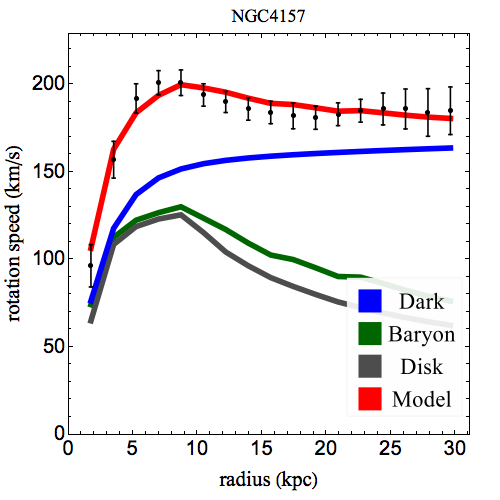} & \includegraphics[scale=0.192]{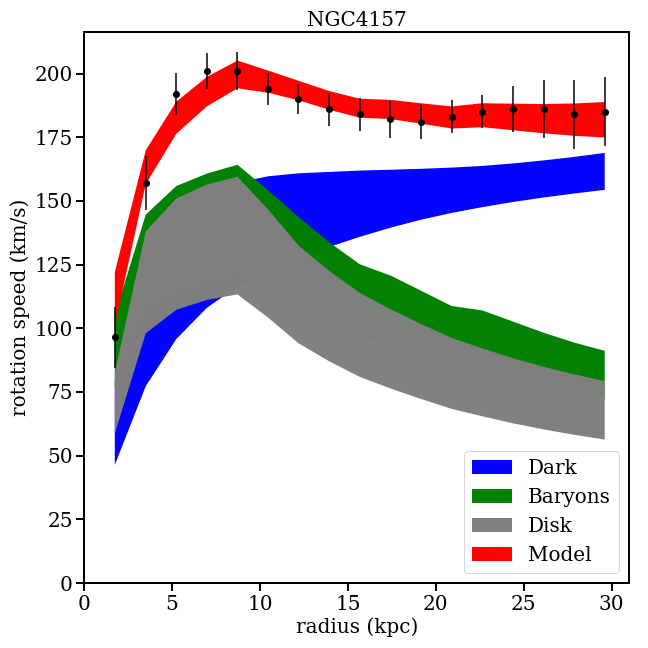} & \includegraphics[scale=0.215]{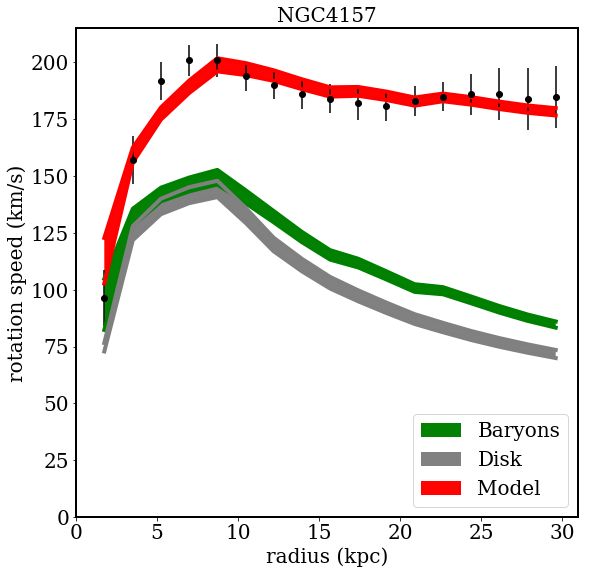} \\
\includegraphics[scale=0.25]{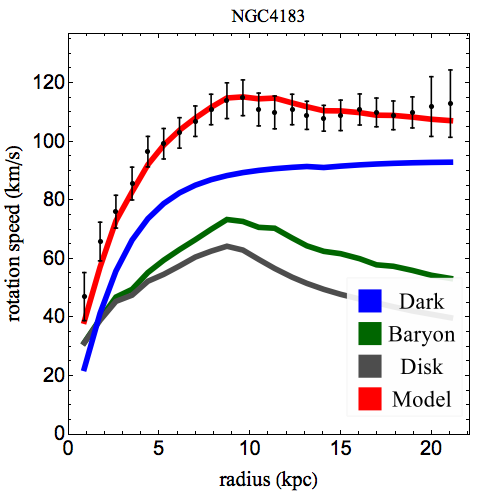} & \includegraphics[scale=0.192]{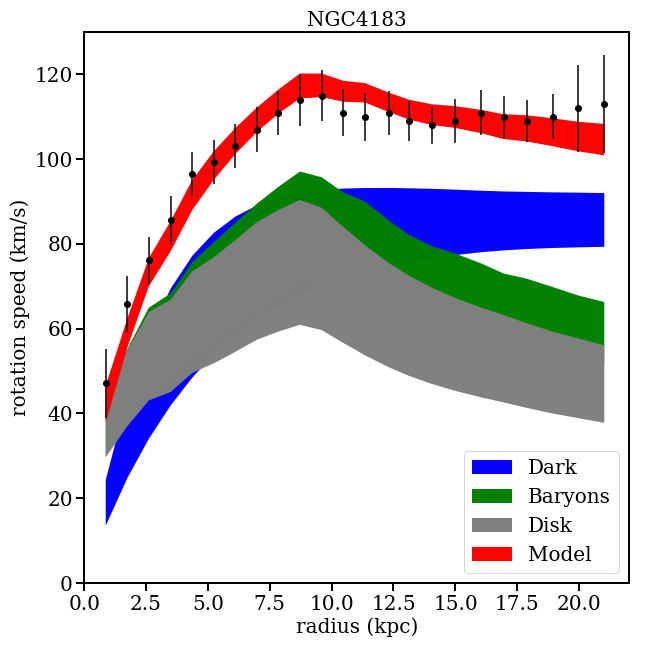} & \includegraphics[scale=0.215]{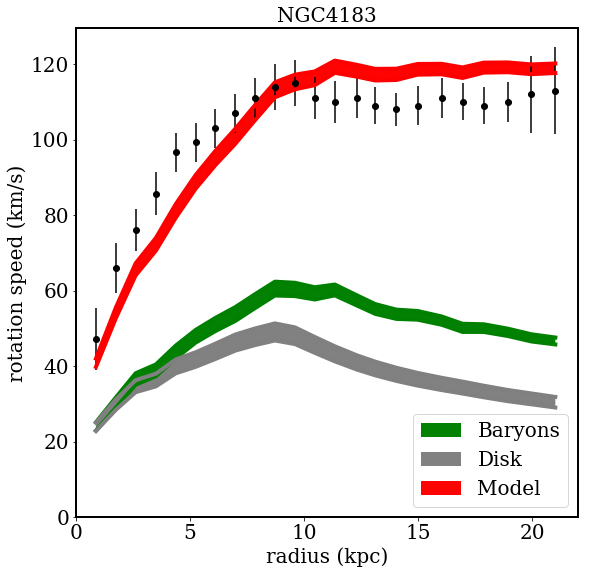} \\
\includegraphics[scale=0.25]{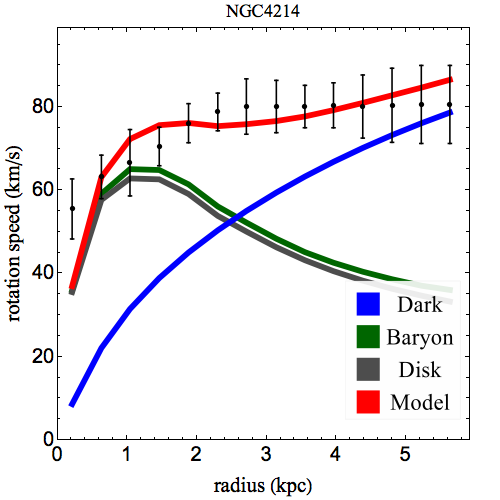} & \includegraphics[scale=0.192]{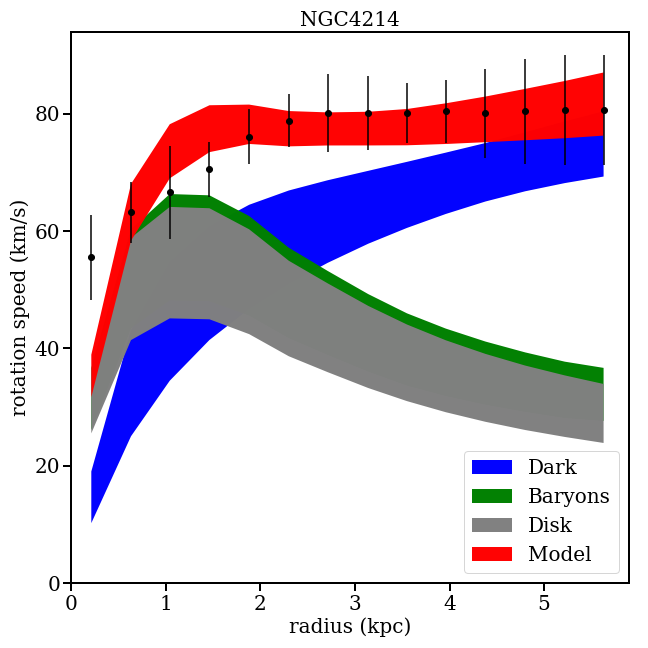} & \includegraphics[scale=0.215]{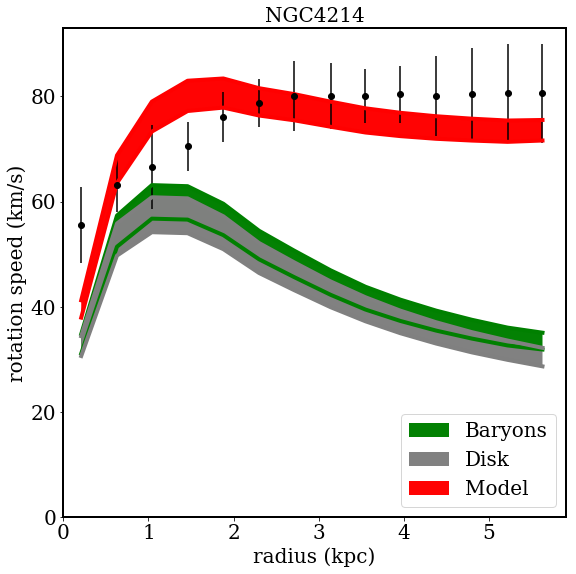} \\
\includegraphics[scale=0.25]{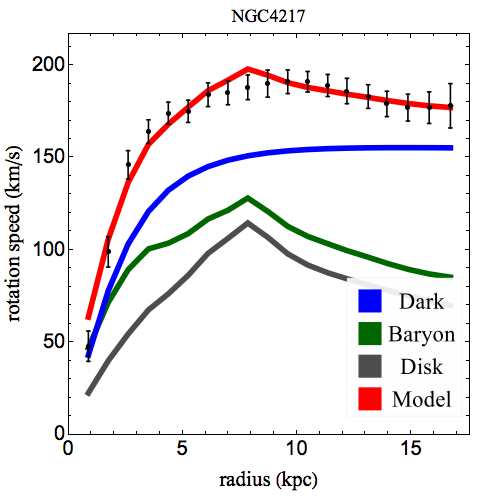} & \includegraphics[scale=0.192]{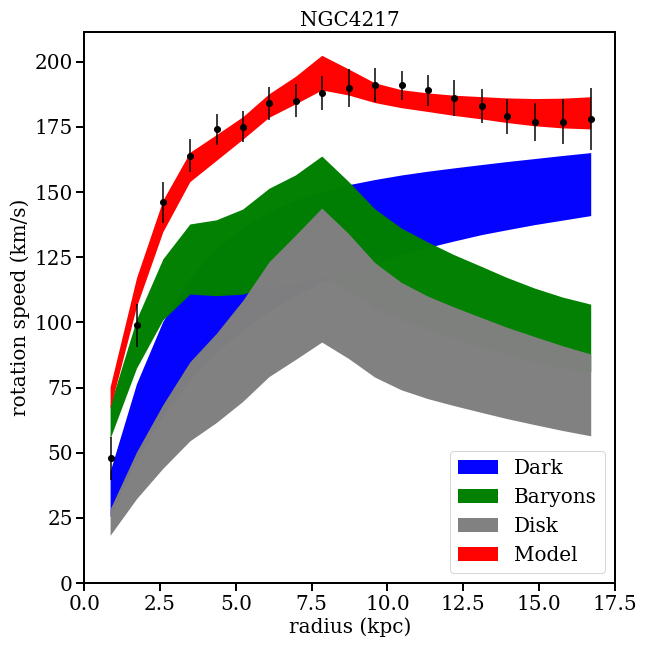} & \includegraphics[scale=0.215]{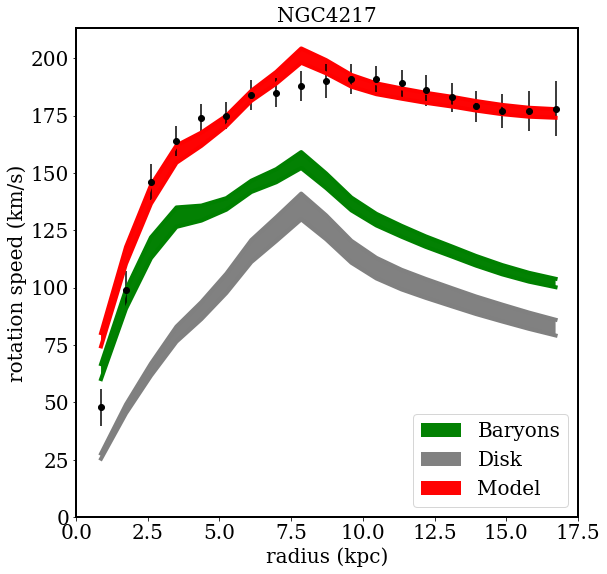} \\
\includegraphics[scale=0.25]{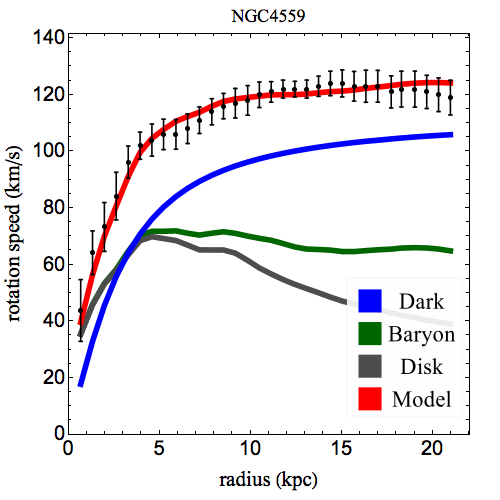} & \includegraphics[scale=0.192]{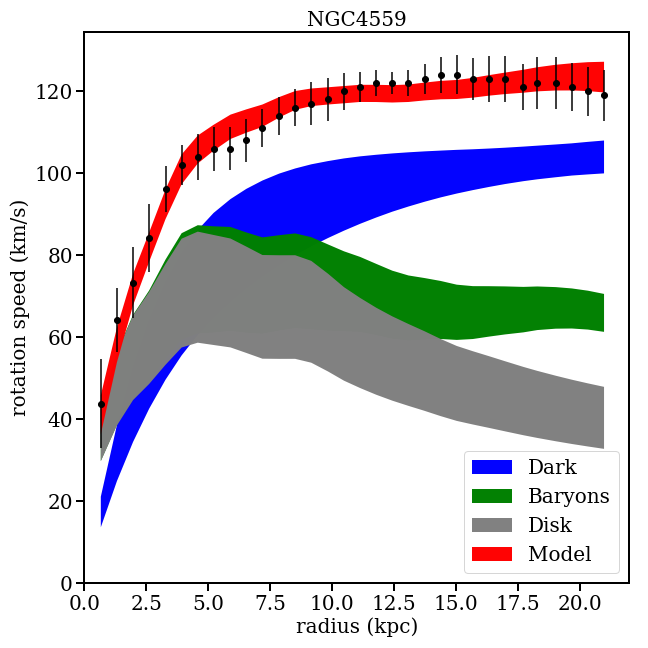} & \includegraphics[scale=0.215]{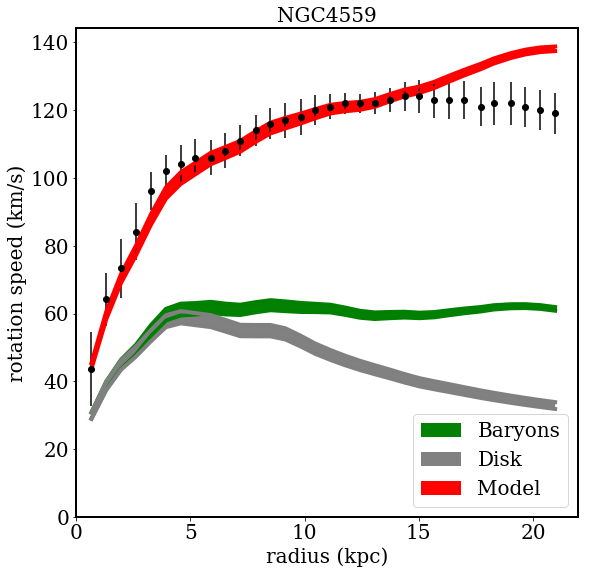} \\
\end{tabular}
\caption{Continued.}
\end{figure}

\addtocounter{figure}{-1}
\begin{figure}[H]
\centering
\begin{tabular}{@{}ccc@{}}
\includegraphics[scale=0.25]{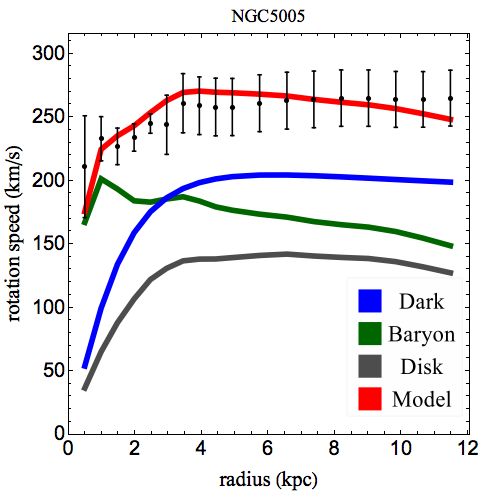} & \includegraphics[scale=0.192]{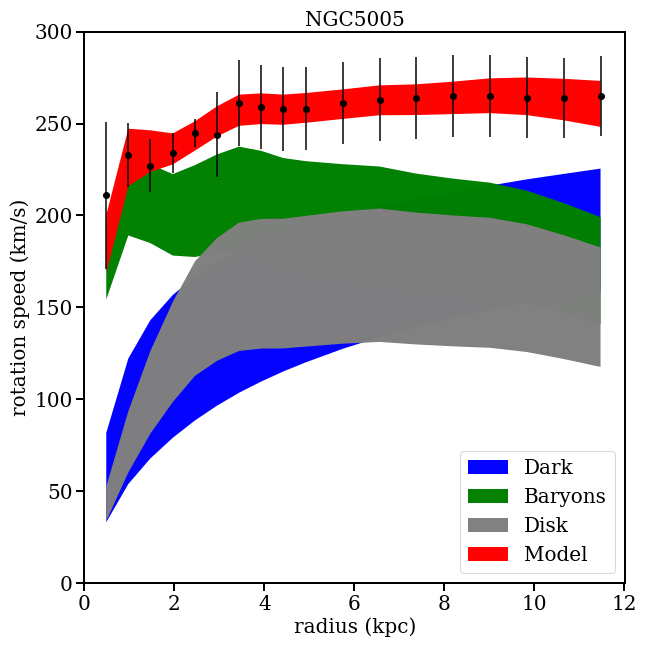} & \includegraphics[scale=0.215]{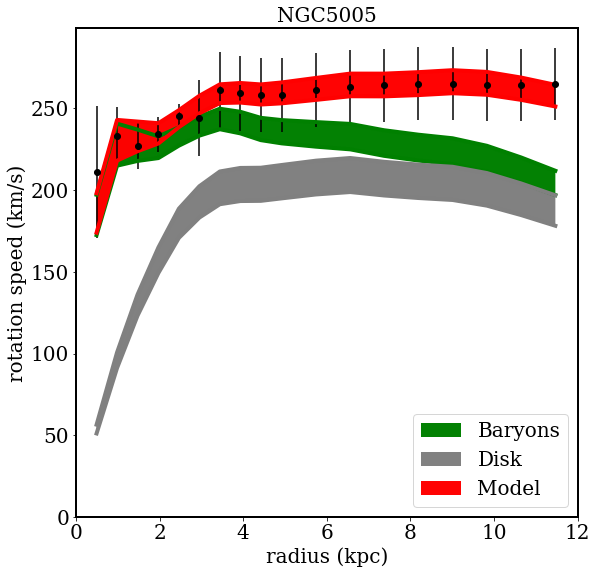} \\
\includegraphics[scale=0.25]{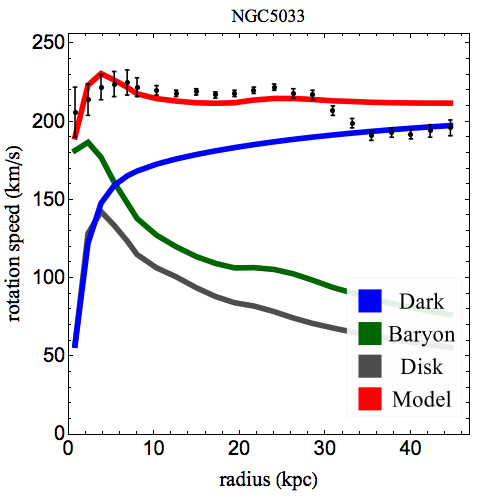} & \includegraphics[scale=0.192]{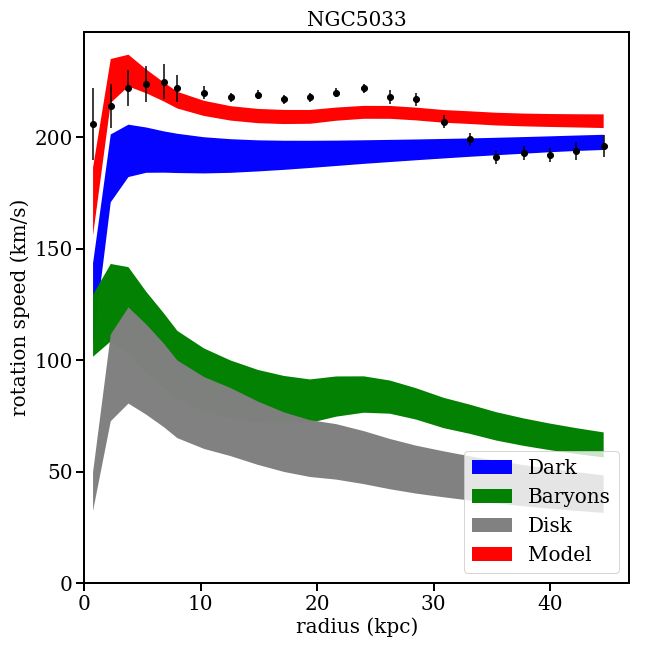} & \includegraphics[scale=0.215]{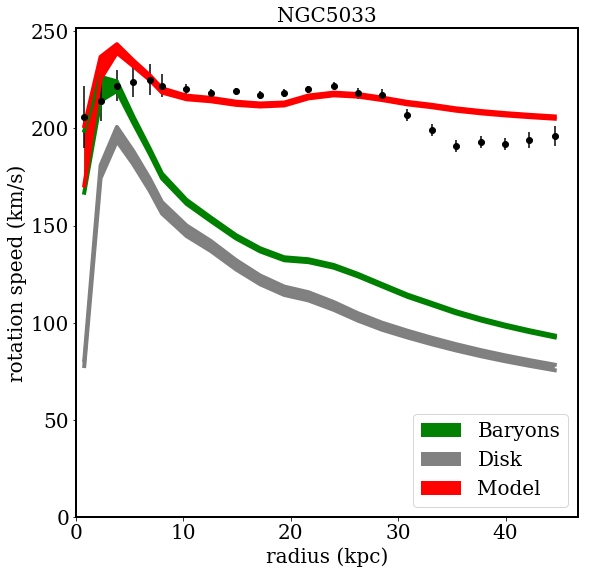} \\
\includegraphics[scale=0.25]{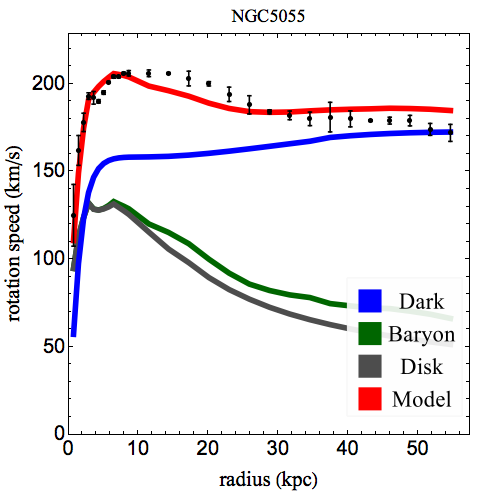} & \includegraphics[scale=0.192]{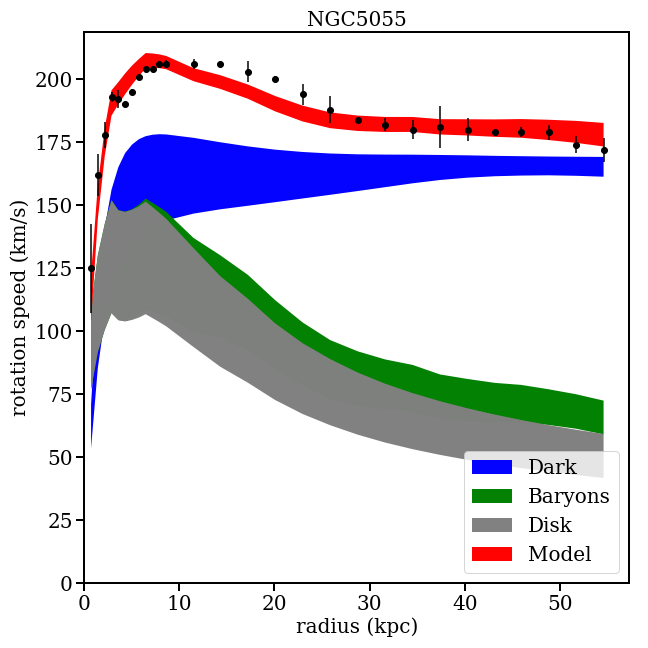} & \includegraphics[scale=0.215]{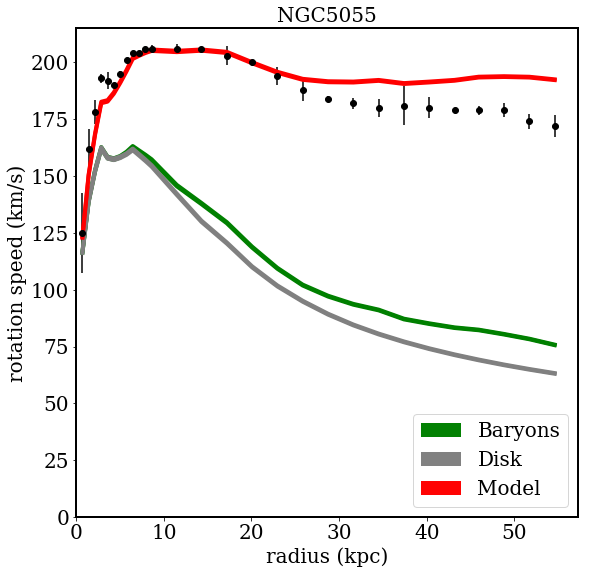} \\
\includegraphics[scale=0.25]{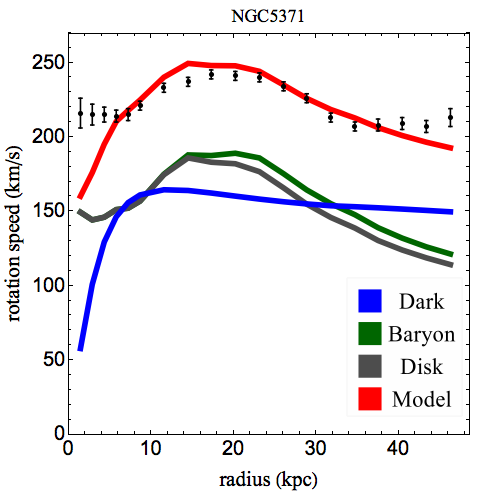} & \includegraphics[scale=0.192]{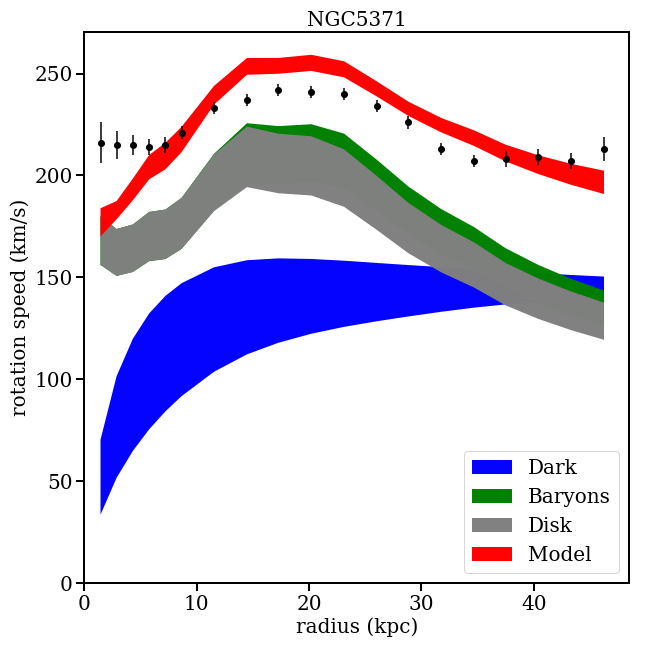} & \includegraphics[scale=0.215]{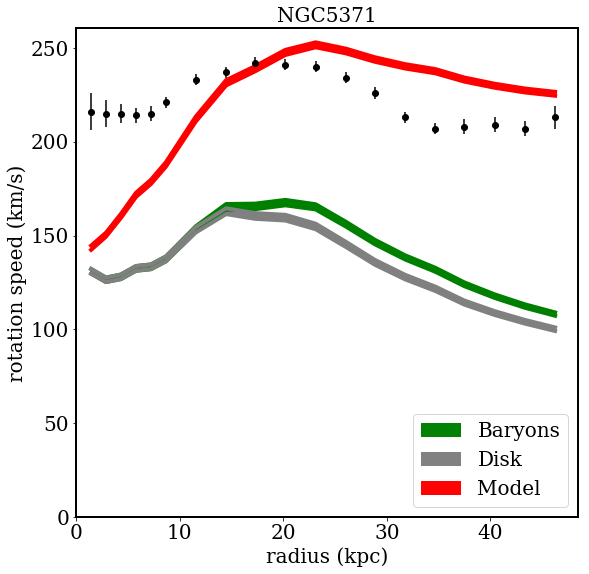} \\
\includegraphics[scale=0.25]{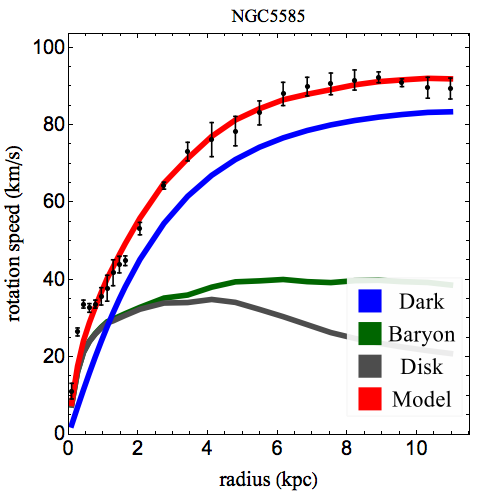} & \includegraphics[scale=0.192]{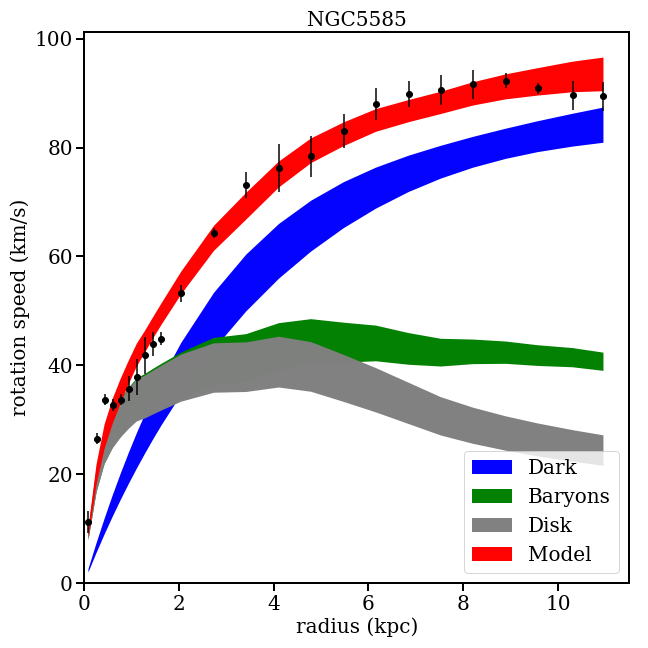} & \includegraphics[scale=0.215]{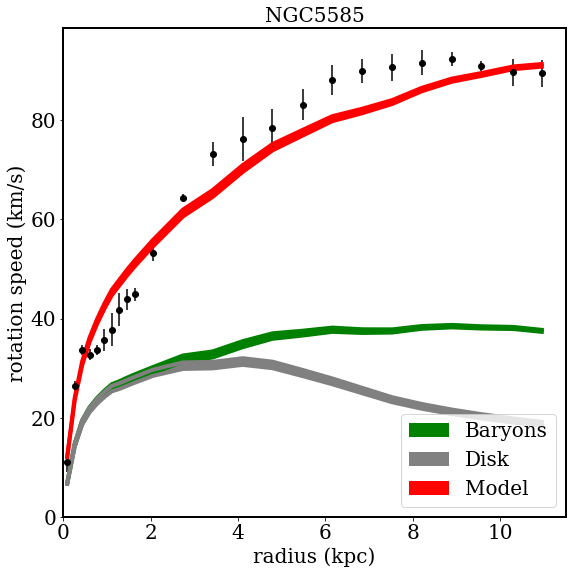} \\
\end{tabular}
\caption{Continued.}
\end{figure}

\addtocounter{figure}{-1}
\begin{figure}[H]
\centering
\begin{tabular}{@{}ccc@{}}
\includegraphics[scale=0.25]{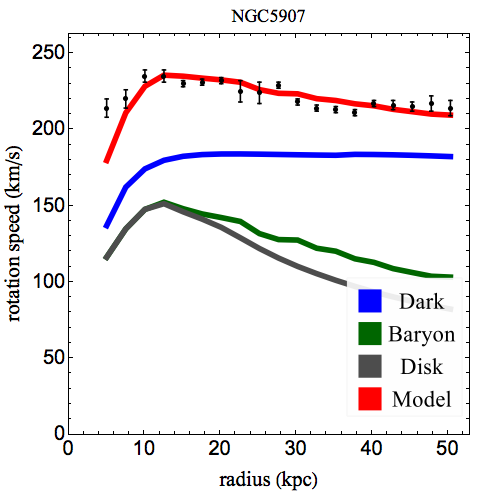} & \includegraphics[scale=0.192]{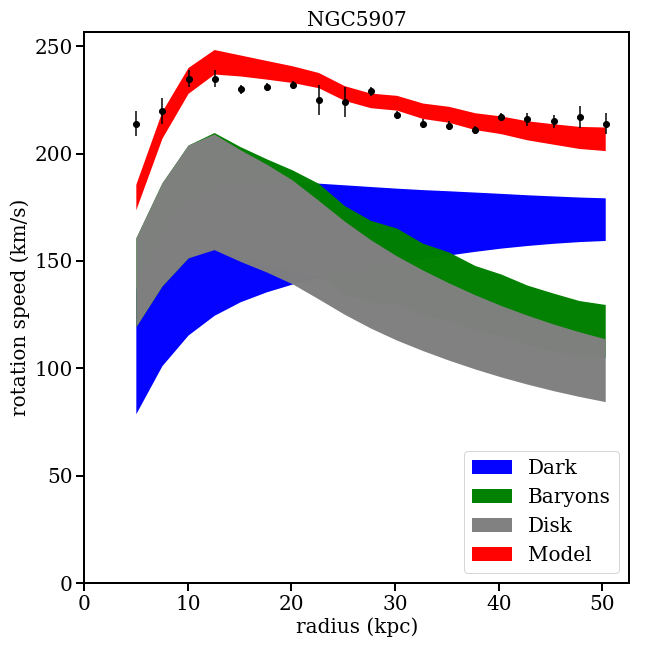} & \includegraphics[scale=0.215]{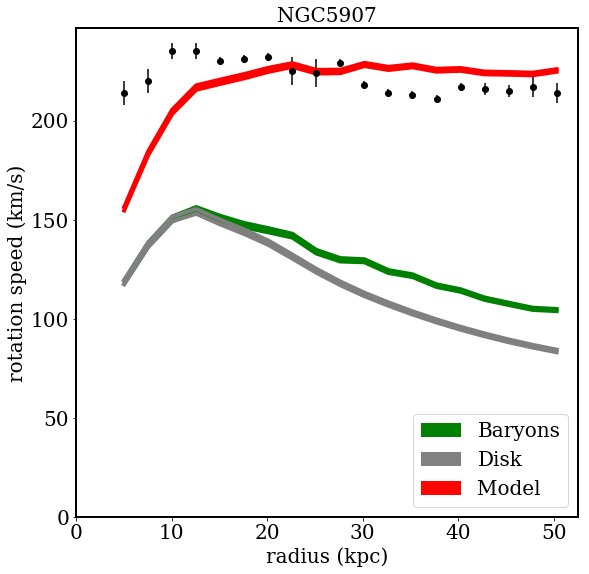} \\
\includegraphics[scale=0.25]{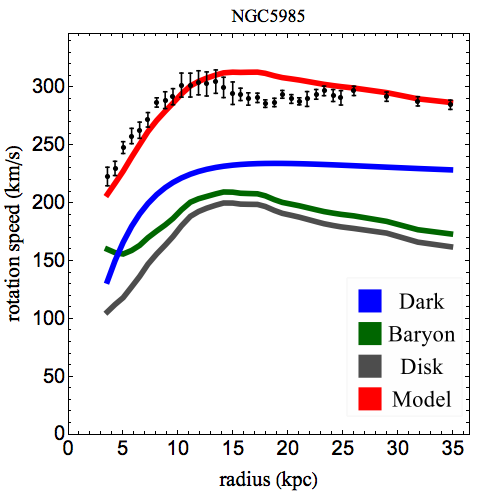} & \includegraphics[scale=0.192]{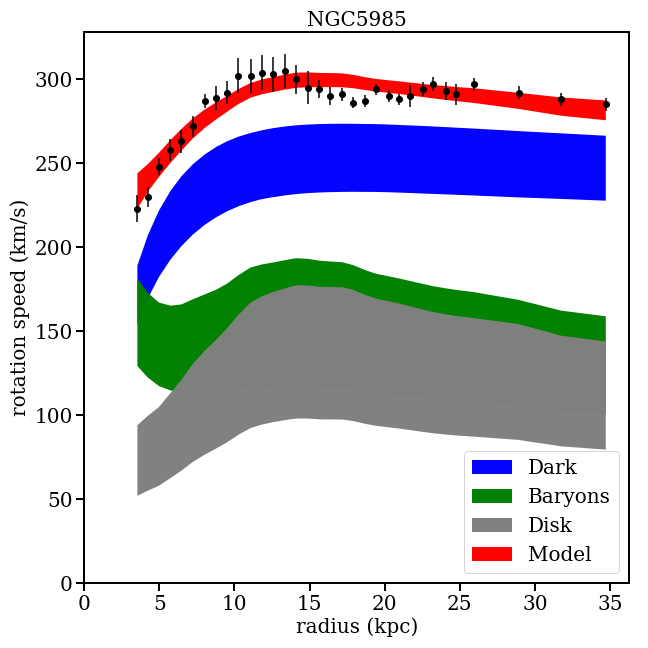} & \includegraphics[scale=0.215]{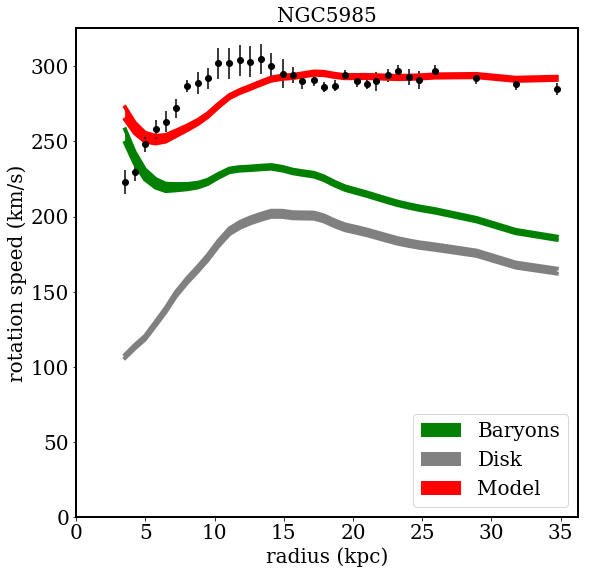} \\
\includegraphics[scale=0.25]{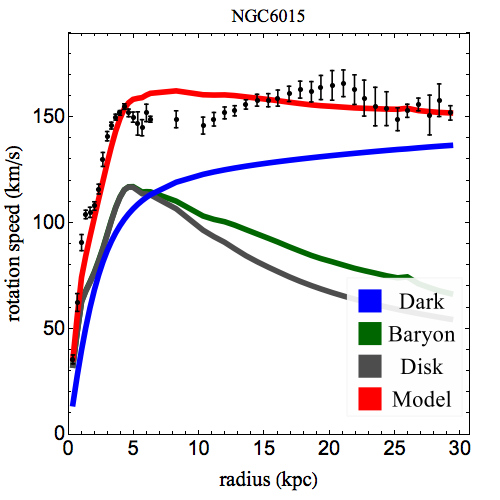} & \includegraphics[scale=0.192]{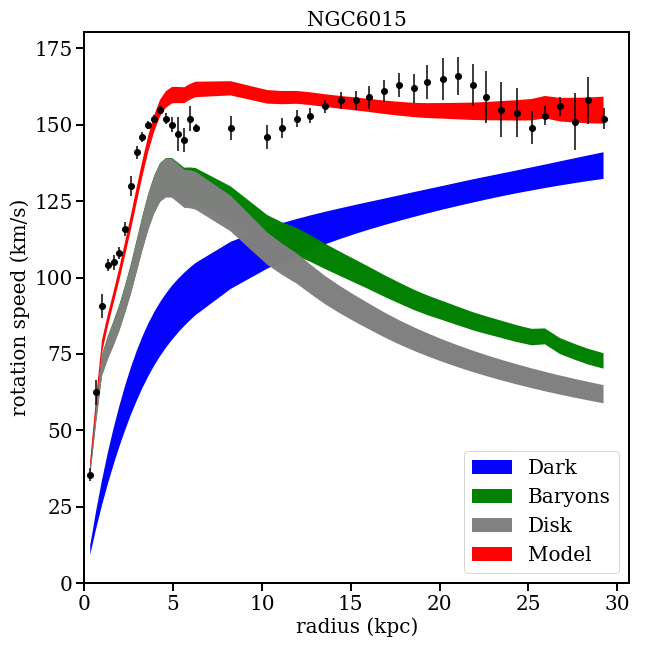} & \includegraphics[scale=0.215]{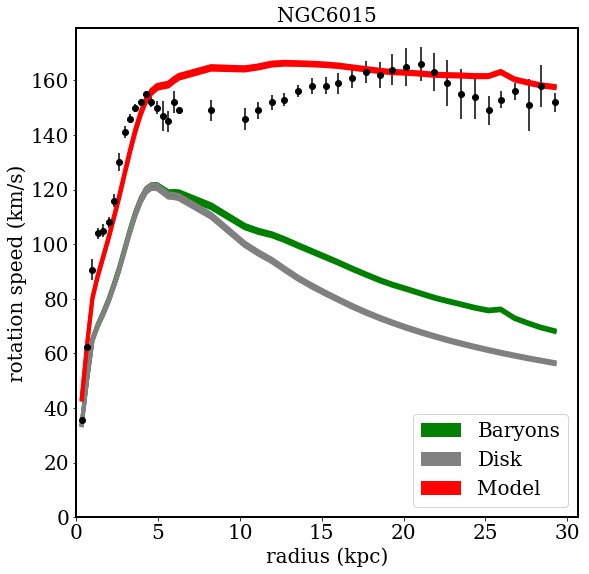} \\
\includegraphics[scale=0.25]{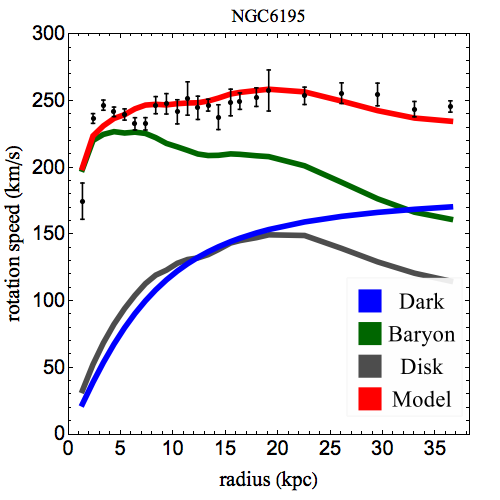} & \includegraphics[scale=0.192]{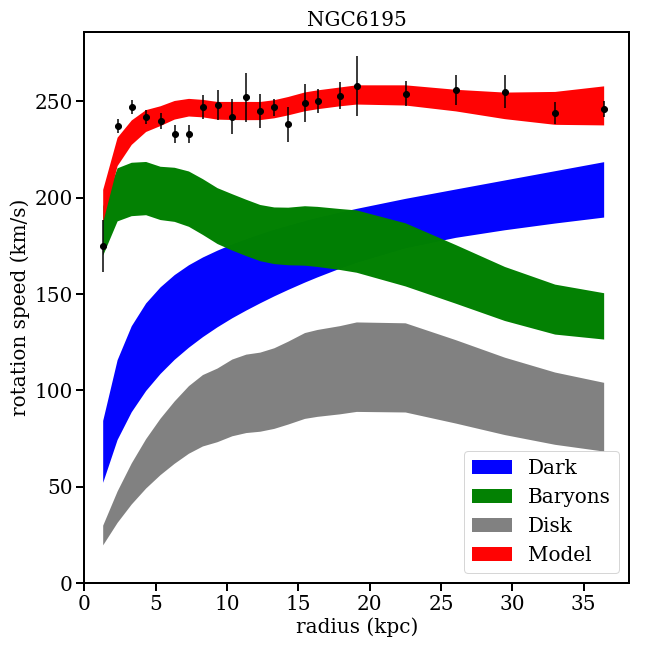} & \includegraphics[scale=0.215]{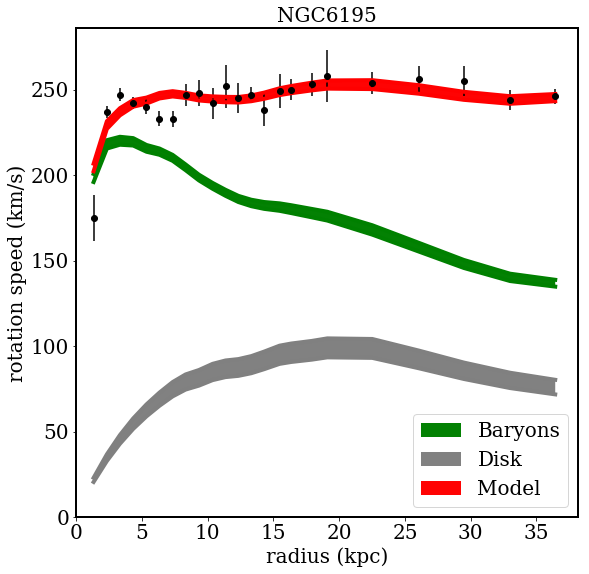} \\
\includegraphics[scale=0.25]{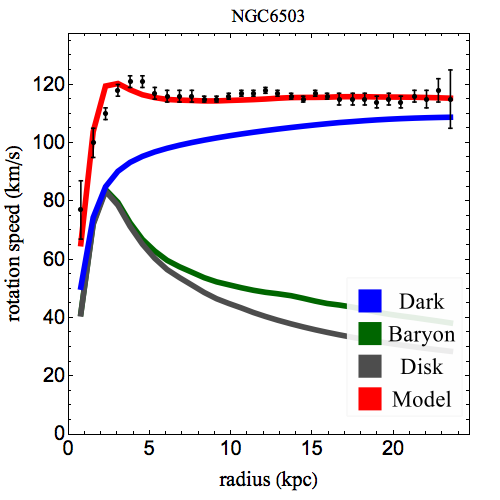} & \includegraphics[scale=0.192]{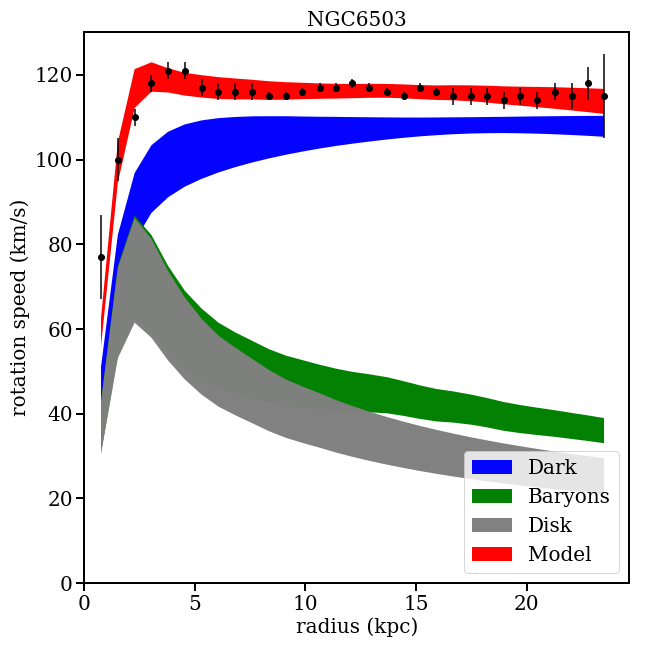} & \includegraphics[scale=0.215]{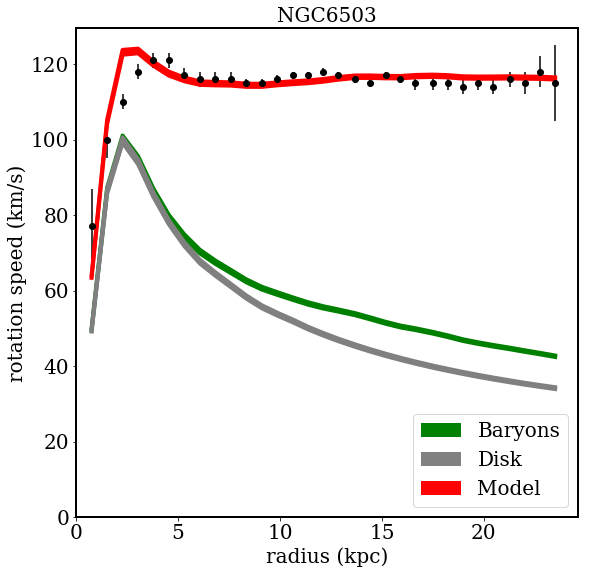} \\
\end{tabular}
\caption{Continued.}
\end{figure}

\addtocounter{figure}{-1}
\begin{figure}[H]
\centering
\begin{tabular}{@{}ccc@{}}
\includegraphics[scale=0.25]{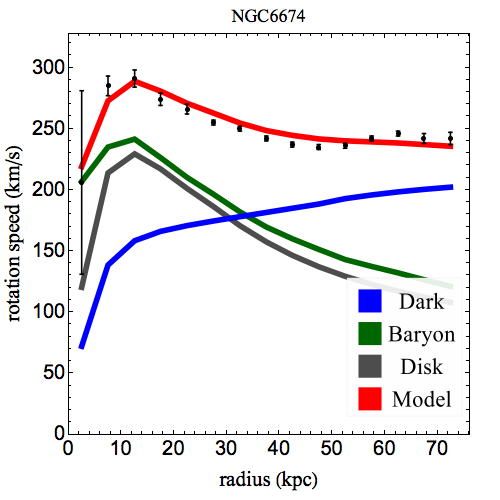} & \includegraphics[scale=0.192]{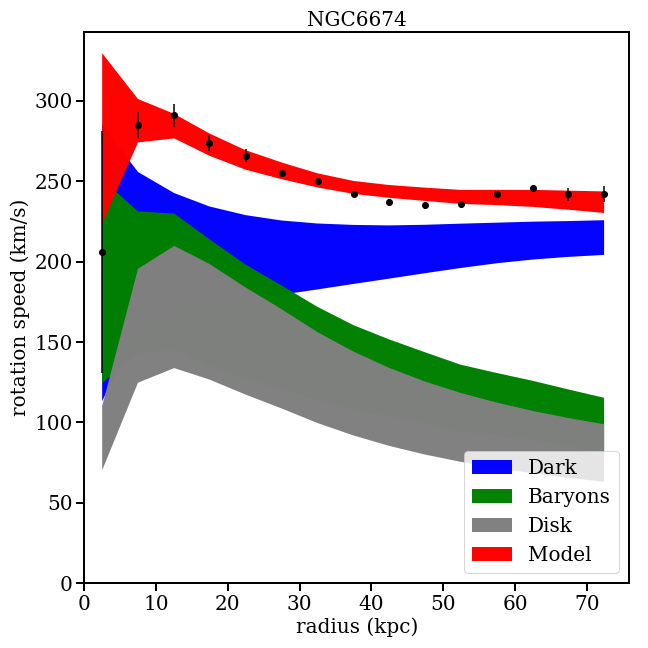} & \includegraphics[scale=0.215]{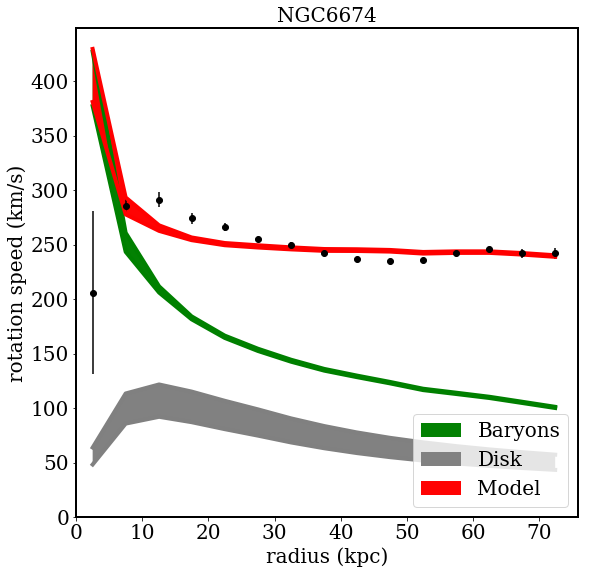} \\
\includegraphics[scale=0.25]{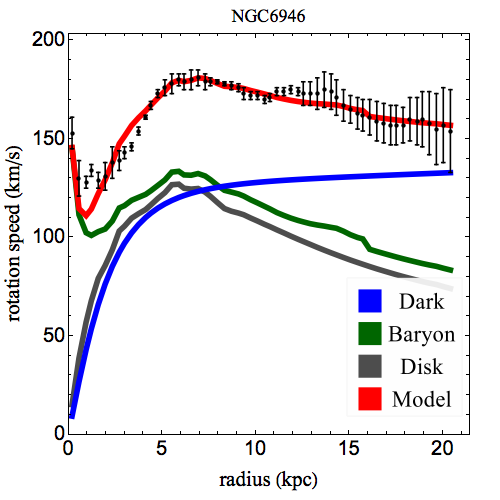} & \includegraphics[scale=0.192]{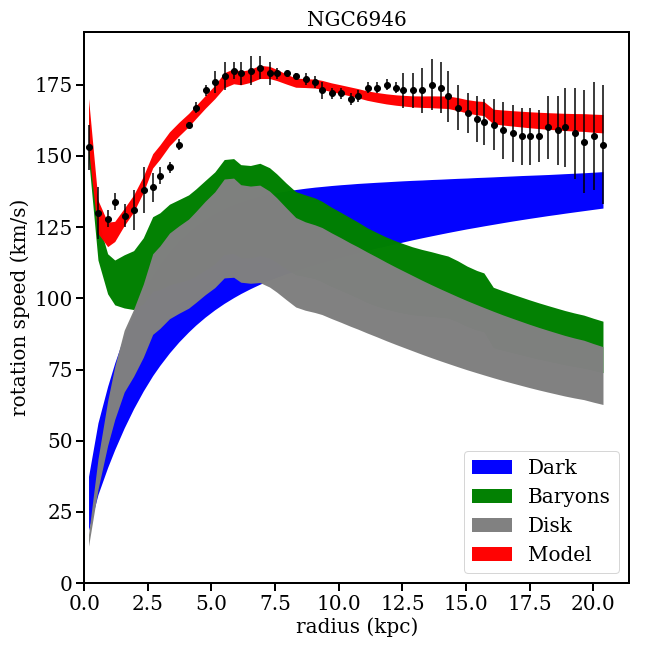} & \includegraphics[scale=0.215]{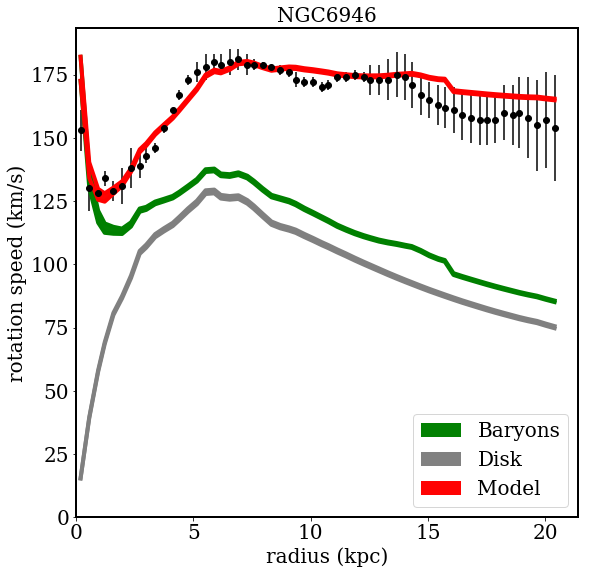} \\
\includegraphics[scale=0.25]{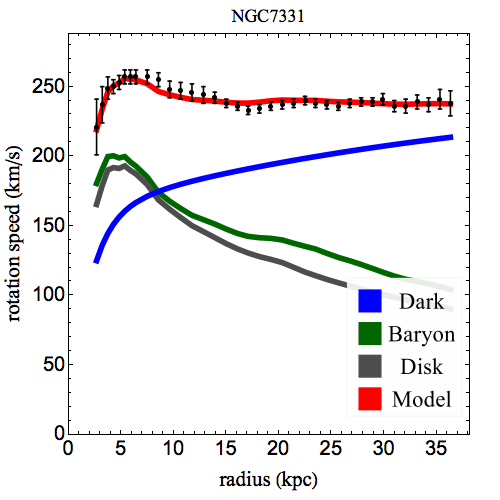} & \includegraphics[scale=0.192]{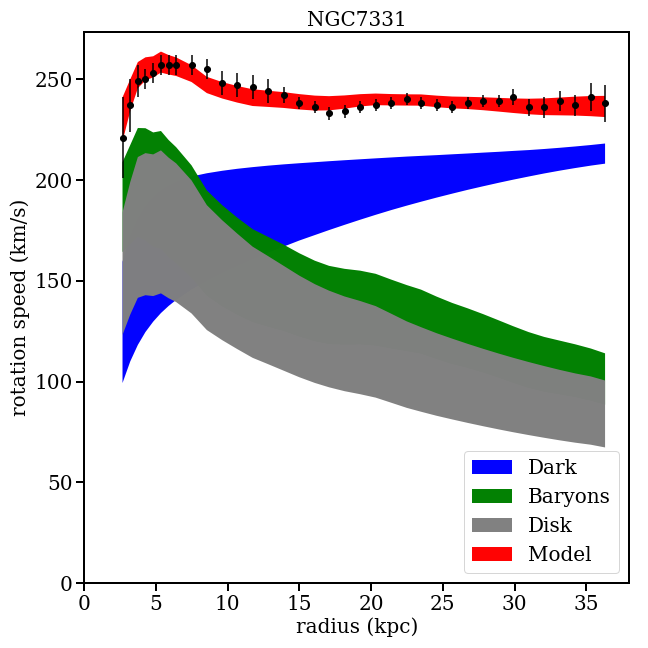} & \includegraphics[scale=0.215]{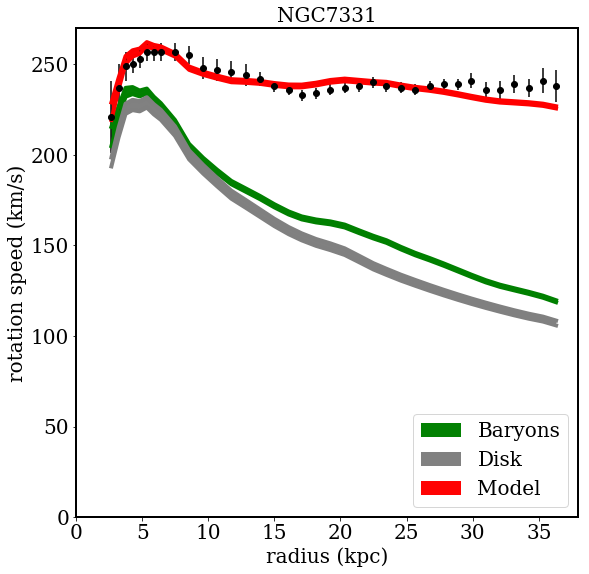} \\
\includegraphics[scale=0.25]{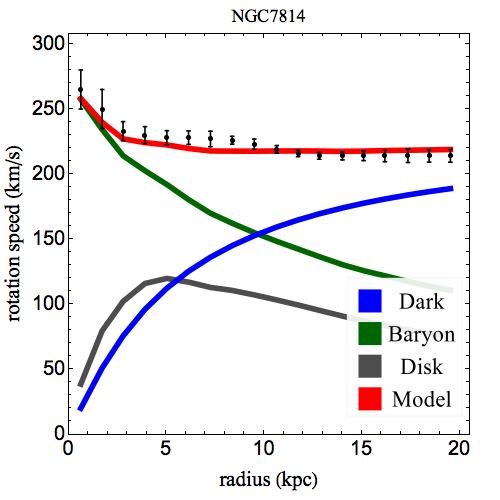} & \includegraphics[scale=0.192]{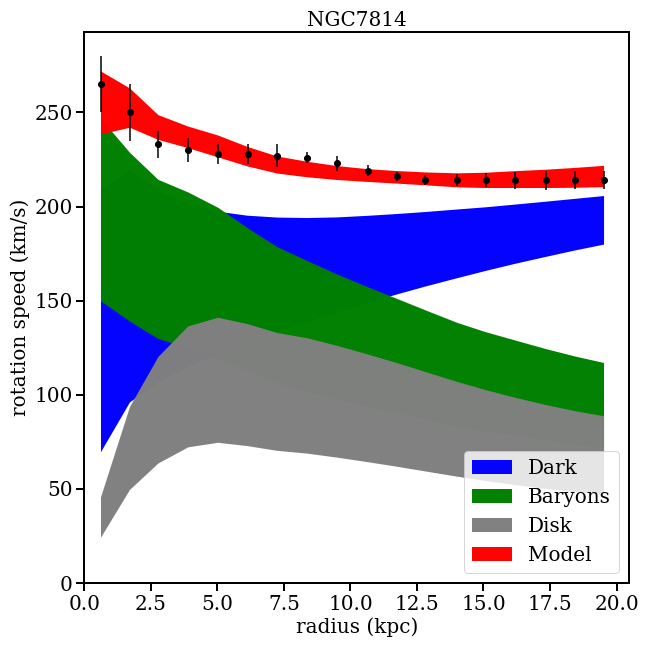} & \includegraphics[scale=0.215]{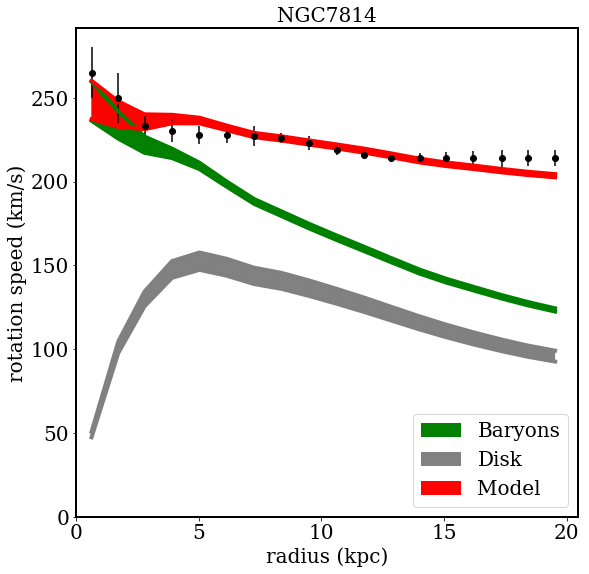} \\
\includegraphics[scale=0.25]{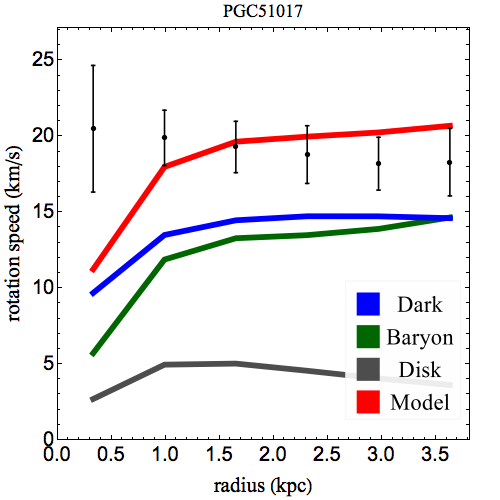} & \includegraphics[scale=0.192]{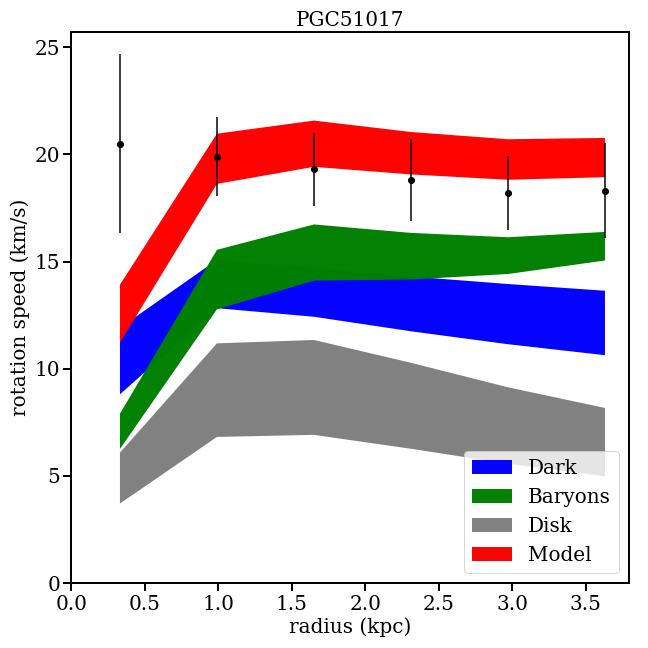} & \includegraphics[scale=0.215]{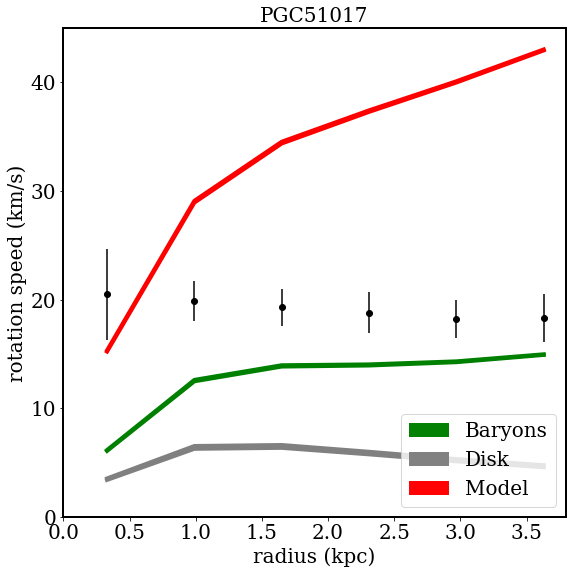} \\
\end{tabular}
\caption{Continued.}
\end{figure}

\addtocounter{figure}{-1}
\begin{figure}[H]
\centering
\begin{tabular}{@{}ccc@{}}
\includegraphics[scale=0.25]{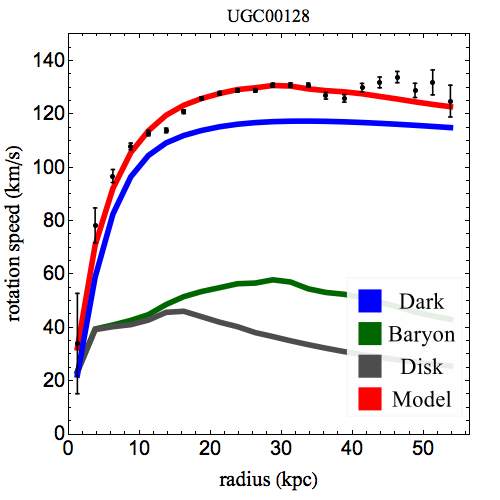} & \includegraphics[scale=0.192]{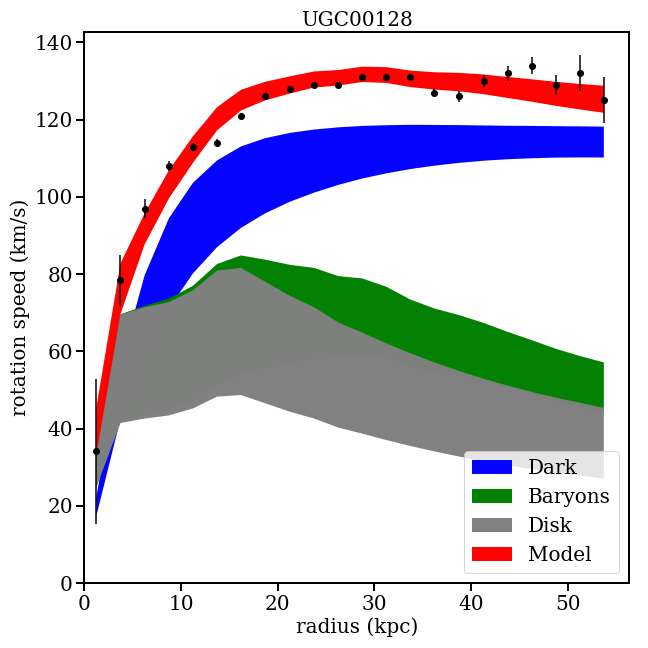} & \includegraphics[scale=0.215]{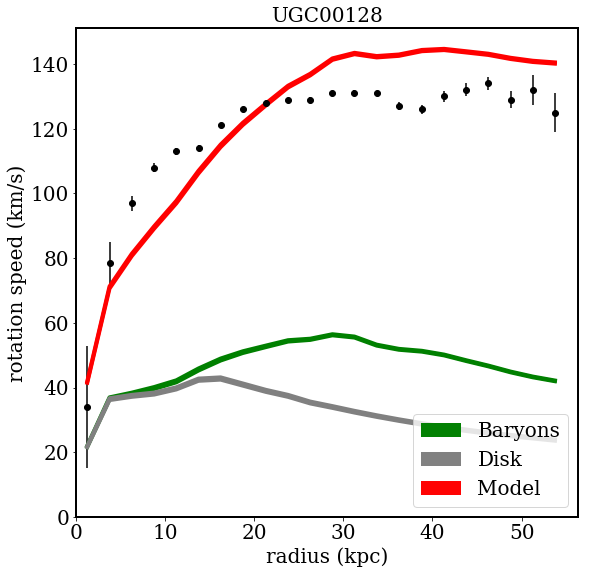} \\
\includegraphics[scale=0.25]{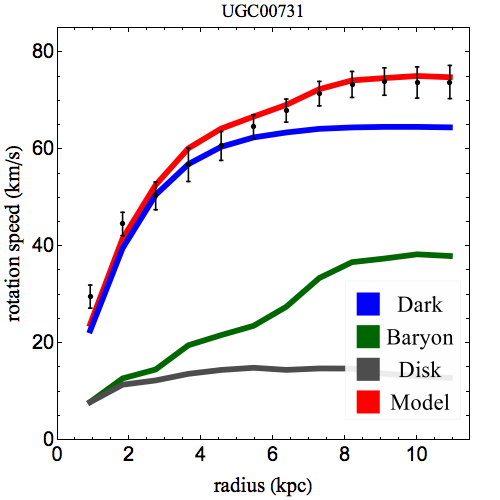} & \includegraphics[scale=0.192]{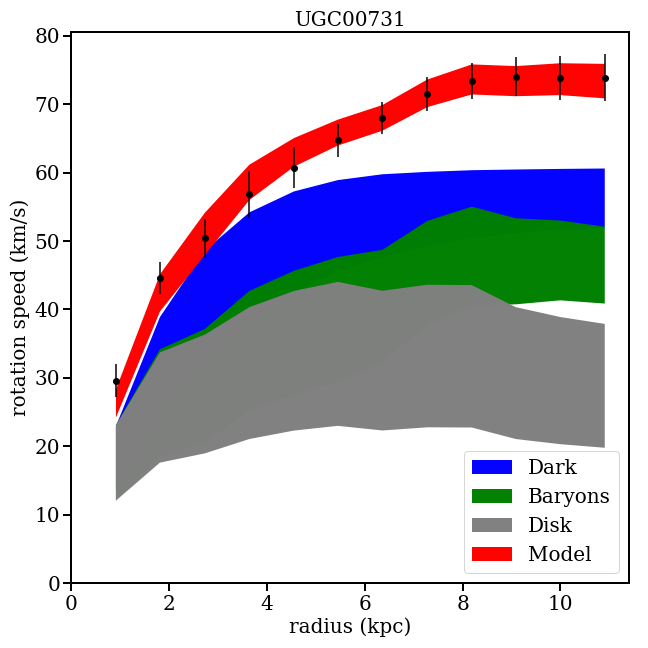} & \includegraphics[scale=0.215]{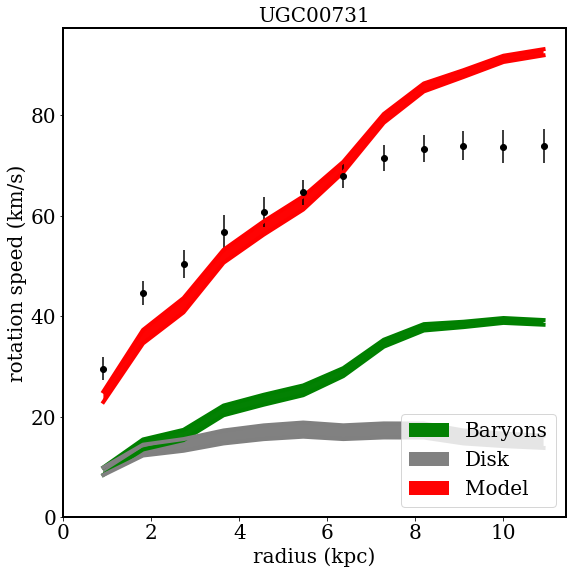} \\
\includegraphics[scale=0.25]{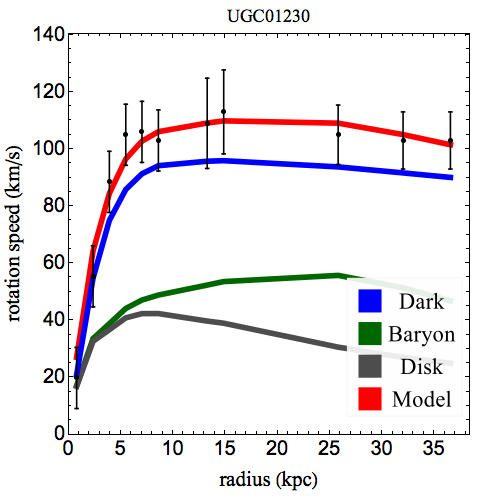} & \includegraphics[scale=0.192]{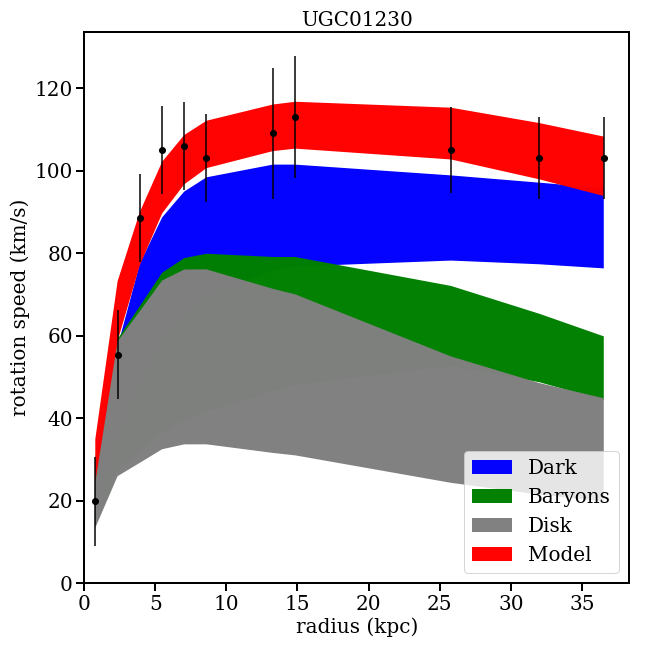} & \includegraphics[scale=0.215]{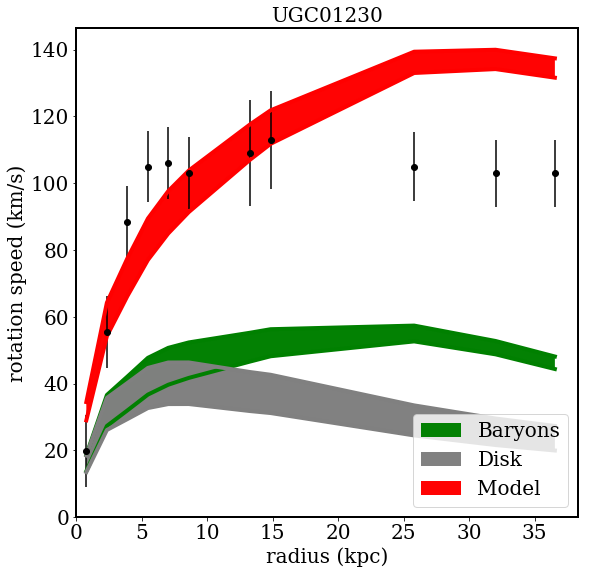} \\
\includegraphics[scale=0.25]{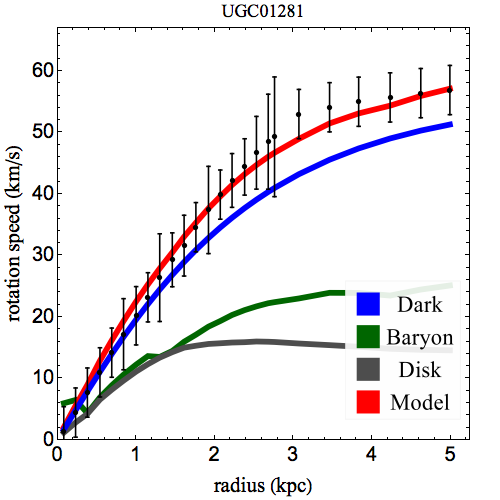} & \includegraphics[scale=0.192]{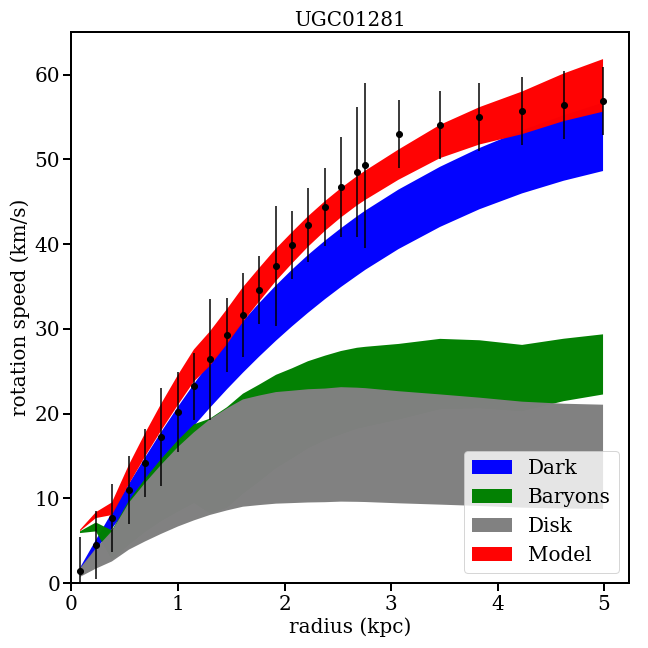} & \includegraphics[scale=0.215]{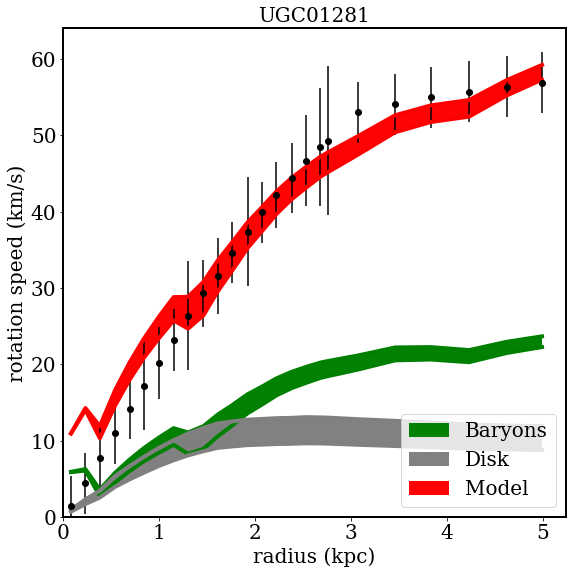} \\
\includegraphics[scale=0.25]{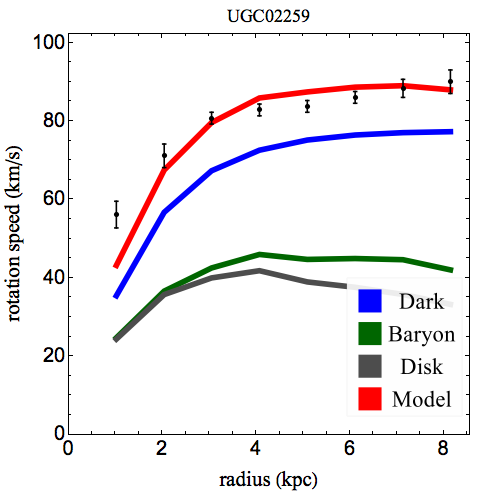} & \includegraphics[scale=0.192]{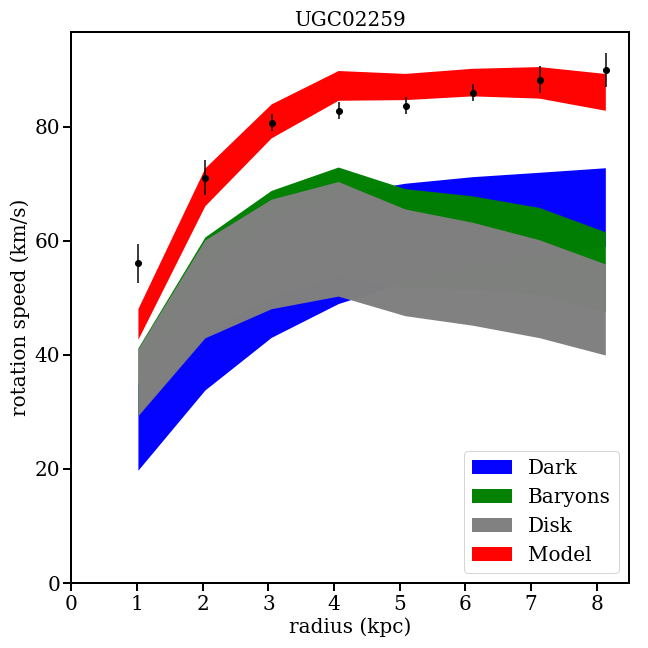} & \includegraphics[scale=0.215]{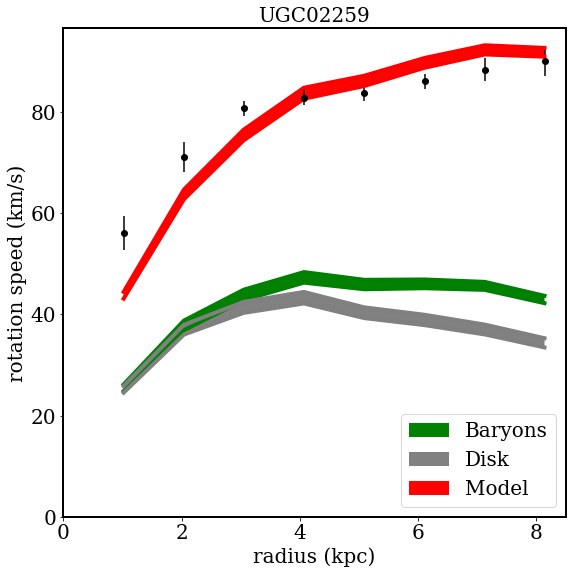} \\
\end{tabular}
\caption{Continued.}
\end{figure}

\addtocounter{figure}{-1}
\begin{figure}[H]
\centering
\begin{tabular}{@{}ccc@{}}
\includegraphics[scale=0.25]{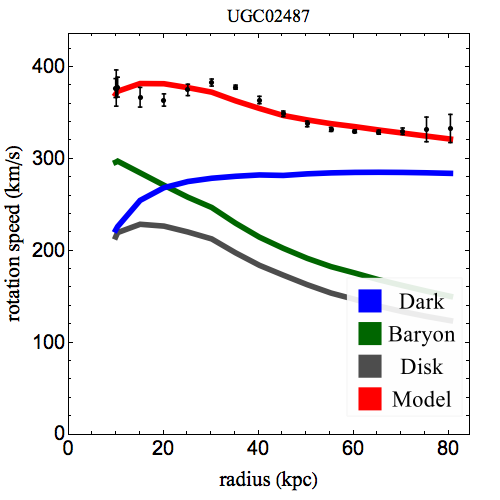} & \includegraphics[scale=0.192]{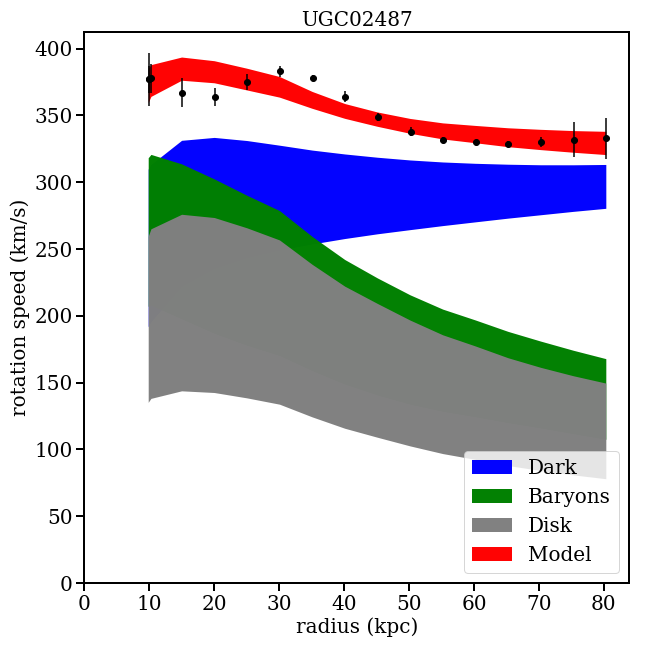} & \includegraphics[scale=0.215]{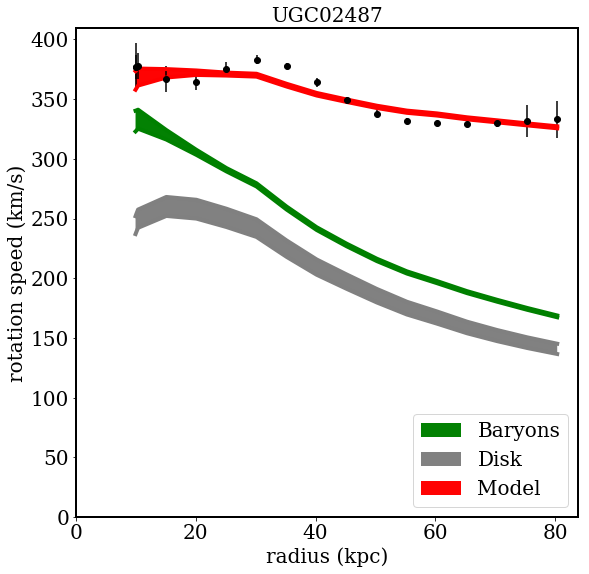} \\
\includegraphics[scale=0.25]{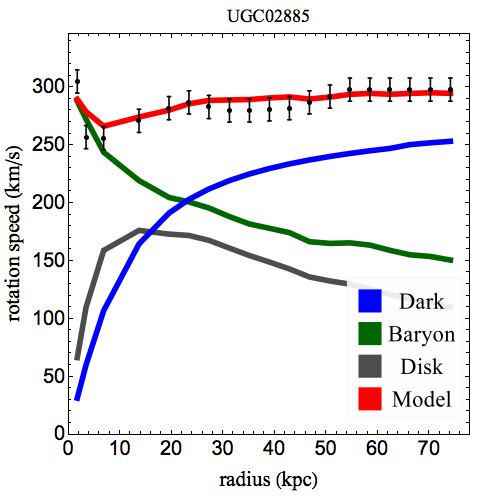} & \includegraphics[scale=0.192]{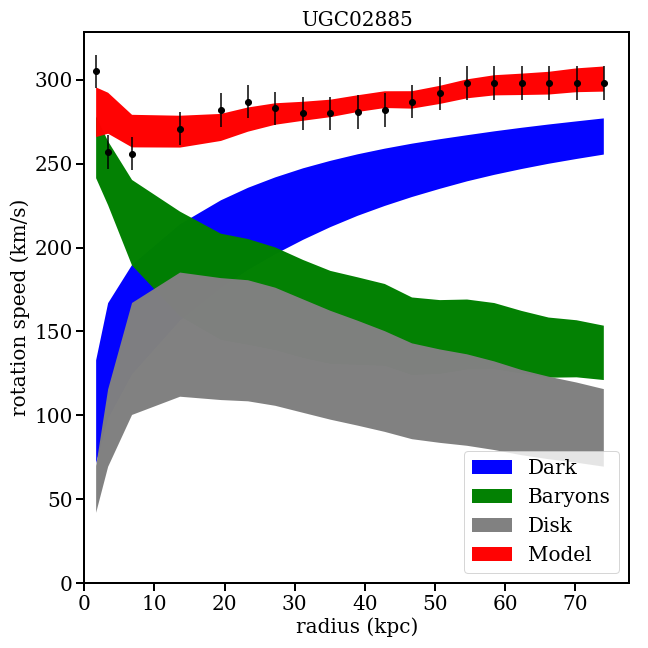} & \includegraphics[scale=0.215]{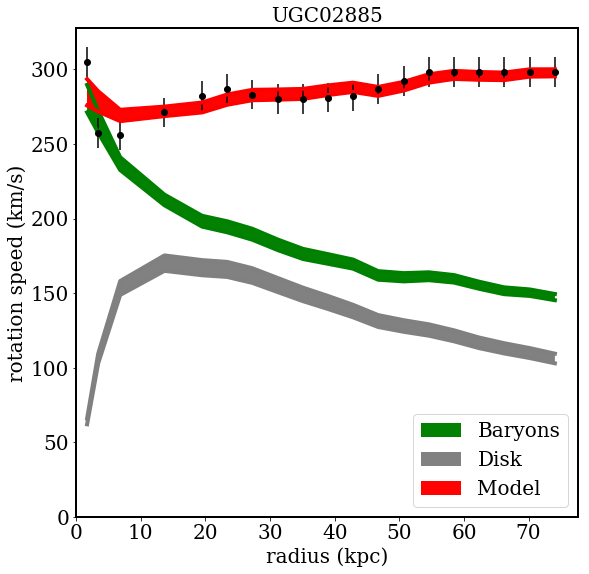} \\
\includegraphics[scale=0.25]{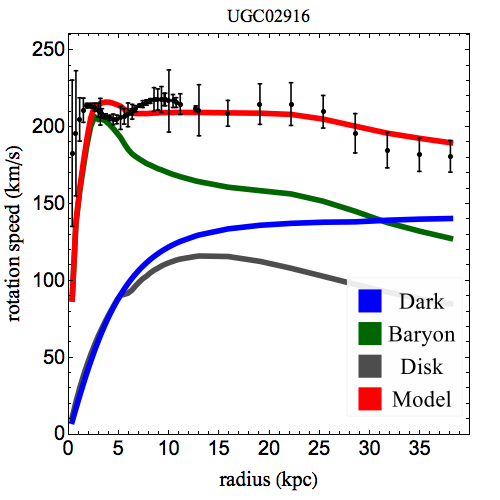} & \includegraphics[scale=0.192]{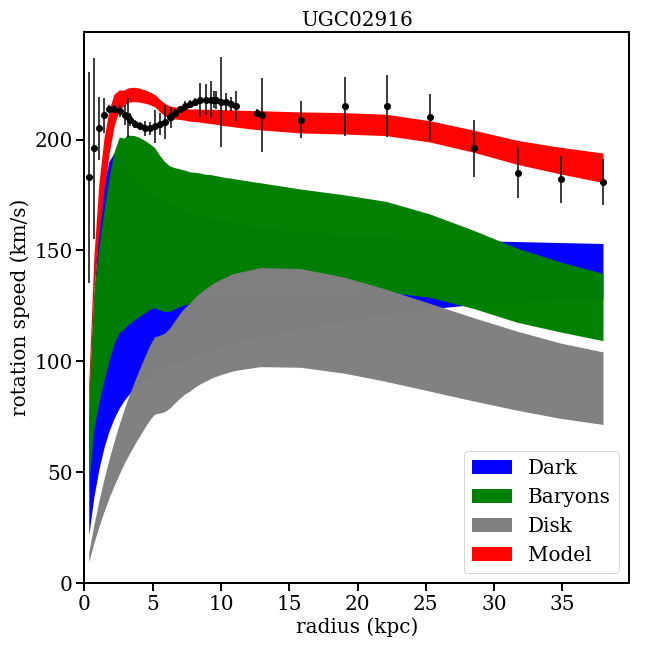} & \includegraphics[scale=0.215]{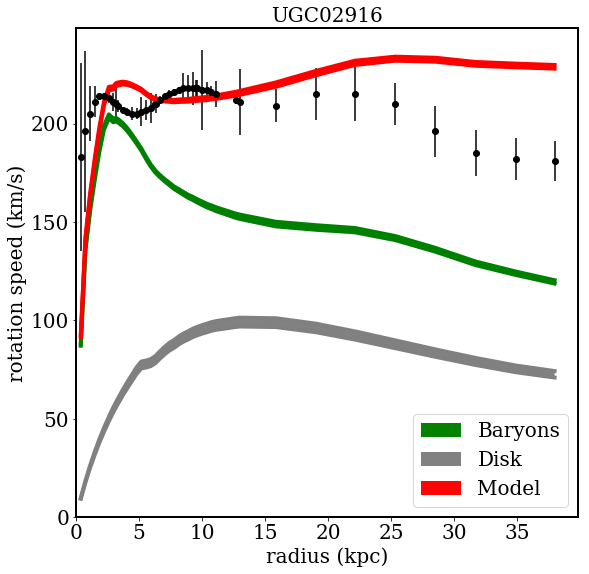} \\
\includegraphics[scale=0.25]{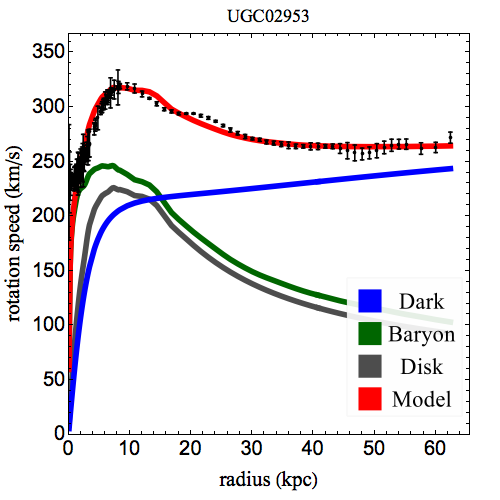} & \includegraphics[scale=0.192]{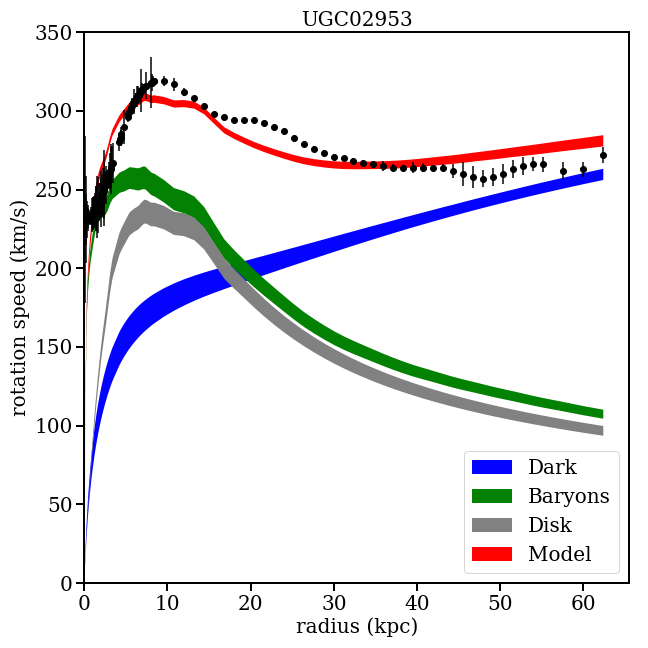} & \includegraphics[scale=0.215]{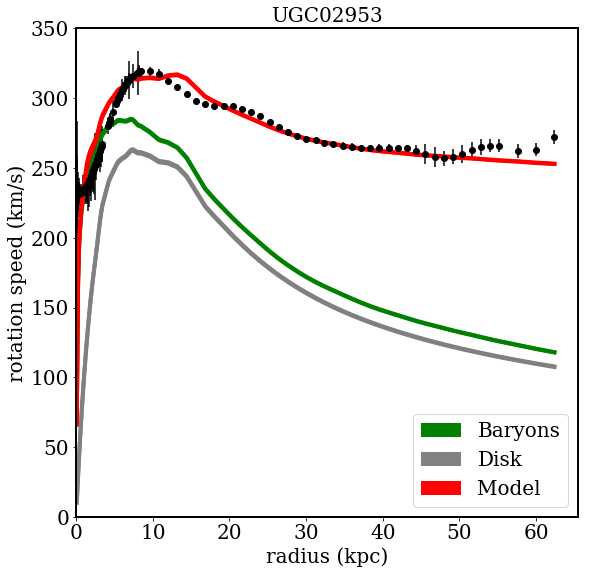} \\
\includegraphics[scale=0.25]{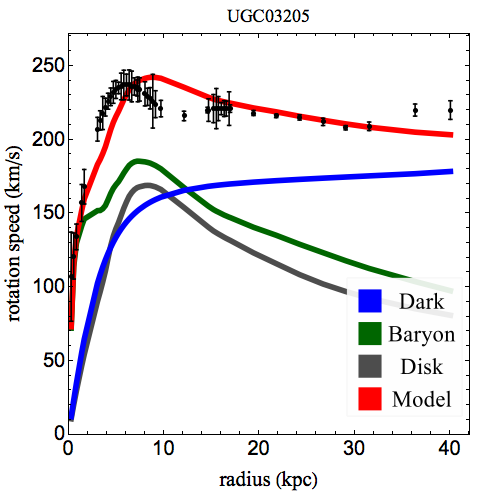} & \includegraphics[scale=0.192]{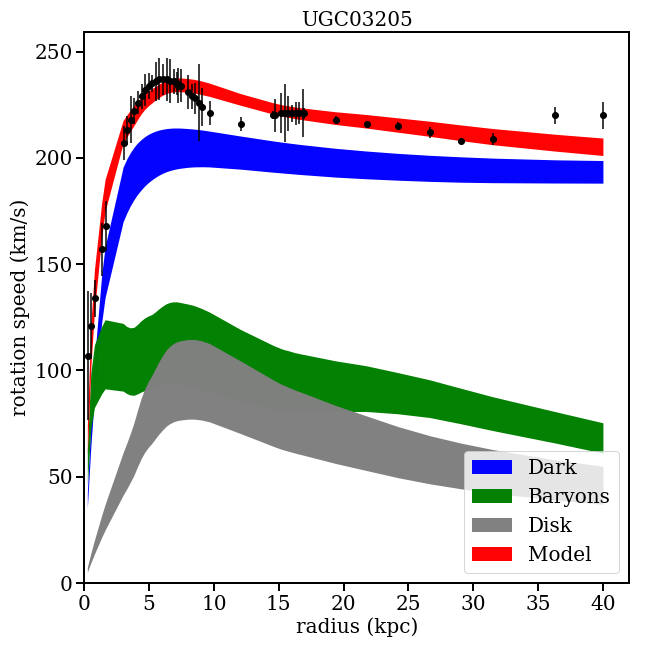} & \includegraphics[scale=0.215]{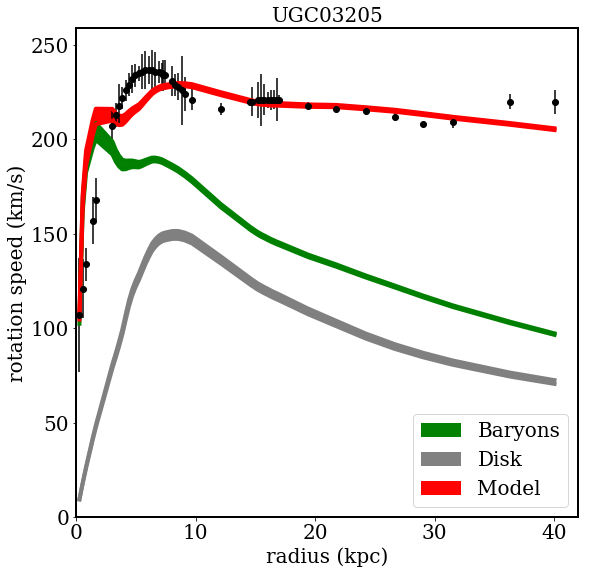} \\
\end{tabular}
\caption{Continued.}
\end{figure}

\addtocounter{figure}{-1}
\begin{figure}[H]
\centering
\begin{tabular}{@{}ccc@{}}
\includegraphics[scale=0.25]{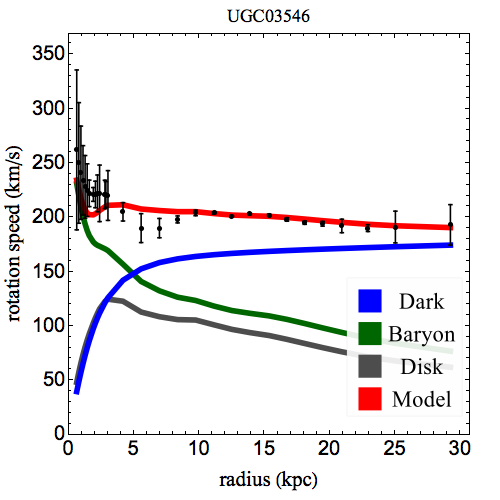} & \includegraphics[scale=0.192]{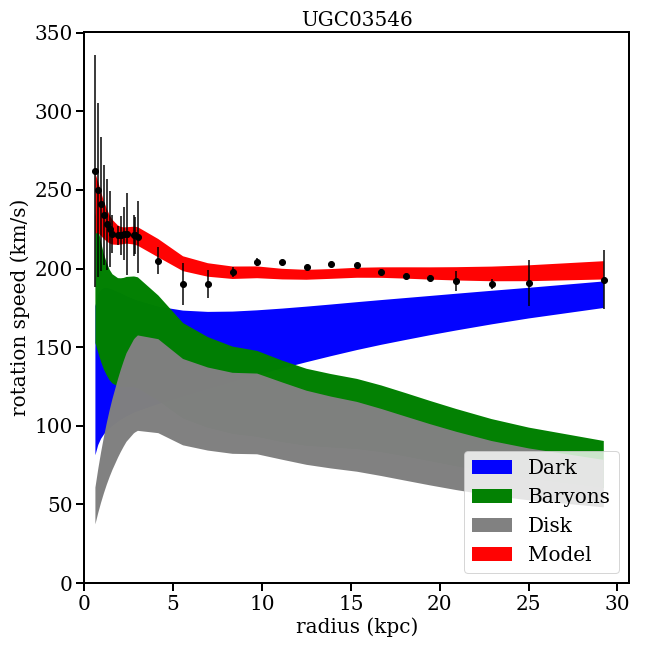} & \includegraphics[scale=0.215]{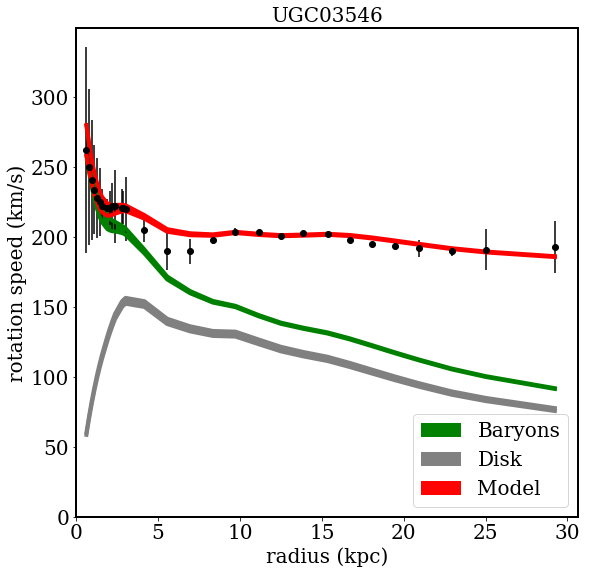} \\
\includegraphics[scale=0.25]{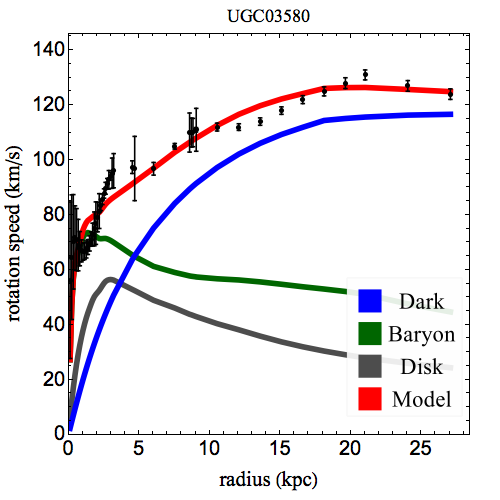} & \includegraphics[scale=0.192]{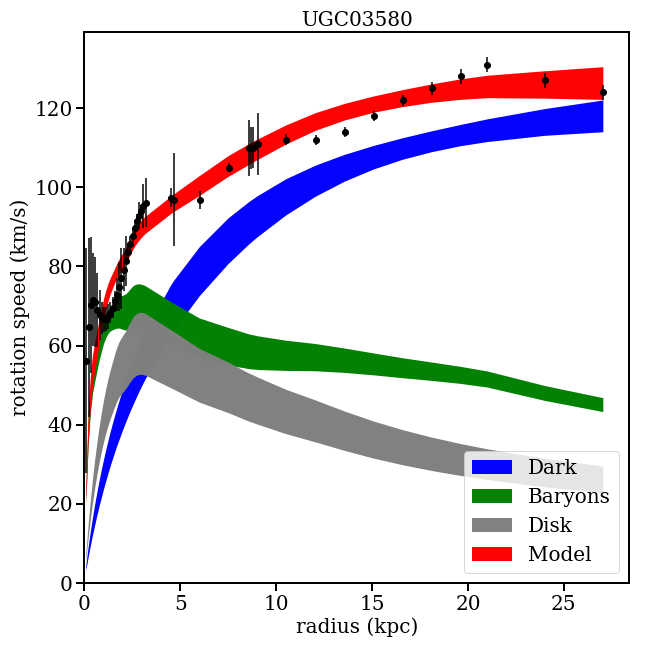} & \includegraphics[scale=0.215]{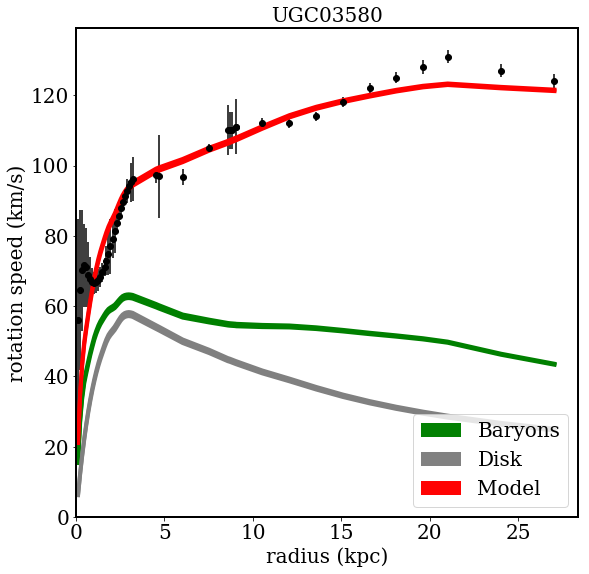} \\
\includegraphics[scale=0.25]{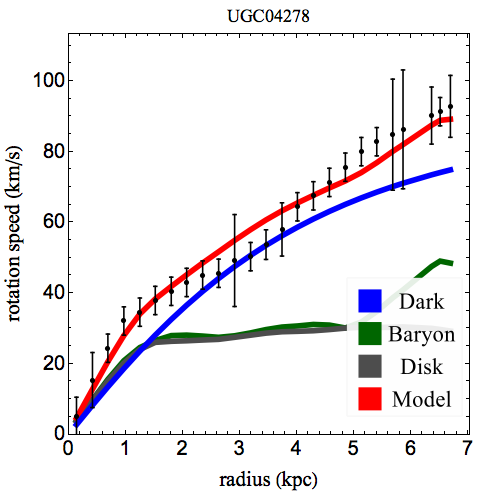} & \includegraphics[scale=0.192]{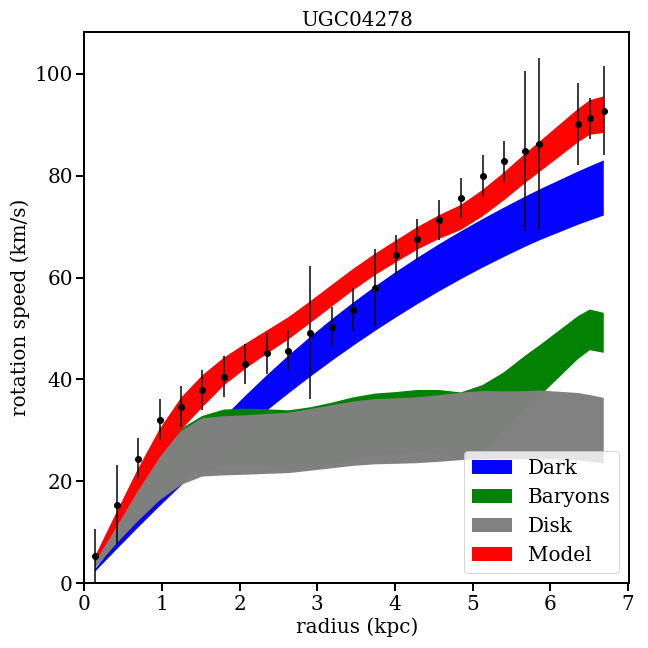} & \includegraphics[scale=0.215]{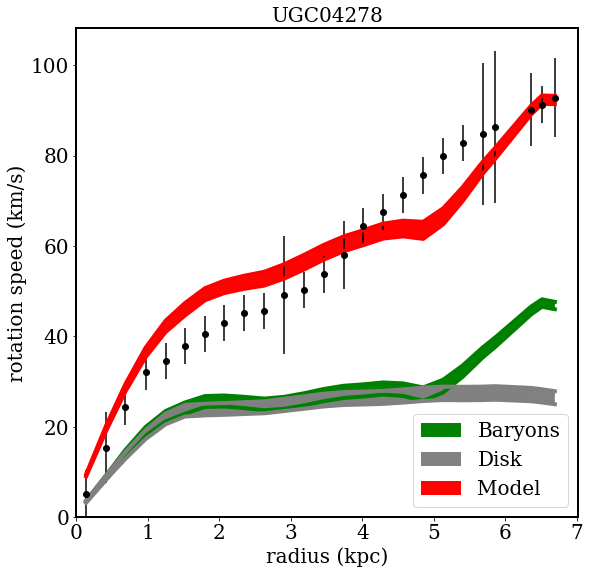} \\
\includegraphics[scale=0.25]{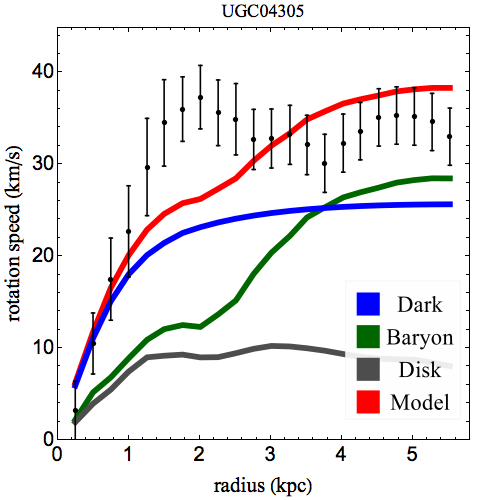} & \includegraphics[scale=0.192]{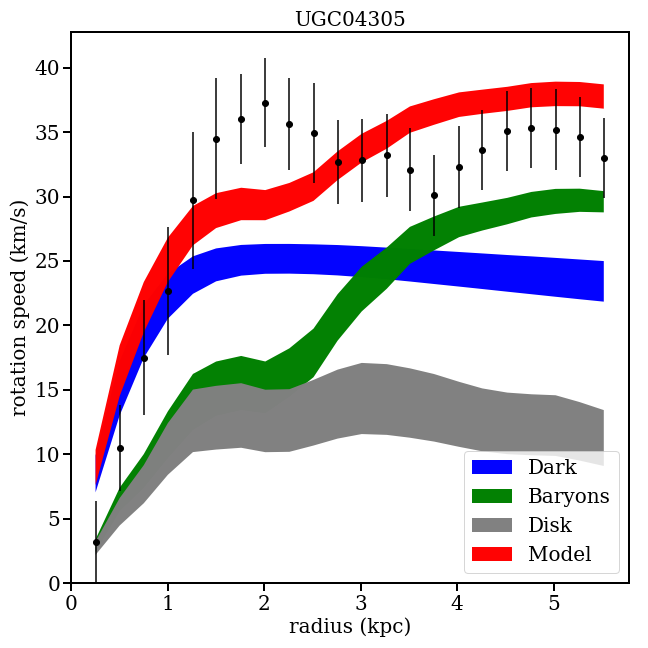} & \includegraphics[scale=0.215]{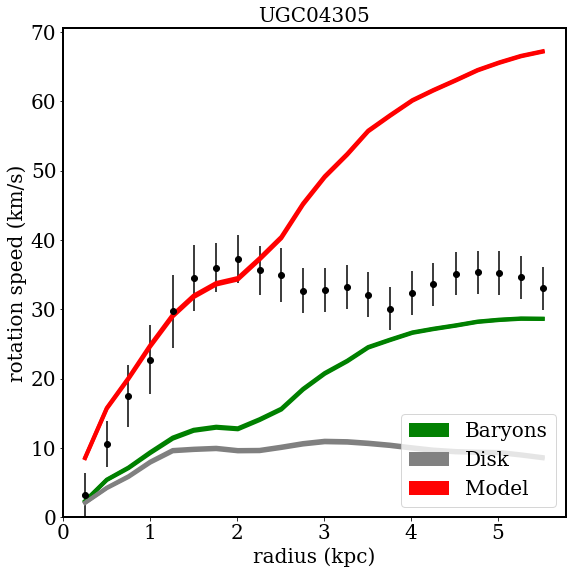} \\
\includegraphics[scale=0.25]{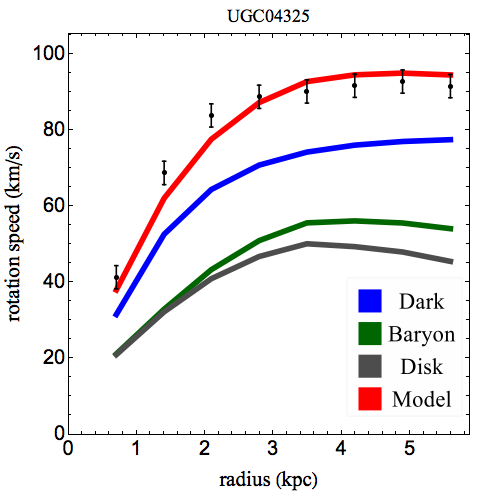} & \includegraphics[scale=0.192]{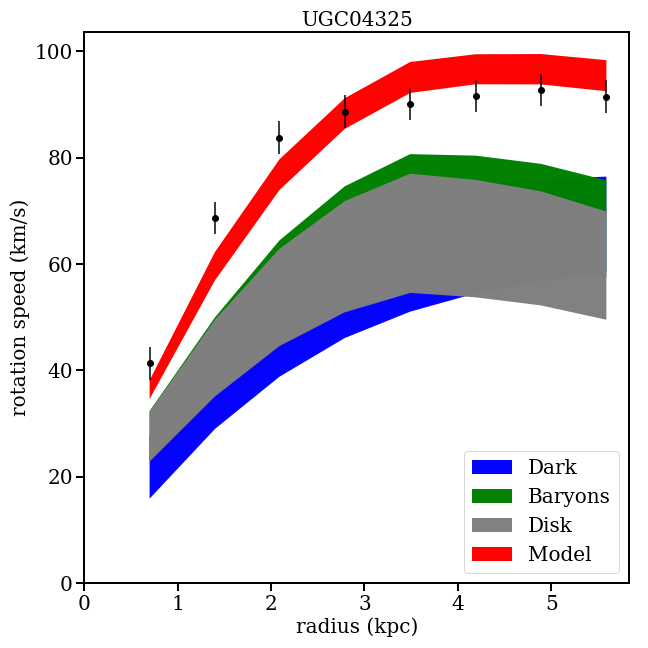} & \includegraphics[scale=0.215]{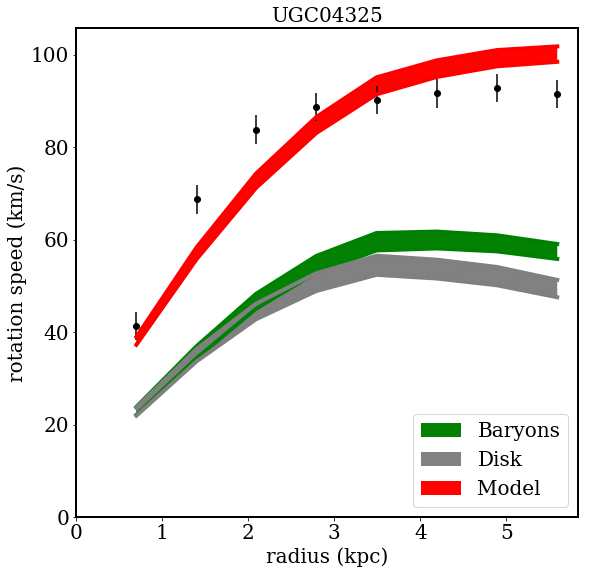} \\
\end{tabular}
\caption{Continued.}
\end{figure}

\addtocounter{figure}{-1}
\begin{figure}[H]
\centering
\begin{tabular}{@{}ccc@{}}
\includegraphics[scale=0.25]{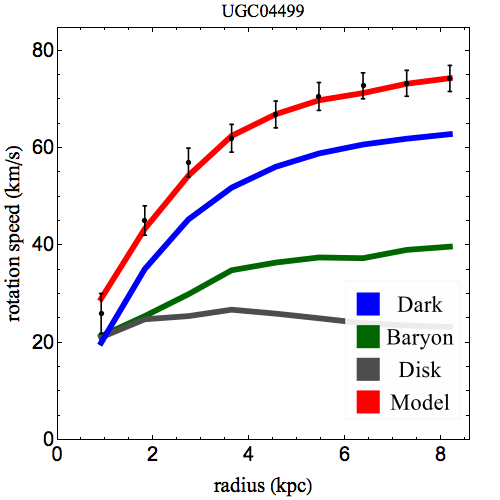} & \includegraphics[scale=0.192]{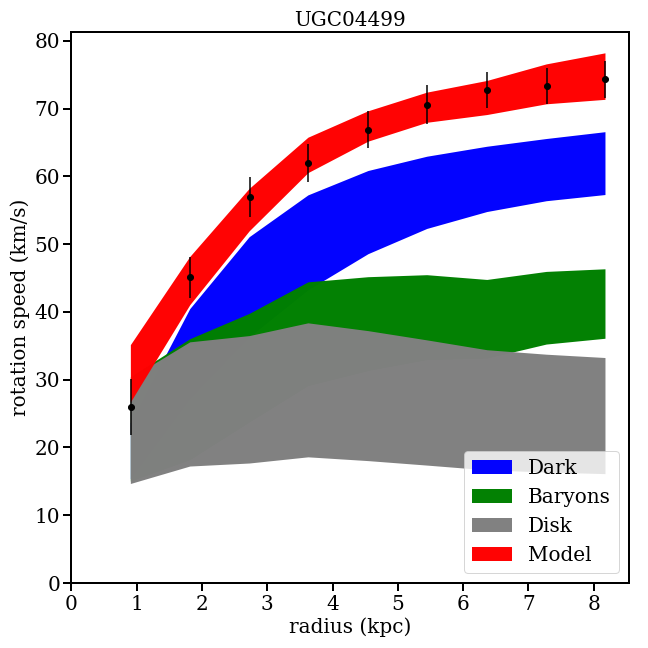} & \includegraphics[scale=0.215]{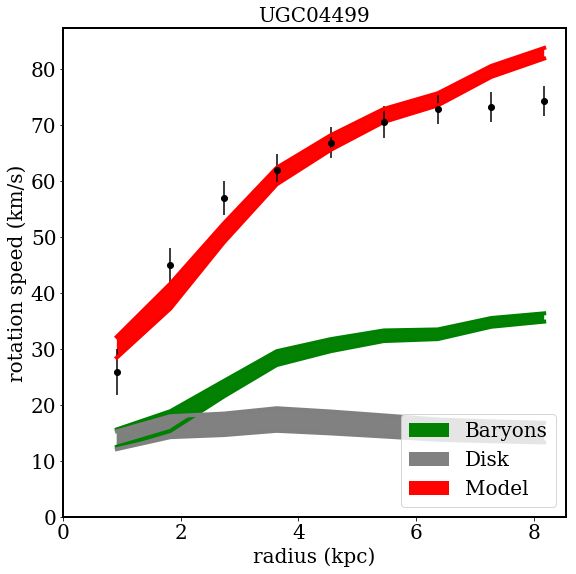} \\
\includegraphics[scale=0.25]{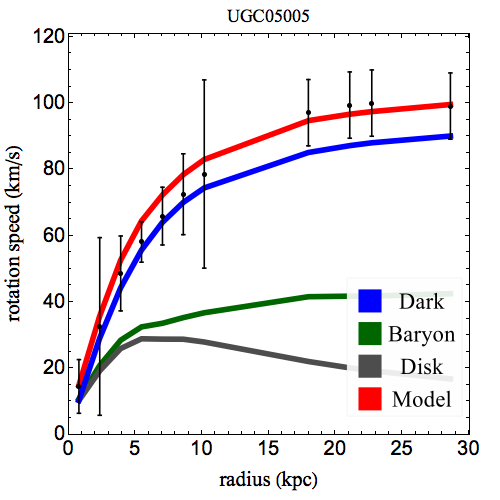} & \includegraphics[scale=0.192]{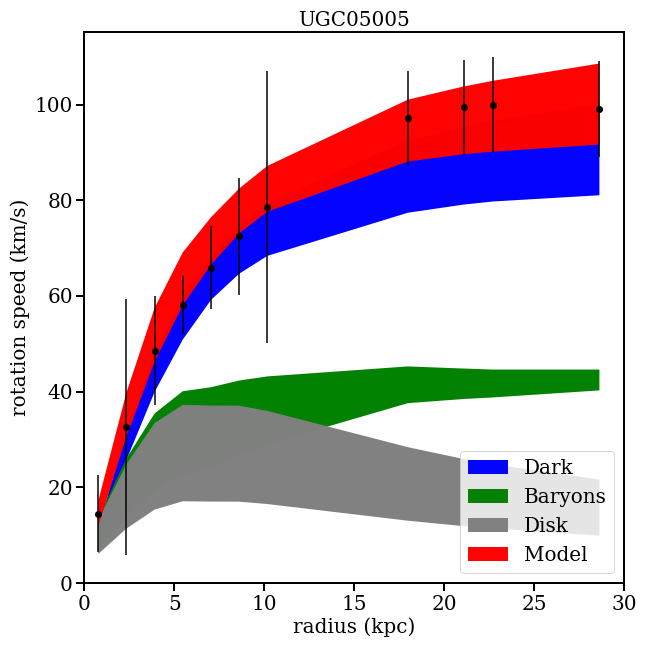} & \includegraphics[scale=0.215]{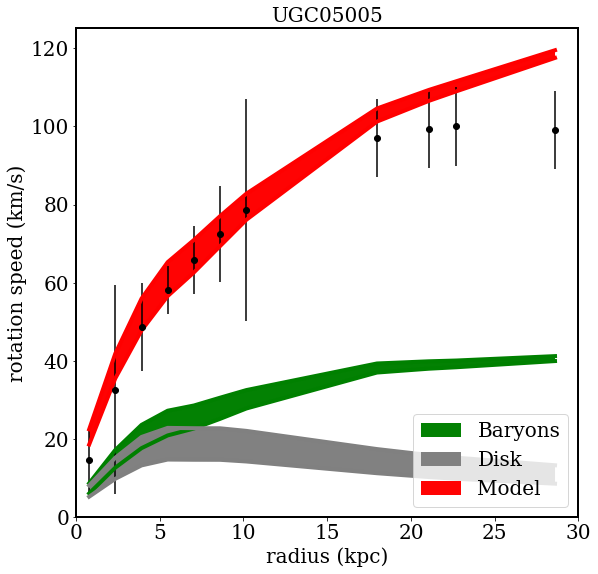} \\
\includegraphics[scale=0.25]{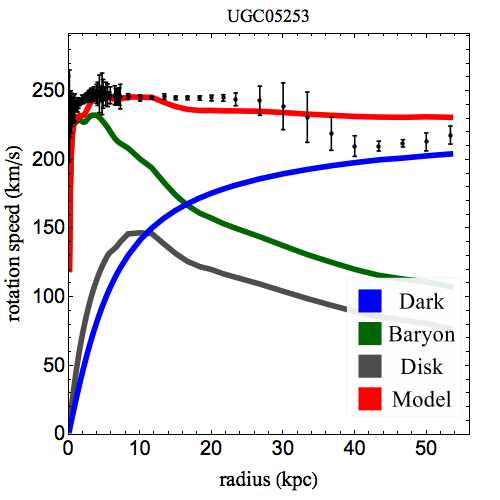} & \includegraphics[scale=0.192]{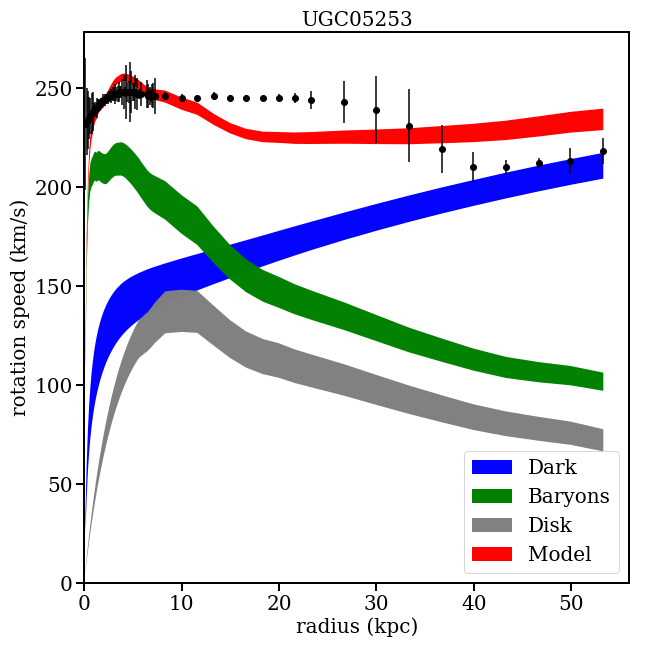} & \includegraphics[scale=0.215]{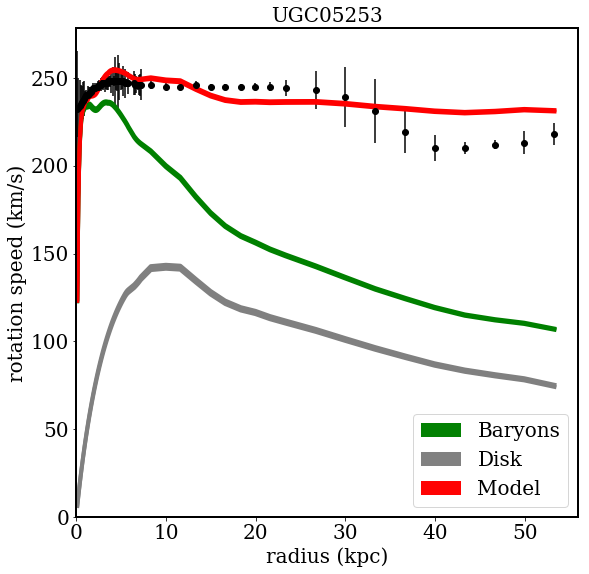} \\
\includegraphics[scale=0.25]{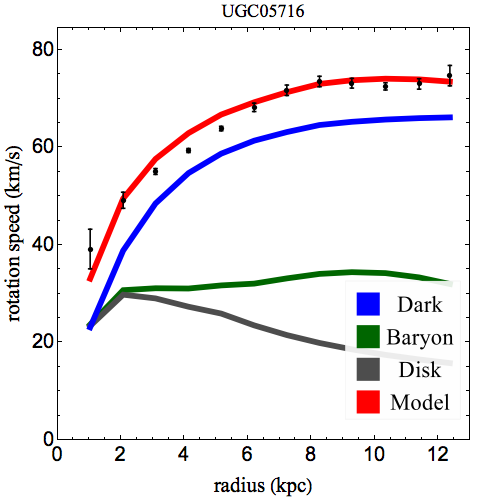} & \includegraphics[scale=0.192]{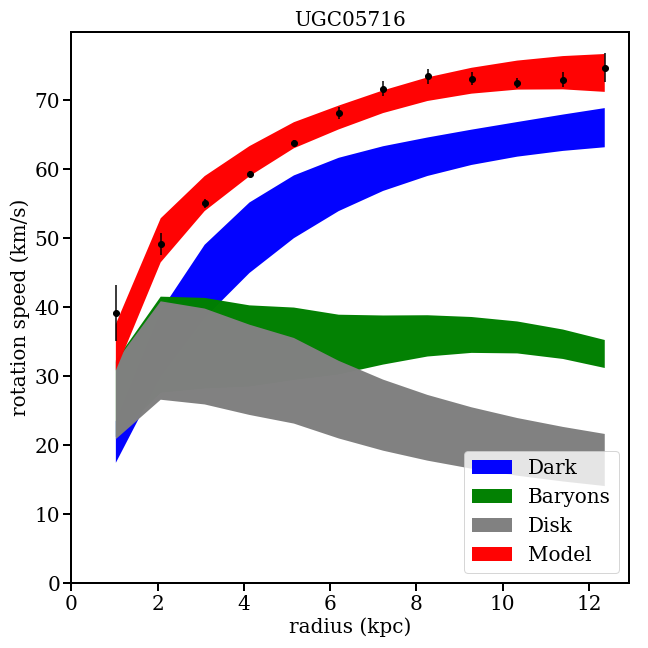} & \includegraphics[scale=0.215]{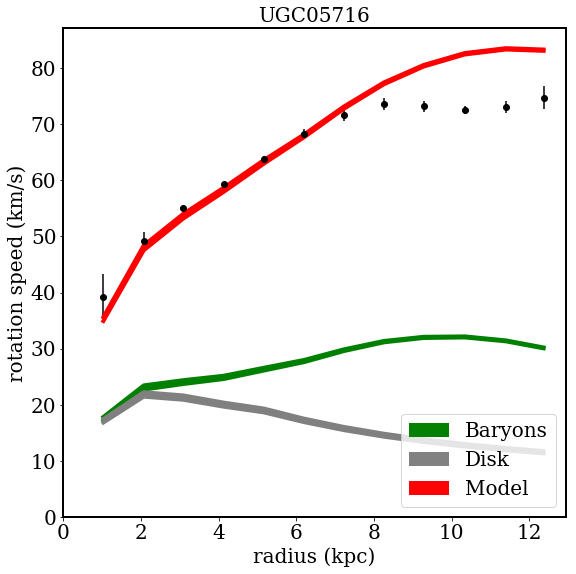} \\
\includegraphics[scale=0.25]{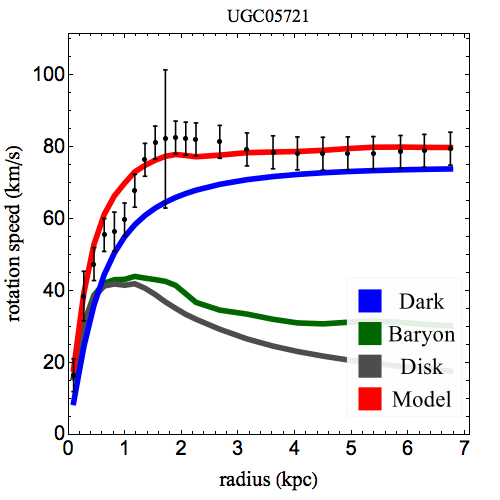} & \includegraphics[scale=0.192]{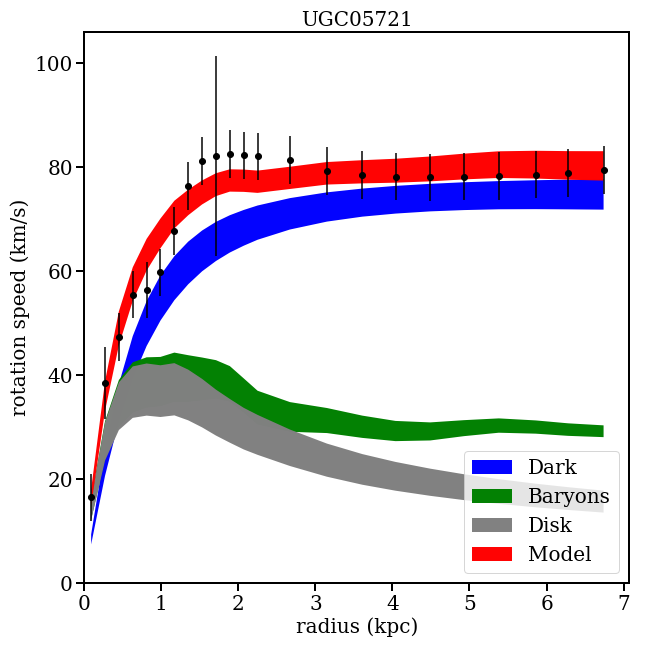} & \includegraphics[scale=0.215]{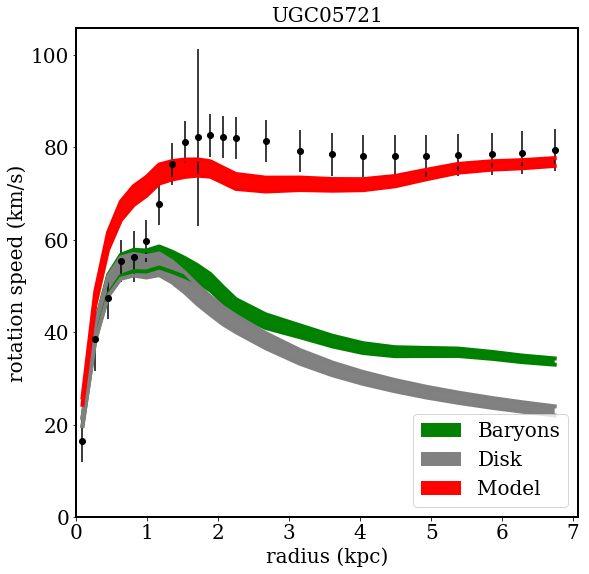} \\
\end{tabular}
\caption{Continued.}
\end{figure}

\addtocounter{figure}{-1}
\begin{figure}[H]
\centering
\begin{tabular}{@{}ccc@{}}
\includegraphics[scale=0.25]{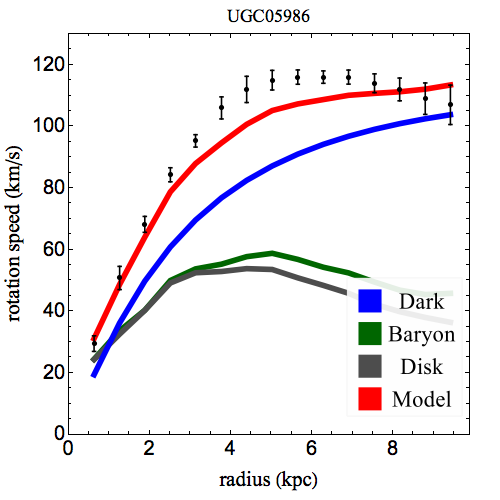} & \includegraphics[scale=0.192]{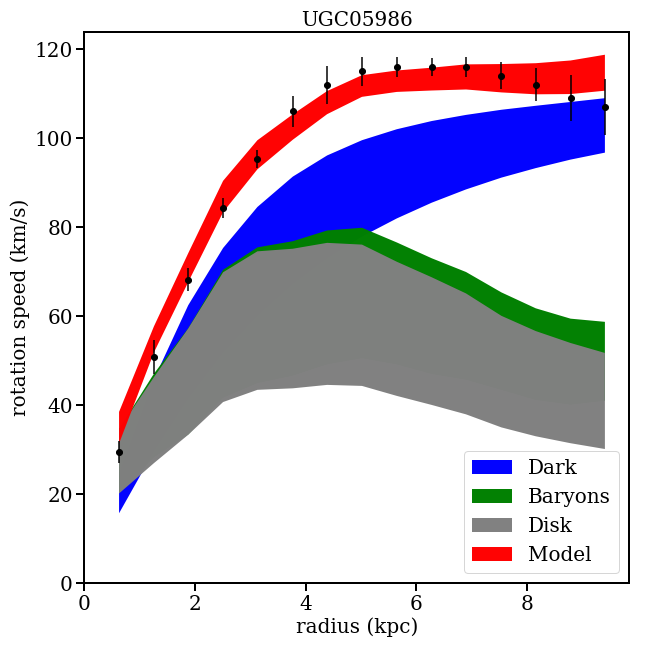} & \includegraphics[scale=0.215]{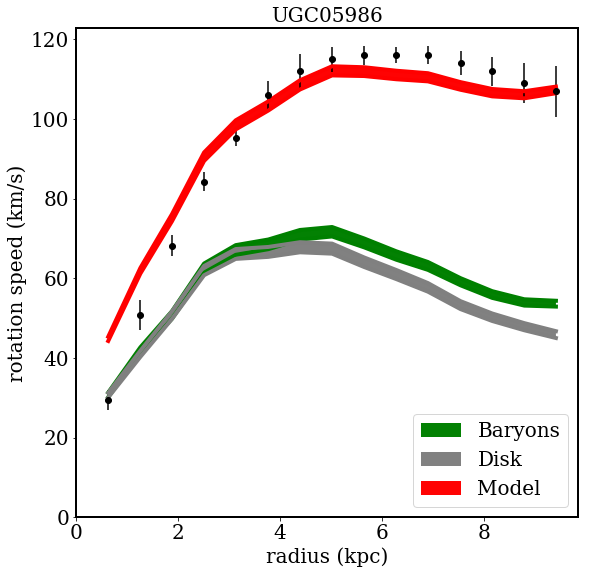} \\
\includegraphics[scale=0.25]{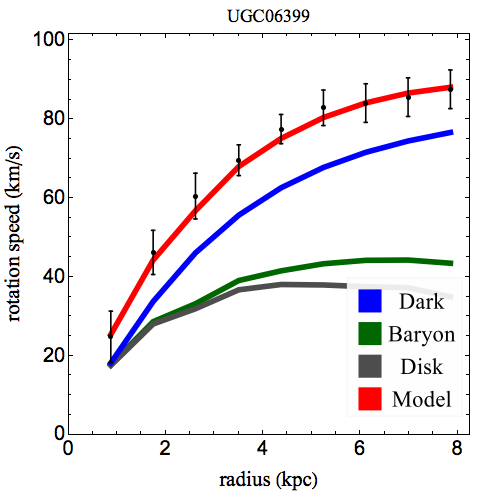} & \includegraphics[scale=0.192]{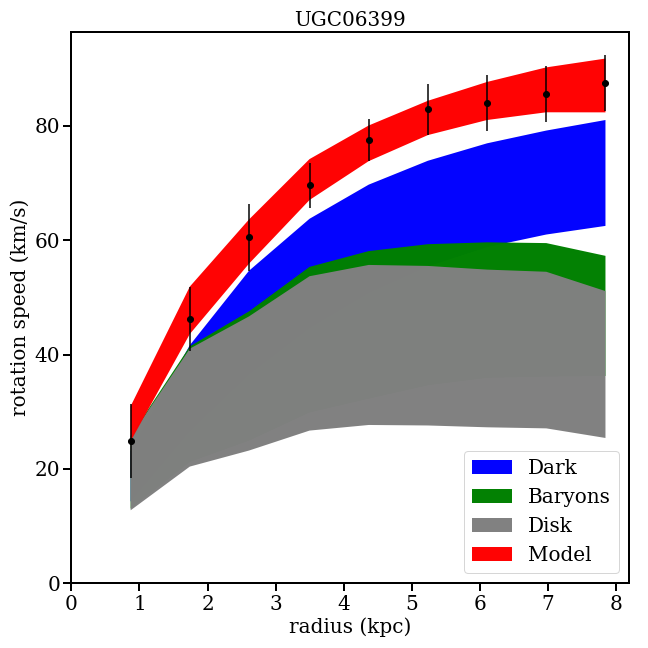} & \includegraphics[scale=0.215]{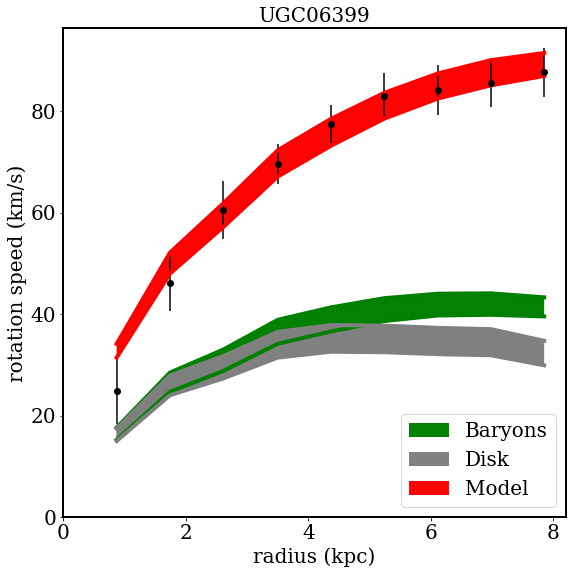} \\
\includegraphics[scale=0.25]{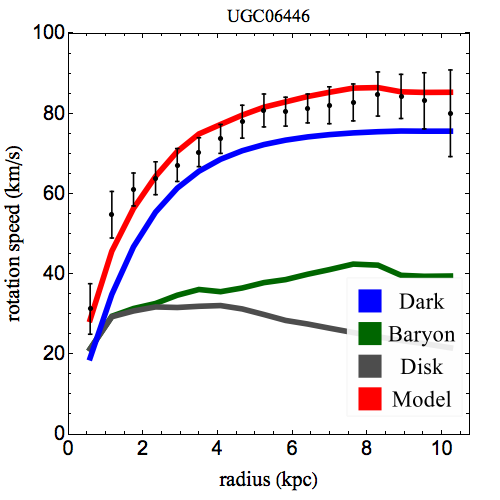} & \includegraphics[scale=0.192]{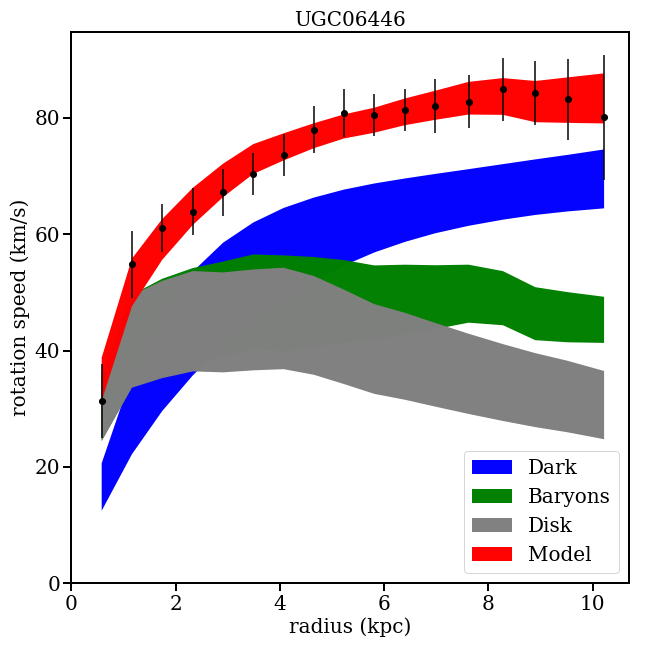} & \includegraphics[scale=0.215]{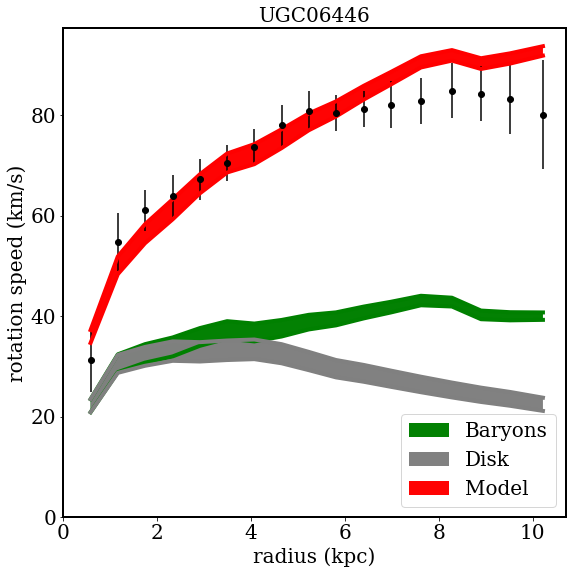} \\
\includegraphics[scale=0.25]{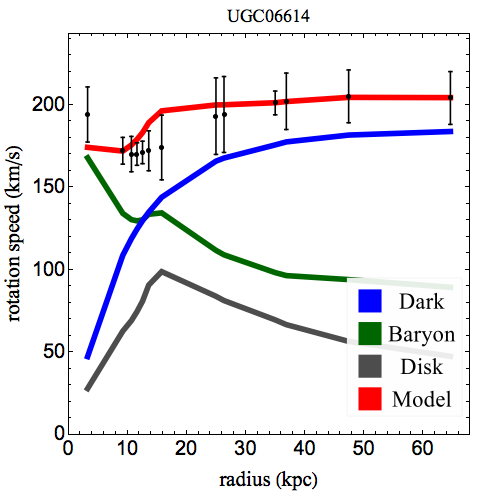} & \includegraphics[scale=0.192]{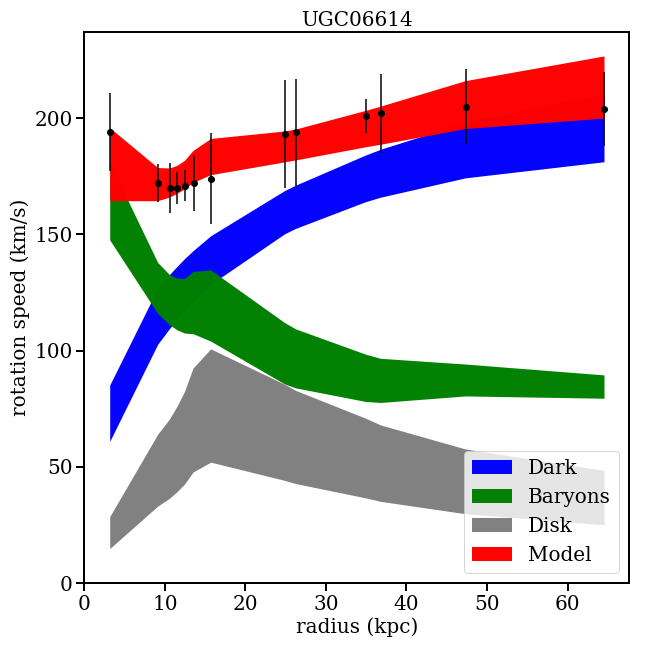} & \includegraphics[scale=0.215]{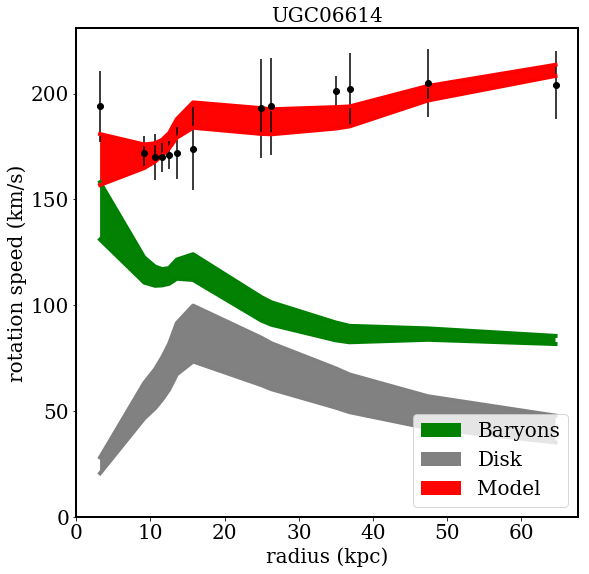} \\
\includegraphics[scale=0.25]{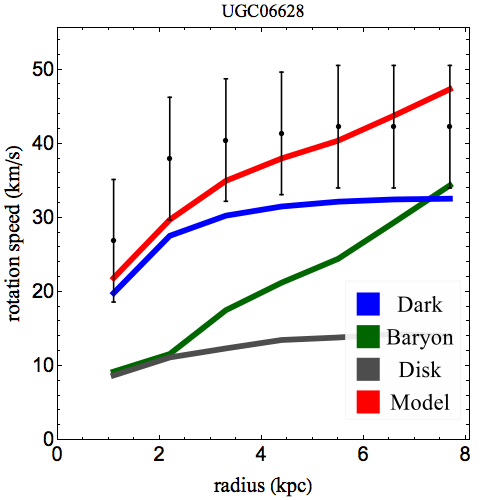} & \includegraphics[scale=0.192]{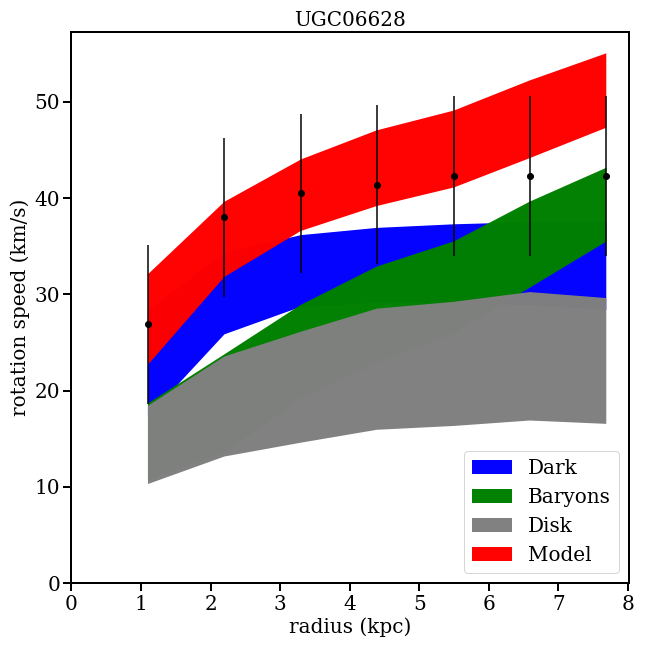} & \includegraphics[scale=0.215]{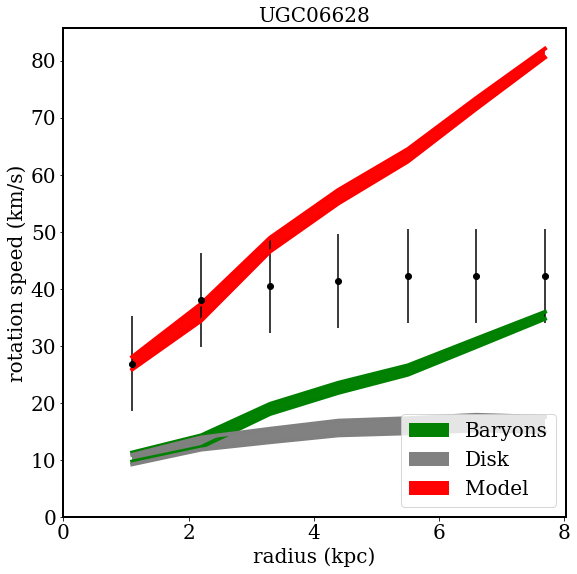} \\
\end{tabular}
\caption{Continued.}
\end{figure}

\addtocounter{figure}{-1}
\begin{figure}[H]
\centering
\begin{tabular}{@{}ccc@{}}
\includegraphics[scale=0.25]{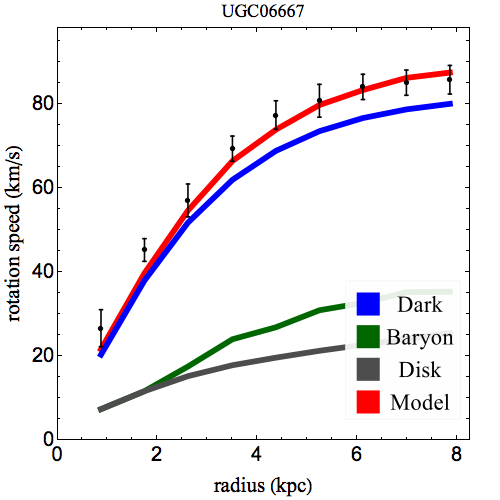} & \includegraphics[scale=0.192]{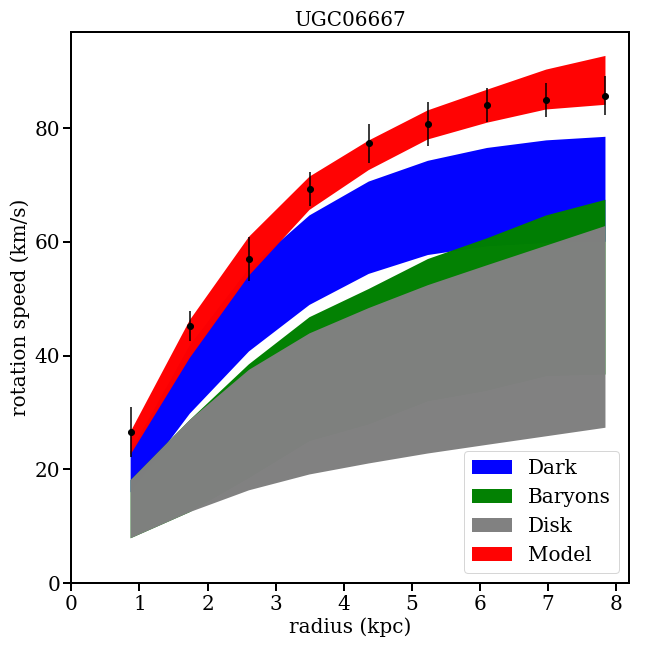} & \includegraphics[scale=0.215]{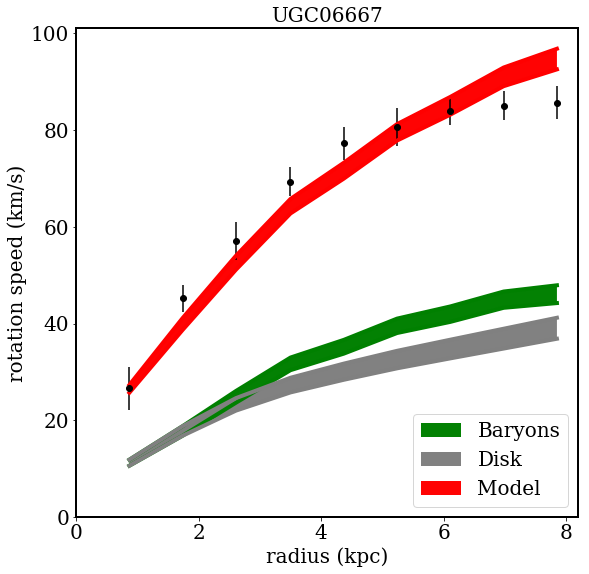} \\
\includegraphics[scale=0.25]{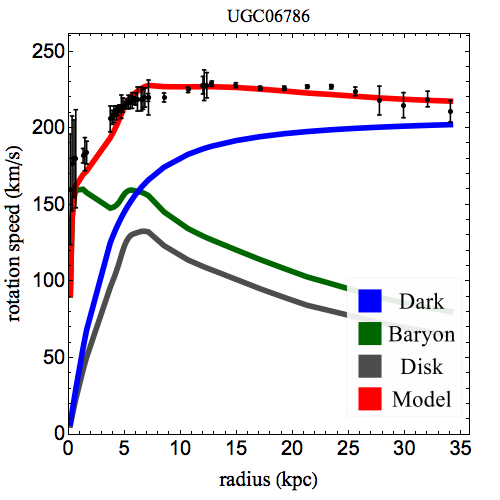} & \includegraphics[scale=0.192]{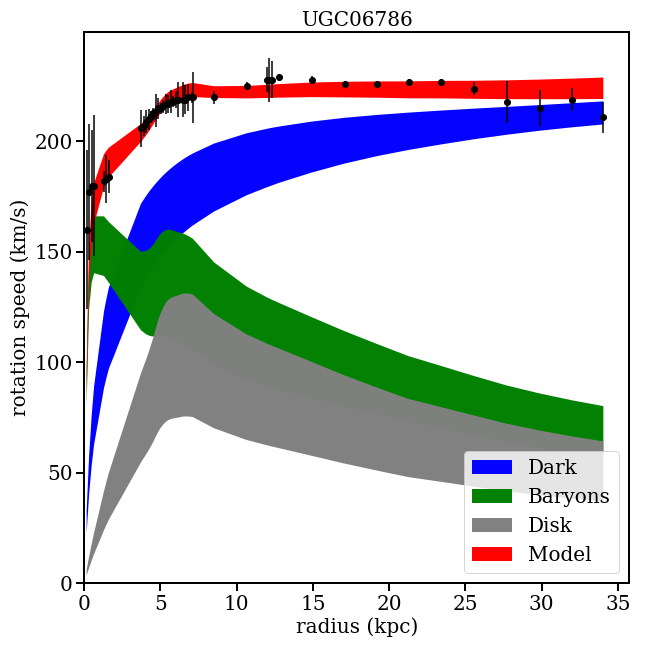} & \includegraphics[scale=0.215]{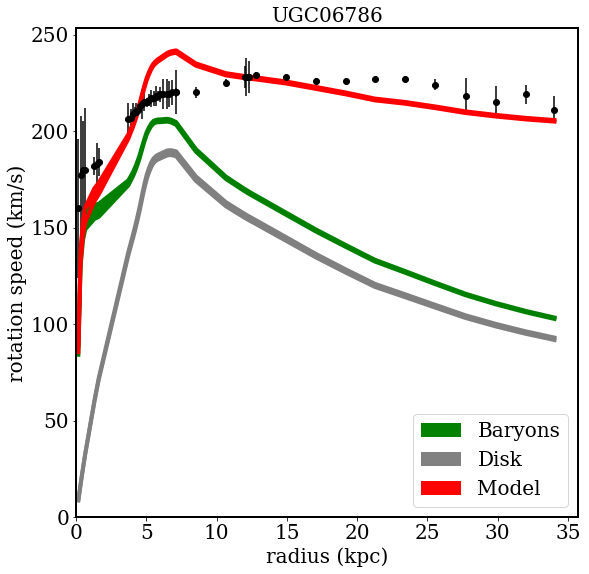} \\
\includegraphics[scale=0.25]{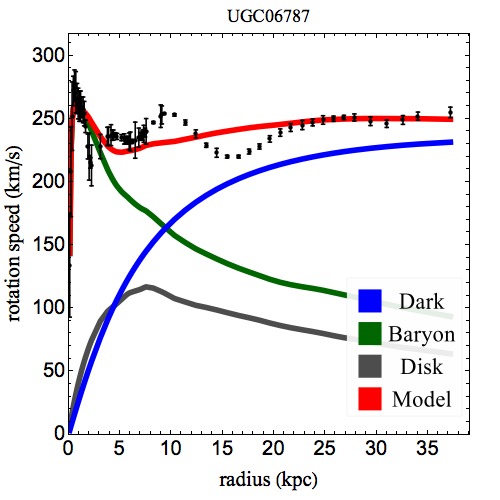} & \includegraphics[scale=0.192]{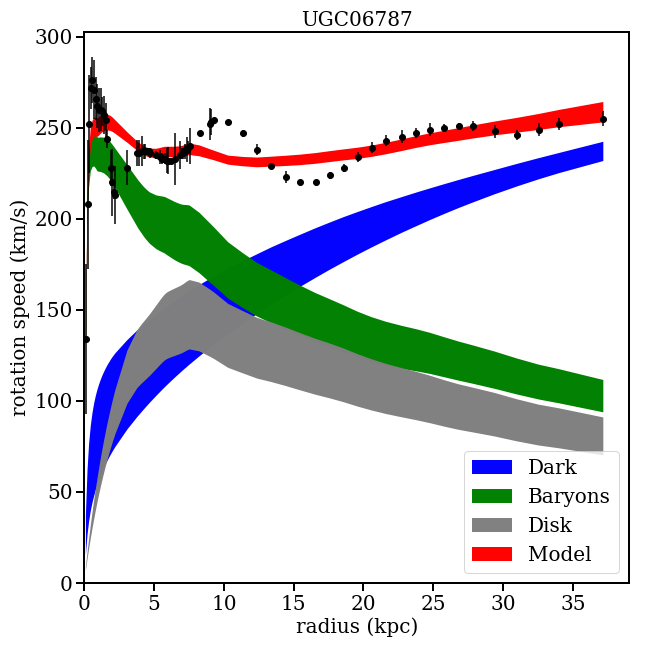} & \includegraphics[scale=0.215]{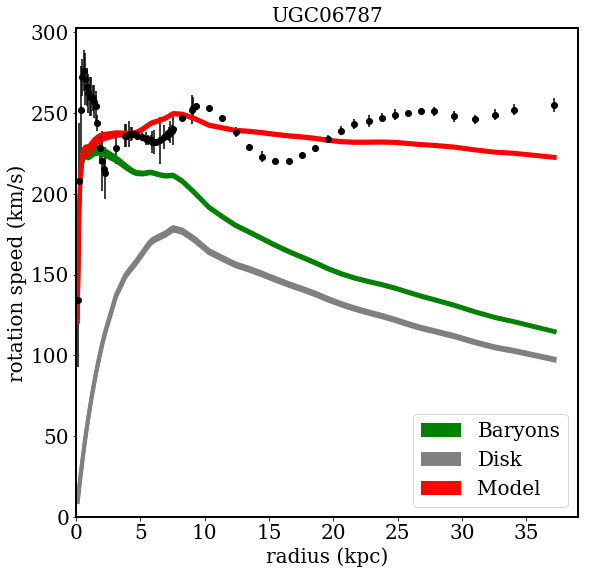} \\
\includegraphics[scale=0.25]{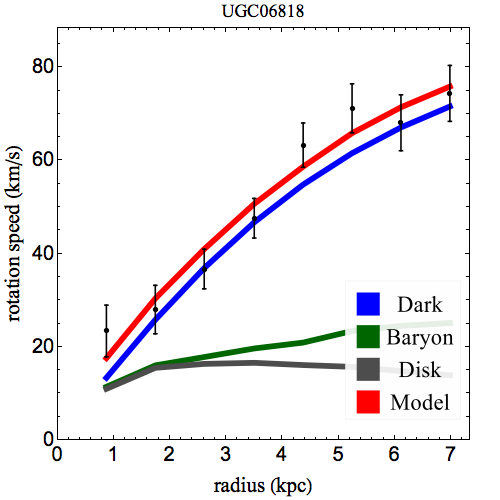} & \includegraphics[scale=0.192]{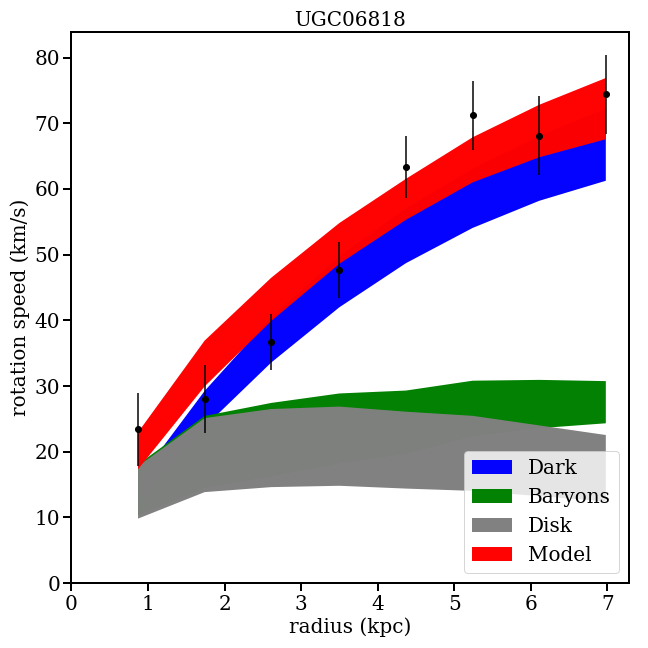} & \includegraphics[scale=0.215]{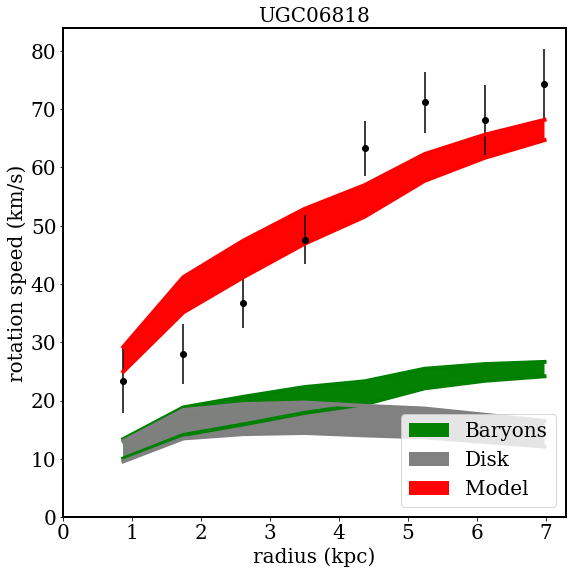} \\
\includegraphics[scale=0.25]{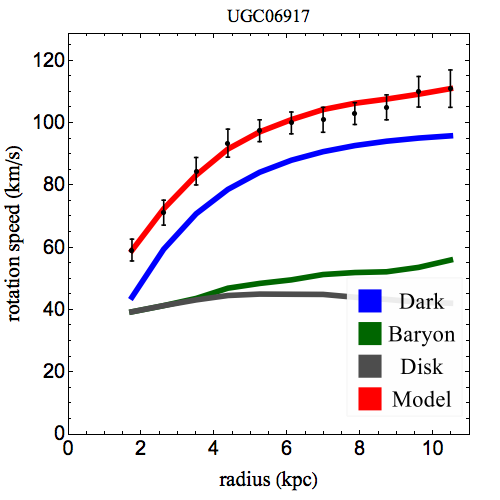} & \includegraphics[scale=0.192]{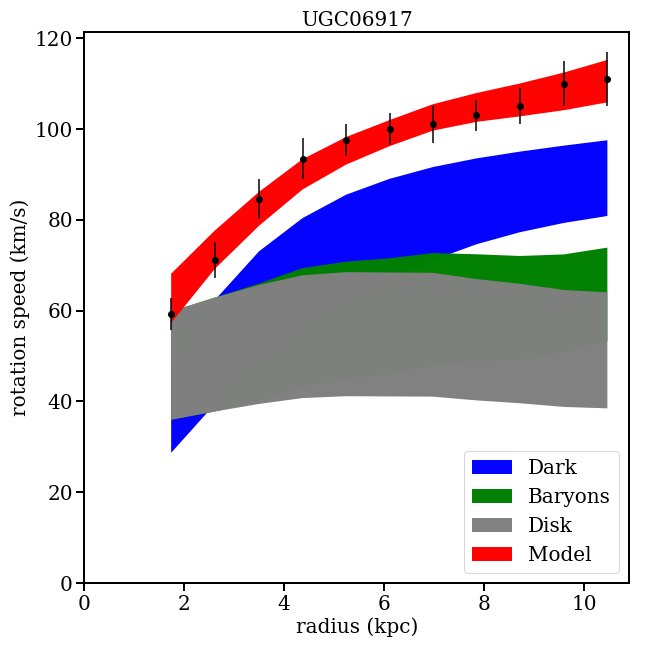} & \includegraphics[scale=0.215]{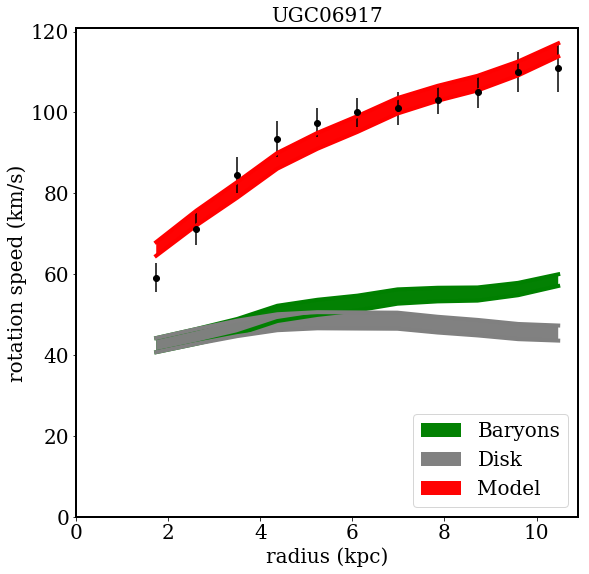} \\
\end{tabular}
\caption{Continued.}
\end{figure}

\addtocounter{figure}{-1}
\begin{figure}[H]
\centering
\begin{tabular}{@{}ccc@{}}
\includegraphics[scale=0.25]{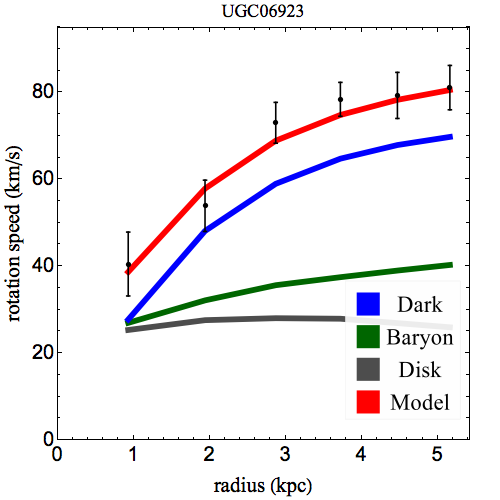} & \includegraphics[scale=0.192]{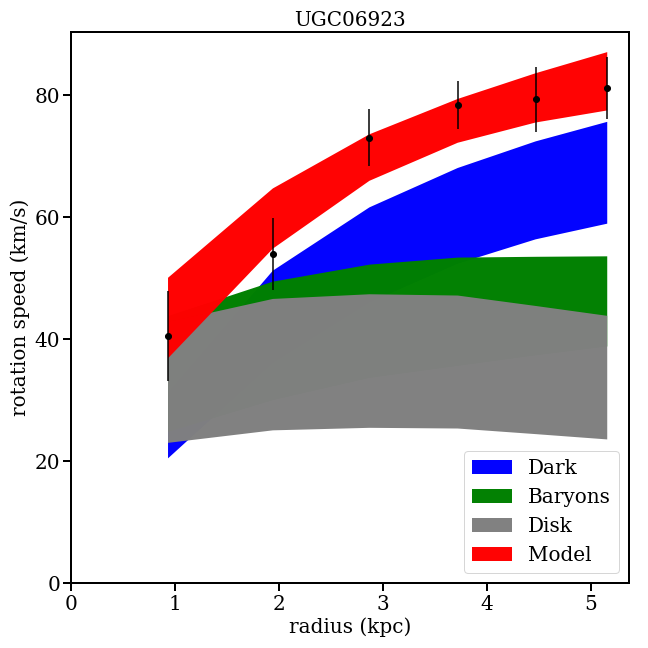} & \includegraphics[scale=0.215]{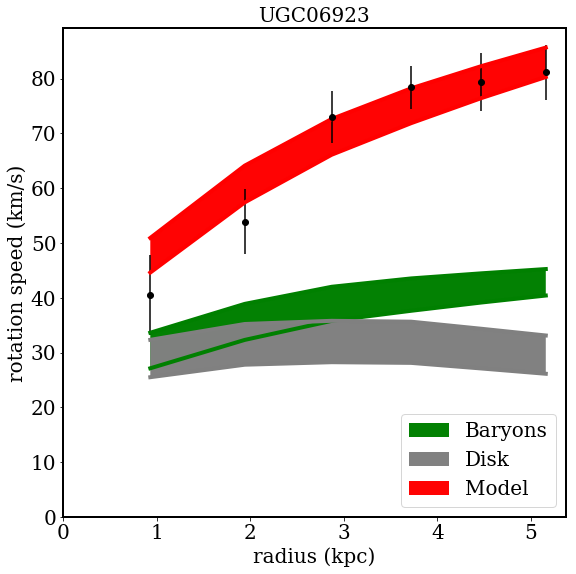} \\
\includegraphics[scale=0.25]{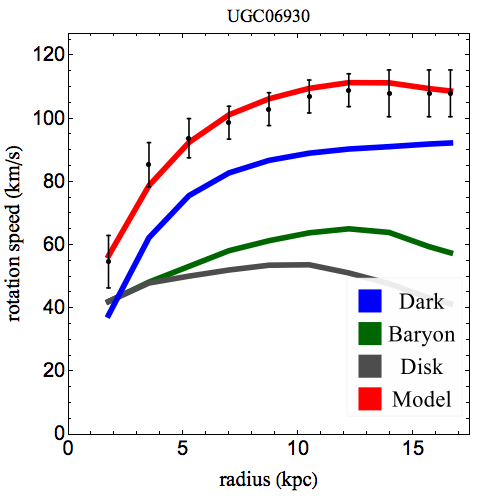} & \includegraphics[scale=0.192]{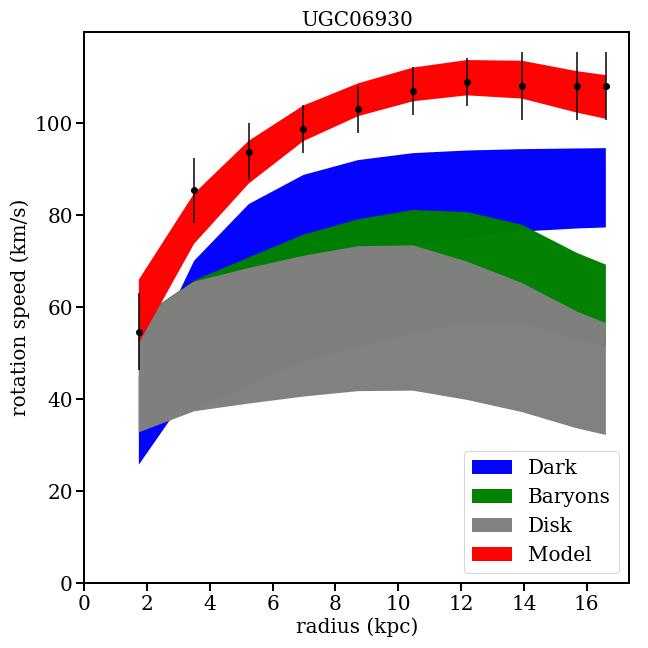} & \includegraphics[scale=0.215]{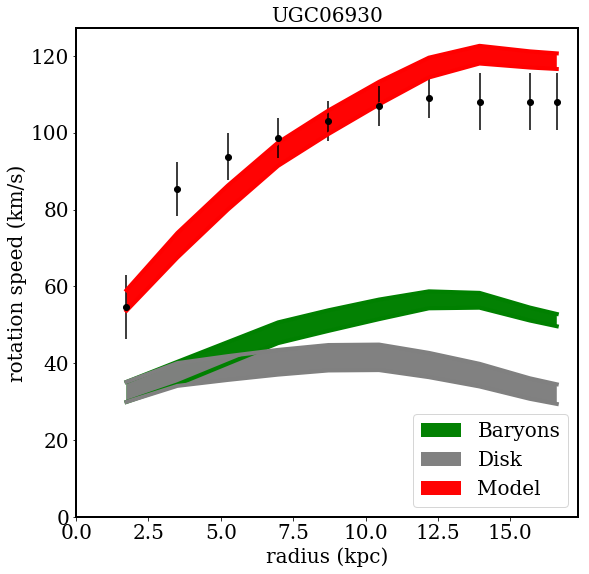} \\
\includegraphics[scale=0.25]{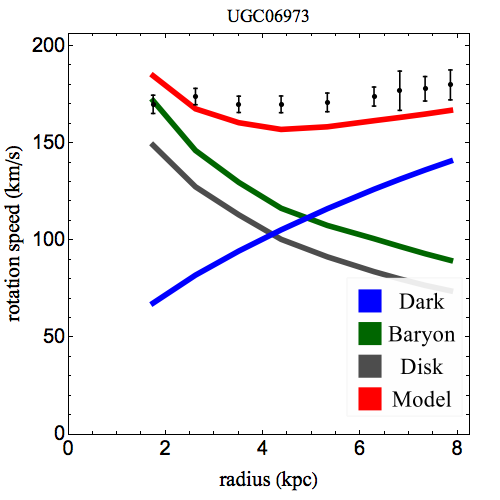} & \includegraphics[scale=0.192]{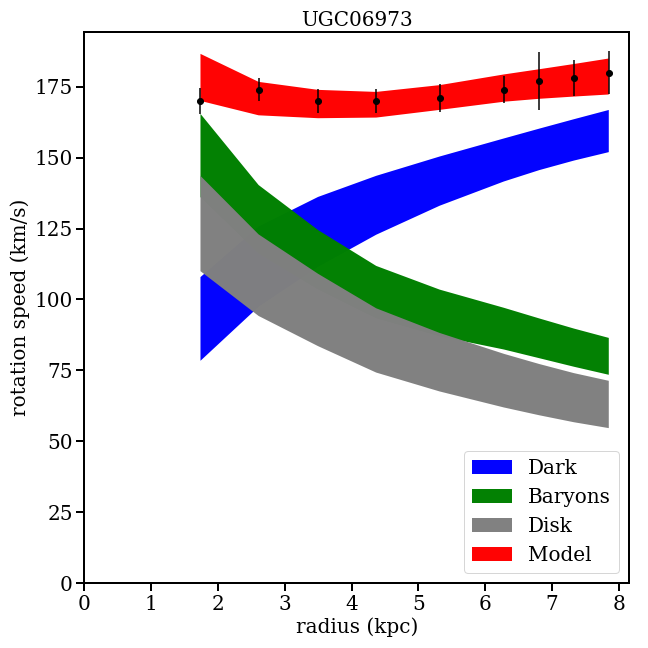} & \includegraphics[scale=0.215]{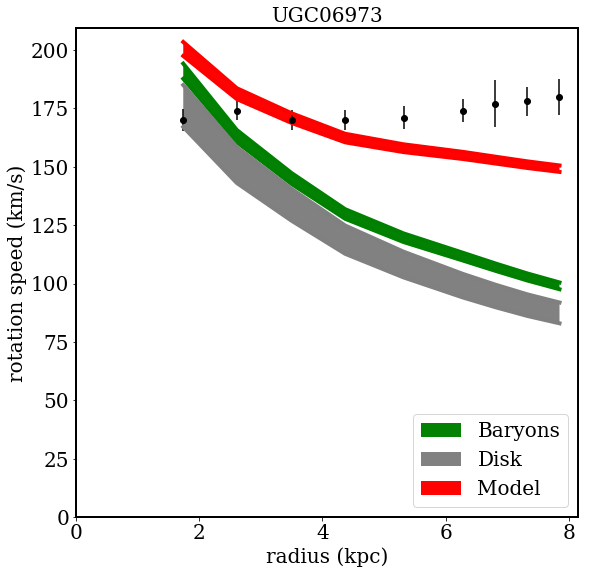} \\
\includegraphics[scale=0.25]{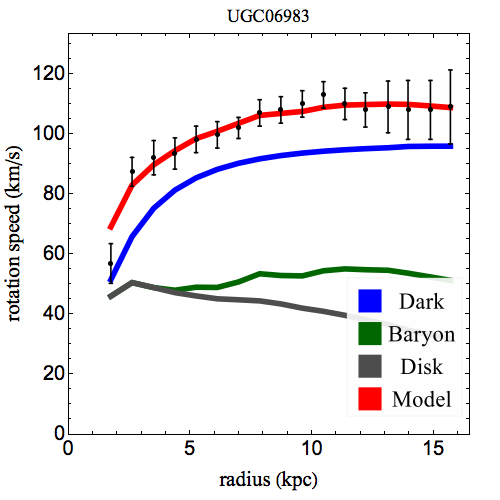} & \includegraphics[scale=0.192]{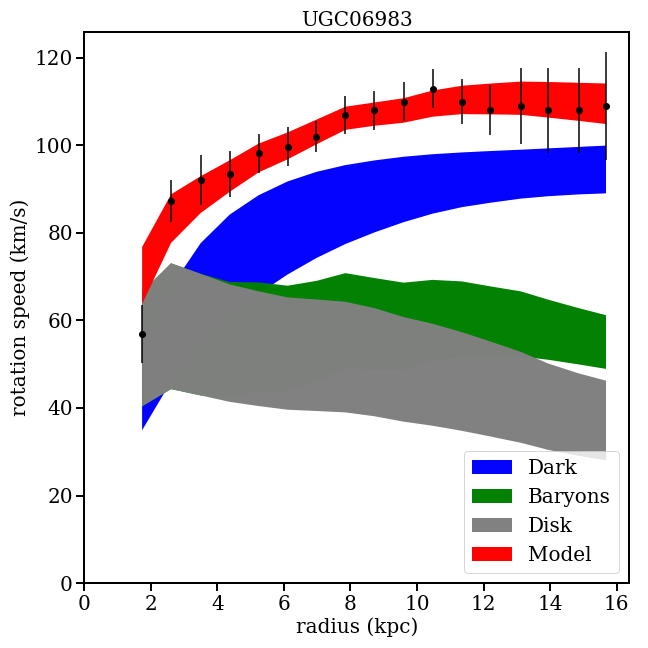} & \includegraphics[scale=0.215]{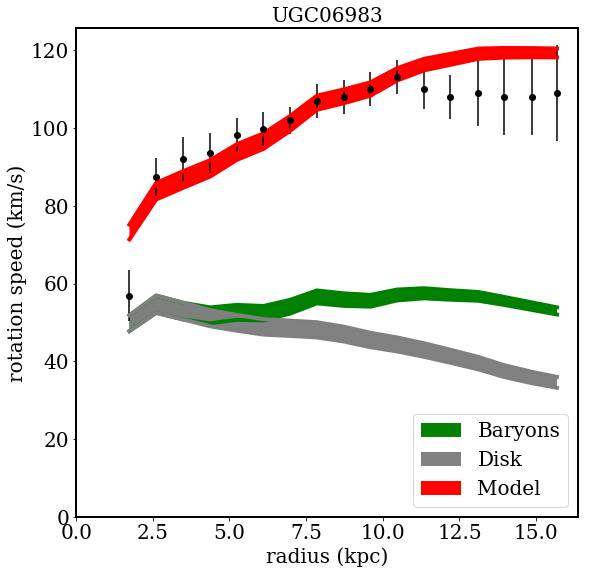} \\
\includegraphics[scale=0.25]{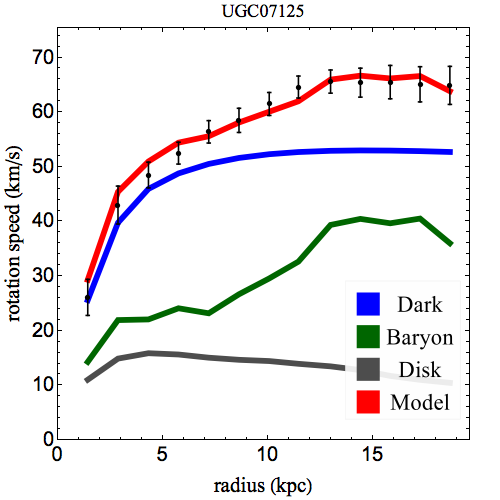} & \includegraphics[scale=0.192]{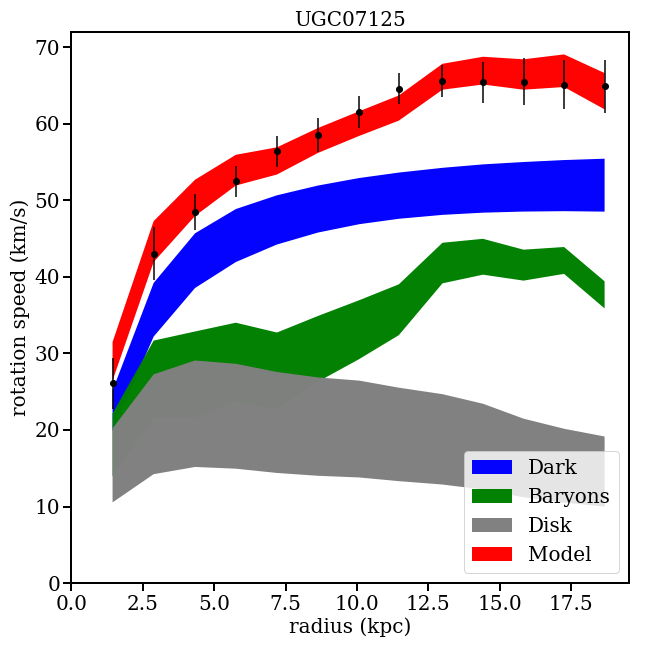} & \includegraphics[scale=0.215]{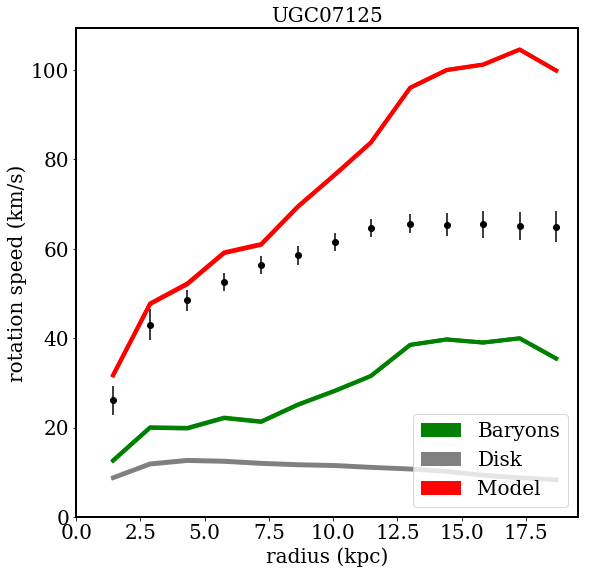} \\
\end{tabular}
\caption{Continued.}
\end{figure}

\addtocounter{figure}{-1}
\begin{figure}[H]
\centering
\begin{tabular}{@{}ccc@{}}
\includegraphics[scale=0.25]{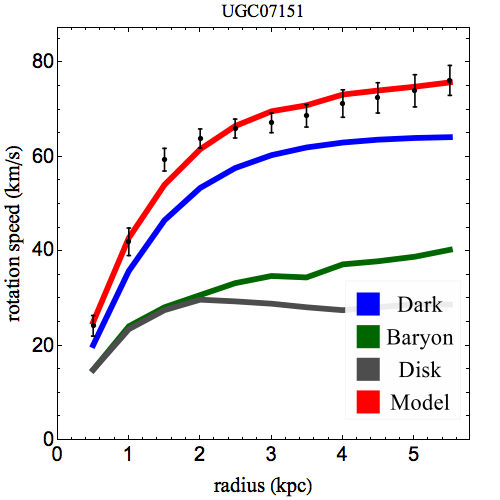} & \includegraphics[scale=0.192]{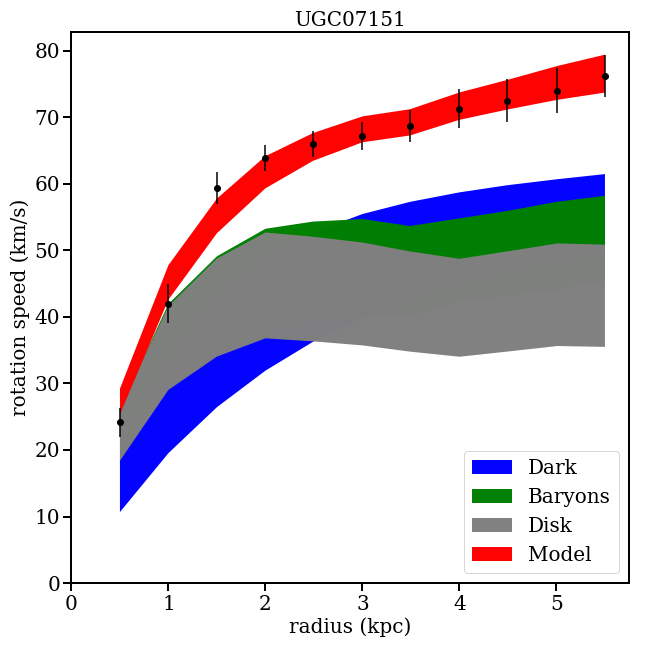} & \includegraphics[scale=0.215]{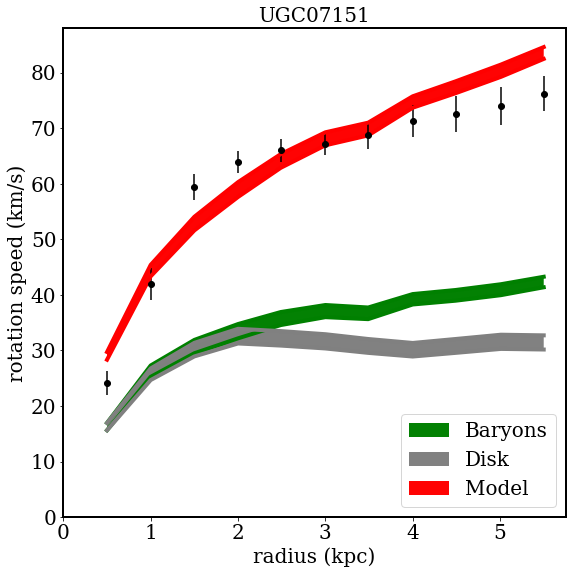} \\
\includegraphics[scale=0.25]{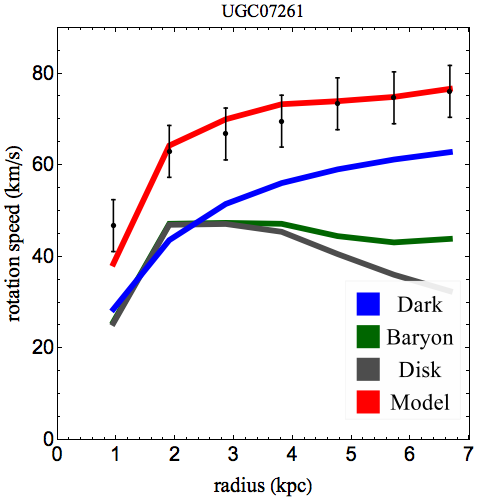} & \includegraphics[scale=0.192]{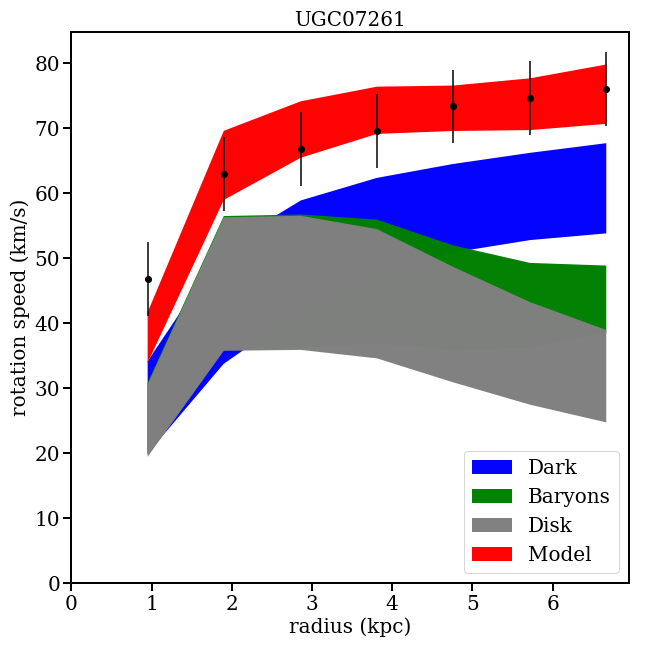} & \includegraphics[scale=0.215]{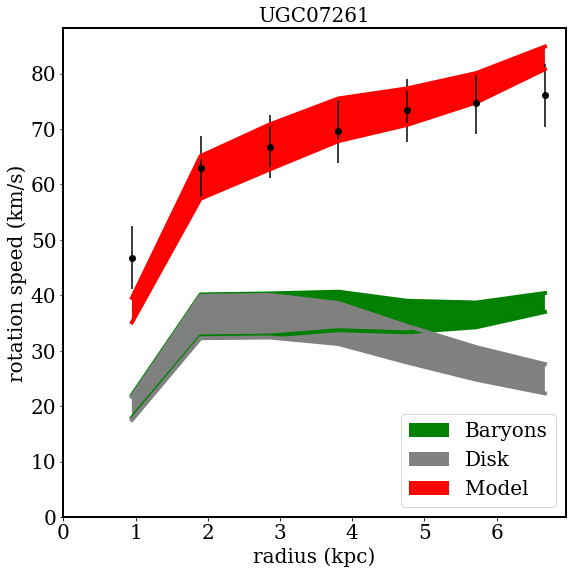} \\
\includegraphics[scale=0.25]{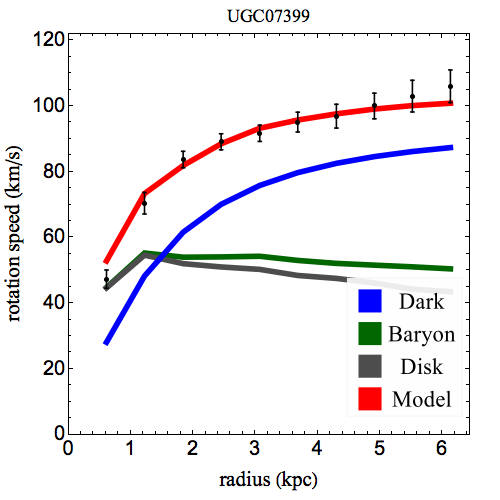} & \includegraphics[scale=0.192]{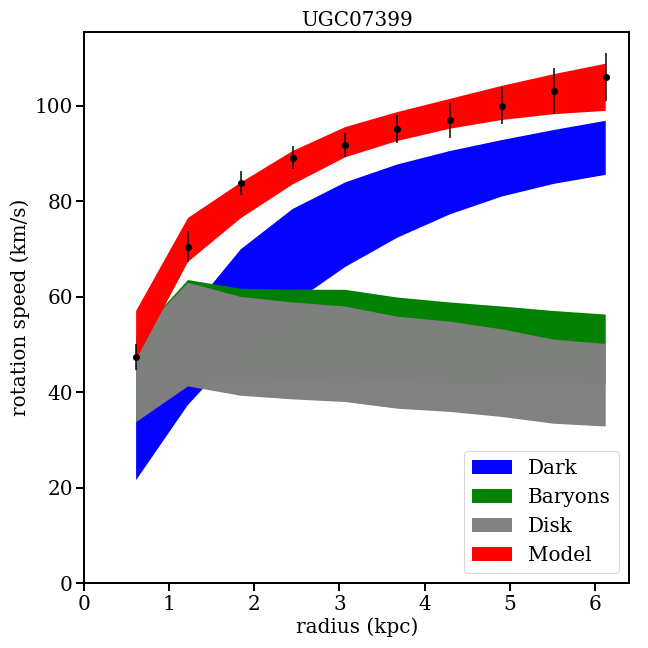} & \includegraphics[scale=0.215]{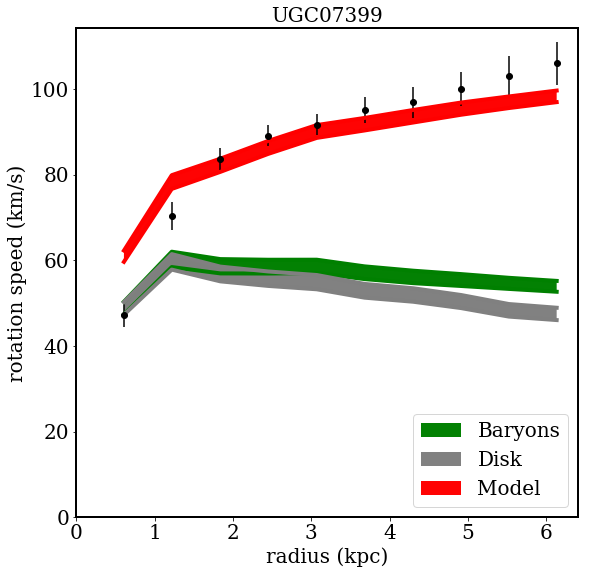} \\
\includegraphics[scale=0.25]{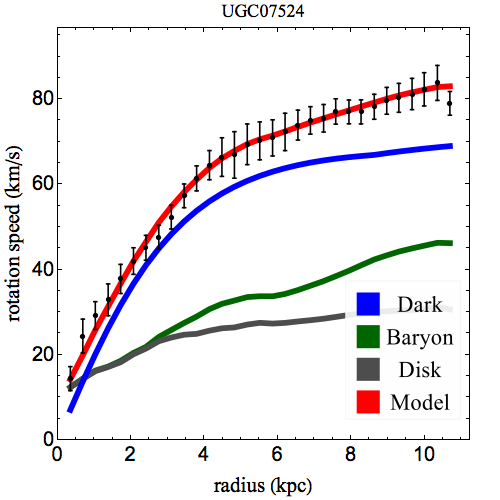} & \includegraphics[scale=0.192]{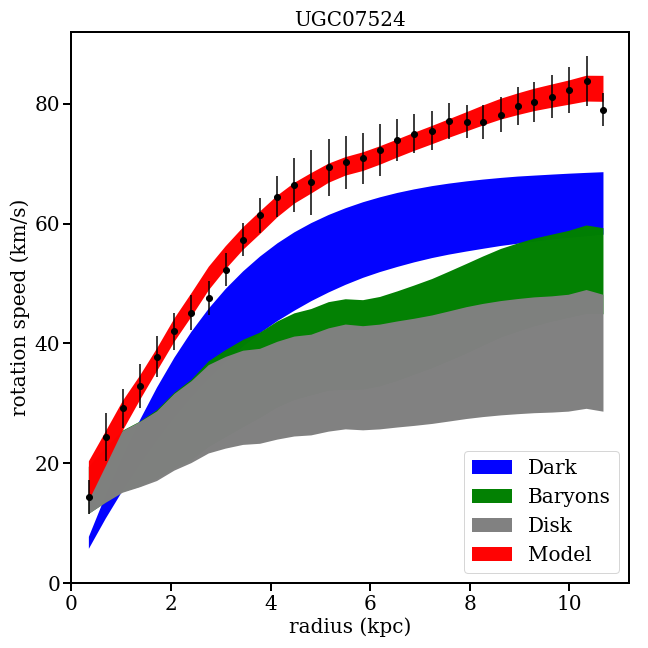} & \includegraphics[scale=0.215]{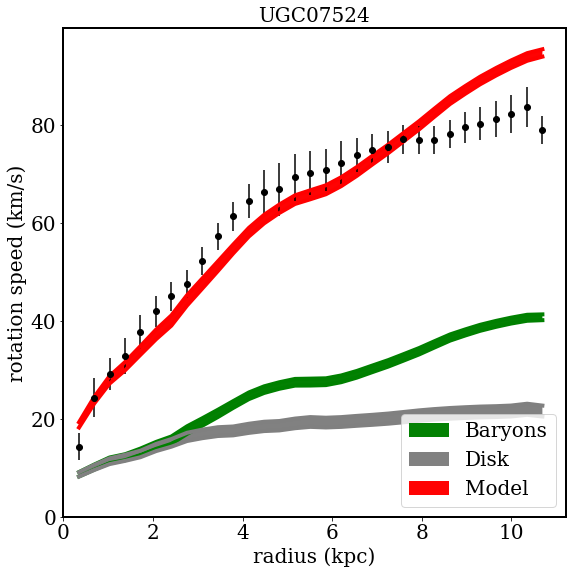} \\
\includegraphics[scale=0.25]{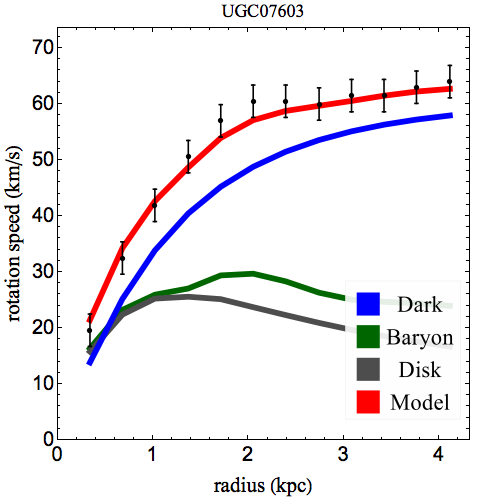} & \includegraphics[scale=0.192]{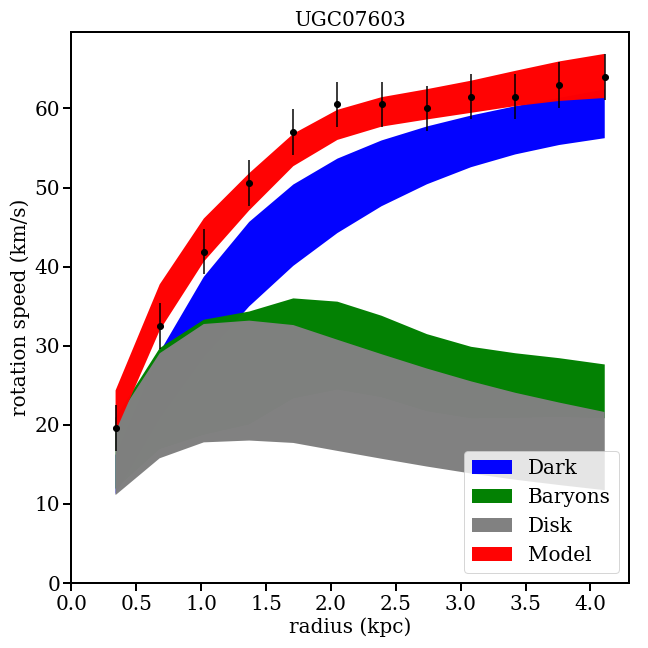} & \includegraphics[scale=0.215]{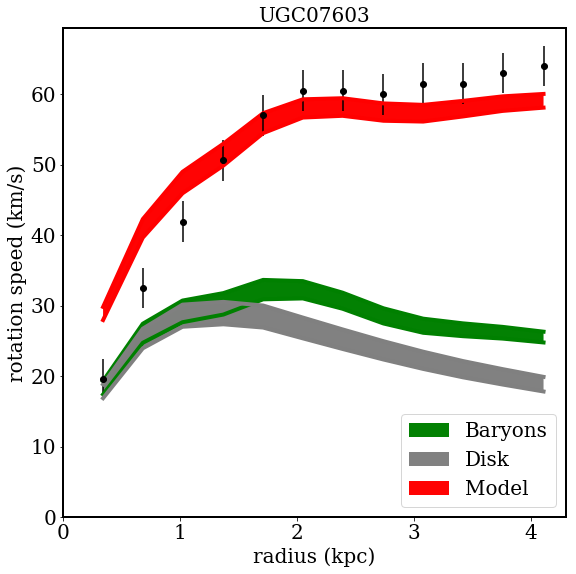} \\
\end{tabular}
\caption{Continued.}
\end{figure}

\addtocounter{figure}{-1}
\begin{figure}[H]
\centering
\begin{tabular}{@{}ccc@{}}
\includegraphics[scale=0.25]{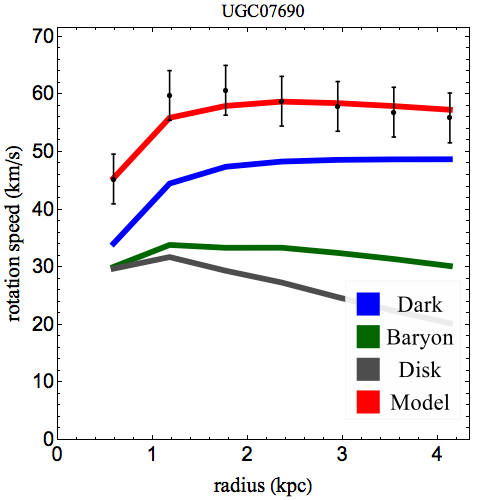} & \includegraphics[scale=0.192]{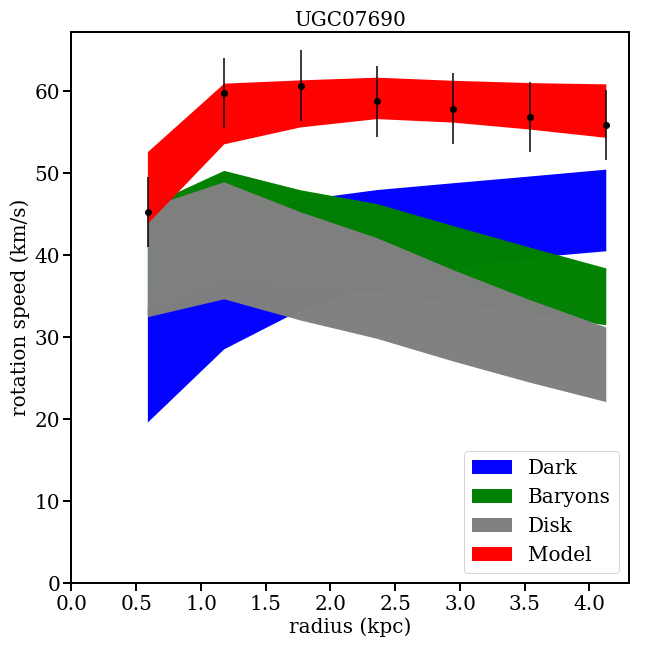} & \includegraphics[scale=0.215]{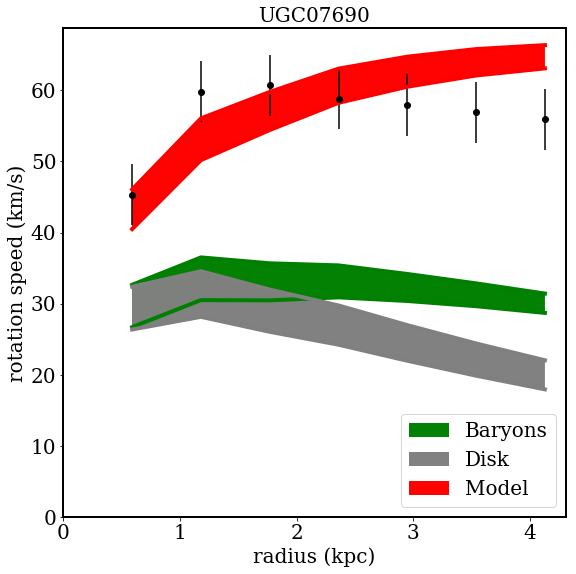} \\
\includegraphics[scale=0.25]{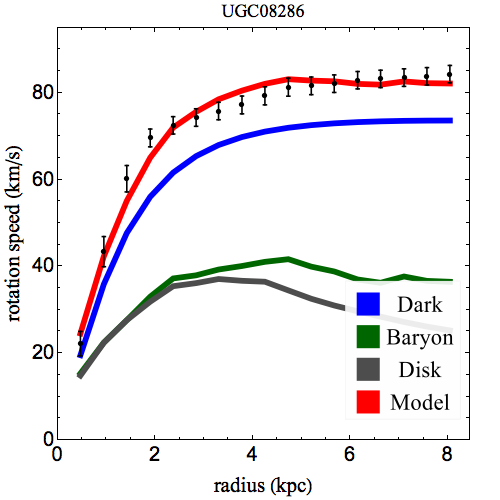} & \includegraphics[scale=0.192]{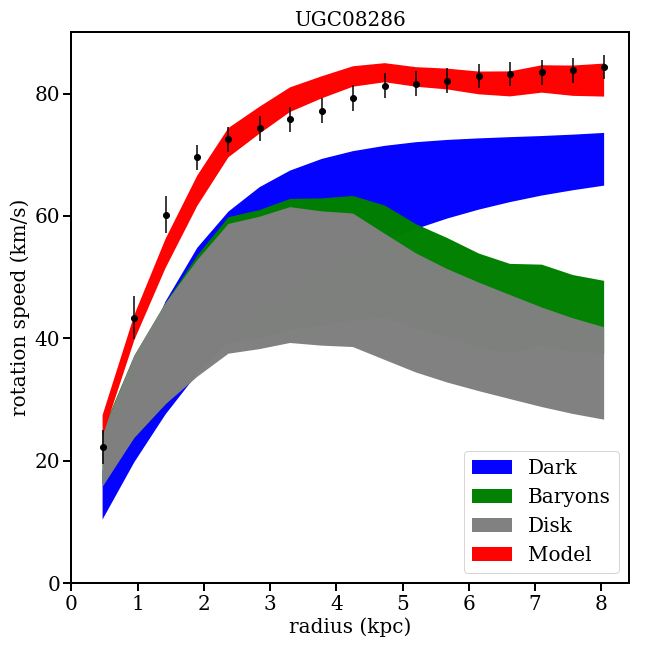} & \includegraphics[scale=0.215]{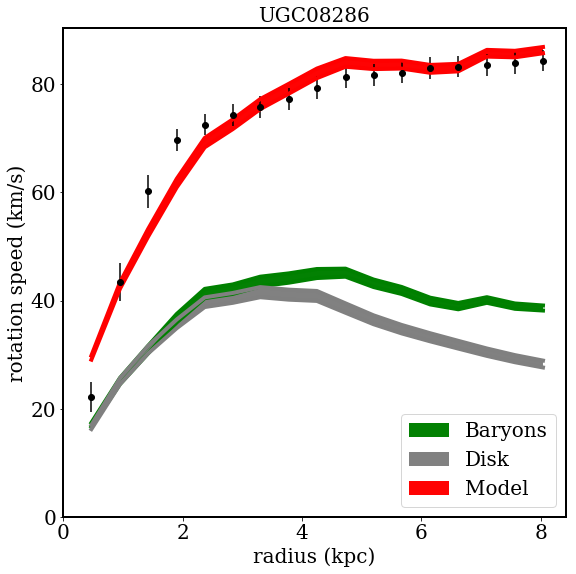} \\
\includegraphics[scale=0.25]{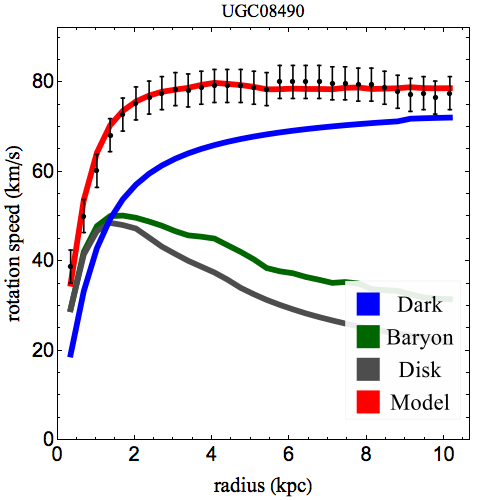} & \includegraphics[scale=0.192]{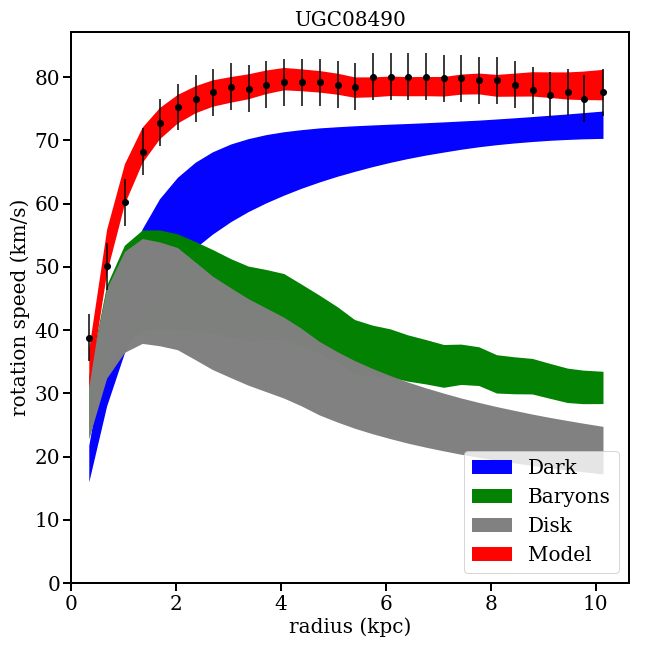} & \includegraphics[scale=0.215]{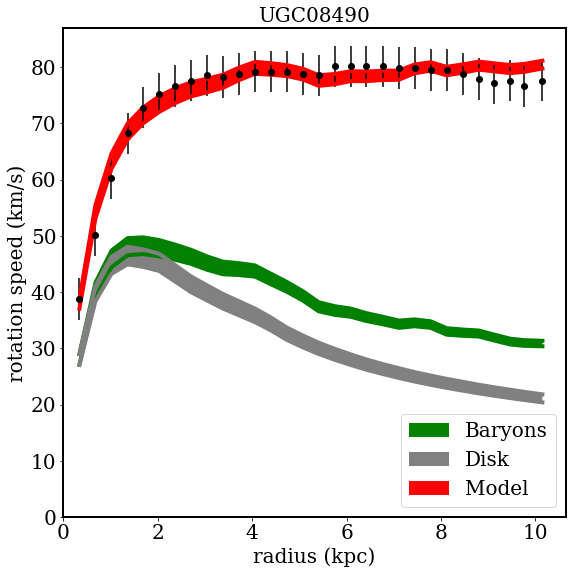} \\
\includegraphics[scale=0.25]{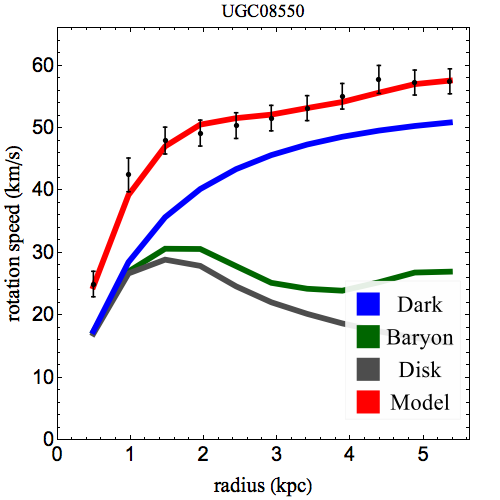} & \includegraphics[scale=0.192]{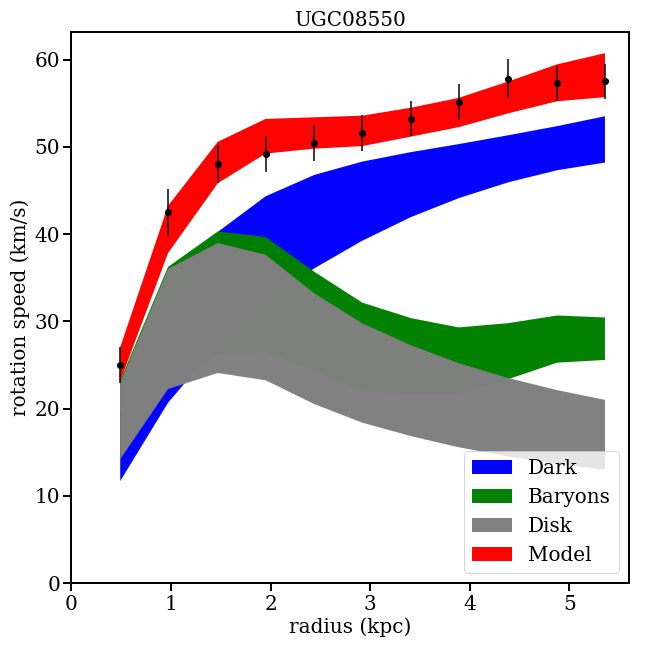} & \includegraphics[scale=0.215]{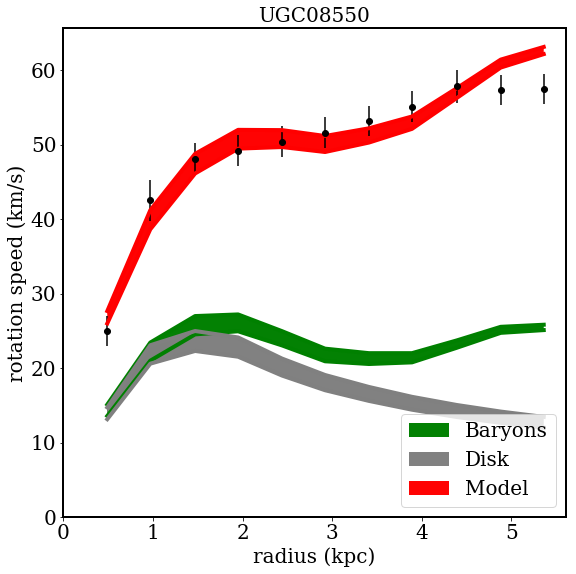} \\
\includegraphics[scale=0.25]{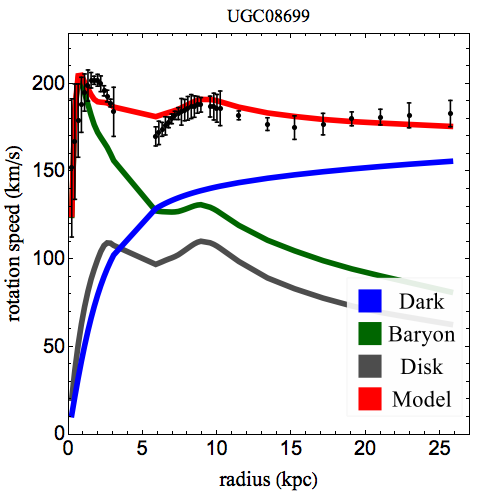} & \includegraphics[scale=0.192]{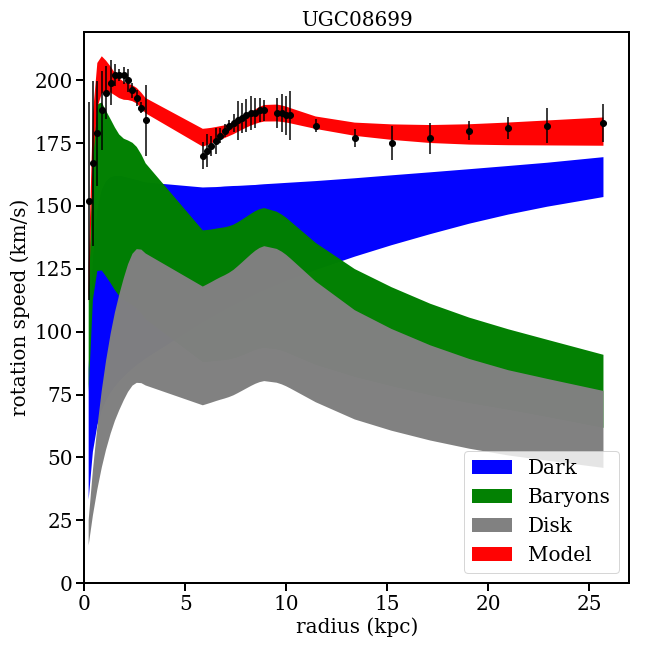} & \includegraphics[scale=0.215]{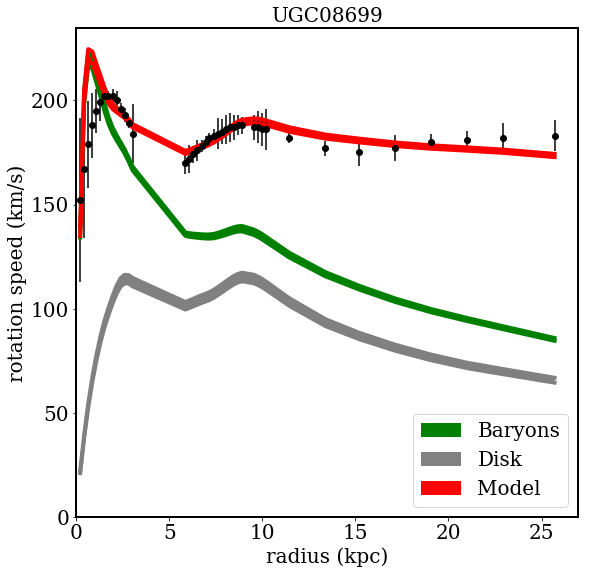} \\
\end{tabular}
\caption{Continued.}
\end{figure}

\addtocounter{figure}{-1}
\begin{figure}[H]
\centering
\begin{tabular}{@{}ccc@{}}
\includegraphics[scale=0.25]{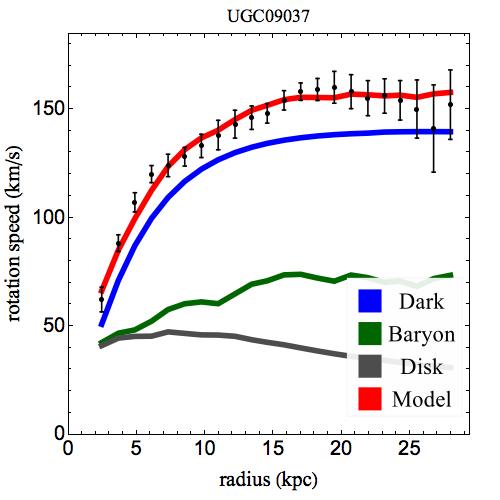} & \includegraphics[scale=0.192]{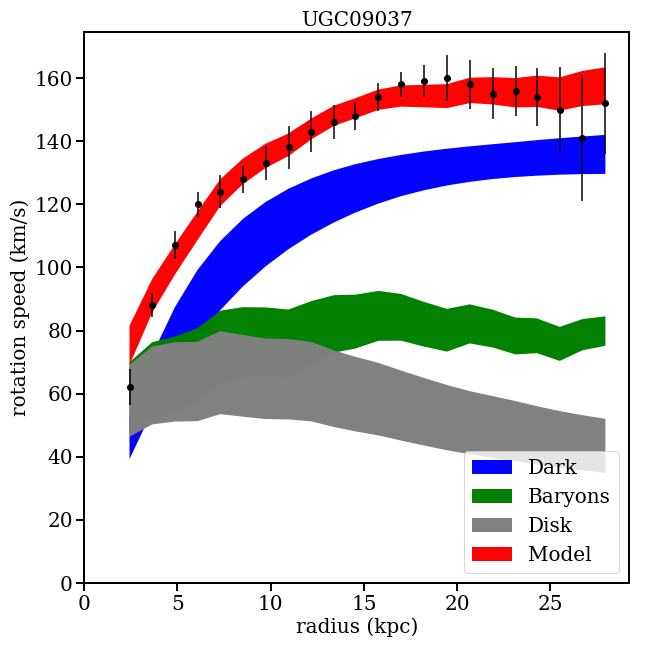} & \includegraphics[scale=0.215]{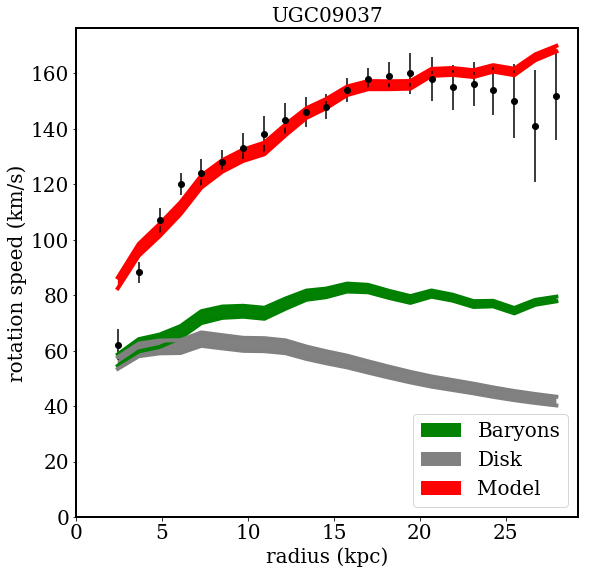} \\
\includegraphics[scale=0.25]{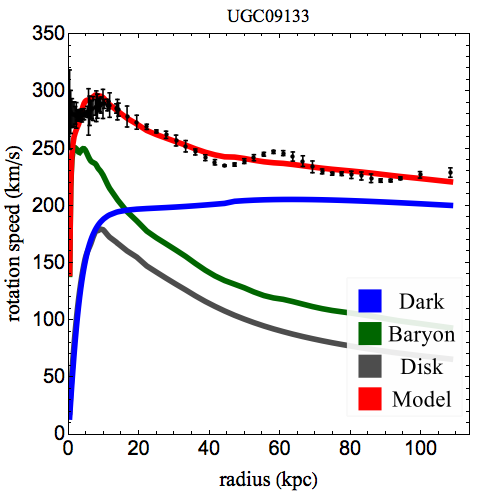} & \includegraphics[scale=0.192]{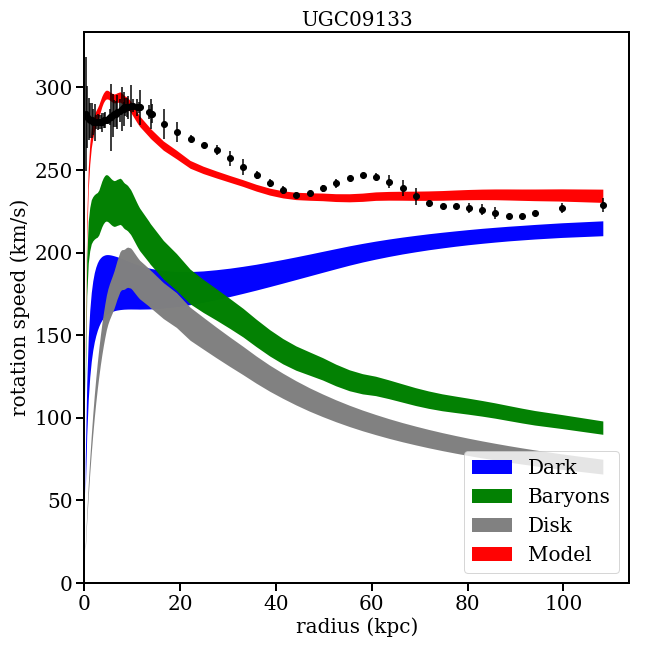} & \includegraphics[scale=0.215]{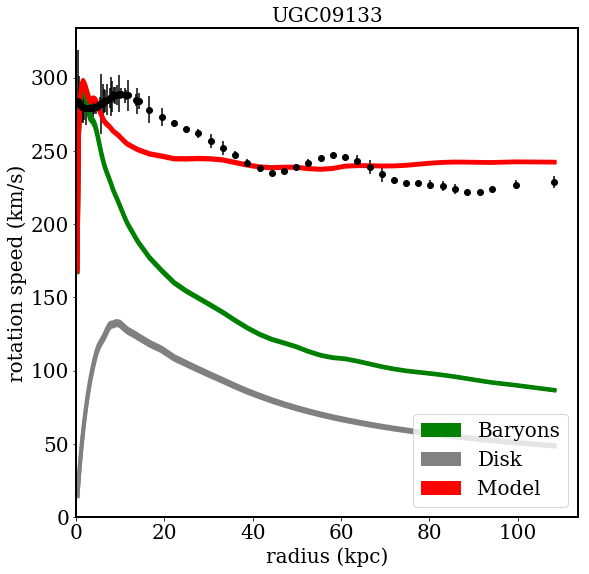} \\
\includegraphics[scale=0.25]{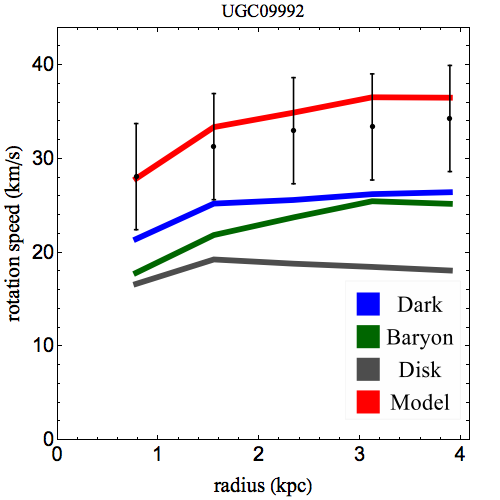} & \includegraphics[scale=0.192]{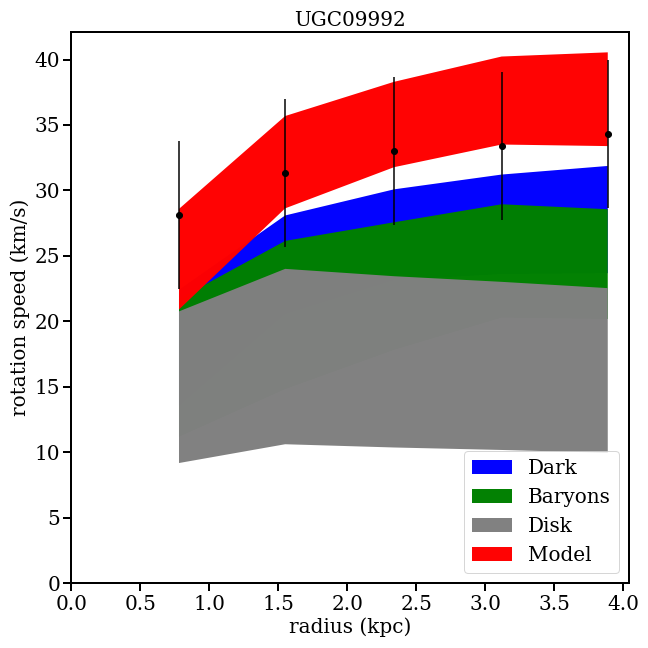} & \includegraphics[scale=0.215]{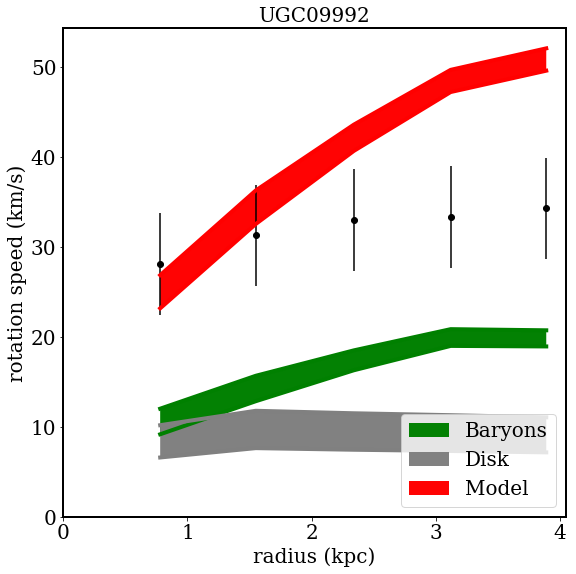} \\
\includegraphics[scale=0.25]{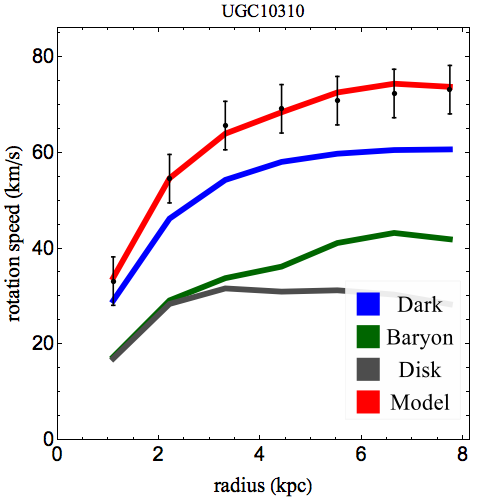} & \includegraphics[scale=0.192]{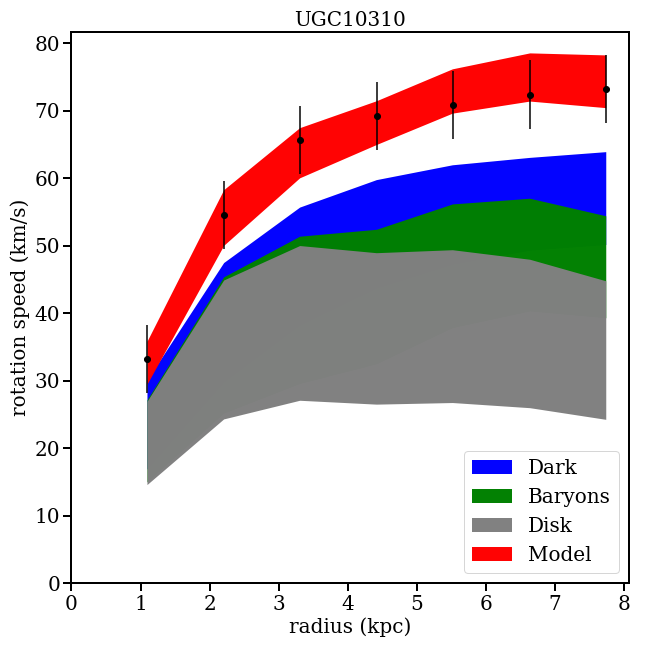} & \includegraphics[scale=0.215]{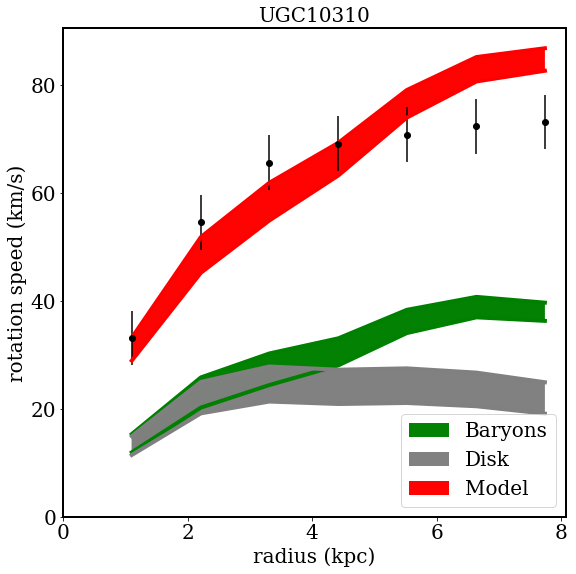} \\
\includegraphics[scale=0.25]{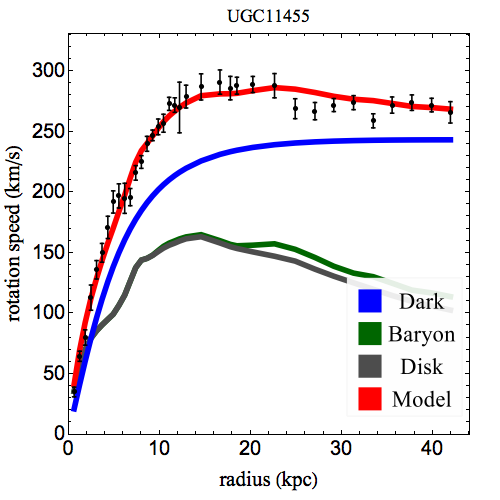} & \includegraphics[scale=0.192]{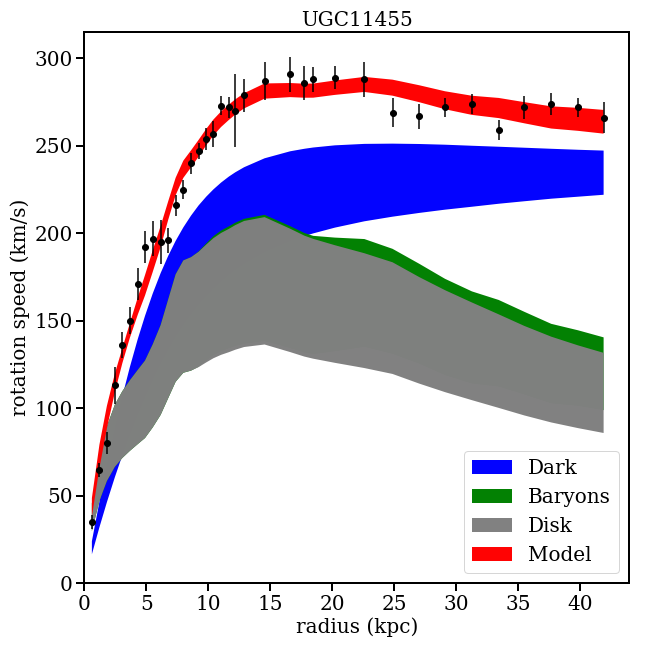} & \includegraphics[scale=0.215]{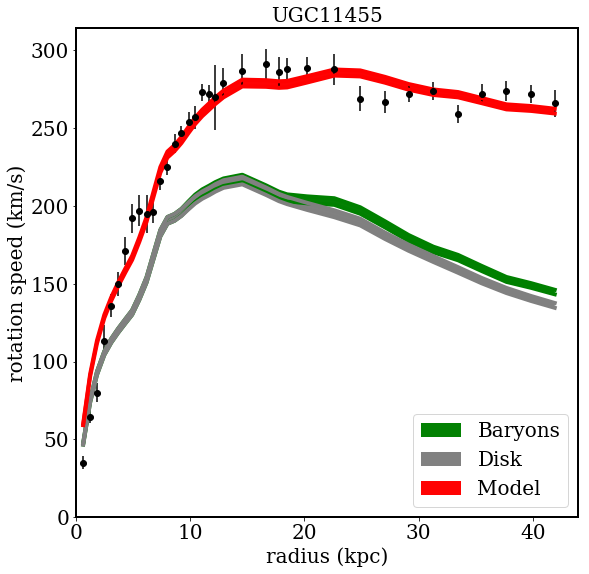} \\
\end{tabular}
\caption{Continued.}
\end{figure}

\addtocounter{figure}{-1}
\begin{figure}[H]
\centering
\begin{tabular}{@{}ccc@{}}
\includegraphics[scale=0.25]{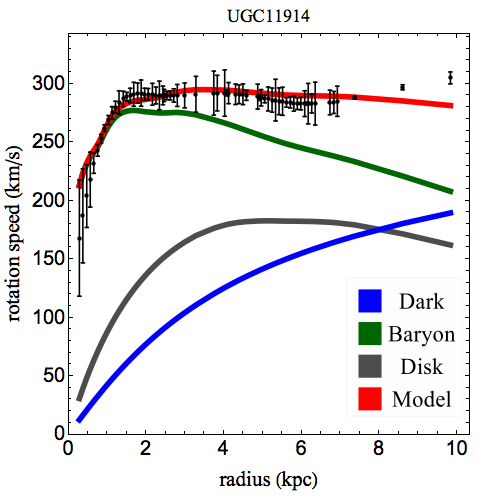} & \includegraphics[scale=0.192]{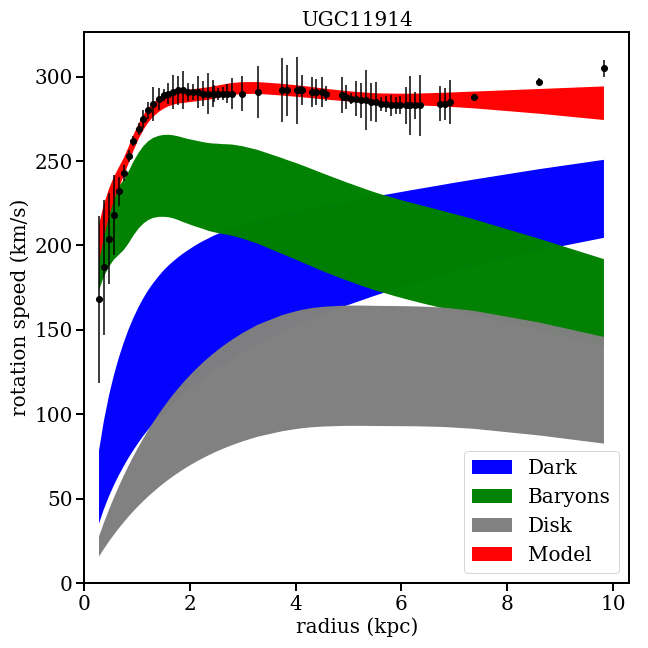} & \includegraphics[scale=0.215]{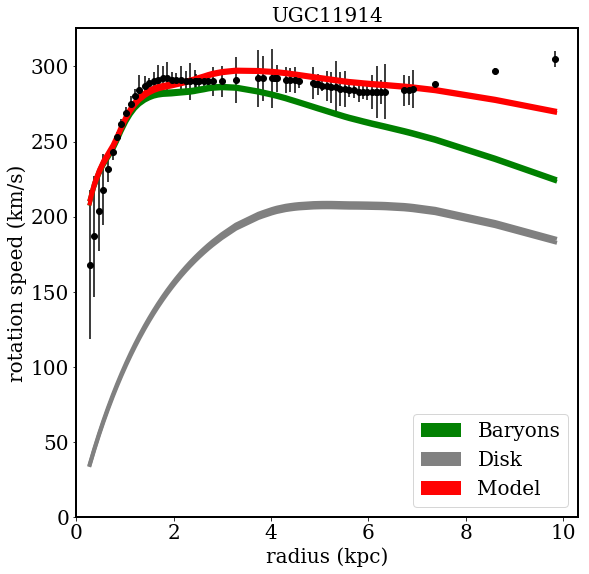} \\
\includegraphics[scale=0.25]{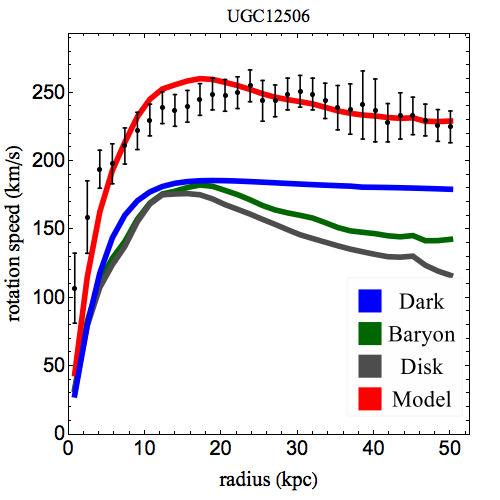} & \includegraphics[scale=0.192]{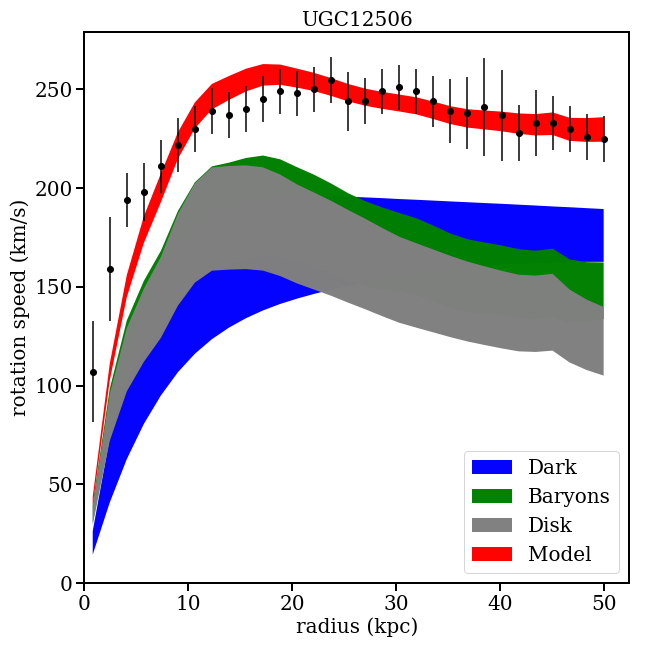} & \includegraphics[scale=0.215]{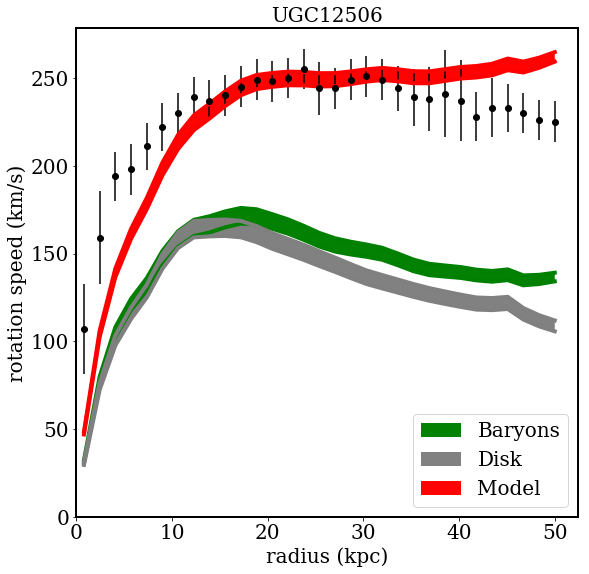} \\
\includegraphics[scale=0.25]{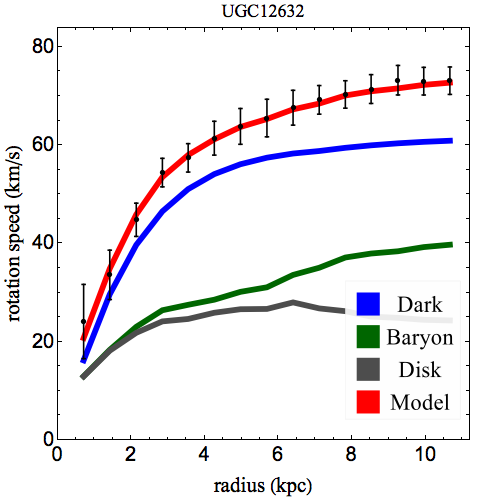} & \includegraphics[scale=0.192]{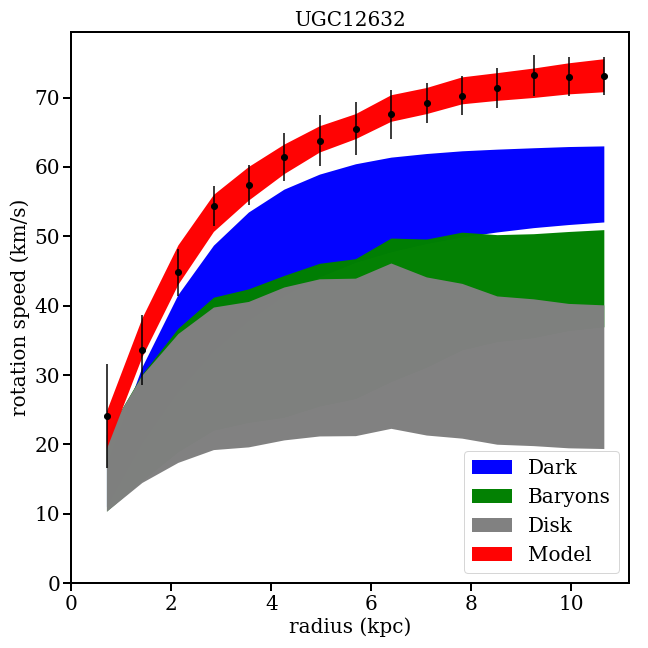} & \includegraphics[scale=0.215]{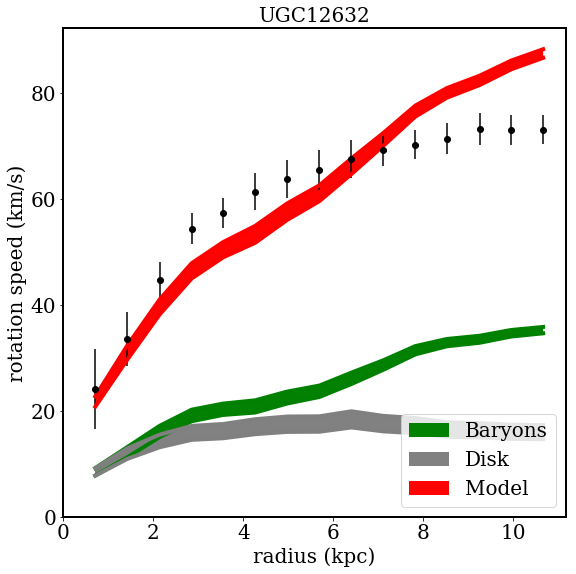} \\
\includegraphics[scale=0.25]{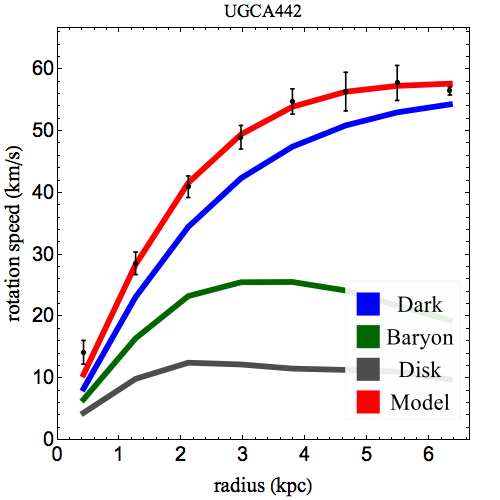} & \includegraphics[scale=0.192]{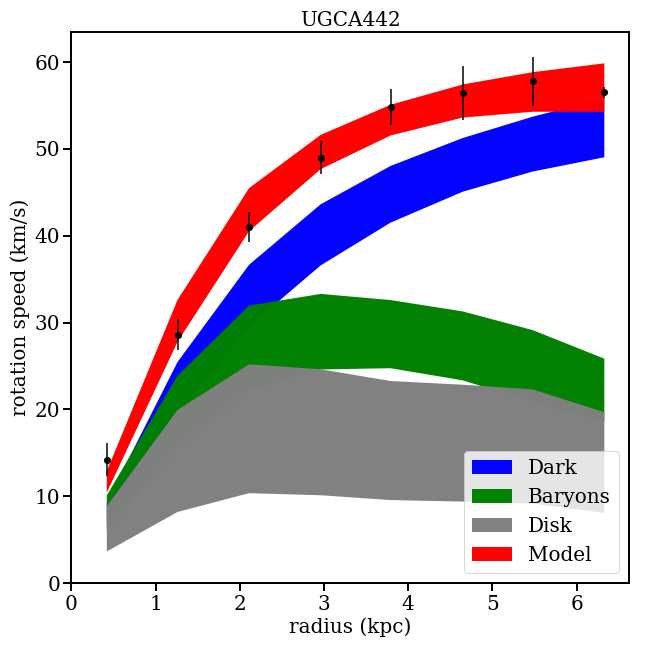} & \includegraphics[scale=0.215]{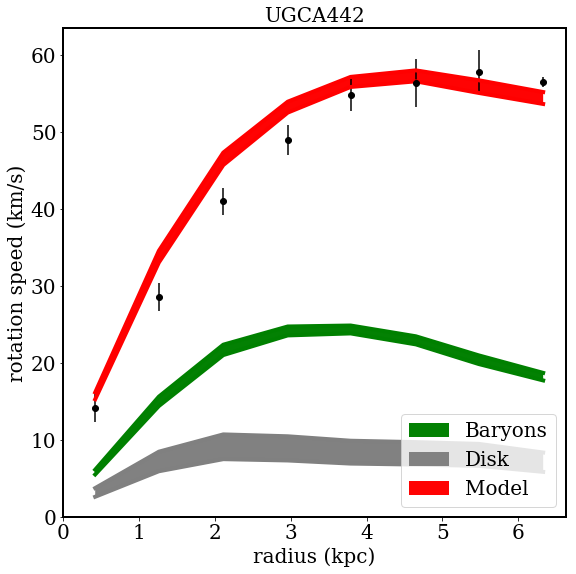} \\
\includegraphics[scale=0.25]{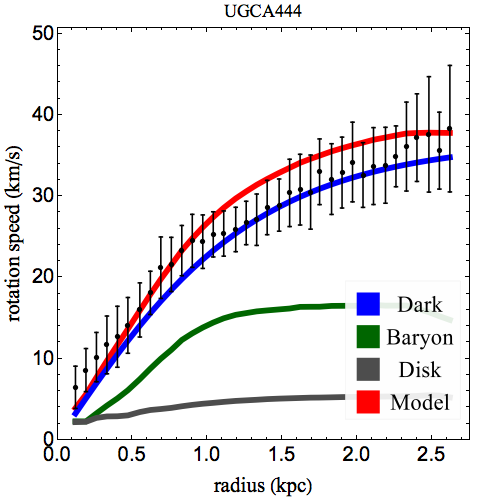} & \includegraphics[scale=0.192]{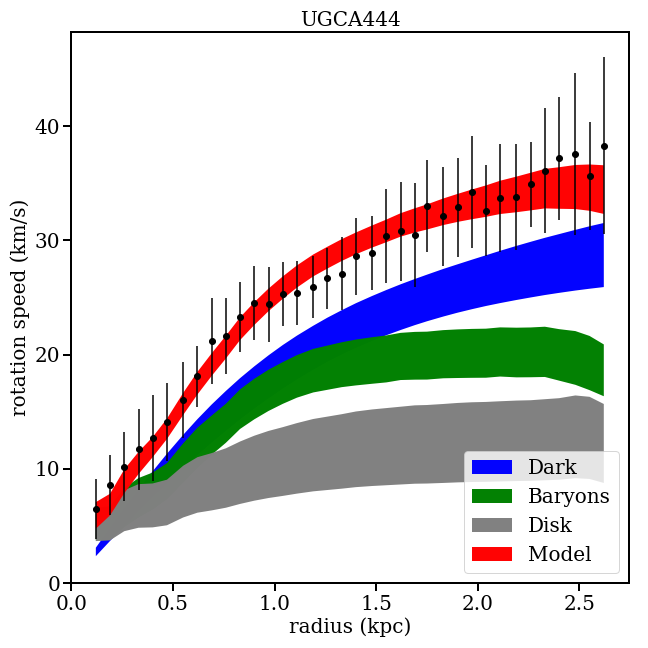} & \includegraphics[scale=0.215]{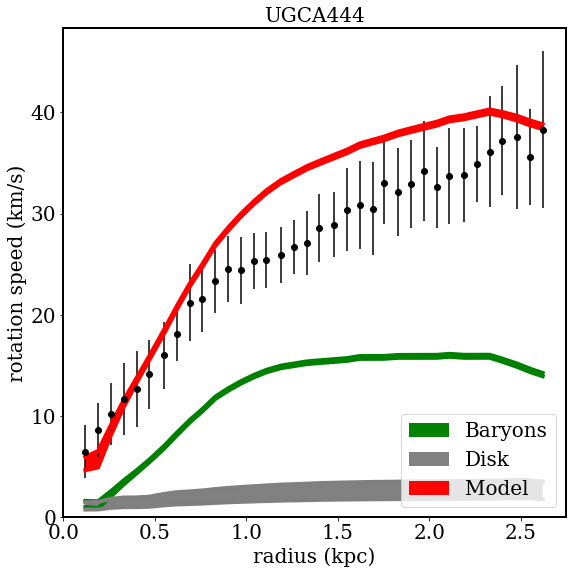} \\
\end{tabular}
\caption{Continued.}
\end{figure}

\centering

\addtocounter{table}{-1}
\text{}
\begin{table}[H]
\begin{longtable}{*{9}{l}}
\hline
Name & ${V}_{\rm max}$ & ${r}_{\rm max}$ & $\rho_{0}\times10^7$ & $\sigma_{0}$ & $-\alpha$ & $\Upsilon_{\star,{\rm disk}}$ &$\Upsilon_{\star,{\rm bulge}}$ & $\chi^2/{\rm{d.o.f.}}$\\ 
& [km/s] & [kpc] & $[{M}_{\odot}/\rm{kpc}^3]$ & [km/s] &  & $[{M}_{\odot}/{L}_{\odot}]$ & $[{M}_{\odot}/{L}_{\odot}]$ &\\ 
(1) & (2) & (3) & (4) & (5) & (6) & (7) & (8) & (9)\\
\hline
D631-7 & 60.96 & 14.56 & 1.10 & 38.15 & $0.30$ & 0.20 & 0.00 & 0.49\\
 & 82.92 & 47.16 & 1.16 & 47.56 & $0.57$ & 0.10 & 0.00 & 0.47\\
 & $75.88_{-7.87}^{+4.56}$ & $40.36_{-9.02}^{+7.01}$ & $1.36_{-0.13}^{+0.22}$ & $44.49_{-3.07}^{+2.17}$ & $0.63_{-0.04}^{+0.04}$ & $0.19_{-0.07}^{+0.18}$ & $0.00_{-0.00}^{+0.00}$ & \\
DDO064 & 54.42 & 11.61 & 3.47 & 35.00 & $0.73$ & 0.72 & 0.00 & 0.59\\
 & 45.76 & 4.37 & 5.45 & 31.69 & $0.64$ & 0.11 & 0.00 & 0.31\\
\hline\hline
\end{longtable}
\caption{Model parameters and $\chi^2/{\rm d.o.f.}$ values for the SIDM fits shown in Fig. S3, with controlled sampling (first row associated with each galaxy) and MCMC sampling with the tophat prior (best-fit value, second row; medium with $1\sigma$ errors, third row). The $\alpha$ value is the logarithmic slope of the dark matter density profile at $r=1.5\% r_{\rm vir}$. Galaxies are listed alphabetically, corresponding to the order in Fig. S3.}
\end{table}

\clearpage
\addtocounter{table}{-2}
\centering
\text{}
\begin{table}[H]
\begin{longtable}{*{9}{l}}
\hline
Name & ${V}_{\rm max}$ & ${r}_{\rm max}$ & $\rho_{0}\times10^7$ & $\sigma_{0}$ & $-\alpha$ & $\Upsilon_{\star,{\rm disk}}$ &$\Upsilon_{\star,{\rm bulge}}$ & $\chi^2/{\rm{d.o.f.}}$\\ 
& [km/s] & [kpc] & $[{M}_{\odot}/\rm{kpc}^3]$ & [km/s] &  & $[{M}_{\odot}/{L}_{\odot}]$ & $[{M}_{\odot}/{L}_{\odot}]$ &\\ 
(1) & (2) & (3) & (4) & (5) & (6) & (7) & (8) & (9)\\
\hline
F563-1 & 100.10 & 27.27 & 3.21 & 68.87 & $0.75$ & 2.25 & 0.00 & 0.73\\
 & 104.81 & 10.30 & 4.57 & 67.19 & $0.63$ & 0.11 & 0.00 & 0.55\\
 & $100.94_{-8.30}^{+9.62}$ & $19.38_{-7.60}^{+31.98}$ & $2.93_{-1.35}^{+1.53}$ & $71.28_{-5.12}^{+8.27}$ & $0.70_{-0.10}^{+0.11}$ & $2.08_{-1.40}^{+1.67}$ & $0.00_{-0.00}^{+0.00}$ & \\
F563-V2 & 102.17 & 20.74 & 9.70 & 75.08 & $1.22$ & 1.96 & 0.00 & 0.13\\
 & 94.30 & 22.72 & 4.38 & 82.37 & $1.01$ & 3.05 & 0.00 & 0.11\\
\hline
\end{longtable}
\caption{Continued}
\end{table}
\clearpage
\addtocounter{table}{-2}
\centering
\text{}
\begin{table}[H]
\begin{longtable}{*{9}{l}}
\hline
Name & ${V}_{\rm max}$ & ${r}_{\rm max}$ & $\rho_{0}\times10^7$ & $\sigma_{0}$ & $-\alpha$ & $\Upsilon_{\star,{\rm disk}}$ &$\Upsilon_{\star,{\rm bulge}}$ & $\chi^2/{\rm{d.o.f.}}$\\ 
& [km/s] & [kpc] & $[{M}_{\odot}/\rm{kpc}^3]$ & [km/s] &  & $[{M}_{\odot}/{L}_{\odot}]$ & $[{M}_{\odot}/{L}_{\odot}]$ &\\ 
(1) & (2) & (3) & (4) & (5) & (6) & (7) & (8) & (9)\\
\hline
KK98-251 & 32.46 & 10.60 & 2.59 & 17.89 & $0.74$ & 0.25 & 0.00 & 1.99\\
 & 45.88 & 26.23 & 1.27 & 26.95 & $0.63$ & 0.11 & 0.00 & 0.28\\
 & $35.25_{-5.55}^{+6.89}$ & $15.01_{-4.44}^{+5.11}$ & $1.52_{-0.12}^{+0.26}$ & $23.59_{-2.94}^{+3.14}$ & $0.54_{-0.03}^{+0.04}$ & $0.36_{-0.21}^{+0.51}$ & $0.00_{-0.00}^{+0.00}$ & \\
NGC0024 & 106.13 & 21.22 & 19.65 & 77.38 & $1.70$ & 1.00 & 0.00 & 1.48\\
 & 154.15 & 120.65 & 4.60 & 97.34 & $1.25$ & 1.79 & 0.00 & 0.37\\
\hline
\end{longtable}
\caption{Continued}
\end{table}
\clearpage
\addtocounter{table}{-2}
\centering
\text{}
\begin{table}[H]
\begin{longtable}{*{9}{l}}
\hline
Name & ${V}_{\rm max}$ & ${r}_{\rm max}$ & $\rho_{0}\times10^7$ & $\sigma_{0}$ & $-\alpha$ & $\Upsilon_{\star,{\rm disk}}$ &$\Upsilon_{\star,{\rm bulge}}$ & $\chi^2/{\rm{d.o.f.}}$\\ 
& [km/s] & [kpc] & $[{M}_{\odot}/\rm{kpc}^3]$ & [km/s] &  & $[{M}_{\odot}/{L}_{\odot}]$ & $[{M}_{\odot}/{L}_{\odot}]$ &\\ 
(1) & (2) & (3) & (4) & (5) & (6) & (7) & (8) & (9)\\
\hline
NGC1090 & 140.44 & 23.86 & 13.59 & 109.15 & $1.31$ & 0.34 & 0.00 & 1.38\\
 & 147.65 & 16.51 & 10.55 & 109.37 & $1.11$ & 0.22 & 0.00 & 0.78\\
 & $141.51_{-7.52}^{+6.36}$ & $21.28_{-3.84}^{+19.24}$ & $8.87_{-3.09}^{+2.21}$ & $113.25_{-3.00}^{+7.15}$ & $1.09_{-0.06}^{+0.07}$ & $0.35_{-0.09}^{+0.16}$ & $0.00_{-0.00}^{+0.00}$ & \\
NGC1705 & 71.66 & 10.21 & 69.18 & 50.87 & $1.88$ & 1.21 & 0.00 & 0.24\\
 & 73.83 & 15.79 & 75.78 & 54.08 & $1.83$ & 1.22 & 0.00 & 0.14\\
\hline
\end{longtable}
\caption{Continued}
\end{table}
\clearpage
\addtocounter{table}{-2}
\centering
\text{}
\begin{table}[H]
\begin{longtable}{*{9}{l}}
\hline
Name & ${V}_{\rm max}$ & ${r}_{\rm max}$ & $\rho_{0}\times10^7$ & $\sigma_{0}$ & $-\alpha$ & $\Upsilon_{\star,{\rm disk}}$ &$\Upsilon_{\star,{\rm bulge}}$ & $\chi^2/{\rm{d.o.f.}}$\\ 
& [km/s] & [kpc] & $[{M}_{\odot}/\rm{kpc}^3]$ & [km/s] &  & $[{M}_{\odot}/{L}_{\odot}]$ & $[{M}_{\odot}/{L}_{\odot}]$ &\\ 
(1) & (2) & (3) & (4) & (5) & (6) & (7) & (8) & (9)\\
\hline
NGC3109 & 68.89 & 16.16 & 1.82 & 42.94 & $0.44$ & 0.72 & 0.00 & 0.27\\
 & 77.74 & 26.29 & 1.56 & 49.99 & $0.51$ & 1.23 & 0.00 & 0.21\\
 & $78.87_{-9.46}^{+10.55}$ & $28.24_{-9.79}^{+12.66}$ & $1.51_{-0.24}^{+0.40}$ & $50.56_{-5.06}^{+5.05}$ & $0.52_{-0.04}^{+0.06}$ & $1.44_{-0.97}^{+1.17}$ & $0.00_{-0.00}^{+0.00}$ & \\
NGC3198 & 134.72 & 39.92 & 6.21 & 102.08 & $1.10$ & 0.49 & 0.00 & 0.93\\
 & 135.97 & 24.37 & 6.42 & 103.86 & $0.79$ & 0.37 & 0.00 & 0.74\\
\hline
\end{longtable}
\caption{Continued}
\end{table}
\clearpage
\addtocounter{table}{-2}
\centering
\text{}
\begin{table}[H]
\begin{longtable}{*{9}{l}}
\hline
Name & ${V}_{\rm max}$ & ${r}_{\rm max}$ & $\rho_{0}\times10^7$ & $\sigma_{0}$ & $-\alpha$ & $\Upsilon_{\star,{\rm disk}}$ &$\Upsilon_{\star,{\rm bulge}}$ & $\chi^2/{\rm{d.o.f.}}$\\ 
& [km/s] & [kpc] & $[{M}_{\odot}/\rm{kpc}^3]$ & [km/s] &  & $[{M}_{\odot}/{L}_{\odot}]$ & $[{M}_{\odot}/{L}_{\odot}]$ &\\ 
(1) & (2) & (3) & (4) & (5) & (6) & (7) & (8) & (9)\\
\hline
NGC3953 & 173.19 & 36.76 & 19.95 & 141.92 & $1.60$ & 0.36 & 0.00 & 0.61\\
 & 172.65 & 21.66 & 17.76 & 139.84 & $1.67$ & 0.26 & 0.00 & 0.07\\
 & $177.20_{-16.06}^{+37.02}$ & $44.51_{-20.45}^{+107.94}$ & $8.59_{-5.07}^{+6.87}$ & $151.02_{-9.29}^{+18.53}$ & $1.47_{-0.11}^{+0.13}$ & $0.42_{-0.16}^{+0.17}$ & $0.00_{-0.00}^{+0.00}$ & \\
NGC3972 & 126.66 & 25.34 & 6.21 & 91.16 & $0.99$ & 0.49 & 0.00 & 1.24\\
 & 117.65 & 12.73 & 7.94 & 82.95 & $0.95$ & 0.24 & 0.00 & 0.80\\
\hline
\end{longtable}
\caption{Continued}
\end{table}
\clearpage
\addtocounter{table}{-2}
\centering
\text{}
\begin{table}[H]
\begin{longtable}{*{9}{l}}
\hline
Name & ${V}_{\rm max}$ & ${r}_{\rm max}$ & $\rho_{0}\times10^7$ & $\sigma_{0}$ & $-\alpha$ & $\Upsilon_{\star,{\rm disk}}$ &$\Upsilon_{\star,{\rm bulge}}$ & $\chi^2/{\rm{d.o.f.}}$\\ 
& [km/s] & [kpc] & $[{M}_{\odot}/\rm{kpc}^3]$ & [km/s] &  & $[{M}_{\odot}/{L}_{\odot}]$ & $[{M}_{\odot}/{L}_{\odot}]$ &\\ 
(1) & (2) & (3) & (4) & (5) & (6) & (7) & (8) & (9)\\
\hline
NGC4157 & 166.13 & 39.68 & 21.54 & 127.03 & $1.81$ & 0.25 & 0.25 & 0.60\\
 & 217.48 & 193.46 & 4.04 & 154.23 & $1.53$ & 0.43 & 0.10 & 0.26\\
 & $177.84_{-15.59}^{+38.45}$ & $87.17_{-52.40}^{+85.28}$ & $7.62_{-3.05}^{+6.80}$ & $142.09_{-11.14}^{+12.17}$ & $1.57_{-0.12}^{+0.11}$ & $0.33_{-0.13}^{+0.07}$ & $0.52_{-0.31}^{+0.56}$ & \\
NGC4183 & 93.14 & 22.30 & 5.58 & 69.74 & $0.90$ & 0.64 & 0.00 & 0.35\\
 & 92.87 & 66.34 & 1.21 & 90.51 & $0.68$ & 1.36 & 0.00 & 0.44\\
\hline
\end{longtable}
\caption{Continued}
\end{table}
\clearpage
\addtocounter{table}{-2}
\centering
\text{}
\begin{table}[H]
\begin{longtable}{*{9}{l}}
\hline
Name & ${V}_{\rm max}$ & ${r}_{\rm max}$ & $\rho_{0}\times10^7$ & $\sigma_{0}$ & $-\alpha$ & $\Upsilon_{\star,{\rm disk}}$ &$\Upsilon_{\star,{\rm bulge}}$ & $\chi^2/{\rm{d.o.f.}}$\\ 
& [km/s] & [kpc] & $[{M}_{\odot}/\rm{kpc}^3]$ & [km/s] &  & $[{M}_{\odot}/{L}_{\odot}]$ & $[{M}_{\odot}/{L}_{\odot}]$ &\\ 
(1) & (2) & (3) & (4) & (5) & (6) & (7) & (8) & (9)\\
\hline
NGC5907 & 183.88 & 37.36 & 12.02 & 147.11 & $1.25$ & 0.42 & 0.00 & 5.82\\
 & 194.34 & 168.66 & 1.58 & 179.58 & $1.04$ & 0.77 & 0.00 & 6.25\\
 & $176.57_{-12.49}^{+12.08}$ & $46.00_{-20.43}^{+80.23}$ & $3.62_{-1.79}^{+2.94}$ & $162.38_{-8.66}^{+15.69}$ & $0.90_{-0.08}^{+0.11}$ & $0.65_{-0.20}^{+0.16}$ & $0.00_{-0.00}^{+0.00}$ & \\
NGC5985 & 225.66 & 44.24 & 15.85 & 187.19 & $1.34$ & 0.81 & 1.16 & 11.33\\
 & 264.71 & 38.63 & 26.44 & 190.84 & $1.81$ & 0.10 & 0.95 & 1.34\\
\hline
\end{longtable}
\caption{Continued}
\end{table}
\clearpage
\addtocounter{table}{-2}
\centering
\text{}
\begin{table}[H]
\begin{longtable}{*{9}{l}}
\hline
Name & ${V}_{\rm max}$ & ${r}_{\rm max}$ & $\rho_{0}\times10^7$ & $\sigma_{0}$ & $-\alpha$ & $\Upsilon_{\star,{\rm disk}}$ &$\Upsilon_{\star,{\rm bulge}}$ & $\chi^2/{\rm{d.o.f.}}$\\ 
& [km/s] & [kpc] & $[{M}_{\odot}/\rm{kpc}^3]$ & [km/s] &  & $[{M}_{\odot}/{L}_{\odot}]$ & $[{M}_{\odot}/{L}_{\odot}]$ &\\ 
(1) & (2) & (3) & (4) & (5) & (6) & (7) & (8) & (9)\\
\hline
UGC00128 & 117.63 & 33.05 & 2.26 & 79.67 & $0.54$ & 0.64 & 0.00 & 5.67\\
 & 116.55 & 87.69 & 1.04 & 92.52 & $0.72$ & 2.04 & 0.00 & 6.05\\
 & $115.88_{-4.80}^{+4.61}$ & $51.16_{-19.82}^{+24.20}$ & $1.42_{-0.28}^{+0.44}$ & $87.05_{-4.84}^{+4.84}$ & $0.66_{-0.12}^{+0.07}$ & $1.47_{-0.76}^{+0.53}$ & $0.00_{-0.00}^{+0.00}$ & \\
UGC00731 & 64.73 & 9.48 & 4.04 & 41.36 & $0.65$ & 1.00 & 0.00 & 1.31\\
 & 52.68 & 25.40 & 1.37 & 47.62 & $0.68$ & 9.93 & 0.00 & 0.17\\
\hline
\end{longtable}
\caption{Continued}
\end{table}
\clearpage
\addtocounter{table}{-2}
\centering
\text{}
\begin{table}[H]
\begin{longtable}{*{9}{l}}
\hline
Name & ${V}_{\rm max}$ & ${r}_{\rm max}$ & $\rho_{0}\times10^7$ & $\sigma_{0}$ & $-\alpha$ & $\Upsilon_{\star,{\rm disk}}$ &$\Upsilon_{\star,{\rm bulge}}$ & $\chi^2/{\rm{d.o.f.}}$\\ 
& [km/s] & [kpc] & $[{M}_{\odot}/\rm{kpc}^3]$ & [km/s] &  & $[{M}_{\odot}/{L}_{\odot}]$ & $[{M}_{\odot}/{L}_{\odot}]$ &\\ 
(1) & (2) & (3) & (4) & (5) & (6) & (7) & (8) & (9)\\
\hline
UGC03546 & 178.92 & 45.52 & 35.21 & 134.67 & $2.10$ & 0.34 & 0.34 & 1.88\\
 & 230.62 & 189.25 & 152.50 & 163.96 & $1.62$ & 0.63 & 0.25 & 1.63\\
 & $213.72_{-13.31}^{+31.98}$ & $110.03_{-36.08}^{+79.05}$ & $498.07_{-324.78}^{+684.56}$ & $155.63_{-6.24}^{+11.85}$ & $1.68_{-0.28}^{+0.22}$ & $0.36_{-0.16}^{+0.18}$ & $0.25_{-0.10}^{+0.10}$ & \\
UGC03580 & 116.82 & 28.65 & 3.11 & 79.29 & $0.86$ & 0.36 & 0.36 & 5.24\\
 & 125.58 & 60.33 & 4.41 & 90.41 & $1.06$ & 0.57 & 0.13 & 2.10\\
\hline
\end{longtable}
\caption{Continued}
\end{table}
\clearpage
\addtocounter{table}{-2}
\centering
\text{}
\begin{table}[H]
\begin{longtable}{*{9}{l}}
\hline
Name & ${V}_{\rm max}$ & ${r}_{\rm max}$ & $\rho_{0}\times10^7$ & $\sigma_{0}$ & $-\alpha$ & $\Upsilon_{\star,{\rm disk}}$ &$\Upsilon_{\star,{\rm bulge}}$ & $\chi^2/{\rm{d.o.f.}}$\\ 
& [km/s] & [kpc] & $[{M}_{\odot}/\rm{kpc}^3]$ & [km/s] &  & $[{M}_{\odot}/{L}_{\odot}]$ & $[{M}_{\odot}/{L}_{\odot}]$ &\\ 
(1) & (2) & (3) & (4) & (5) & (6) & (7) & (8) & (9)\\
\hline
UGC05986 & 110.35 & 18.18 & 7.36 & 76.00 & $1.10$ & 0.56 & 0.00 & 7.18\\
 & 109.44 & 11.01 & 9.34 & 74.92 & $1.05$ & 0.26 & 0.00 & 0.80\\
 & $111.74_{-7.66}^{+25.20}$ & $22.91_{-10.25}^{+32.95}$ & $6.40_{-2.80}^{+2.57}$ & $80.59_{-5.49}^{+13.63}$ & $1.13_{-0.09}^{+0.10}$ & $0.74_{-0.35}^{+0.40}$ & $0.00_{-0.00}^{+0.00}$ & \\
UGC06399 & 84.29 & 21.44 & 3.21 & 57.29 & $0.71$ & 0.74 & 0.00 & 0.25\\
 & 82.14 & 8.83 & 4.40 & 54.32 & $0.63$ & 0.13 & 0.00 & 0.03\\
\hline
\end{longtable}
\caption{Continued}
\end{table}
\clearpage
\addtocounter{table}{-2}
\centering
\text{}
\begin{table}[H]
\begin{longtable}{*{9}{l}}
\hline
Name & ${V}_{\rm max}$ & ${r}_{\rm max}$ & $\rho_{0}\times10^7$ & $\sigma_{0}$ & $-\alpha$ & $\Upsilon_{\star,{\rm disk}}$ &$\Upsilon_{\star,{\rm bulge}}$ & $\chi^2/{\rm{d.o.f.}}$\\ 
& [km/s] & [kpc] & $[{M}_{\odot}/\rm{kpc}^3]$ & [km/s] &  & $[{M}_{\odot}/{L}_{\odot}]$ & $[{M}_{\odot}/{L}_{\odot}]$ &\\ 
(1) & (2) & (3) & (4) & (5) & (6) & (7) & (8) & (9)\\
\hline
UGC06923 & 73.15 & 9.47 & 6.92 & 49.03 & $0.94$ & 0.25 & 0.00 & 0.72\\
 & 75.77 & 8.74 & 6.80 & 53.60 & $0.92$ & 0.17 & 0.00 & 0.46\\
 & $80.80_{-13.62}^{+21.95}$ & $18.62_{-10.26}^{+19.50}$ & $4.57_{-1.50}^{+3.16}$ & $59.43_{-7.31}^{+10.40}$ & $0.97_{-0.15}^{+0.17}$ & $0.42_{-0.22}^{+0.29}$ & $0.00_{-0.00}^{+0.00}$ & \\
UGC06930 & 93.14 & 22.30 & 4.30 & 67.49 & $0.78$ & 0.64 & 0.00 & 0.33\\
 & 92.28 & 10.36 & 4.55 & 66.56 & $0.75$ & 0.41 & 0.00 & 0.20\\
\hline
\end{longtable}
\caption{Continued}
\end{table}
\clearpage
\addtocounter{table}{-2}
\centering
\text{}
\begin{table}[H]
\begin{longtable}{*{9}{l}}
\hline
Name & ${V}_{\rm max}$ & ${r}_{\rm max}$ & $\rho_{0}\times10^7$ & $\sigma_{0}$ & $-\alpha$ & $\Upsilon_{\star,{\rm disk}}$ &$\Upsilon_{\star,{\rm bulge}}$ & $\chi^2/{\rm{d.o.f.}}$\\ 
& [km/s] & [kpc] & $[{M}_{\odot}/\rm{kpc}^3]$ & [km/s] &  & $[{M}_{\odot}/{L}_{\odot}]$ & $[{M}_{\odot}/{L}_{\odot}]$ &\\ 
(1) & (2) & (3) & (4) & (5) & (6) & (7) & (8) & (9)\\
\hline
UGC07690 & 48.85 & 5.14 & 48.60 & 36.49 & $2.05$ & 0.42 & 0.00 & 0.35\\
 & 47.38 & 3.50 & 33.64 & 38.71 & $1.96$ & 0.46 & 0.00 & 0.18\\
 & $49.37_{-5.91}^{+13.77}$ & $9.91_{-5.80}^{+17.34}$ & $15.64_{-7.27}^{+13.97}$ & $42.95_{-3.55}^{+5.16}$ & $1.73_{-0.31}^{+0.25}$ & $0.75_{-0.25}^{+0.25}$ & $0.00_{-0.00}^{+0.00}$ & \\
UGC08286 & 73.63 & 8.91 & 13.18 & 51.96 & $1.28$ & 0.85 & 0.00 & 1.34\\
 & 74.35 & 6.20 & 9.97 & 52.78 & $1.03$ & 0.67 & 0.00 & 1.68\\
\hline
\end{longtable}
\caption{Continued}
\end{table}
\clearpage
\addtocounter{table}{-2}
\centering
\text{}
\begin{table}[H]
\begin{longtable}{*{9}{l}}
\hline
Name & ${V}_{\rm max}$ & ${r}_{\rm max}$ & $\rho_{0}\times10^7$ & $\sigma_{0}$ & $-\alpha$ & $\Upsilon_{\star,{\rm disk}}$ &$\Upsilon_{\star,{\rm bulge}}$ & $\chi^2/{\rm{d.o.f.}}$\\ 
& [km/s] & [kpc] & $[{M}_{\odot}/\rm{kpc}^3]$ & [km/s] &  & $[{M}_{\odot}/{L}_{\odot}]$ & $[{M}_{\odot}/{L}_{\odot}]$ &\\ 
(1) & (2) & (3) & (4) & (5) & (6) & (7) & (8) & (9)\\
\hline
UGC11914 & 276.49 & 68.51 & 16.85 & 191.90 & $1.60$ & 0.53 & 0.99 & 2.02\\
 & 331.63 & 87.51 & 363.29 & 229.96 & $1.59$ & 0.10 & 0.85 & 0.70\\
 & $337.51_{-37.08}^{+46.05}$ & $104.72_{-12.28}^{+27.77}$ & $261.23_{-154.72}^{+292.28}$ & $232.68_{-21.08}^{+26.55}$ & $1.51_{-0.28}^{+0.28}$ & $0.25_{-0.11}^{+0.18}$ & $0.86_{-0.20}^{+0.10}$ & \\
UGC12506 & 181.13 & 36.58 & 10.96 & 156.50 & $1.07$ & 1.00 & 0.00 & 1.00\\
 & 195.24 & 23.37 & 6.71 & 159.72 & $0.89$ & 0.76 & 0.00 & 0.96\\
\hline
\end{longtable}
\caption{Continued}
\end{table}

\end{document}